\documentclass[a4paper,11pt]{article}

\usepackage{jheppub}
\usepackage{amsmath}
\usepackage{amsfonts}
\usepackage{amssymb}

\usepackage{graphicx}
\usepackage{dcolumn}
\usepackage{bm}
\usepackage{epsfig}

\newcommand{\dhd}{{\textstyle d}
	\lower.03ex\hbox{\kern-0.38em$^{\scriptstyle-}$}\kern-0.05em{}}
\newcommand{\dbar}{{\textstyle \delta}
	\lower.03ex\hbox{\kern-0.38em$^{\scriptstyle-}$}\kern-0.05em{}}
\newcommand{\half}{{1\over 2}}

\newcommand{\cala}{{\cal A}}
\newcommand{\calb}{{\cal B}}

\newcommand{\calf}{{\cal F}}

\newcommand{\calh}{{\cal H}}
\newcommand{\cali}{{\cal I}}

\newcommand{\calo}{{\cal O}} 
 
\newcommand{\calq}{{\cal Q}}

\newcommand{\calu}{{\cal U}}

\newcommand{\calx}{{\cal X}} 
 
\newcommand{\calz}{{\cal Z}}

\newcommand{\barG}{{\bar G}}

\newcommand{\barpsi}{{\bar \psi}} 
\newcommand{\bpsi}{\bar{\psi}}

\newcommand{\hatp}{{\hat p}}

\newcommand{\hatO}{{\hat O}} 
\newcommand{\hatX}{{\hat X}} 
\newcommand{\hatY}{{\hat Y}}

\newcommand{\tildeq}{{\tilde q}}

\newcommand{\tildeG}{{\tilde G}}

\newcommand{\tildeQ}{{\tilde Q}}

\newcommand \ket [1] {|{#1}\rangle}
\newcommand \bra [1] {\langle {#1}|}

\newcommand{\ketx}{\ket{x}}
\newcommand{\kety}{\ket{y}}

\newcommand{\brax}{\bra{x}}
\newcommand{\bray}{\bra{y}}

\newcommand{\ketxp}{\ket{x_\perp}}
\newcommand{\ketyp}{\ket{y_\perp}}
\newcommand{\ketzp}{\ket{z_\perp}}

\newcommand{\ketop}{\ket{\omega_\perp}}

\newcommand{\braxp}{\bra{x_\perp}}
\newcommand{\brayp}{\bra{y_\perp}}
\newcommand{\brazp}{\bra{z_\perp}}

\newcommand \sslash [1] {\slash\hspace{-0.2cm}{#1}}
\newcommand{\ssp}{\sslash{p}}

\newcommand{\ssq}{\sslash{q}}

\newcommand{\ssx}{\sslash{x}}
\newcommand{\ssy}{\sslash{y}}
\newcommand{\ssz}{\sslash{z}}

\newcommand{\slashd}{\sslash{\partial}}

\newcommand \Slash [1] {\slash\hspace{-0.23cm}{#1}}
\newcommand{\Sa}{\Slash{A}}
\newcommand{\Sd}{\Slash{D}}
\newcommand{\Sp}{\Slash{P}}
\newcommand{\Sx}{\Slash{X}}
\newcommand{\Sy}{\sslash{Y}}

\newcommand{\Tra}{{\rm Tr}_{\rm a}} 
\newcommand{\Tr}{{\rm Tr}} 
\newcommand{\tr}{{\rm tr}}

\begin{document}
	
	\title{High-energy Operator Product Expansion at sub-eikonal level}

\author[]{Giovanni Antonio Chirilli}
\affiliation[]{Institut f\"ur Theoretische Physik, Universit\"at Regensburg,\\ D-93040 Regensburg, Germany}
\emailAdd{giovanni.chirilli@ur.de}

\abstract{The high energy Operator Product Expansion for the product of
	two electromagnetic currents is extended to the sub-eikonal level in a rigorous way. I
	calculate the impact factors for polarized and unpolarized structure functions, define new distribution functions,
	and derive the evolution equations for unpolarized and polarized structure functions in the flavor singlet and non-singlet case.
	
}

\maketitle

\section{Introduction}

In view of future collider experiments, in which hadronic matter will be probed at unprecedented 
kinematic regimes, there has been, in recent years, an ever growth of interest in understanding 
the fundamental properties of hadrons, spin and mass, from their constituents, quarks, and gluons
\cite{Boer:2011fh, Nocera:2014gqa, deFlorian:2008mr, deFlorian:2009vb, 
Kovchegov:2015pbl, Kovchegov:2017lsr, Boussarie:2019icw, Tarasov:2020cwl, Hatta:2016aoc, Hatta:2020riw, Hatta:2020ltd}.

At high energy (Regge limit) scattering amplitudes are dominated by gluon dynamics, in particular, the cross section of 
Deep Inelastic Scattering (DIS) processes, is dominated, in the unpolarized case, by the gluon structure function.
Within the Leading Log Approximation (LLA), the resummation of log of energy through BFKL \cite{Kuraev:1977fs, Balitsky:1978ic}
formalism predicts a steep rise of the DIS cross-section that has been observed experimentally at HERA experiments 
\cite{Aaron:2009aa, Abramowicz:2015mha}.

In perturbative quantum chromodynamics (pQCD) a standard technique to study the asymptotic behavior 
of cross-sections is the operator product expansions (OPE).
In deep inelastic scattering (DIS), in the Bjorken limit, the $T$-product of two electromagnetic currents
is expanded in terms of coefficient functions, perturbatively calculable, and matrix elements of non-local operators that
encode the non perturbative information of the process. Evolution equations of the non-local matrix elements
with respect to the factorization scale $\mu_f$ provide information on the scaling behavior of the 
parton distributions \cite{Balitsky:1987bk}.  

About twenty-five years ago, the high-energy Operator Product Expansion was introduced for the first time
as a method to study the asymptotic behavior of high-energy structure functions
from a gauge-invariant formalism\cite{Balitsky:1995ub}. Since then, the high-energy OPE
proved to be a successful method to systematically study high-energy scattering amplitudes
\cite{Balitsky:2008zza, Balitsky:2009xg, Balitsky:2010ze, Balitsky:2012bs}.
Within the high-energy OPE the scattering amplitude is factorized 
in rapidity space in coefficient functions (also called impact factors) and matrix elements of Wilson lines. 
The evolution equation of Wilson lines with respect to the rapidity parameter
provides the energy dependence of the amplitude: each step in rapidity generates a new Wilson line at a different
point in impact parameter space thus obtaining the non-linear 
Balitsky-JIMWLK evolution equation 
\cite{Balitsky:1995ub}\cite{JalilianMarian:1997gr, Ferreiro:2001qy, Iancu:2000hn}  which in the dipole approximation takes the 
form of the Balitsky-Kovchegov equation \cite{Kovchegov:1999yj, Kovchegov:1999ua}
(see Ref. \cite{Balitsky:2001gj, Kovchegov:2012mbw} for a review).

Contrary to the OPE in the Bjorken limit, in the high-energy OPE
the coefficient functions and the matrix elements both receive perturbative and non-perturbative contributions. The latter, however, are screened
by the saturation scale, which being much larger than the scale of the confinement region, justifies the
applicability of perturbative methods.

The unpolarized DIS structure functions at small-x are  known at next-to-leading order (NLO)
\cite{Balitsky:2010ze, Balitsky:2012bs} (see also \cite{Beuf:2017bpd}) 
in $\alpha_s$ and at next-leading-logarithmic accuracy \cite{Balitsky:2008zza, Fadin:1998py}. 
The NLO pomeron intercept, through which the log of energy are resummed at next-to-leading log approximation (NLL), 
has been calculated long ago \cite{Fadin:1998py} and later confirmed
through the linearizion of the NLO Balitsky-Kovchegov equation \cite{Balitsky:2008zza}.
Moreover, the pomeron residue, which is related to the impact factor
is now available in an analytic form in coordinate and momentum space at NLO \cite{Balitsky:2010ze, Balitsky:2012bs}. These two
ingredients, the NLO pomeron intercept and the NLO pomeron residue allow one to study the behavior of the small-x structure function
at NLO accuracy in $\alpha_s$ and at NLL approximation. A first fit of the dipole model to HERA data
using the full NLO impact factor and Balitsky-Kovchegov evolution
has been performed in Ref. \cite{Beuf:2020dxl}.

Unlike the unpolarized ones, the behavior of the polarized DIS structure functions at small-x is not yet well understood.
In the polarized case, the pomeron intercept  receives an extra complication due to the presence of double logarithm of energy
$\alpha_s\ln^2x_B^{-1}$ \cite{Bartels:1995iu, Bartels:1996wc}.  Such double logarithms of energy
are also present in the unpolarized quark structure function \cite{Kirschner:1983di,Ermolaev:1995fx}, although
they are suppressed by one power of energy with respect to the gluon one.
In the polarized case, instead, quark structure functions are not suppressed thus their behavior
at small-$x_B$ is as relevant as the gluon structure functions. 

What makes the study of polarized structure functions even more challenging is the fact that, the 
double logarithm of energy, $\alpha_s\ln^2x_B^{-1}$, cannot be reached 
from the LO anomalous dimension of twist two operators.
Indeed, if on one side, the relation at the \textit{non-physical} point $n=1$ ($n$ being the moment index),
between the BFKL and DGLAP was established
through the anomalous dimension of twist two operators \cite{Jaroszewicz:1982gr, BALITSKY:2014zza}, on the other,
such relation cannot be easily obtained in the polarized case due to the presence of double log of energy contributions
which are absent in the LO \cite{Altarelli:1977zs} and NLO polarized splitting functions \cite{Mertig:1995ny, Vogelsang:1996im}. 
However, the double log of energy contribution, $\alpha_s\ln^2x_B^{-1}$, appears at there-loop 
polarized splitting function \cite{Moch:2014sna, Behring:2019tus, Blumlein:1995jp, Blumlein:1996hb}
and the result is in agreement with the one originally found in Ref. \cite{Bartels:1995iu, Bartels:1996wc}.

The plan is to bring the knowledge of the unpolarized and polarized small-x structure function at the same level.
To this end, first we need to extend the high-energy OPE of the $T$-product of two-electromagnetic currents to
include terms that are not symmetric in the exchange of the two Lorentz indexes of the DIS hadronic tensor.
Indeed, in the polarized DIS
both the leptonic tensor and the hadronic one have to be antisymmetric. To obtain antisymmetric contributions
in the high-energy OPE, it is necessary to relax the eikonal approximation and allow sub-eikonal terms to enter into
the game. As we will show in the subsequent sections, the OPE with sub-eikonal contributions 
will be given in terms of new operators and the task will be to calculate their evolution equations.
An important ingredient to perform the high-energy OPE with sub-eikonal corrections is 
the quark and gluon propagators with sub-eikonal corrections. 
In Ref. \cite{Chirilli:2018kkw} such corrections have been all calculated as deviation from the shock-wave approximation: the 
corrections are suppressed by the large Lorentz boost parameter. Corrections to the eikonal approximation
can be also  included by assuming that the fields of the background shock-wave do not have the same transverse
momenta, rather they are ordered. In this case the sub-eikonal corrections are included as light-cone expansion of the
background shock-wave \cite{Balitsky:2015qba, Balitsky:2016dgz}. However, these corrections will be
irrelevant for our analysis since here we are interested in small-x dynamics rather than studying the overlapping kinematic regime
where these corrections will play a central role.

In section \ref{sec: opehe} we will review the main idea behind 
the high-energy OPE in the unpolarized case. We will derive the leading order (LO) impact factor
and the associated Wilson-line operators together with their evolution equations. 
This will serve as a smooth transition to the OPE at sub-eikonal level
in section \ref{sec: sub-opehe} where we will give a new expression 
(equivalent to the one obtained before in Ref. \cite{Chirilli:2018kkw})
of the quark propagator with sub-eikonal corrections in the background of gluon fields, and
will identify the relevant sub-eikonal corrections which will be used, in section \ref{sec: OPEsubeik},
to calculate the impact factors for polarized and unpolarized structure functions. 
In section \ref{sec: parametri-matrixele} we summarize the parametrization of the matrix elements
of operators found through the OPE, and identify 10 new distribution functions.
In section \ref{sec: subeikeq} we will derive the evolution equations for the 
operators associated to the sub-eikonal impact factors using the propagators calculated
in previous work, Ref. \cite{Chirilli:2018kkw}.
In the same section we will calculate, for the first time, the quark-to-gluon propagator and use it to calculate diagrams 
that have not been 
calculated before in the contest of spin at small-x.
Summary of the evolution equations for singlet and non-singlet case will be presented in section \ref{sec: summaryevo}.
In the last section we will summarize our findings and compare them with other results that have been obtained in the same 
direction. We will argue that the result we obtained although agree in some limiting case, they actually differ from the ones calculated in Refs. 
\cite{Kovchegov:2015pbl, Kovchegov:2018znm, Kovchegov:2016zex} because of the way the quark and gluon operators are treated
under one loop evolution.

\section{Operator Product Expansion at High-Energy}
\label{sec: opehe}

Before calculating the sub-eikonal corrections to the high-energy OPE for DIS, let us first briefly review
the high-energy OPE in the eikonal approximation.

The inclusive differential DIS cross-section in the laboratory frame
for detecting the final lepton in the solid angle $d\Omega$ with final energy within $[E', E'+dE']$ is
\begin{eqnarray}
{d^2\sigma\over d\Omega\,dE'} = {\alpha^2\over Mq^4}{E'\over E}L_{\mu\nu}W^{\mu\nu}\,.
\end{eqnarray}
Here the hadronic target of mass $M$ has momentum $P^\mu=p_2^\mu + {M^2\over s}p_1^\mu$, and  the virtual photon
has momentum $q^\mu = p_1^\mu - x_Bp^\mu_2$ with $p_1^\mu, p_2^\mu$ light-cone vectors such that
$p_1^\mu p_{2\mu} = {s\over 2}$ and $x_B = {-q^2\over s}\ll1$. 
The leptonic tensor is denoted by $L^{\mu\nu}$ and the hadronic one by $W^{\mu\nu}$.

In strong and electromagnetic interactions parity is conserved, thus
the hadronic tensor can be expanded in terms of the unpolarized structure functions $F_1$ and $F_2$ and the
polarized structure functions $g_1$ and $g_2$
\begin{eqnarray}
\hspace{-1cm}W_{\mu\nu}=\!\!&& \left(- g_{\mu\nu} + {q_\mu q_\nu\over q^2}\right)F_1(x, Q^2)
+ \left(P_\mu - q_\mu {q\cdot P\over q^2}\right) \left(P_\nu - q_\nu {q\cdot P\over q^2}\right){F_2(x, Q^2)\over P\cdot q}
\nonumber\\
\hspace{-1cm}&& + i\,\varepsilon_{\mu\nu\lambda\sigma}\,q^\lambda S^\sigma {M\over P\cdot q}\,g_1(x,Q^2)
+  i\,\varepsilon_{\mu\nu\lambda\sigma}q^\lambda\left(S^\sigma  - P^\sigma {q\cdot S\over q\cdot P}\right){M\over P\cdot q}\,g_2(x,Q^2)
\label{Htensor}
\end{eqnarray}
where $S^\mu$ is the spin of the target that satisfies  $S^2 = -1$ and $S\!\cdot\!P=0$.

To extract the polarized structure functions $g_1$ and $g_2$, we need the antisymmetric part of the leptonic tensor.
This means that both the incoming lepton and the hadronic target have to be polarized.

With the help of the optical theorem, 
the hadronic tensor is related to the imaginary part of the Fourier transform of the $T$-product of two electromagnetic currents 
\begin{eqnarray}
&&W_{\mu\nu} = {1\over \pi}{\rm Im} T_{\mu\nu}\,.
\end{eqnarray}
where
\begin{eqnarray}
&&T_{\mu\nu} = i\!\int d^4 x \, e^{iq\cdot x}
\langle P,S|{\rm T}\{j_{\mu}(x)j_\nu(0)\}|P,S\rangle
\end{eqnarray}
To study the polarized structure functions $g_1$ and $g_2$ at high-energy 
we need to extract the antisymmetric part of the $T_{\mu\nu}$ tensor.
This is what we will do in the next section.

The T-product of two electromagnetic currents is considered in the background of gluon field. As we will see later,
in order to calculate sub-eikonal corrections it will be necessary to consider the OPE in the background
of gluons and quarks as well. However for the moment we consider a background made only of gluons.

In the spectator frame the background field reduces to a shock wave (see Appendix \ref{sec: notation} and \cite{Balitsky:2001gj}
for review), and the virtual photon, which  mediates the interactions
between the lepton and the nucleon (or nucleus) in DIS processes, 
splits into a quark anti-quark pair long before the interaction with the target. In the eikonal approximation
the propagation of the 
quark anti-quark pair in the shock wave background, reduces to two infinite Wilson lines.  
Although with less probability, the virtual photon
may fluctuate in a quark, an anti-quark and a gluon before interacting with the target. In this case, the 
number of Wilson lines increases. Consequently, at high energy, the T-product of two electromagnetic 
currents is expanded in terms of infinite Wilson lines as
\begin{eqnarray}
\hspace{-1.3cm}&&
T\{\bar{\hat{\psi}}(x)\gamma^\mu\hat{\psi}(x)\bar{\hat{\psi}}(y)\gamma^\nu\hat{\psi}(y)\}
\nonumber\\
\hspace{-1.3cm}&&=\int\! d^2z_1d^2z_2~I_{\rm LO}^{\mu\nu}(z_1,z_2,x,y)
{\rm Tr}\{\hat{U}^\eta_{z_1}\hat{U}^{\dagger\eta}_{z_2}\}
\nonumber\\
\hspace{-1.3cm}&&
+\int\! d^2z_1d^2z_2d^2z_3~I_{\rm NLO}^{\mu\nu}(z_1,z_2,z_3,x,y)
[{\rm tr}\{\hat{U}^\eta_{z_1}\hat{U}^{\dagger\eta}_{z_3}\}{\rm tr}\{\hat{U}^\eta_{z_3}\hat{U}^{\dagger\eta}_{z_2}\}
-N_c{\rm tr}\{\hat{U}^\eta_{z_1}\hat{U}^{\dagger\eta}_{z_2}\}] + \cdots
\label{OPEexp}
\end{eqnarray}
where with the symbol $\,\hat{}\,$ over the fields we indicate operators, 
$U_x = {\rm Pexp}\{ig\int dx^+A^-(x^+ +x_\perp)\}$ is the Wilson line and $A^-$ is the 
background gluon field. The first term of expansion  (\ref{OPEexp}), which is proportional to the LO impact factor $I_{\rm LO}$, 
corresponds to the probability for the virtual photon, to split in a quark-anti-quark pair.
The second term, proportional to the NLO impact factor $I_{NLO}$, 
and which has been calculated in Ref. \cite{Balitsky:2012bs, Balitsky:2010ze}, corresponds to the probability that the virtual photon has
to split in a quark-anti-quark and a gluon before scattering with the target.
Here we use the light-cone variables 
$x^\pm = {x^0\pm x^3\over \sqrt{2}}$. In Appendix \ref{sec: notation} we provide further details on the notation
used throughout this paper.

Let us see how to get expansion (\ref{OPEexp}) in a bit more detail. We know that to get the DIS cross-section we need to evaluate
the T-product of two electromagnetic currents in the nucleus or nucleon target state $|P,S\rangle$
\begin{eqnarray}
\langle P,S| {\rm T}\{\bar{\hat{\psi}}(x)\gamma^\mu\hat{\psi}(x)\bar{\hat{\psi}}(y)\gamma^\nu\hat{\psi}(y)\}|P,S\rangle
\label{Tjj}
\end{eqnarray}

\begin{figure}[htb]
	\begin{center}
	\includegraphics[width=2.0in]{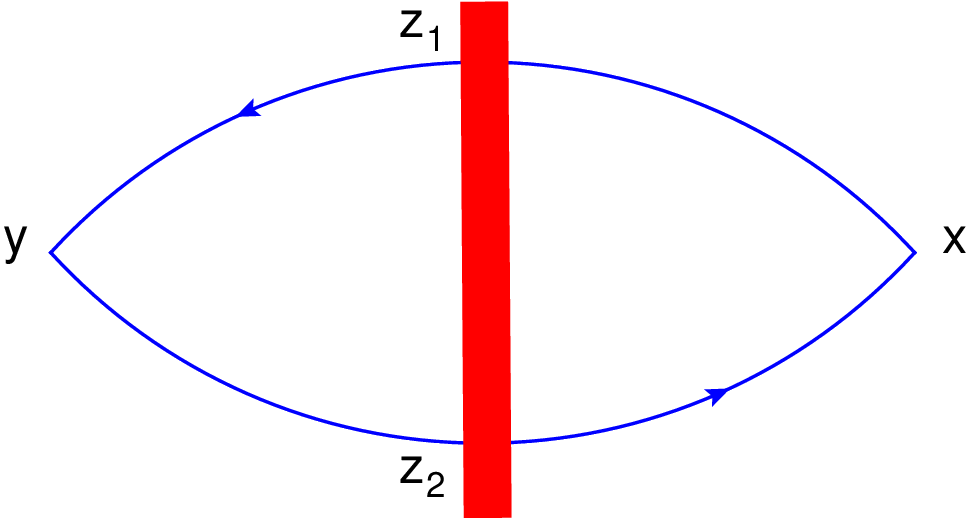}
	\caption{Diagram for the LO impact factor in the eikonal approximation.
		As usual, we will indicate in blue the quantum fields and in red the classical background ones.}
	\label{loif}
\end{center}
\end{figure}

Since we do not know exactly the proton or nucleus state in terms of quarks and gluons, we are forced to
make approximations which are suitable for the kinematic regime under consideration. As anticipated before, 
at high-energy (Regge limit), the target state, made by quarks and gluons, represents the background field
which, in the conveniently chosen spectator frame, shrinks into a shock wave. In first (eikonal) approximation the shock-wave
is made only by gluon field. In light of these considerations we have 
\begin{eqnarray}
\langle P, S| {\rm T}\{\bar{\hat{\psi}}(x)\gamma^\mu\hat{\psi}(x)\bar{\hat{\psi}}(y)\gamma^\nu\hat{\psi}(y)\}|P, S\rangle
~~\rightarrow ~~ \langle{\rm T}\{\bar{\psi}(x)\gamma^\mu\psi(x)\bar{\psi}(y)\gamma^\nu\psi(y)\}\rangle_A
\label{TjjA}
\end{eqnarray}
where the subscript $A$ indicates that the matrix element is evaluated in the background of gluon field generated by the target.
The transition from the LHS to the RHS of eq. (\ref{TjjA}) is similar to the analysis that one performs in
the usual local OPE in which one considers the target in terms of its partonic content, quarks or gluons. 
In Ref. \cite{Balitsky:1987bk} it was shown that the local OPE can be reformulated in terms of 
non-local operators using the background field method.
It turns out that also at high-energy (Regge) limit it is more convenient to perform the OPE analysis 
using the background field method \cite{Balitsky:1995ub}. 
Once the relevant operators are obtained, and we will see they are infinite Wilson lines, 
they will be evaluated in the target state again. 
We now perform functional integration over the spinor fields. Considering only the fully connected diagrams, we have
(see Fig. \ref{loif})
\begin{eqnarray}
&&\hspace{-2cm}\langle {\rm T}\{\bar{\psi}(x)\gamma^\mu\psi(x)\bar{\psi}(y)\gamma^\nu\psi(y)\}\rangle_A
= {\rm tr}\Big\{\brax {1\over \Sp + i\epsilon}\kety \gamma^\nu \bray{1\over \Sp + i\epsilon}\ketx\gamma^\mu\Big\}
\label{Tjjcontracted}
\end{eqnarray}
where $P^\mu = p^\mu + g A^\mu$ and $A^\mu$ is the background gluon field. In eq. (\ref{Tjjcontracted})
we have used the Schwinger notation for the quark propagator which in the eikonal approximation is
\begin{eqnarray}
\langle{\rm T}\psi(x)\bar{\psi}(y)\rangle 
\stackrel{x^+>0>y^+}{=}\int d^4 z\delta(z^+){\ssx - \ssz\over 2\pi^2(x-z)^4}\ssp_2U_z{\ssz-\ssy\over 2\pi^2(y-z)^4}
\label{qprop-eikona}
\end{eqnarray}
with $x^\mu\gamma_\mu = \ssx$.
In Fig. \ref{loif} the blue lines represent the quark and anti-quark fields, while the red band represents the background shock-wave field.

Using propagator (\ref{qprop-eikona}) in eq. (\ref{Tjjcontracted}) arrive at
\begin{eqnarray}
\hspace{-1cm}&&\langle {\rm T}\{\bar{\psi}(x)\gamma^\mu\psi(x)\bar{\psi}(y)\gamma^\nu\psi(y)\}\rangle_A
\nonumber\\
&&~~~~\stackrel{x^+>0>y^+}{=} - {1\over 4\,\pi^6(x_*y_*)^4}\int\!\!d^2z_1 d^2 z_2
\langle{\rm Tr}\{U(z_{1\perp})U^\dagger(z_{2\perp})\}\rangle_A
\nonumber\\
 &&
 ~~~~~~~~~~~~\times{\tr\{(\ssx-\ssz_1)\ssp_2(\ssy-\ssz_1)\gamma^\nu(\ssy-\ssz_2)\ssp_2(\ssx-\ssz_2)\gamma^\mu\}
\over [\calz_1 + i\epsilon ]^3[\calz_2 + i\epsilon ]^3}
+ \dots
\label{LOIF}
\end{eqnarray}
where the $\dots$ stand for higher order corrections in $\alpha_s$ or sub-eikonal corrections.
In eq. (\ref{LOIF}) we defined
\begin{eqnarray}
\calz_i \equiv {(x-z_i)^2_\perp\over x_*} -  {(y-z_i)^2_\perp\over y_*} - {4\over s}(x_\bullet - y_\bullet) 
\label{calzi}
\end{eqnarray}
where we use the short-hand notation (see Appendix \ref{sec: notation} 
for further details on the notations used through out the paper) 
\begin{eqnarray}
&&p_1^\mu x_\mu = x_\bullet = \sqrt{s\over 2}\,x^- = \sqrt{s\over 2}\,x_+\,,\\
&&p_2^\mu x_\mu = x_* = \sqrt{s\over 2}\,x^+ = \sqrt{s\over 2}\,x_-\,.
\end{eqnarray}
 
Using the identity
\begin{eqnarray}
&&{\rm tr}\{(\ssx-\ssz_1)\ssp_2(\ssy-\ssz_1)\gamma^\nu(\ssy-\ssz_2)\ssp_2(\ssx-\ssz_2)\gamma^\mu\}
\nonumber\\
&&~~~~~~= 4x_*^2y_*^2{\partial^2\over \partial x_\mu\partial y_\nu}\Big(- \calz_1 \calz_2 
+ {z_{12\perp}^2\over x_* y_*}(x-y)^2\Big)
\end{eqnarray}
we define the LO impact factor as
\begin{eqnarray}
\cali_{LO}^{\mu\nu}(z_{1\perp},z_{2\perp};x,y) \equiv 
{1\over \pi^6(x_*y_*)^2}{\big(\calz_1 + i\epsilon \big)^{-3}\over \big(\calz_2 + i\epsilon\big)^3}
{\partial^2\over \partial x_\mu\partial y_\nu}\Big(\calz_1 \calz_2 - {z_{12\perp}^2\over x_* y_*}(x-y)^2\Big)
\label{LOcoeff}
\end{eqnarray}
and the high-energy OPE takes the form
\begin{eqnarray}
\hspace{-0.7cm}{\rm T}\{\bar{\hat{\psi}}(x)\gamma^\mu\hat{\psi}(x)\bar{\hat{\psi}}(y)\gamma^\nu\hat{\psi}(y)\}
= \int\!d^2z_1d^2z_2\,\cali_{LO}^{\mu\nu}(z_{1\perp},z_{2\perp};x,y) {\rm Tr}\{\hat{U}(z_{1\perp})\hat{U}^\dagger(z_{2\perp})\} + \dots
\label{LO-OPE-1}
\end{eqnarray}
Eq. (\ref{LO-OPE-1}) tell us that in the high-energy limit
the operators standing to the left of the equation are approximated by the ones standing to the right.

The LO impact factor enjoys two nice properties: electromagnetic gauge invariance 
\begin{eqnarray}
{\partial\over \partial x^\mu} \cali^{\mu\nu}_{LO}(z_{1\perp},z_{2\perp};x,y) = 0
\end{eqnarray}
and M\"obius $SL(2,C)$ conformal invariance
\begin{eqnarray}
\int\! d^2z_1d^2 z_2\,\cali^{\mu\nu}_{LO}(z_{1\perp},z_{2\perp};x,y) 
\stackrel{{\rm inv.}}{\longrightarrow} \int\! d^2z_1d^2 z_2\,\cali^{\mu\nu}_{LO}(z_{1\perp},z_{2\perp};x,y)
\end{eqnarray}
where the symbol $\stackrel{{\rm inv.}}{\longrightarrow}$ means that we perform the inversion transformation $x^\mu\over x^2$
to all coordinates.

If we try to calculate one loop correction either to the coefficient function (the impact factor) or to the 
matrix element of the Wilson-line operators, we will find divergences which are identified by rapidity divergences 
as a remnant of the fact that the parameter we use to discriminate between background (or classical) field 
from the quantum field is indeed the rapidity. 
As explained in the Introduction, at high-energy fields are ordered in their rapidity. 

The rapidity divergences we just mentioned represent the log of energy which are usually resummed through an evolution equation.
The easiest way to identify such log of energy is to consider one loop correction to the matrix element of Wilson lines, rather then
to the coefficient function. 
In Ref. \cite{Balitsky:2012bs, Balitsky:2010ze} a one loop correction to the coefficient function (NLO impact factor) was calculated,
and it was shown that the BK evolution equation can also be obtained from the one loop correction to the coefficient
function (impact factor). One may observe the similarity with the non-local OPE where DGLAP evolution kernel
can be obtained either from the one loop evolution of the non-local operator or from the NLO coefficient function
\cite{Balitsky:1987bk}.

\begin{figure}[htb]
	\begin{center}
		\includegraphics[width=2.5in]{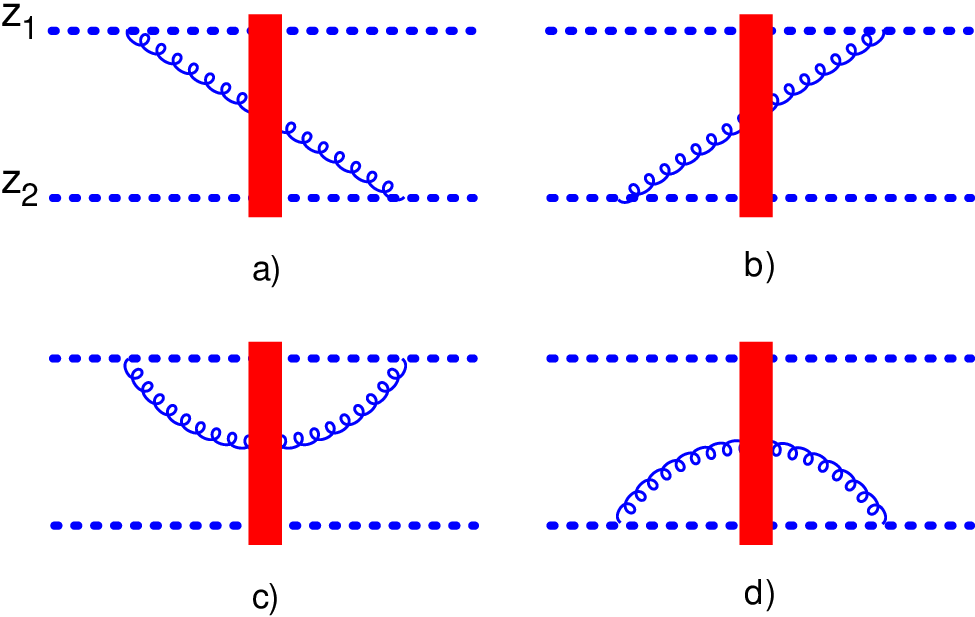}
	\end{center}
	\caption{Sample of diagrams (real ones) for the LO BK equation.}
	\label{lobk}
\end{figure}

The evolution of the ${\rm Tr}\{\hat{U}(z_\perp)\hat{U}^\dagger(z'_\perp)\}$ with respect to rapidity is the BK equation.
The diagrams (except the virtual ones) contributing to the kernel of the evolution equation are given in Fig. \ref{lobk}.
The evolution equation is \cite{Balitsky:1995ub}
\begin{eqnarray}
\langle\Tr\{U_{z_1}U^\dagger_{z_2}\}\rangle =\!\!
&& {\alpha_s\over 2\pi}\int{d\alpha\over \alpha}
\int d^2z\,{(z_1-z_2)^2\over (z_1-z)^2(z_2-z)^2}
\nonumber\\
&&\times\Big[
\Tr\{U_{z_1}U^\dagger_z\}{\rm Tr}\{U_zU^\dagger_{z_2}\} - N_c{\rm Tr}\{U_{z_1}U^\dagger_{z_2}\}\Big]\,.
\end{eqnarray}

The evolution equation is obtained in the background fileld method by integrating over the infinitesimal step in rapidity
where $\int_{\eta_2}^{\eta_1}{d\alpha\over \alpha} \to \Delta\eta$. Taking derivative with respect to rapidity
we get the evolution equation
\begin{eqnarray}
{d\over d\eta}\hat{\calu}^\eta_{z_1z_2} = {\alpha_sN_c\over 2\pi}
\int d^2z\,{(z_1-z_2)^2\over (z_1-z)^2(z_2-z)^2}
\Big[\hat{\calu}^\eta_{z_1z} + \hat{\calu}^\eta_{zz_2} - \hat{\calu}^\eta_{z_1z_2} - 
\hat{\calu}^\eta_{z_1z}\hat{\calu}^\eta_{zz_2}\Big]\,.
\label{balitskyeq}
\end{eqnarray}
where we indicate the rapidity dependence of the operators by the subscript $\eta$ and we defined the operator
\begin{eqnarray}
\hat{\calu}(x_\perp,y_\perp) = 1 - {1\over N_c}\Tr\{\hat{U}(x_\perp)\hat{U}^\dagger(y_\perp)\}\,.
\end{eqnarray}

Equation (\ref{balitskyeq}) is the Balitsky evolution equation. The operator we started with, 
$\Tr\{\hat{U}_{z_1}\hat{U}^\dagger_{z_2}\}$, after one loop evolution became a sum of two operators.
One is clearly the same as the original one (before evolution), the other one,
$\Tr\{\hat{U}_{z_1}\hat{U}^\dagger_z\}{\rm Tr}\{\hat{U}_z\hat{U}^\dagger_{z_2}\} $ is a new operator.
To solve the evolution equation one 
should, in principle, find the evolution equation of the new operator as well. However, this will actually generate 
a further new operator. This process generates a hierarchy of evolution equations known as Balitsky-hierarchy.
It is also known that the Balitsky-hierarchy is equivalent to the JIMWLK formalism \cite{JalilianMarian:1997gr, Ferreiro:2001qy, Iancu:2000hn} 
and for this reason, in the literature, they are mentioned together as B-JIMWLK equation. 

In the large $N_c$ approximation, the matrix element of the 
new operator factorizes into a product of operators equal to the one we started with, thus breaking the 
Balitsky-hierarchy to the non-linear Balitsky-Kovchegov \cite{Balitsky:1995ub, Kovchegov:1999ua, Kovchegov:1999yj} 
equation
\begin{eqnarray}
\hspace{-1cm}{d\over d\eta}\langle\calu^\eta_{z_1z_2}\rangle = {\alpha_sN_c\over 2\pi}
\int {d^2z(z_1-z_2)^2\over (z_1-z)^2(z_2-z)^2}
\Big[\langle\calu^\eta_{z_1z}\rangle + \langle\calu^\eta_{zz_2}\rangle - \langle\calu^\eta_{z_1z_2}\rangle - 
\langle\calu^\eta_{z_1z}\rangle\langle\calu^\eta_{zz_2}\rangle\Big]\,.
\end{eqnarray}

If sub-eikonal corrections are included, as we will see, we will obtain new evolution equations which will generate 
different type of new operators.

\section{Sub-eikonal corrections to the high-energy OPE}
\label{sec: sub-opehe}

Now we are going to include the sub-eikonal corrections to the high-energy OPE. 
To this end we need the quark propagator with sub-eikonal corrections which was calculated in Ref. \cite{Chirilli:2018kkw}.

\subsection{Quark propagator in the background of gluon}

The quark propagator in the background of gluon field with sub-eikonal corrections was calculated in Ref. \cite{Chirilli:2018kkw}.
In Appendix \ref{sec: propagators}, we provide a new derivation of the quark propagator which confirms the 
one obtained in the previous publication \cite{Chirilli:2018kkw}. In the same Appendix, we also put the result in a new form which facilitates
the analysis of the matrix elements with sub-eikonal corrections. The new form of the quark propagator in the background
of gluon fields up to subeikonal terms is
\begin{eqnarray}
\hspace{-1cm}&&\brax{i\over \hat{\Sp} +i\epsilon}\kety
\nonumber\\
&&= \left[\int_0^{+\infty}\!\!{\dhd \alpha\over 2\alpha}\theta(x_*-y_*) - 
\int_{-\infty}^0\!\!{\dhd\alpha\over 2\alpha}\theta(y_*-x_*) \right] e^{-i\alpha(x_\bullet - y_\bullet)}
{1\over \alpha s}\,
\nonumber\\
&&
\times\braxp\,e^{-i{\hatp^2_\perp\over \alpha s}x_*}\Bigg\{
\hat{\ssp}\,\ssp_2\,[x_*,y_*]\,\hat{\ssp}
+ \hat{\ssp}\,\ssp_2\,\hat{\mathcal{O}}_1(p_\perp; x_*,y_*)\,\hat{\ssp}
\nonumber\\
&&
+ {i\,\alpha s\over 2} \epsilon^{ij}\gamma^5\gamma_i\hat{\mathcal{O}}_j(p_\perp;x_*,y_*) 
- \half \ssp_2\big[\hat{p}^j,\hat{\mathcal{O}}_j(p_\perp; x_*,y_*)\big]
- {i\over 2}\epsilon^{ij}\gamma^5\ssp_2\{\hat{p}_i,\hat{\mathcal{O}}_j(p_\perp;x_*,y_*)\} 
\nonumber\\
&& +{i\,\alpha s\over 2}\epsilon^{ij}\gamma^5\gamma_j\{p_i,\calo_{\bullet *}(x_*,y_*)\}  
- \half\ssp_2[\hat{p}^2_\perp,\hat{\mathcal{O}}_{\bullet*}(x_*,y_*)]
\Bigg\}e^{i{\hatp^2_\perp\over \alpha s}y_*}\ketyp + O(\lambda^{-2})\,.
\label{ojobulletstar}
\end{eqnarray}
where operators $\calo_1$, $\calo_j$ and $\calo_{\bullet *}$ are defined in eqs. (\ref{O1}), (\ref{Oj}), and (\ref{Obulletstar}).
This is a new expression (equivalent to the one obtained before in Ref. \cite{Chirilli:2018kkw})
of the quark propagator with sub-eikonal corrections in the background of gluon fields.

As it is shown in Appendix \ref{sec: matrielements},  from all the terms present in the quark propagator (\ref{ojobulletstar}),
the only one that contributes to the polarized $g_1$ structure function is
\begin{eqnarray}
\hspace{-1cm}&&\langle{\rm T}\{\psi(x)\bar{\psi}(y)\}\rangle_{\psi,\bar{\psi}} 
\nonumber\\
&&\hspace{-1cm}\stackrel{x_*>0>y_*}{\ni}
\int_0^{+\infty}\!{\dhd\alpha\over 4s\alpha^3}\,e^{-i\alpha(x_\bullet-y_\bullet)}\int d^2z
\braxp \ssp \,e^{-i{\hatp^2_\perp\over \alpha s}x_*}\ketzp
\nonumber\\
\hspace{-1cm}&&\times ig\int_{-\infty}^{+\infty}\!d{2\over s}z_*\,\,\ssp_2[\infty p_1,z_*]_z
\half \sigma^{ij}F_{ij}(z_*,z_\perp)[z_*,-\infty p_1]_z\brazp\ssp\,e^{i{\hatp^2\over \alpha s}y_*}\ketyp
\label{qpropFijterm}
\end{eqnarray}
where we used $\sigma^{ij} = {i\over 2}(\gamma^i\gamma^j-\gamma^j\gamma^i)$, 
and the $\hbar$-inspired notation $\dhd^n \alpha = {d^n\alpha\over (2\pi)^n}$, and
we defined the gauge link at fixed transverse position $z_\perp$ as
\begin{eqnarray}
{\rm Pexp} \left\{ig{2\over s}\!\! \int_{y_*}^{x_*}\!\!\!dz_*\, 
A_\bullet(z_*,z_\perp)\right\}\ \equiv [x_*p_1 + z_\perp, y_*p_1+z_\perp] \equiv [x_*,y_*]_z\,.
\end{eqnarray}

We will work again in coordinate space so, performing the Fourier transform of (\ref{qpropFijterm}),
we have
\begin{eqnarray}
&&\langle\Tr\{\psi(x)\bar{\psi}(y)\}\rangle_{\psi,\bar{\psi}} 
\label{Fijterm-coord}\\
&&\stackrel{x_*>0>y_*}{\ni}
\!\! - {1\over 16\pi^3 x_*^2y_*^2}\int {d^2 z\over [\calz+i\epsilon]^2}
\big({2\over s}x_*\ssp_1 + \Sx_\perp\big)
{1\over s}\ssp_2\gamma^5\calf(z_\perp)\big({2\over s}x_*\ssp_1+\Sy_\perp\big)
\nonumber
\end{eqnarray}
with
\begin{eqnarray}
\calz \equiv {(x-z)^2_\perp\over x_*} -  {(y-z)^2_\perp\over y_*} - {4\over s}(x_\bullet - y_\bullet) 
\end{eqnarray}
and where we have introduced the short-hand notation $X^\mu_\perp = (x-z)^\mu_\perp$ and similarly for $Y^\mu_\perp$
and defined, for later convenience, the operator
\begin{eqnarray}
\calf(z_\perp) \equiv ig{s\over 2}\int_{-\infty}^{+\infty}\!\!dz_*[\infty p_1,z_*]_z
\,\epsilon^{ij}F_{ij}\big(z_*,z_{1\perp}\big)[z_*,-\infty p_1]_z\,.
\label{calfoperator}
\end{eqnarray}
We use the symbol $\ni$ to indicate that we are considering only one of the possible terms that make up the quark propagator.
Moreover, the superscript $x_*>0>y_*$ indicates that we are considering only the contribution 
in which the quark goes through the shock-wave from positive values of light-cone coordinate, $x_*>0$, to negative ones,
$y_*<0$.

Throughout this paper we use the convention 
$\gamma^5 = i\gamma^0\gamma^1\gamma^2\gamma^3 = - {i\over 4!}\epsilon_{\mu\nu\rho\sigma}$ and $\epsilon^{0123}=1$.
The two dimensional antisymmetric tensor is $p_{1\mu}p_{2\nu}\epsilon^{\mu\nu ij} = {s\over 2}\epsilon^{ij}$
where, as usual, Latin indexes take values $1,2$ and $\epsilon^{12}=-\epsilon^{21} = 1$.

\subsection{Quark propagator in the background of quark fields}

The quark propagator in the background of quark fields is \cite{Chirilli:2018kkw}
\begin{eqnarray}
\hspace{-1cm}\langle{\rm T}\{\psi(x)\bar{\psi}(y)\}\rangle_{\psi,\bar{\psi}} 
&&\stackrel{x_*>0>y_*}{\ni} g^2\int_0^{+\infty}\!{\dhd\alpha\over 2\alpha}
\int_{-\infty}^{+\infty}\!\!\! dz_*\! \int_{-\infty}^{z_*}\!\!\!dz'_*
e^{-i\alpha(x_\bullet - y_\bullet)}
\nonumber\\
&&
\times { 1\over\alpha^4 s^4}
\braxp e^{-i{\hatp^2_\perp\over \alpha s}x_*}\ssp\ssp_2
\ssp\,[\infty p_1,z_*] \gamma^\mu t^a\psi(z_*)\left(\delta^\xi_\mu - {p_{2\mu}\over p_*}p^\xi\right)[z_*,z'_*]^{ab}
\nonumber\\
&&
\times\!\left(g_{\xi\nu} - p_\xi {p_{2\nu}\over p_*}\right)
\bar{\psi}(z'_*)t^b\gamma^\nu [z'_*,-\infty p_1]\,\ssp 
\ssp_2\ssp \,e^{i{\hatp^2_\perp\over \alpha s}y_*}\ketyp
\label{qprpinquark}
\end{eqnarray}
Performing the Fourier transform we have
\begin{eqnarray}
&&\langle{\rm T}\{\psi(x)\bar{\psi}(y)\}\rangle_{\psi,\bar{\psi}} 
\nonumber\\
&&\stackrel{x_*>0>y_*}{\ni} - {g^2\over 16\pi^3x_*^2y_*^2}\!\int_{-\infty}^{+\infty}\!\!\! dz_*\! \int_{-\infty}^{z_*}\!\!dz'_*
\int {d^2z\over (\calz^2+i\epsilon)^2}
\big({2\over s} x_*\ssp_1 + \Sx_\perp\big)
\nonumber\\
&&\hspace{-1cm} 
\times[\infty p_1,z_*]_z t^a \Big(\gamma^\mu_\perp \psi(z_*,x_\perp)[z_*,z'_*]^{ab}_z
\bar{\psi}(z'_*,z_\perp)\gamma_\mu^\perp\Big) t^b[z'_*,-\infty p_1]_z
\big({2\over s} y_*\ssp_1 + \Sy_\perp\big)
\label{qprop-backqua-coord}
\end{eqnarray}
To highlight the structure of the propagator (\ref{qprop-backqua-coord}), let us define the following operator
\begin{eqnarray}
&&Q^{\alpha\beta}_{ij}(x_\perp) 
\nonumber\\
&&\equiv  g^2\!\int_{-\infty}^{+\infty}\!\!\! dz_*\! \int_{-\infty}^{z_*}\!\!dz'_*
\Big([\infty p_1,z_*]_xt^a  \psi^\alpha(z_*,x_\perp)[z_*,z'_*]^{ab}_x
\bar{\psi}^\beta(z'_*,x_\perp) t^b[z'_*,-\infty p_1]_x\Big)_{ij}
\label{Qoper}
\end{eqnarray}
where $\alpha$, and $\beta$ are spinor indexes, and $i, j$ are color indexes in the fundamental representation.
Thus, the quark propagator (\ref{qprop-backqua-coord}) becomes
\begin{eqnarray}
\hspace{-1cm}&&\langle{\rm T}\{\psi(x)\bar{\psi}(y)\}\rangle_{\psi,\bar{\psi}} 
\nonumber\\
&&\stackrel{x_*>0>y_*}{\ni}
- {1\over 16\pi^3x_*^2y_*^2}
\int {d^2z\over (\calz+i\epsilon)^2}
\big({2\over s} x_*\ssp_1 + \Sx_\perp\big)
\gamma^\mu_\perp Q(z_\perp)\gamma_\mu^\perp\big({2\over s} y_*\ssp_1 + \Sy_\perp\big)\,.
\label{qpropglu}
\end{eqnarray}

\subsection{Quark propagator relevant for polarized high-energy DIS}

\begin{figure}[htb]
	\begin{center}
	\includegraphics[width=4.0 in]{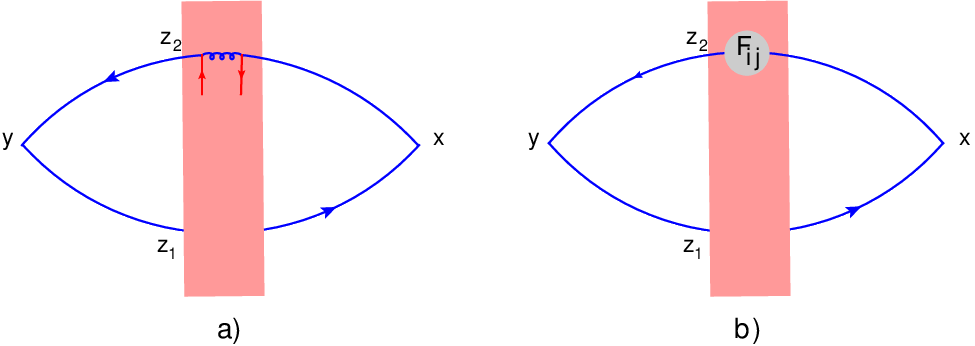}
	\caption{Diagram of the LO impact factor with quark-sub-eikonal correction.}
	\label{subeikonalif}
	\end{center}
\end{figure}

The quark propagator we need for polarized DIS at high-energy is the sum of the eikonal 
propagator (\ref{qprop-eikona}), the sub-eikonal correction due to the background of gluon field (\ref{Fijterm-coord}),
and the sub-eikonal correction due to the background made by quark fields eq. (\ref{qpropglu}). 
Putting all these contributions together we have
\begin{eqnarray}
&&\langle{\rm T}\{\psi(x)\bar{\psi}(y)\}\rangle_{A,\psi,\bar{\psi}} 
\nonumber\\
&&\stackrel{x_*>0>y_*}{\ni}
- {1\over 2\pi^3x_*^2y_*^2}
\int\!\! {d^2z\over (\calz+i\epsilon)^3}\Big({2\over s}x_*\ssp_1 + \Sx_\perp\Big)
\nonumber\\
&&~~~\times 
\bigg\{i\,\ssp_2U(z_\perp) + {\calz\over 8}
\Big({1\over s}\ssp_2\gamma^5\calf(z_\perp) + \gamma^\mu_\perp Q(z_\perp)\gamma_\mu^\perp
\Big)\bigg\}\Big({2\over s}y_*\ssp_1 + \Sy_\perp\Big) 
\label{eik-noeik}
\end{eqnarray}
Propagator (\ref{eik-noeik}) will be used to calculate the impact factor in Fig. \ref{subeikonalif}.

\section{Operator Product Expansion with sub-eikonal terms}
\label{sec: OPEsubeik}

\subsection{OPE with Quark sub-eikonal terms}

Now that we obtained the quark propagator with the necessary sub-eikonal corrections,
we are ready to extend the expansion of the T-product of two electromagnetic currents to the polarized DIS case.

The diagram we need to calculate are given in Fig. \ref{subeikonalif}.
As before, we make functional integration over the spinor field
\begin{eqnarray}
&&\hspace{-2cm}\langle {\rm T}\{j^{\mu}(x) j^{\nu}(y)\}\rangle_{A,\psi,\bar{\psi}}
= {\rm tr}\Big\{\brax {1\over \Sp + i\epsilon}\kety \gamma^\nu \bray{1\over \Sp + i\epsilon}\ketx\gamma^\mu\Big\}
\label{spinor-contraction}
\end{eqnarray}
where for each quark propagator in eq. (\ref{spinor-contraction})
we should use eq. (\ref{eik-noeik}). We will consider operators $\calf(z_\perp)$ and $Q(z_\perp)$ separately starting, in this section, with
$Q(z_\perp)$. 

The product of two quark propagators given in eq. (\ref{eik-noeik}) will result in four terms. The first term is a product of two eikonal 
propagators which we considered in section \ref{sec: opehe}. The second and the third terms are obtained by the product of one eikonal term and
one sub-eikonal term; we will consider them in this section.  The last term is the product of two sub-eikonal terms,
and since it will contribute as a sub-sub-eikonal correction it will be disregarded.

As it is illustrated in Fig. \ref{subeikonalif}, if we do not want 
to exceed our precision, we need a sub-eikonal correction to only one of the two
interactions between the quark and the shock-wave target while keeping eikonal the other one.
Diagram \ref{subeikonalif} a, is
\begin{eqnarray}
\hspace{-1cm}\langle {\rm T}\{j^{\mu}(x) j^{\nu}(y)\}\rangle_{\small{\rm Fig.} \ref{subeikonalif}a}
 \stackrel{x_*>0>y_*}{\ni}\!\!\!&&
 {i\over 32\pi^6 x_*^4y_*^4} \!\!\int {d^2z_1 d^2z_2\over [\calz_1+i\epsilon]^3[\calz_2+i\epsilon]^2}
\label{OPEquark}\\
&&\times\!
\Bigg[{\rm Tr\,tr}\{\gamma^\mu \Sx_2 
\gamma^\rho_\perp Q(z_{2\perp})\gamma_\rho^\perp\,\Sy_2\gamma^\nu\,\Sy_1\,\ssp_2U^\dagger_{z_1} \,\Sx_1\}
\nonumber\\
&&
~~~~- \Big({\rm Tr\,tr}\{\gamma^\mu \Sx_2 \gamma^\rho_\perp Q(z_{2\perp})\gamma_\rho^\perp
\,\Sy_2\gamma^\nu\,\Sy_1\,\ssp_2U^\dagger_{z_1} \,\Sx_1\}\Big)^\dagger
\Bigg]
\nonumber
\end{eqnarray}
where we used again the short-hand notation
$X^\mu_i = {2\over s}x_*p_1^\mu + X^\mu_{i\perp}$  with $X^\mu_{i\perp} = x^\mu_\perp - z^\mu_{i\perp}$ and $i=1,2$
and similarly for $Y_i$.
Moreover, we use $\tr$ for trace over spinor indexes while $\Tr$ over color indexes
(we refer the reader to Appendix \ref{sec: qsub-correctionIF} for the details of the derivation of eq. (\ref{OPEquark})).

Our task is now the evaluation of the double trace 
\begin{eqnarray}
i\,{\rm Tr\,tr}\{\gamma^\mu \Sx_2 \gamma^\rho_\perp Q(z_{2\perp})\gamma_\rho^\perp
\,\Sy_2\gamma^\nu\,\Sy_1\,\ssp_2U^\dagger_{z_1} \,\Sx_1\}\,.
\end{eqnarray}

After some lengthy algebra, and using the rescaling of the spinor field given in eq. (\ref{spinorboost}), 
one can show that the trace over the spinor indexes can be cast in the following form
\begin{eqnarray}
&&
i\,\bar{\psi}(z'_*,z_{2\perp})\gamma^\perp_\rho \hatY_2\gamma^\nu\hatY_1\ssp_2\hatX_1\gamma^\mu\hatX_2\gamma^\rho_\perp
\psi(z_*,z_{2\perp})
\nonumber\\
&&~~~~= {8\over s}\bar{\psi}(z'_*,z_{2\perp})\,i\,\ssp_1\psi(z_*,z_{2\perp})I^{\mu\nu}_1(x_*,y_*;z_{1\perp},z_{2\perp})
\nonumber\\
&&~~~~~~ - {8\over s}\bar{\psi}(z'_*,z_{2\perp})\gamma^5\ssp_1\psi(z_*,z_{2\perp})I^{\mu\nu}_5(x_*,y_*;z_{1\perp},z_{2\perp})
+ O(\lambda^{-1})
\label{spinor-trace}
\end{eqnarray}
where
\begin{eqnarray}
I^{\mu\nu}_1(x,y;z_1,z_2) = \half 
x_*^2y_*^2{\partial^2\over \partial x_\mu\partial y_\nu}\left(\calz_1\calz_2 - z_{12\perp}^2 {(x-y)^2\over x_*y_*}\right)\,,
\label{I1}
\end{eqnarray}
with $\calz_i$ defined in eq. (\ref{calzi}), and
\begin{eqnarray}
\hspace{-0.5cm}
I^{\mu\nu}_5(x,y;z_1,z_2) = \big(x_*\partial^\mu_x - p_2^\mu\big)\big(y_*\partial^\nu_y - p_2^\nu\big)
\big[(\vec{Y}_1\!\times\!\vec{Y}_2) X_1\!\cdot\! X_2 - (\vec{X}_1\!\times\!\vec{X}_2)Y_1\!\cdot\! Y_2\big]\,,
\label{I5}
\end{eqnarray}
where we used the notation for vector product in two dimensions 
\begin{eqnarray}
\vec{x}\times\vec{y} = \epsilon^{ij} x_iy_j\,.
\end{eqnarray}
The explicit expressions of $I^{\mu\nu}_1$ and $I^{\mu\nu}_5$ can be found in Appendix \ref{sec: explicitIFs}.

In result (\ref{spinor-trace}) we obtained two operators with definite parity. 
The symmetric tensor $I_1^{\mu\nu}$ is proportional to the parity even operator 
$\barpsi(z'_*,z_{2\perp})\ssp_1\psi(z_*,z_{2\perp})$, while the operator proportional to
the antisymmetric tensor $I^{\mu\nu}_5$ is proportional to the parity odd operator
$\barpsi(z'_*,z_{2\perp})\gamma^5\ssp_1\psi(z_*,z_{2\perp})$.

Using result (\ref{spinor-trace}) in the eq. (\ref{OPEquark}), we arrive at
\begin{eqnarray}
\hspace{-0.8cm}\langle {\rm T}\{j^{\mu}(x) j^{\nu}(y)\}\rangle_{A,\psi,\bar{\psi}}
\stackrel{x_*>0>y_*}{\ni}\!\!\!&&
{1\over 4\pi^6 s\, x_*^4y_*^4} \!\!\int
{ d^2z_1 d^2z_2\over [\calz_1+i\epsilon]^3[\calz_2+i\epsilon]^2}
\label{OPEQ1Q5}\\
&&
\times\!\Bigg[
\Big(\Tr\{\calq_{1z_2}U^\dagger_{z_1}\} + \Tr\{\calq^\dagger_{1z_2}U_{z_1}\}\Big)I^{\mu\nu}_1(x,y;z_1,z_2)
\nonumber\\
&&
~~~~- \Big(\Tr\{\calq_{5z_2}U^\dagger_{z_1}\} + \Tr\{\calq^\dagger_{5z_2}U_{z_1}\}\Big)I_5^{\mu\nu}(x,y;z_1,z_2)
\Bigg]
\nonumber
\end{eqnarray}
where we defined the operators $\calq_{1z}$ and $\calq_{5z}$ as
\begin{eqnarray}
&&\hspace{-1.2cm}
\calq_1(z_{\perp})\! \equiv\!
g^2\!\!\int^{+\infty}_{-\infty}\!\!\!dz_*\!\!\int_{-\infty}^{z_*}\!\!\!dz'_*
[\infty p_1,z_*]_{z}t^a\tr\{i\,\ssp_1\psi(z_*,z_{\perp})
\nonumber\\
&&~~~\times[z_*,z'_*]^{ab}_{z}
\bar{\psi}(z'_*,z_{\perp})\}
 t^b[z'_*,-\infty p_1]_{z}\,,
\label{calq1}
\\
&&\hspace{-1.2cm}
\calq_5(z_\perp) \!\equiv\! g^2\!\!\int^{+\infty}_{-\infty}\!\!\!dz_*\!\!\int_{-\infty}^{z_*}\!\!\!dz'_*
[\infty p_1,z_*]_{z}t^a\tr\{\gamma^5\ssp_1\psi(z_*,z_{\perp})
\nonumber\\
&&~~~\times
[z_*,z'_*]^{ab}_{z}
\bar{\psi}(z'_*,z_{\perp})\}t^b[z'_*,-\infty p_1]_{z}\,.
\label{calq5}
\end{eqnarray}

To put result (\ref{OPEQ1Q5}) in the same form as eq. (\ref{LO-OPE-1}), we define the quark-sub-eikonal impact factors as
\begin{eqnarray}
&&\cali^{\mu\nu}_1(x,y;z_1,z_2) = {1\over 4\pi^6 x_*^4y_*^4}
{I^{\mu\nu}_1(x,y;z_1,z_2)\over [\calz_1+i\epsilon]^3[\calz_2+i\epsilon]^2}
\label{I1definition}
\\
&&\cali^{\mu\nu}_5(x,y;z_1,z_2) = - {1\over 4\pi^6 x_*^4y_*^4}
{I^{\mu\nu}_5(x,y;z_1,z_2)\over [\calz_1+i\epsilon]^3[\calz_2+i\epsilon]^2}
\label{I5definition}
\end{eqnarray}
thus the high-energy OPE with quark sub-eikonal contributions is
\begin{eqnarray}
&&
T\{\bar{\hat{\psi}}(x)\gamma^\mu\hat{\psi}(x)\bar{\hat{\psi}}(y)\gamma^\nu\hat{\psi}(y)\}
\nonumber\\
&& \stackrel{x_*>0>y_*}{\ni}
{1\over s}\int d^2z_1 d^2z_2\Bigg[
\Big(\Tr\{\hat{\calq}_{1z_2}\hat{U}^\dagger_{z_1}\} + \Tr\{\hat{\calq}^\dagger_{1z_2}\hat{U}_{z_1}\}\Big)\cali^{\mu\nu}_1(x,y;z_1,z_2)
\nonumber\\
&&\hspace{4cm}
 + \Big(\Tr\{\hat{\calq}_{5z_2}\hat{U}^\dagger_{z_1}\} + \Tr\{\hat{\calq}^\dagger_{5z_2}\hat{U}_{z_1}\}\Big)
\cali_5^{\mu\nu}(x,y;z_1,z_2)
\Bigg]
\label{OPEQ1Q5operator}
\end{eqnarray}
We notice that the sub-eikonal quark contribution to the quark propagator, eq. (\ref{Qoper}), gave rise to two
different impact factors and to two different operators. The impact factor (coefficient function) $\cali_1$ which is associated to the operator 
$\Tr\{\hat{\calq}_{1z_2}\hat{U}^\dagger_{z_1}\} + \Tr\{\hat{\calq}^\dagger_{1z_2}\hat{U}_{z_1}\}$
is a sub-eikonal correction to eikonal unpolarized case eq. (\ref{LO-OPE-1}) because it is symmetric under the
exchange $\mu\leftrightarrow \nu$, $x\leftrightarrow y$. 
Moreover, we notice that there is a relation between the LO impact factor $\cali_{LO}^{\mu\nu}$
and the impact factor $\cali_1^{\mu\nu}$. Indeed, using definitions (\ref{LOcoeff}) and (\ref{I1definition}) we can write
\begin{eqnarray}
\cali_1^{\mu\nu} = {\calz_2\over 8}\cali_{LO}^{\mu\nu}
\end{eqnarray}

The impact factor $\cali_5$ and its associated operator $\Tr\{\hat{\calq}_{5z_2}\hat{U}^\dagger_{z_1}\} + \Tr\{\hat{\calq}^\dagger_{5z_2}\hat{U}_{z_1}\}$.

Like the LO impact factor, one can check that the coefficients $\cali^{\mu\nu}_1$, and $\cali^{\mu\nu}_5$ 
in eqs. (\ref{I1definition}) and (\ref{I5definition}) satisfy 
the electromagnetic gauge invariance
\begin{eqnarray}
\partial_\mu^x{1\over \calz^2_2\calz_1^3x_*^4y_*^4}I^{\mu\nu}_1 = 
{1\over x_*^5y_*^4\calz^3_2\calz_1^4}\Big[\!\!\!\!
&& x_*\calz_1\calz_2\partial^x_\mu I^{\mu\nu}_1 - I_1^{\mu\nu}
\Big(4\calz_1\calz_2p_{2\mu} 
\nonumber\\
&& + 3 x_*\calz_2\partial^x_\mu\calz_1 + 2x_*\calz_1\partial^x_\mu\calz_2\Big)\Big] = 0
\end{eqnarray}
and
\begin{eqnarray}
\partial_\mu^x{1\over \calz^2_2\calz_1^3x_*^4y_*^4}I^{\mu\nu}_5 = 
{1\over x_*^5y_*^4\calz^3_2\calz_1^4}\Big[\!\!\!\!
&& x_*\calz_1\calz_2\partial^x_\mu I^{\mu\nu}_5 - I_5^{\mu\nu}
\Big(4\calz_1\calz_2p_{2\mu} 
\nonumber\\
&& + 3 x_*\calz_2\partial^x_\mu\calz_1 + 2x_*\calz_1\partial^x_\mu\calz_2\Big)\Big]
= 0\,.
\end{eqnarray}
One may also easily check that impact factors (\ref{I1definition}) and (\ref{I5definition}) 
are M\"obius $SL(2,C)$ invariant.

Before proceeding with the calculation of the evolution equations, we observe that, working out the color algebra
we can rewrite operator $Q^{\alpha\beta}_{ij}$ defined in (\ref{Qoper}) as
\begin{eqnarray}
Q^{\alpha\beta}_{ij}(x_\perp) =\!\!&& - g^2\!\int_{-\infty}^{+\infty}\!\!\! dz_*\! \int_{-\infty}^{z_*}\!\!dz'_*
\bigg[\half\,U_x^{ij} \, \barpsi^\alpha(z'_*,x_\perp)[z'_*,z_*]_x\psi^\beta(z_*,x_\perp)
\label{Q2lightray}\\
&&\hspace{3cm}+ {1\over 2N_c}\big([\infty p_1,z_*]_x \psi^\beta(z_*,x_\perp)\barpsi^\alpha(z'_*,x_\perp)[z'_*,-\infty]\big)_{ij}
\bigg]
\nonumber
\end{eqnarray}
From (\ref{Q2lightray}), we see that in the large $N_c$ limit, the $Q^{\alpha\beta}_{ij}$ operator reduces
to the product of an infinite Wilson line times
the usual light-cone quark operator when multiplied by 
$\ssp_1 = \sqrt{s\over 2}\gamma_+$ (recall that $\gamma_+=\gamma^-$
is what is call the \textit{good component})
or to the parity odd quark operator when multiplied by $\gamma^5\ssp_1$.

\begin{figure}[thb]
	\begin{center}
		\includegraphics[width=4.5in]{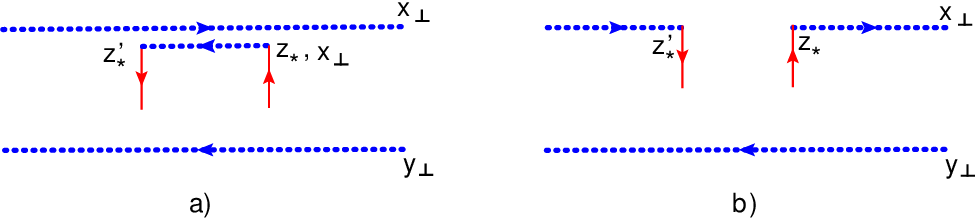}
		\caption{Figure a) represents the operators $\Tr\{U_xU^\dagger_y\}Q_{1x}$, and $\Tr\{U_xU^\dagger_y\}Q_{5x}$,
			 while Figure b) represents operators $\Tr\{\tildeQ_{1x}U^\dagger_y\}$, and $\Tr\{\tildeQ_{5x}U^\dagger_y\}$.}
		\label{Q-tildeQ-operators}
	\end{center}
\end{figure}

We define the light-ray quark operators as 
\begin{eqnarray}
Q_{1x} =&& -g^2\!\!\int_{-\infty}^{+\infty}\!\!dz_*\!\!\int_{-\infty}^{z_*}\!\!dz'_*\,
\bar{\psi}(z'_*,x_\perp)\,i\,\ssp_1[z'_*,z_*]_x\psi(z_*,x_\perp)\,,
\label{Q1}
\\
Q^\dagger_{1x} =&&  g^2\!\!\int_{-\infty}^{+\infty}\!\!dz_*\!\!\int_{-\infty}^{z_*}\!\!dz'_*\,
\bar{\psi}(z_*,x_\perp)\,i\,\ssp_1[z_*,z'_*]_x\psi(z'_*,x_\perp)\,,
\label{Qd1}
\\
Q_{5x} \equiv&& - g^2\!\!\int_{-\infty}^{+\infty}\!\!dz_*\!\!\int_{-\infty}^{z_*}\!\!dz'_*\,
\bar{\psi}(z'_*,x_\perp)\,\gamma^5\ssp_1[z'_*,z_*]_x \psi(z_*,x_\perp)\,,
\label{Q5}
\\
Q^\dagger_{5x} \equiv&& g^2\!\!\int_{-\infty}^{+\infty}\!\!dz_*\!\!\int_{-\infty}^{z_*}\!\!dz'_*\,
 \bar{\psi}(z_*,x_\perp)\,\gamma^5\ssp_1[z_*,z'_*]_x \psi(z'_*,x_\perp)\,,
\label{Qd5}
\end{eqnarray}
and
\begin{eqnarray}
&&\hspace{-1.3cm}\tilde{Q}_{1ij}(x_\perp) \equiv g^2\!\!\int_{-\infty}^{+\infty}\!\!dz_*\!\!\int_{-\infty}^{z_*}\!\!dz'_*
\big([\infty p_1,z_*]_x {\rm tr}\{\psi(z_*,x_\perp)\bar{\psi}(z'_*,x_\perp)\,i\,\ssp_1\}[z'_*, -\infty p_1]\big)_{ij}\,,
\label{Qt1}
\\
&&\hspace{-1.3cm}\tilde{Q}^\dagger_{1ij}(x_\perp) \equiv - g^2\!\!\int_{-\infty}^{+\infty}\!\!dz_*\!\!\int_{-\infty}^{z_*}\!\!dz'_*
\big([-\infty p_1,z'_*]_x {\rm tr}\{\psi(z'_*,x_\perp)\bar{\psi}(z_*,x_\perp)\,i\,\ssp_1\}[z_*, \infty p_1]\big)_{ij}\,,
\label{Qtd1}
\\
&&\hspace{-1.3cm}\tilde{Q}_{5ij}(x_\perp) \equiv g^2\!\!\int_{-\infty}^{+\infty}\!\!dz_*\!\!\int_{-\infty}^{z_*}\!\!dz'_*
\big([\infty p_1,z_*]_x {\rm tr}\{\psi(z_*,x_\perp)\bar{\psi}(z'_*,x_\perp)\,\gamma^5\ssp_1\}[z'_*, - \infty p_1]\big)_{ij}\,,
\label{Qt5}
\\
&&\hspace{-1.3cm}\tilde{Q}^\dagger_{5ij}(x_\perp) \equiv g^2\!\!\int_{-\infty}^{+\infty}\!\!dz_*\!\!\int_{-\infty}^{z_*}\!\!dz'_*
\big([-\infty p_1,z'_*]_x {\rm tr}\{\psi(z'_*,x_\perp)\bar{\psi}(z_*,x_\perp)\,\gamma^5\ssp_1\}[z_*, \infty p_1]\big)_{ij}\,.
\label{Qtd5}
\end{eqnarray}
where $i, j$ are color indexes in the fundamental representation.
So, we can rewrite the operator $\Tr\{\hat{\calq}_{1x}\hat{U}^\dagger_y\}$, and 
$\Tr\{\hat{\calq}_{5x}\hat{U}^\dagger_y\}$, and their adjoint conjugated as
\begin{eqnarray}
&&\Tr\{\hat{\calq}_{1x}\hat{U}^\dagger_y\}
= \half \Tr\{\hat{U}^\dagger_y \hat{U}_x\}\hat{Q}_{1\,x} - 
{1\over 2N_c} \Tr\{\hat{U}^\dagger_y \hat{\tilde{Q}}_{1x}\}
\label{calq1toq1}
\\
&&\Tr\{\hat{\calq}^\dagger_{1x}\hat{U}_y\}
= \half \Tr\{\hat{U}_y \hat{U}^\dagger_x\}\hat{Q}_{1x}^\dagger - 
{1\over 2N_c}\Tr\{\hat{U}_y \hat{\tilde{Q}}^\dagger_{1x}\}
\label{calq1toq1dag}
\\
&&\Tr\{\hat{\calq}_{5x}\hat{U}^\dagger_y\}
= \half \Tr\{\hat{U}^\dagger_y \hat{U}_x\}\hat{Q}_{5x} - 
{1\over 2N_c} \Tr\{\hat{U}^\dagger_y \hat{\tilde{Q}}_{5x}\}
\label{calq5toq5}
\\
&&\Tr\{\hat{\calq}^\dagger_{5x}\hat{U}_y\}
= \half \Tr\{\hat{U}_y \hat{U}^\dagger_x\}\hat{Q}_{5x}^\dagger - 
{1\over 2N_c}\Tr\{\hat{U}_y \hat{\tilde{Q}}^\dagger_{5x}\}
\label{calq5toq5dag}
\end{eqnarray}
Operators (\ref{calq1toq1})-(\ref{calq5toq5dag}) are the new operators of which we will calculate the evolution equations and in provide
the parametrization through new distribution functions. They are represented in Fig. \ref{Q-tildeQ-operators}.

The OPE (\ref{OPEQ1Q5operator}) in terms of operators (\ref{calq1toq1})-(\ref{calq5toq5dag}) is
\begin{eqnarray}
&&T\{\bar{\hat{\psi}}(x)\gamma^\mu\hat{\psi}(x)\bar{\hat{\psi}}(y)\gamma^\nu\hat{\psi}(y)\}
\nonumber\\
&& \stackrel{x_*>0>y_*}{\ni}
{1\over 2s}\int d^2z_1 d^2z_2\bigg[
\cali^{\mu\nu}_1(x,y;z_1,z_2)
\Big(\Tr\{\hat{U}^\dagger_{z_1} \hat{U}_{z_2}\}\hat{Q}_{1z_2} +  \Tr\{\hat{U}_{z_1} \hat{U}^\dagger_{z_2}\}\hat{Q}_{1z_2}^\dagger
\nonumber\\
&&\hspace{6cm}
-  {1\over N_c} \Tr\{\hat{U}^\dagger_{z_1} \hat{\tilde{Q}}_{1z_2}\}
- {1\over N_c}\Tr\{\hat{U}_{z_1} \hat{\tilde{Q}}^\dagger_{1z_2}\}\Big)
\nonumber\\
&& \hspace{4cm}+ \cali_5^{\mu\nu}(x,y;z_1,z_2)\Big(\Tr\{\hat{U}^\dagger_{z_1} \hat{U}_{z_2}\}\hat{Q}_{5z_2} 
 +  \Tr\{\hat{U}_{z_1} \hat{U}^\dagger_{z_2}\}\hat{Q}_{5z_2}^\dagger
\nonumber\\
&&\hspace{6cm} -  {1\over N_c} \Tr\{\hat{U}^\dagger_{z_1} \hat{\tilde{Q}}_{5z_2}\}
- {1\over N_c}\Tr\{\hat{U}_{z_1} \hat{\tilde{Q}}^\dagger_{5z_2}\}\Big)
\bigg]
\label{OPEQ1Q5Qt1Qt5}
\end{eqnarray}

\subsection{OPE with Gluon sub-eikonal terms}

The contribution to the high-energy OPE of the T-product of two electromagnetic currents
due to sub-eikonal terms in the background of gluon fields, given in eq. (\ref{Fijterm-coord}), is obtained 
again by performing the spinor contractions (see Fig. \ref{subeikonalif}b)
\begin{eqnarray}
&&\langle\bar{\psi}(x)\gamma^\mu\psi(x)\bar{\psi}(y)\gamma^\nu\psi(y)\rangle_A
\nonumber\\
&&= {\rm tr}\Big\{\brax {1\over \Sp + i\epsilon}\kety \gamma^\nu \bray{1\over \Sp + i\epsilon}\ketx\gamma^\mu\Big\}
\nonumber\\
&&\stackrel{x_*>0>y_*}{\ni}
{g\over 64\pi^6}\int\!\!d^2z_1 d^2 z_2\,
{ [\calz_2 + i\epsilon ]^{-2}\over [\calz_1 + i\epsilon ]^3}
{\rm tr}\{\Sx_1\ssp_2\Sy_1\gamma^\nu\Sy_2\ssp_2\sigma^{\rho\sigma}_\perp\Sx_2\gamma^\mu\}
\nonumber\\
&&~~~~\times
\Bigg[ \int_{-\infty}^{+\infty}\!\!\!d\omega_*
{\rm Tr}\{[\infty p_1,\omega_*]_{z_2} F^\perp_{\rho\sigma}[\omega_*,-\infty p_1]_{z_2}U^\dagger_{z_1}\}
\nonumber\\
&&~~~~~~~~ - \int^{+\infty}_{-\infty}\!\!\!d\omega_*
{\rm Tr}\{U_{z_1}[-\infty p_1,\omega_*]_{z_2} F^\perp_{\rho\sigma}[\omega_*,\infty p_1]_{z_2}\}
\Bigg]\,.
\label{OPE-F}
\end{eqnarray}
After a lengthy algebra, the calculation of the gamma matrices in (\ref{OPE-F}) gives
\begin{eqnarray}
&&\tr\{\Sx_1\ssp_2\Sy_1\gamma^\nu\Sy_2\ssp_2\sigma^{\alpha\beta}_\perp\Sx_2\gamma^\mu\}
\nonumber\\
&&~~~~~= 8i\epsilon^{\alpha\beta}\big(x_*\partial^\mu_x - p_2^\mu\big)\big(y_*\partial^\nu_y - p_2^\nu\big)
\big[(\vec{X}_1\times\vec{X}_2)Y_1\cdot Y_2 - (\vec{Y}_1\times\vec{Y}_2) X_1\cdot X_2\big]
\nonumber\\
&&~~~~~\equiv 8\,i\, \epsilon^{\alpha\beta} \, I^{\mu\nu}_{\calf}\,.
\label{Fij-impfact}
\end{eqnarray}
So, it turns out that $I_\calf^{\mu\nu} = - I_5^{\mu\nu}$. 
This should not be surprising because the two operators, $\calq_{5x}$ and $\calf_x$ are both
parity odd.

Using the operator $\calf_z$ defined in eq. (\ref{calfoperator}), eq. (\ref{OPE-F}) becomes
\begin{eqnarray}
\hspace{-1.4cm}&&\langle\bar{\psi}(x)\gamma^\mu\psi(x)\bar{\psi}(y)\gamma^\nu\psi(y)\rangle_A
\label{OPE-F1}\\
\hspace{-1.4cm}&&= \tr\Big\{\brax {1\over \Sp + i\epsilon}\kety \gamma^\nu \bray{1\over \Sp + i\epsilon}\ketx\gamma^\mu\Big\}
\nonumber\\
\hspace{-1.4cm}&&\stackrel{x_*>0>y_*}{\ni}
{1\over 4\pi^6s\,x_*^4y_*^4}\int\!\!d^2z_1 d^2 z_2\,
{ [\calz_2 + i\epsilon ]^{-2}\over [\calz_1 + i\epsilon ]^3}
I^{\mu\nu}_{\calf}(z_1,z_1;x,y)
\Big[\Tr\big\{U^\dagger_{z_1}\calf_{z_2}\big\}
+ {\rm Tr}\big\{U_{z_1}\calf_{z_2}^\dagger\big\}\Big]\,,
\nonumber
\end{eqnarray}
and using the definition of $\cali_5^{\mu\nu}$, (\ref{OPE-F1}) can be written as
\begin{eqnarray}
&&{\rm T}\{\bar{\hat{\psi}}(x)\gamma^\mu\psi(x)\bar{\hat{\psi}}(y)\gamma^\nu\hat{\psi}(y)\}
\label{OPE-F2}\\
&&~~~~~\stackrel{x_*>0>y_*}{\ni}
{1\over s}\int\!\!d^2z_1 d^2 z_2\,
\cali_5^{\mu\nu}(z_{1\perp},z_{2\perp};x,y)
\Big[{\rm Tr}\big\{\hat{U}^\dagger_{z_1}\hat{\calf}_{z_2}\big\}
+ \Tr\big\{\hat{U}_{z_1}\hat{\calf}_{z_2}^\dagger\big\}\Big]\,,
\nonumber
\end{eqnarray}
therefore, the coefficient $\cali^{\mu\nu}_5$ is also the impact-factor associated to the operator
$\Tr\big\{\hat{U}^\dagger_{z_1}\hat{\calf}(z_{2\perp})\big\}
+ \Tr\big\{\hat{U}_{z_1}\hat{\calf}^\dagger(z_{2\perp})\big\}$.

\subsection{OPE with sub-eikonal corrections: flavor singlet}
\label{sec: OPEfs}

We can now add together the contributions of the quark, eq. (\ref{OPEQ1Q5operator}),
and gluon, eq. (\ref{OPE-F2}), sub-eikonal corrections. We have
\begin{eqnarray}
\hspace{-1cm}&&{\rm T}\{\bar{\hat{\psi}}(x)\gamma^\mu\psi(x)\bar{\hat{\psi}}(y)\gamma^\nu\hat{\psi}(y)\}
\nonumber\\
\hspace{-1cm}&&= \int\!dz_1dz_2\,\cali_{LO}^{\mu\nu}(z_{1\perp},z_{2\perp};x,y) 
\bigg[{\rm Tr}\{\hat{U}_{z_1}\hat{U}^\dagger_{z_2}\} 
+ {\calz_2\over 8\,s}
\Big(\Tr\{\hat{U}^\dagger_{z_1}\hat{\calq}_{1z_2}\} + \Tr\{\hat{U}_{z_1}\hat{\calq}^\dagger_{1z_2}\}\Big)\bigg]
\nonumber\\
\hspace{-1cm}&&
~~~+ {1\over s}\int\!\!d^2z_1 d^2 z_2\,
\cali_5^{\mu\nu}(z_{1\perp},z_{2\perp};x,y)
\bigg[\Tr\{\big(\hat{\calq}_{5z_2}+\hat{\calf}_{z_2}\big)\hat{U}^\dagger_{z_1}\}
+ \Tr\{\big(\hat{\calq}^\dagger_{5z_2}+\hat{\calf}^\dagger_{z_2}\big)\hat{U}_{z_1}\}
\bigg] 
\nonumber\\
\hspace{-1cm}&&~~~+ \calo(\alpha_s) + \calo(\lambda^{-2})\,.
\label{OPEtot-calq}
\end{eqnarray}
The same expansion can be written in terms of the operators $\hat{Q}_1(x_\perp)$ and $\hat{Q}_5(x_\perp)$,
$\hat{\tildeQ}_1(x_\perp)$ and $\hat{\tildeQ}_5(x_\perp)$ and their adjoint conjugated as
\begin{eqnarray}
\hspace{-1cm}&&{\rm T}\{\bar{\hat{\psi}}(x)\gamma^\mu\psi(x)\bar{\hat{\psi}}(y)\gamma^\nu\hat{\psi}(y)\}
\nonumber\\
&&= \int\!dz_1dz_2\,\cali_{LO}^{\mu\nu}(z_{1\perp},z_{2\perp};x,y) 
\bigg[{\rm Tr}\{\hat{U}^\dagger_{z_1}\hat{U}^\dagger_{z_2}\} 
\nonumber\\
\hspace{-1cm}&&
~~~+ {\calz_2\over 16\,s}
\Big(\Tr\{\hat{U}^\dagger_{z_1} \hat{U}_{z_2}\}\hat{Q}_{1z_2} +  \Tr\{\hat{U}_{z_1} \hat{U}^\dagger_{z_2}\}\hat{Q}_{1z_2}^\dagger
-  {1\over N_c} \Tr\{\hat{U}^\dagger_{z_1} \hat{\tilde{Q}}_{1z_2}\}
- {1\over N_c}\Tr\{\hat{U}_{z_1} \hat{\tilde{Q}}^\dagger_{1z_2}\}\Big)\bigg]
\nonumber\\
\hspace{-1cm}&&
~~~+ {1\over s}\int\!\!d^2z_1 d^2 z_2\,
\cali_5^{\mu\nu}(z_{1\perp},z_{2\perp};x,y)
\bigg[\Tr\{\hat{U}^\dagger_{z_1} \hat{U}_{z_2}\}\hat{Q}_{5z_2} +  \Tr\{\hat{U}_{z_1} \hat{U}^\dagger_{z_2}\}\hat{Q}_{5z_2}^\dagger
\nonumber\\
\hspace{-1cm}&&
~~~-  {1\over N_c} \Tr\{\hat{U}^\dagger_{z_1} \big(\hat{\tilde{Q}}_{5z_2} - 2N_c\hat{\calf}_{z_2}\big)\}
 - {1\over N_c}\Tr\{\hat{U}_{z_1}\big(\hat{\tilde{Q}}^\dagger_{5z_2} - 2N_c\hat{\calf}^\dagger_{z_2}\big)\}
\bigg] 
\nonumber\\
\hspace{-1cm}&&
~~~+ \calo(\alpha_s) + \calo(\lambda^{-2})
\label{OPEtot}
\end{eqnarray}
Equations (\ref{OPEtot}) and (\ref{OPEtot-calq}) are equivalent, and
are the high-energy OPE with the sub-eikonal corrections that come in with a ${1\over s}$ factor.
In principle, if we consider the
full quark propagator in the background of gluon field, (\ref{ojobulletstar}), 
there might be other sub-eikonal contributions to the unpolarized and polarized high-energy OPE
(see Appendix \ref{sec: matrielements} for more details). In other words, if we use the full quark propagator (\ref{ojobulletstar}),
from diagram in Fig \ref{subeikonalif}b
we will get other terms besides the one we obtained using only the term proportional to $\epsilon^{ij}F_{ij}$. However
the new terms will not contribute to $g_1$ structure function.

Considering that both $\hat{\calq}_{5x}$ and $\hat{\calf}_x$ are parity odd operators
and that their impact factors results, which came from two independent and different calculations, are equal,
we may consider this as an indirect proof of the validity of the entire result.

At small-x, the unpolarized quark structure function is known to be energy suppressed 
with respect to the gluon structure function. Indeed, this can be observed from our OPE result
(\ref{OPEtot}), where the sub-eikonal impact factors $\cali_1^{\mu\nu}$ and $\cali_5^{\mu\nu}$
are proportional to $1\over s$.

In the large $N_c$ approximation we have 
\begin{eqnarray}
&&{\rm T}\{\bar{\hat{\psi}}(x)\gamma^\mu\psi(x)\bar{\hat{\psi}}(y)\gamma^\nu\hat{\psi}(y)\}
\nonumber\\
&&= \int\!dz_1dz_2\bigg\{\cali_{LO}^{\mu\nu}(z_{1\perp},z_{2\perp};x,y) 
{\rm Tr}\{\hat{U}_{z_1}\hat{U}^\dagger_{z_2}\} 
\nonumber\\
&&
~~~+ {1\over 2s}\cali^{\mu\nu}_1(z_{1\perp},z_{2\perp};x,y)
\Big(\Tr\{\hat{U}^\dagger_{z_1} \hat{U}_{z_2}\}\hat{Q}_{1z_2} 
+ \Tr\{\hat{U}_{z_1} \hat{U}^\dagger_{z_2}\}\hat{Q}^\dagger_{1z_2}\Big)
\nonumber\\
&&
~~~+ {1\over 2s}\cali^{\mu\nu}_5(z_{1\perp},z_{2\perp};x,y)
\Big(\Tr\{\hat{U}^\dagger_{z_1} \hat{U}_{z_2}\}\hat{Q}_{5z_2} 
+ \Tr\{\hat{U}_{z_1} \hat{U}^\dagger_{z_2}\}\hat{Q}^\dagger_{5z_2}
\nonumber\\
&&~~~ + 2\Tr\{U^\dagger_{z_1}\calf_{z_2}\}  + 2\Tr\{U_{z_1}\calf_{z_2}^\dagger\}\Big)\bigg\} 
+ \calo(\alpha_s) + \calo(\lambda^{-2}) + \calo(1/N_c)\,,
\label{OPEunpLargN}
\end{eqnarray}
(recall that $\cali_1^{\mu\nu} = {\calz_2\over 8}\cali_{LO}^{\mu\nu}$).

\subsection{OPE with sub-eikonal contributions: flavor non-singlet}
\label{sec: OPEnonfs}

In the flavor non-singlet case, the high-energy OPE will be the same as in (\ref{OPEtot}) with the exception that
the operator $\hat{\calf}(x_\perp)$ will be absent because it does not allow exchange of flavor with the target. So we have
\begin{eqnarray}
&&{\rm T}\{\bar{\hat{\psi}}(x)\gamma^\mu\psi(x)\bar{\hat{\psi}}(y)\gamma^\nu\hat{\psi}(y)\}
\nonumber\\
&&=\int\!dz_1dz_2\,\cali_{LO}^{\mu\nu}(z_{1\perp},z_{2\perp};x,y) 
\bigg[{\rm Tr}\{\hat{U}_{z_1}\hat{U}^\dagger_{z_2}\} 
+ {\calz_2\over 8\,s}
\Big(\Tr\{U^\dagger_{z_1}\calq_{1z_2}\} + \Tr\{U_{z_1}\calq^\dagger_{1z_2}\}\Big)\bigg]
\nonumber\\
&&
~~~+ {1\over s}\int\!\!d^2z_1 d^2 z_2\,
\cali_5^{\mu\nu}(z_{1\perp},z_{2\perp};x,y)
\Big[\Tr\{\hat{\calq}_{5z_2}\hat{U}^\dagger_{z_1}\}
+ \Tr\{\hat{\calq}^\dagger_{5z_2}\hat{U}_{z_1}\}
\Big] 
\nonumber\\
&&~~~+ \calo(\alpha_s) + \calo(\lambda^{-2})
\end{eqnarray}
As for the singlet case, the same expansion can be written in terms of the operators $\hat{Q}_1(x_\perp)$ and $\hat{Q}_5(x_\perp)$,
$\hat{\tildeQ}_1(x_\perp)$ and $\hat{\tildeQ}_5(x_\perp)$ and their adjoint conjugated as
\begin{eqnarray}
&&{\rm T}\{\bar{\hat{\psi}}(x)\gamma^\mu\psi(x)\bar{\hat{\psi}}(y)\gamma^\nu\hat{\psi}(y)\}
\nonumber\\
&&= \int\!dz_1dz_2\bigg\{\cali_{LO}^{\mu\nu}(z_{1\perp},z_{2\perp};x,y)
+ {\calz_2\over 16\,s}\cali_{LO}^{\mu\nu}(z_{1\perp},z_{2\perp};x,y)
\Big(\Tr\{\hat{U}^\dagger_{z_1} \hat{U}_{z_2}\}\hat{Q}_{1z_2} 
\nonumber\\
&&
\hspace{5cm}+  \Tr\{\hat{U}_{z_1} \hat{U}^\dagger_{z_2}\}\hat{Q}_{1z_2}^\dagger 
-  {1\over N_c} \Tr\{\hat{U}^\dagger_{z_1} \hat{\tilde{Q}}_{1z_2}\}
- {1\over N_c}\Tr\{\hat{U}_{z_1} \hat{\tilde{Q}}^\dagger_{1z_2}\}\Big)
\nonumber\\
&& ~~~+ {1\over s}
\cali_5^{\mu\nu}(z_{1\perp},z_{2\perp};x,y)
\Big(\Tr\{\hat{U}^\dagger_{z_1} \hat{U}_{z_2}\}\hat{Q}_{5z_2} +  \Tr\{\hat{U}_{z_1} \hat{U}^\dagger_{z_2}\}\hat{Q}_{5z_2}^\dagger
\nonumber\\
&&\hspace{3.5cm} -  {1\over N_c} \Tr\{\hat{U}^\dagger_{z_1} \hat{\tilde{Q}}_{5z_2}\}
- {1\over N_c}\Tr\{\hat{U}_{z_1} \hat{\tilde{Q}}^\dagger_{5z_2}\}\Big) 
 + \calo(\alpha_s) + \calo(\lambda^{-2})
\bigg\}
\label{OPEtot-nonsing}
\end{eqnarray}

In the large $N_c$ approximation we can simplify to
\begin{eqnarray}
&&{\rm T}\{\bar{\hat{\psi}}(x)\gamma^\mu\psi(x)\bar{\hat{\psi}}(y)\gamma^\nu\hat{\psi}(y)\}
\nonumber\\
&&= \int\!dz_1dz_2\bigg\{\cali_{LO}^{\mu\nu}(z_{1\perp},z_{2\perp};x,y)
\nonumber\\
&&
\hspace{2cm} + {\calz_2\over 16\,s}\cali_{LO}^{\mu\nu}(z_{1\perp},z_{2\perp};x,y)
\Big(\Tr\{\hat{U}^\dagger_{z_1} \hat{U}_{z_2}\}\hat{Q}_{1z_2} +  \Tr\{\hat{U}_{z_1} \hat{U}^\dagger_{z_2}\}\hat{Q}_{1z_2}^\dagger\Big)
\nonumber\\
&&\hspace{2cm} + {1\over s}
\cali_5^{\mu\nu}(z_{1\perp},z_{2\perp};x,y)
\Big(\Tr\{\hat{U}^\dagger_{z_1} \hat{U}_{z_2}\}\hat{Q}_{5z_2} +  \Tr\{\hat{U}_{z_1} \hat{U}^\dagger_{z_2}\}\hat{Q}_{5z_2}^\dagger\Big) 
\nonumber\\
&&\hspace{2cm} + \calo(\alpha_s) + \calo(\lambda^{-2}) + \calo(1/N_c)
\bigg\}
\label{OPEtot-nonsingLargeN}
\end{eqnarray}
We will find the evolution of these operators in the next sections.

\section{Parametrization of the forward matrix elements}
\label{sec: parametri-matrixele}

In this section we give an account of the operators that we have found and provide 
their parametrization through new distribution functions (all of dimensions $[M^{-2}]$).

We denote by $S^\mu_L$ the longitudinal spin of the hadronic target. In the DIS kinematics we have
$MS^\mu_L \simeq \lambda P^\mu$ so, we may write $S^\mu \simeq {\lambda\over M}P^\mu + S_\perp^\mu$
with helicity $\lambda$.

To consider forward matrix elements, we need to define the reduced matrix elements.
Consider an operator $\hat{O}(x_\perp,y_\perp)$, function of two transverse distances,
which can be one of the dipole type of operators we are 
going to consider in this section. Then, we define the reduced matrix elements as
\begin{eqnarray}
\langle P,S|\hatO(k_\perp)|P',S\rangle = 2\pi{s\over 2}\delta((P'-P)\!\cdot\!p_1)\langle\langle P,S|\hat{O}(k_\perp)|P,S\rangle\rangle
\end{eqnarray}
with $P'^\mu  = P^\mu + \beta\,p_2^\mu$ and
\begin{eqnarray}
\int d^2\Delta\, e^{-i(k,\Delta)_\perp}\langle\langle P,S |\hat{O}(x_\perp,y_\perp)|P,S\rangle\rangle
= \langle\langle P,S|\hat{O}(k_\perp)|P,S\rangle\rangle
\end{eqnarray}
where we defined $\Delta^\mu_\perp \equiv (x-y)_\perp^\mu$.
The delta function $\delta((P'-P)\!\cdot\!p_1)$ takes into account that forward matrix elements of dipole type operators contain
unrestricted integration along $p_1$.

Let us start considering the relevant matrix elements. Consider matrix element with operator $\hat{Q}_1(x_\perp)$
\begin{eqnarray}
	&& 
	\int \!d^2\Delta \, e^{i(\Delta,k)}\langle\langle P,S|\Big[Q_1(x_\perp)\Tr\{ U_x U^\dagger_y\} 
	+ {\rm a.c.}\Big]|P,S\rangle\rangle 
	\nonumber\\
	&&=
	-ig^2\!\!\int \!d^2\Delta \, e^{i(\Delta,k)}\int_{-\infty}^{+\infty}\!\! dz_*\!\!\int _{-\infty}^{z_*}\!\! dz'_*
	\nonumber\\
	&&~~~\times
	\langle\langle P,S|\Big[\barpsi(z'_*,x_\perp)\ssp_1[z'_*, z_*]_x\psi(z_*,x_\perp)\Tr\{ U_xU^\dagger_y\} + {\rm a.c}\Big]|P,S\rangle\rangle 
	\nonumber\\
	&&= {s\over 2}\Big( q_1(k^2_\perp,x) + {\vec{S}\times \vec{k}\over M} q_{1T}(k^2_\perp,x)\Big)\,.
	\label{U1paramet}
\end{eqnarray}
with $\epsilon^{ij}S_i k_j = \vec{S}\times \vec{k}$. (recall $z_*, z'_*$ are dimensionless)
and ${\rm a.c}$ stands for adjoint conjugated.

Matrix element with operator $\hat{\tildeQ}_1(x_\perp)$
\begin{eqnarray}
\hspace{-1cm}&& 
\int \!d^2\Delta \, e^{i(\Delta,k)}\langle\langle P,S|\Big[\Tr\{\tildeQ_1(x_\perp)U^\dagger_y \}
+ {\rm a.c.} \Big]|P,S\rangle\rangle
\nonumber\\
\hspace{-1cm}&&=
ig^2\!\!\int \!d^2\Delta \, e^{i(\Delta,k)}\int_{-\infty}^{+\infty}\!\!dz_*\!\!\int_{-\infty}^{z_*}\!\!dz'_*\!\!
\nonumber\\
\hspace{-1cm}&&~~~\times 
\langle\langle P,S|\Big[\Tr\{[\infty p_1,z_*]_x {\rm tr}\{\psi(z_*,x_\perp)\bar{\psi}(z'_*,x_\perp)\ssp_1\}
[z'_*, -\infty p_1] U^\dagger_y \} + {\rm a.c.}\Big]|P,S\rangle\rangle 
\nonumber\\
\hspace{-1cm}&&=  {s\over 2}\Big(\tilde{q}_1(k^2_\perp,x) + {\vec{S}\times \vec{k}\over M} 
\tilde{q}_{1T}(k^2_\perp,x)\Big)\,.
\label{tildeU1paramet}
\end{eqnarray}

Matrix element with operator $\hat{Q}_5(x_\perp)$
\begin{eqnarray}
\hspace{-1cm}&& 
\int \!d^2\Delta \, e^{i(\Delta,k)}\langle\langle P,S|\Big[Q_5(x_\perp)\Tr\{U_xU^\dagger_y \} + {\rm a.c}\Big]|P,S\rangle\rangle
\nonumber\\
\hspace{-1cm}&& =
-g^2\!\!\int \!d^2\Delta \, e^{i(\Delta,k)}\int_{-\infty}^{+\infty}\!\! dz_*\!\!\int _{-\infty}^{z_*}\!\! dz'_*
\nonumber\\
&&~~~\times
\langle\langle P,S|\Big[\barpsi(z'_*,x_\perp)\gamma^5\ssp_1[z'_*, z_*]_x\psi(z_*,x_\perp) \Tr\{U_xU^\dagger_y\}+ {\rm a.c}\Big]|P,S\rangle\rangle 
\nonumber\\
\hspace{-1cm}&& =
{s\over 2}\Big( \lambda q_{5L}(k^2_\perp,x) - {(S,k)_\perp\over M}q_{5T}(k^2_\perp,x)\Big)\,.
\label{U5paramet}
\end{eqnarray}

Matrix element with operator $\hat{\tildeQ}_5(x_\perp)$
\begin{eqnarray}
	\hspace{-1cm}&& 
	\int \!d^2\Delta \, e^{i(\Delta,k)}\langle\langle P,S|\Big[\Tr\{\tildeQ_5(x_\perp)U^\dagger_y\}+ {\rm a.c}\Big]|P,S\rangle\rangle
	\nonumber\\
	\hspace{-1cm}&&=
	g^2\!\!\int \!d^2\Delta \, e^{i(\Delta,k)}\int_{-\infty}^{+\infty}\!\!dz_*\!\!\int_{-\infty}^{z_*}\!\!dz'_*\!\!
	\nonumber\\
	\hspace{-1cm}&&~~~\times 
	\langle\langle P,S|\Big[\Tr\{[\infty p_1,z_*]_x {\rm tr}\{\psi(z_*,x_\perp)\bar{\psi}(z'_*,x_\perp)\gamma^5\ssp_1\}
	[z'_*, -\infty p_1] U^\dagger_y \}+ {\rm a.c}\Big]|P,S\rangle\rangle
	\nonumber\\
	\hspace{-1cm}&&=
	{s\over 2}\Big(\lambda\tildeq_{5L}(k^2_\perp,x) - {(S,k)_\perp\over M}\tildeq_{5T}(k^2_\perp,x)\Big)\,.
	\label{tildeU5paramet}
\end{eqnarray}

We can also parametrize the matrix elements with the operators $\calq_{1x}$ and $\calq_{5x}$ as
\begin{eqnarray}
&&
\int \!d^2\Delta \, e^{i(\Delta,k)}\langle\langle P,S|\Big[\Tr\{\calq_1(x_\perp)U^\dagger_y \}+ {\rm a.c}\Big]|P,S\rangle\rangle
\nonumber\\
&&~~~~={s\over 2}\Big(Q_1(k^2_\perp,x) + {\vec{S}\times \vec{k}\over M} Q_{1T}(k^2_\perp,x)\Big)\,.
\end{eqnarray}
and
\begin{eqnarray}
&&\int \!d^2\Delta \, e^{i(\Delta,k)}\langle\langle P,S|\Tr\{\calq_5(x_\perp)U^\dagger_y \}+ {\rm a.c}|P,S\rangle\rangle
\nonumber\\
&&~~~~= {s\over 2}\Big(\lambda Q_{5L}(k^2_\perp,x) + {(S,k)_\perp\over M} Q_{1T}(k^2_\perp,x)\Big)\,.
\end{eqnarray}

Using eqs. (\ref{calq1toq1})-(\ref{calq5toq5dag}) we can find the following relations
\begin{eqnarray}
&&2\, Q_1(k^2_\perp,x) = q_1(k^2_\perp,x) - {1\over N_c}\tildeq_1(k^2_\perp,x)\,,\\
&&2\, Q_{1T}(k^2_\perp,x) = q_{1T}(k^2_\perp,x) - {1\over N_c}\tildeq_{1T}(k^2_\perp,x)\,,\\
&&2\, Q_{5L}(k^2_\perp,x) = q_{5L}(k^2_\perp,x) - {1\over N_c}\tildeq_{5L}(k^2_\perp,x)\,,\\
&&2\, Q_{5T}(k^2_\perp,x) = q_{5T}(k^2_\perp,x) - {1\over N_c}\tildeq_{5T}(k^2_\perp,x)\,.
\end{eqnarray}

The matrix element with the $\epsilon^{ij}F_{ij}(x_*,x_\perp)$ term is
\begin{eqnarray}
&&\int \!d^2\Delta e^{i(\Delta,k)_\perp}\!\!\int_{-\infty}^{+\infty}\!\! d z_*
\langle\langle P,S|\Big[\Tr\{[\infty p_1, z_*]_x \,ig{s\over 2}\epsilon^{ij}F_{ij}(z_*,x_\perp)[z_*,-\infty p_1]_x U^\dagger_y\}
+ {\rm a.c}\Big]|P,S\rangle\rangle
\nonumber\\
&& = {s\over 2}\Big[\lambda {k^2_\perp\over M^2}
G_L(k^2_\perp,x)  + {(S,k)_\perp\over M} G_T(k^2_\perp,x) \Big]\,,
\label{GLGT}
\end{eqnarray}
with helicity $\lambda= \pm \half$.
In (\ref{GLGT}), $G_L(k^2_\perp,x)$ and $G_T(k^2_\perp,x)$ are the polarized longitudinal and transverse gluon distributions
of dimension $[M^{-2}]$ (see Appendix \ref{sec: matrielements} for details).
Note that the parameterizations (\ref{U1paramet})-(\ref{GLGT}) are similar to the
standard parameterizations for TMD \cite{Mulders:1995dh, Goeke:2005hb, Bacchetta:2006tn}.

\section{Evolution equation of sub-eikonal corrections}
\label{sec: subeikeq}

In the eikonal approximation, which was presented in section \ref{sec: opehe}, we first derived the LO impact factor and its associated dipole operator $\Tr\{U(x_\perp)U^\dagger(y_\perp)\}$, eq. (\ref{LO-OPE-1}), and then we proceeded with the calculation of the evolution of the
dipole operator thus obtaining the BK equation. The plan is to repeat the same steps at sub-eikonal level.
In the previous sections we derived the sub-eikonal impact factors and their associated operators. Our task is now the calculation of the
evolution equations of these operators.

We use again the background field method. We have to separate fields in quantum and classical and perform the 
functional integration over the quantum fields leaving untouched the classical one. As a result of this procedure
we will obtain a relation between the operator at the starting renormalization point $\eta_1$ and
the new operators at the end renormalization point $\eta_2$ convoluted with
a coefficient, the evolution kernel,  which is the result of the functional integration over the quantum fields living in
the infinitesimal rapidity interval $\eta_1-\eta_2=\Delta\eta$. This is nothing but the application of the Wilsonian renormalization group.
Here the evolution parameter $\eta$
is the rapidity of the fields which we use to discriminate between classical and quantum fields. Indeed, at high energies,
fields are ordered in rapidity space therefore it is natural to use it as the evolution parameter and as 
factorization parameter for scattering amplitudes. This logic is equal to the one adopted in the Bjorken limit, where
the fields are ordered according to their transverse momentum and the factorization parameter $\mu_f$ discriminates 
between classical and quantum fields according to their transverse momenta.

A feature of the BK equation (and of BFKL equation) is that it is free of ultra violet (UV) and infra red (IR) divergences.
Moreover, in the limit of vanishing dipole size unitarity is restored. What we will observe in the
evolution equations of sub-eikonal operators, is the absence of this property  which can be translated into a
double log of energy contribution of the type $\alpha_s\ln^2{\alpha_1\over \alpha_2}$ where
$\alpha_1$ and $\alpha_2$ are the longitudinal  momenta of the fields within the infinitesimal step in rapidity
where the quantum fields live. 

\begin{figure}[thb]
\begin{center}
	\includegraphics[width=4.0in]{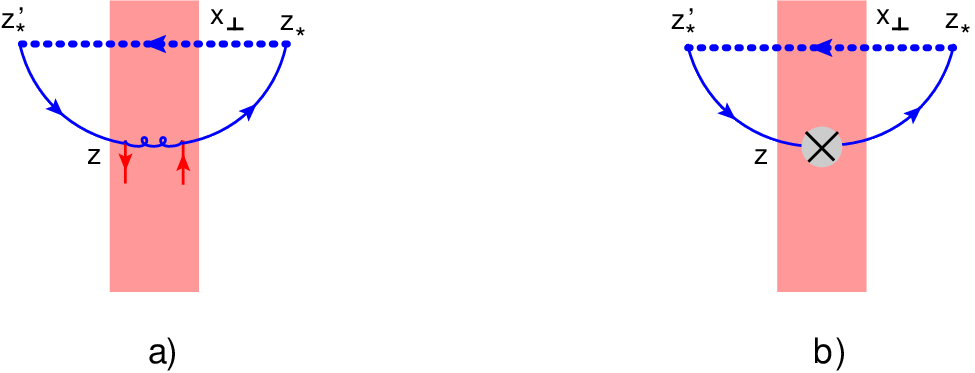}
	\caption{Diagrams with $\hat{Q}_{1x}$ and $\hat{Q}_{5x}$ quantum.}
	\label{LO-psibarpsiadj4}
\end{center}
\end{figure}

To find evolution equation, the first step is to separate the gluon and the quark files in quantum and classical:
$A_\mu\to A^{cl}_\mu+A^q_\mu$ and $\psi\to \psi^{cl}+\psi^q$. 
With this separation, the operator, for example, $\Tr\{\hat{\calq}_{1z_2}\hat{U}^\dagger_{z_1}\}$
will generate several terms which will turn into Feynman diagrams through functional integration over the quantum fields.
If the fields of the sub-eikonal corrections are classical, they will be in the shock-wave and we have to use the eikonal
quark and gluon propagators. If, instead, the fields of the 
sub-eikonal operators are quantum, then they will be outside the shock-wave and will be functionally integrated.
In this case, the sub-eikonal correction will be provided by the quark and gluon propagators with sub-eikonal corrections.

The relation we are looking for is of the type
\begin{eqnarray}
\Tr\{\hat{\calq}_{1z_2}\hat{U}^\dagger_{z_1}\}^{\eta_1} = 
\Delta\eta K\otimes \hat{\calo}^{\eta_2}
\end{eqnarray}
where $\hat{\calo}^{\eta_2}$ is the, in principle yet unknown, operator which we will obtain after one loop evolution.
The \textit{new} operator is convoluted with the kernel $K$ which is the result of functional integration.
If one is able to solve the evolution equation, then we can convolute the solution with the impact factor
with suitable initial conditions.  This procedure gives the DIS structure functions and their behavior at high energy.

One of the operators we need to consider is 
\begin{eqnarray}
\Tr\{\hat{\calq}_{1x}\hat{U}^\dagger_y\} = \half \hat{Q}_{1x}\Tr\{\hat{U}_x\hat{U}^\dagger_y\} 
- {1\over 2N_c}\Tr\{\hat{\tildeQ}_x \hat{U}^\dagger_y\}
\end{eqnarray}
We will calculate the evolution of the LHS in the Appendix, while in the next section we will consider separately
the evolution of the two terms in the RHS. Similarly will be done for the operator with $\hat{\calq}_5(x_\perp)$. 
The reason we calculate the evolution equation of the RHS for the two terms will be clear when 
we consider diagrams with $\hat{\calf}(x_\perp)$ quantum (see Figs. \ref{outshock-2} and \ref{outshock-3}).

\subsection{Diagrams with $\hat{Q}_{1x}$ and $\hat{Q}_{5x}$ quantum}

As we explained in the previous section, we have to split all field in quantum and classical. Here we consider the case
in which the operators $Q_{1x}$ and $Q_{5x}$ are quantum. The diagrams are give in Fig. \ref{LO-psibarpsiadj4}.
Let us consider operator $\Tr\{U^\dagger_yU_x\}Q_{1x}$ and diagram \ref{LO-psibarpsiadj4}a where there are
quark fields in the background. First notice that
\begin{eqnarray}
\langle \Tr\{U^\dagger_yU_x\}Q_{1x}\rangle_{\small{\rm Fig.}\ref{LO-psibarpsiadj4}}
=  \Tr\{U^\dagger_yU_x\}\langle Q_{1x}\rangle_{\small{\rm Fig.}\ref{LO-psibarpsiadj4}}
\end{eqnarray}
So for the moment we will consider only $Q_{1x}$.

As usual we indicate the quantum fields with the superscript $q$. We have 
\begin{eqnarray}
&&\hspace{-2cm}\langle Q_{1x}\rangle_{\small{\rm Fig.}\ref{LO-psibarpsiadj4}a} =  
g^2\!\!\int_{-\infty}^{+\infty}\!\!dz_*\!\!\int_{-\infty}^{z_*}\!\!dz'_*
\big\langle\tr\{i\,\ssp_1\psi^q(z_*,x_\perp)[z_*,z'_*]_x^\dagger\barpsi^q(z'_*,x_\perp)\} \big\rangle
\end{eqnarray}
We need the quark propagator in the background of quark filed given in eq. (\ref{qprop-backqua-coord}) and get
\begin{eqnarray}
\langle Q_{1x}\rangle_{\small{\rm Fig.}\ref{LO-psibarpsiadj4}a} 
=\!\!&& {\alpha_s\over 4\pi^2}\int_0^{+\infty}\!{d\alpha\over \alpha}\int d^2z\,g^2\!\!\int_{-\infty}^{+\infty}\!dz_{1*}
\!\!\int_{-\infty}^{z_{1*}}\!d{z_{2*}}
\nonumber\\
&&\times\bigg[
\half\Tr\{U^\dagger_x U_z\} \tr\bigg\{i\,\ssp_1\,{(\ssx-\ssz)_\perp \gamma^\mu_\perp\over (x-z)^2_\perp}
\psi(z_{1*},z_\perp)[z_{1*},z_{2*}]^\dagger_z\barpsi(z_{2*},z_\perp)
{\gamma^\mu_\perp(\ssz-\ssx)_\perp\over (x-z)^2_\perp}\bigg\}
\nonumber\\
&& - {1\over 2N_c}{\rm Tr}\bigg\{U^\dagger_x [\infty p_1,z_{1*}]_z
{\rm tr}\bigg\{i\,\ssp_1\,{(\ssx-\ssz)_\perp \gamma^\mu_\perp\over (x-z)^2_\perp}
\psi(z_{1*},z_\perp)\bar{\psi}(z_{2*},z_\perp)
\nonumber\\
&&\hspace{4.5cm}\times{\gamma_\perp^\mu(\ssz-\ssx)_\perp\over (x-z)^2_\perp}\bigg\}[z_{2*},-\infty p_2]_z\bigg\}
\bigg]\,.
\end{eqnarray}
Now, use 
\begin{eqnarray}
{\rm tr}\{\ssp_1(\ssx-\ssz)_\perp \gamma^\mu_\perp \psi(z_{1*})\bar{\psi}(z_{2*})\gamma_\mu^\perp(\ssz-\ssx)_\perp\}
= 2(x-z)^2_\perp\tr\{\ssp_1 \psi(z_{1*})\bar{\psi}(z_{2*})\}
\end{eqnarray}
and arrive at
\begin{eqnarray}
\hspace{-1cm}\langle Q_{1x}\rangle_{\small{\rm Fig.}\ref{LO-psibarpsiadj4}a} 
=&& {\alpha_s\over 4\pi^2}\int_0^{+\infty}\!{d\alpha\over \alpha}\int
{d^2z\over (x-z)^2_\perp}\,g^2\!\!\int_{-\infty}^{+\infty}\!dz_{1*}
\!\!\int_{-\infty}^{z_{1*}}\!d{z_{2*}}
\nonumber\\
&&
\times\bigg[
\Tr\{U^\dagger_x U_z\} \tr\{i\,\ssp_1
\psi(z_{1*},z_\perp)[z_{1*},z_{2*}]^\dagger_z\bar{\psi}(z_{2*},z_\perp)\}
\nonumber\\
&&~~~- {1\over N_c}\Tr\{U^\dagger_x [\infty p_1,z_{1*}]_z
\tr\{i\,\ssp_1\,
\psi(z_{1*},z_\perp)\bar{\psi}(z_{2*},z_\perp)\}[z_{2*},-\infty p_2]_z\}
\bigg]
\end{eqnarray}
Using the definition of operators $\hat{Q}_{1x}$ and $\hat{\tildeQ}_{1x}$ we obtain
\begin{eqnarray}
\hspace{-1cm}\langle Q_{1x}\rangle_{\small{\rm Fig.}\ref{LO-psibarpsiadj4}a} 
= {\alpha_s\over 4\pi^2}\int_0^{+\infty}\!{d\alpha\over \alpha}\int
{d^2z\over (x-z)^2_\perp}\,\bigg[
{\rm Tr}\{U^\dagger_x U_z\}\,Q_{1z}
- {1\over N_c}{\rm Tr}\{U^\dagger_x \tilde{Q}_{1z}\}
\bigg]
\label{Qquantum-diagram-a}
\end{eqnarray}
The same diagram calculated for the polarized quark operator $Q_{5x}$ is
\begin{eqnarray}
\hspace{-2cm}\langle Q_{5x}\rangle_{\small{\rm Fig.}\ref{LO-psibarpsiadj4}a}
=&&  g^2\!\!\int_{-\infty}^{+\infty}\!\!dz_*\!\!\int_{-\infty}^{z_*}\!\!dz'_*
\langle{\rm tr}\{\gamma^5\ssp_1\,\psi(z_*,x_\perp)[z_*,z'_*]_x^\dagger\bar{\psi}(z'_*,x_\perp)\} \rangle
\nonumber\\
=&& {\alpha_s\over 4\pi^2}\int_0^{+\infty}\!{d\alpha\over \alpha}\int 
{d^2z\over (x-z)^2_\perp}\,\bigg[
{\rm Tr}\{U^\dagger_x U_z\}\,Q_{5z}
- {1\over N_c}{\rm Tr}\{U^\dagger_x \tilde{Q}_{5z}\}\bigg]
\label{Q5quantum-diagram-a}
\end{eqnarray}
where this time we used
\begin{eqnarray}
\tr\{\gamma^5\ssp_1(\ssx-\ssz)_\perp \gamma^\mu_\perp \psi(z_{1*})\barpsi(z_{2*})\gamma_\mu^\perp(\ssz-\ssx)_\perp\}
= 2(x-z)^2_\perp\tr\{\gamma^5\ssp_1 \psi(z_{1*})\bar{\psi}(z_{2*})\}
\end{eqnarray}

Now we consider diagram \ref{LO-psibarpsiadj4}b and start again with $Q_{1x}$. We have
\begin{eqnarray}
\hspace{-2cm}\langle Q_{1x}\rangle_{\small{\rm Fig.}\ref{LO-psibarpsiadj4}b} 
=&&  g^2\!\!\int_{-\infty}^{+\infty}\!\!dz_*\!\!\int_{-\infty}^{z_*}\!\!dz'_*
\langle{\rm tr}\{i\,\ssp_1\psi(z_*,x_\perp)[z_*,z'_*]_x^\dagger\bar{\psi}(z'_*,x_\perp)\} \rangle
\nonumber\\
=&& {\alpha_s\over 2\pi^2}\int_0^{+\infty}\!{d\alpha\over \alpha}\int d^2z\,\Tr\{U^\dagger_x \calf_z\}
\tr\{i\,\ssp_1{(\ssx-\ssz)_\perp\over (x-z)^2_\perp}\ssp_2\gamma^5
{(\ssz-\ssx)_\perp\over (z-\omega)^2_\perp}\} = 0
\end{eqnarray}
where we used $\tr\{\ssp_1(\ssx-\ssz)_\perp\ssp_2\gamma^5
(\ssz-\ssx)_\perp\} = (x-z)^2_\perp \tr\{\ssp_1\ssp_2\gamma^5\} = 0$.
So, we see that operators of different parity do not mix under evolution.
For operator $\hat{Q}_{5x}$ we have, instead
\begin{eqnarray}
&&\hspace{-2cm}\langle Q_{5x}\rangle_{\small{\rm Fig.}\ref{LO-psibarpsiadj4}b}
=  g^2\!\!\int_{-\infty}^{+\infty}\!\!dz_*\!\!\int_{-\infty}^{z_*}\!\!dz'_*
\langle{\rm tr}\{\gamma^5\ssp_1\psi(z_*,x_\perp)[z_*,z'_*]_x^\dagger\bar{\psi}(z'_*,x_\perp)\} \rangle
\nonumber\\
&&
={\alpha_s\over 2\pi^2}\int_0^{+\infty}\!{d\alpha\over \alpha}
\int  {d^2z \over (x-z)^2_\perp}\,\Tr\{U^\dagger_x \calf_z\}
\end{eqnarray}
So, for diagrams in Fig. \ref{LO-psibarpsiadj4} we have
\begin{eqnarray}
\langle \Tr\{U^\dagger_y U_x\}Q_{1x}\rangle_{\small{\rm Fig.}\ref{LO-psibarpsiadj4}} = &&
{\alpha_s\over 4\pi^2}\int_0^{+\infty}\!{d\alpha\over \alpha}\int d^2z
{\Tr\{U^\dagger_y U_x\}\over (x-z)^2_\perp}
\nonumber\\
&&\times\bigg[
\Tr\{U^\dagger_x U_z\}\,Q_{1z}
- {1\over N_c}{\rm Tr}\{U^\dagger_x \tilde{Q}_{1z}\}\bigg]
\label{Q1quantum-diagram-aball}
\end{eqnarray}
and 
\begin{eqnarray}
\hspace{-1cm}\langle \Tr\{U^\dagger_y U_x\}Q_{5x}\rangle_{\small{\rm Fig.}\ref{LO-psibarpsiadj4}} 
=\!\!&&{\alpha_s\over 4\pi^2}\int_0^{+\infty}\!{d\alpha\over \alpha}\int d^2z
{\Tr\{U^\dagger_y U_x\}\over (x-z)^2_\perp}
\nonumber\\
&&\times\bigg[
\Tr\{U^\dagger_x U_z\}\,Q_{5z} 
- {1\over N_c}{\rm Tr}\{U^\dagger_x \big(\tilde{Q}_{5z} - 2N_c\calf_z\big)\}\bigg]
\label{Q5quantum-diagram-aball}
\end{eqnarray}

\subsection{Diagrams with $\hat{\tildeQ}_{1x}$ and $\hat{\tildeQ}_{5x}$ quantum}

\begin{figure}[thb]
		\begin{center}
		\includegraphics[width=4.5in]{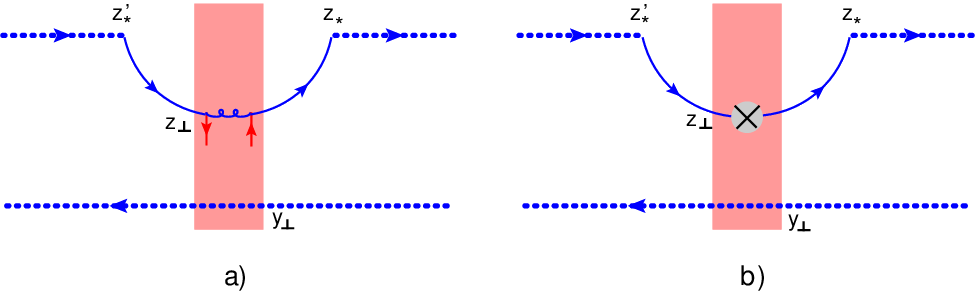}
		\caption{Diagrams with $\tildeQ_{1x}$ and $\tildeQ_{5x}$ quantum.}
		\label{LO-psibarpsiadj5}
	    \end{center}
\end{figure}

The diagrams with operators $\hat{\tildeQ}_{1x}$ and $\hat{\tildeQ}_{5x}$ are shown in Fig. \ref{LO-psibarpsiadj5}.
Let us start with diagram with quark in the background. The calculation is similar to the diagrams in the previous section.
We have
\begin{eqnarray}
\hspace{-0.7cm}\langle {\rm Tr}\{U^\dagger_y \tilde{Q}_{1x}\} \rangle_{\small{\rm Fig.}\ref{LO-psibarpsiadj5}a}
=\!\!&&  g^2\!\int^{+\infty}_{-\infty}\!dz_*\!\!\int_{-\infty}^{z_*}\!dz'_*
\nonumber\\
&&\times\big\langle\Tr\{U^\dagger_y[\infty p_1,z_*]_x 
\tr\{i\,\ssp_1\psi(z_*,x_\perp)\barpsi(z'_*,x_\perp)\}[z'_*,-p_1\infty]_x\} \big\rangle
\nonumber\\
=\!\!&&  g^2\!\int^{+\infty}_{-\infty}\!dz_*\!\!\int_{-\infty}^{z_*}\!dz'_*
\,{\rm Tr}\{U^\dagger_z
\tr\{i\,\ssp_1\big\langle\psi^q(z_*,x_\perp)\barpsi^q(z'_*,x_\perp)\big\rangle\}\} 
\end{eqnarray}
Now, using quark propagator in the background eq. (\ref{qpropglu}) and performing Dirac algebra we get
\begin{eqnarray}
\hspace{-1cm}\langle {\rm Tr}\{U^\dagger_y \tilde{Q}_{1x}\} \rangle_{\small{\rm Fig.}\ref{LO-psibarpsiadj5}a}
= {\alpha_s\over 4\pi^2}\!\int^{+\infty}_0\!{d\alpha\over \alpha}\!\int {d^2z\over (x-z)^2_\perp}
\Big[\Tr\{U^\dagger_y U_z\}Q_{1z} - {1\over N_c}{\rm Tr}\{U^\dagger_y\tilde{Q}_{1z}\}\Big]
\end{eqnarray}
We se that after one loop we get mixing of operators $Q_{1x}$ and $\tildeQ_{1x}$.

For operator $\tildeQ_{5x}$ we get the same result
\begin{eqnarray}
\hspace{-1cm}\langle {\rm Tr}\{U^\dagger_y \tilde{Q}_{5x}\} \rangle_{\small{\rm Fig.}\ref{LO-psibarpsiadj5}a}
=\!\!&& {\alpha_s\over 4\pi^2}\!\int^{+\infty}_0\!{d\alpha\over \alpha}\!\int {d^2z\over (x-z)^2_\perp}
\Big[\Tr\{U^\dagger_y U_z\}Q_{5z} - {1\over N_c}\Tr\{U^\dagger_y\tilde{Q}_{5z}\}\Big]
\end{eqnarray}

Next diagram is \ref{LO-psibarpsiadj5}b. Starting with $\tildeQ_{1x}$ we have
\begin{eqnarray}
\hspace{-1cm}&&\langle \Tr\{U^\dagger_y \tilde{Q}_{1x}\} \rangle_{\small{\rm Fig.}\ref{LO-psibarpsiadj5}b}
\nonumber\\
\hspace{-1cm}&&=  g^2\!\int^{+\infty}_{-\infty}\!dz_*\!\!\int_{-\infty}^{z_*}\!dz'_*
\,\langle\Tr\{U^\dagger_z[\infty p_1,z_*]_x 
\tr\{i\,\ssp_1\psi(z_*,x_\perp)\bar{\psi}(z'_*,x_\perp)\}[z'_*,-p_1\infty]_x\} \rangle
\nonumber\\
\hspace{-1cm}&&=  g^2\!\int^{+\infty}_{-\infty}\!dz_*\!\!\int_{-\infty}^{z_*}\!dz'_*
\,\Tr\{U^\dagger_y
\tr\{i\,\ssp_1\langle\psi(z_*,x_\perp)\bar{\psi}(z'_*,x_\perp)\rangle\}\} 
\nonumber\\
\hspace{-1cm}&&= {\alpha_s\over 8\pi^2}\!\int^{+\infty}_0\!{d\alpha\over \alpha}\!\int {d^2z\over (x-z)^4_\perp}{2\over s}
\Tr\{U^\dagger_y \calf_z\}\tr\{i\,\ssp_1(\ssx-\ssz)_\perp\ssp_2\gamma^5(\ssz-\ssx)_\perp\} = 0
\end{eqnarray}
As before, we get again no mixing between operators of different parity.

For operator $\tildeQ_{5x}$ we have
\begin{eqnarray}
\hspace{-0.5cm}\langle \Tr\{U^\dagger_y \tilde{Q}_{5x}\} \rangle_{\small{\rm Fig.}\ref{LO-psibarpsiadj5}b}
=\!\!&&  g^2\!\int^{+\infty}_{-\infty}\!dz_*\!\!\int_{-\infty}^{z_*}\!dz'_*
\,\Tr\{U^\dagger_y
\tr\{\gamma^5\ssp_1\langle\psi^q(z_*,x_\perp)\barpsi^q(z'_*,x_\perp)\rangle\}\} 
\nonumber\\
\hspace{-0.5cm}=\!\!&& {\alpha_s\over 8\pi^2}\!\int^{+\infty}_0\!{d\alpha\over \alpha}\!\int d^2z
\,{\Tr\{U^\dagger_y \calf_z\}\over (x-z)^4_\perp}\,{2\over s}
\,\tr\{\gamma^5\ssp_1(\ssx-\ssz)_\perp\ssp_2\gamma^5(\ssz-\ssx)_\perp\} 
\nonumber\\
\hspace{-0.5cm}=\!\!&& {\alpha_s\over 2\pi^2}\!\int^{+\infty}_0\!{d\alpha\over \alpha}\!\int {d^2z\over (x-z)^2_\perp}
\Tr\{U^\dagger_y \calf_z\}
\end{eqnarray}
We get, after one loop evolution, mixing between operator $\tildeQ_{5x}$ and $\calf_x$.

So, we conclude that
\begin{eqnarray}
\hspace{-1cm}\langle {\rm Tr}\{U^\dagger_y \tilde{Q}_{1x}\} \rangle_{\small{\rm Fig.}\ref{LO-psibarpsiadj5}}
=\!\!&& {\alpha_s\over 4\pi^2}\!\int^{+\infty}_0\!{d\alpha\over \alpha}\!\int {d^2z\over (x-z)^2_\perp}
\Big[{\rm Tr}\{U^\dagger_y U_z\}Q_{1z} - {1\over N_c}{\rm Tr}\{U^\dagger_y\tilde{Q}_{1z}\}\Big]
\label{LOpsibarpsiadj5sum-tildeq1}
\end{eqnarray}
and
\begin{eqnarray}
\langle {\rm Tr}\{U^\dagger_y \tilde{Q}_{5x}\} \rangle_{\small{\rm Fig.}\ref{LO-psibarpsiadj5}}
=\!\!&& {\alpha_s\over 4\pi^2}\!\int^{+\infty}_0\!{d\alpha\over \alpha}\!\int {d^2z\over (x-z)^2_\perp}
\nonumber\\
&&\times\Big[{\rm Tr}\{U^\dagger_y U_z\}Q_{5z} 
- {1\over N_c}{\rm Tr}\{U^\dagger_y\big(\tilde{Q}_{5z} - 2 N_c\calf_z\big)\}\Big]
\label{LOpsibarpsiadj5sum-tildeq5}
\end{eqnarray}
\begin{figure}[htb]
	\begin{center}
		\includegraphics[width=2.7in]{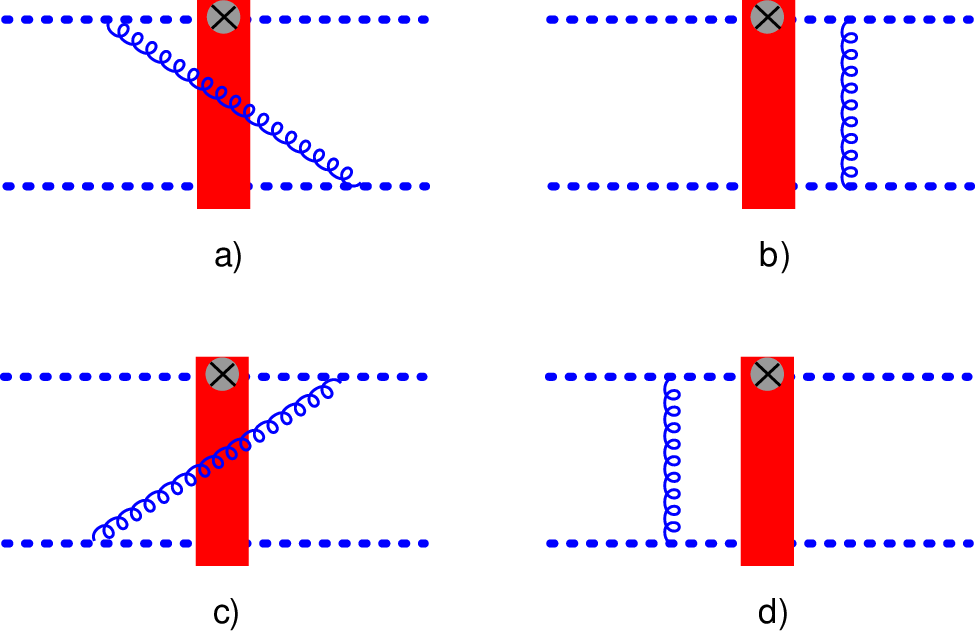}
		\caption{Diagrams with sub-eikonal operator in the shock-wave. The gray circle with a cross on it
			represents the operator being treated as classical thus is situated in the red-band shock-wave.}
		\label{BK-type-diagrams}
	\end{center}
\end{figure}

\subsection{BK-type diagrams}

The diagrams in Fig. \ref{BK-type-diagrams} and \ref{bktype-self} are similar to the BK diagrams in Fig. \ref{lobk}
with the only exception that in the shock-wave there is located a sub-eikonal correction that can be either 
$Q_1$, $\tildeQ_1$, $Q_5$, $\tildeQ_5$ (and their adjoint conjugated) or $\epsilon^{ij}F_{ij}$ 
(and in the Appendix we will consider also $\calq_1$, or $\calq_5$). 

As we will shortly see, the diagrams with $\tildeQ_{1x}$, $\tildeQ_{1x}$ (and their adjoint conjugated)
and $\calf_x$, are actually different then those with $Q_{1x}$, and $Q_{5x}$ (and their adjoint conjugated).
So, we will calculate them separately.

\subsubsection{BK-type diagrams for $Q_{1x}$ and $Q_{5x}$}

The BK-type of diagram for evolution for $\Tr\{U^\dagger_y U_x\}Q_{1x}$ (or $\Tr\{U^\dagger_y U_x\}Q_{5x}$) is
\begin{eqnarray}
\langle Q_{1x}\Tr\{ U_xU^\dagger_y\}\rangle_{\small{\rm Figs.}\ref{BK-type-diagrams}+\ref{bktype-self}} 
=\!\!&& Q_{1x} \langle\Tr\{U_xU^\dagger_y \}\rangle_{\small{\rm Figs.}\ref{BK-type-diagrams}+\ref{bktype-self}}
\nonumber\\
=\!\!&& {\alpha_s\over 2\pi^2}\int_0^{+\infty}\!{d\alpha\over\alpha}\int d^2z {(x-y)^2_\perp\over (x-z)^2_\perp(y-z)^2_\perp}
\nonumber\\
&&\times\Big[{\rm Tr}\{U_x U^\dagger_z \}{\rm Tr}\{U_zU^\dagger_y\} 
- N_c{\rm Tr}\{U_xU^\dagger_y\}\Big]Q_{1x}
\label{Q1bkdiagrams}
\end{eqnarray} 
We get the same result if we replace $Q_{1x}$ with $Q_{5x}$.
In the liner regime, evolution equation (\ref{Q1bkdiagrams}) gives the resummation in the LLA for the light-cone quark operator.
However, in the double log of energy approximation, where 
$\alpha_s\ln^2(1/x_B)\sim 1$ and $\alpha_s\ln(1/x_B)\ll 1$, contribution to the evolution equation 
$Q_{1x}$ and $Q_{5x}$ coming from (\ref{Q1bkdiagrams}) can be neglected.

\subsubsection{BK-type diagrams with $\tildeQ_{1x}$, $\tildeQ_{5x}$, $\calf_x$}

Let us start with the real diagrams in Fig. \ref{BK-type-diagrams} in which the
quantum gluon (in blue) goes through the shock-wave. The calculation is the same as the diagram for 
BK equation. 
Considering, for example, operator $\tildeQ_{1x}$, we have
\begin{eqnarray}
&&\hspace{-2cm}
\langle \Tr\{\tildeQ_{1x}U^\dagger_z\}\rangle_{\small{\rm Fig.}\ref{BK-type-diagrams}a}
=  {\alpha_s\over \pi^2}\int_0^{+\infty}\!{d\alpha\over \alpha}\int d^2z \,
U^{ba}_{z}{\rm Tr}\{\tildeQ_{1x} t^a U^\dagger_y t^b\}\, 
{(y-z,z-x)_\perp\over (y-z)^2_\perp(z-x)^2_\perp}
\label{diagram-aBKtype}
\end{eqnarray}
In eq. (\ref{diagram-aBKtype}) we notice the divergence in the longitudinal momentum component ${1\over \alpha}$. 
Similarly to what we did for the evolution of the trace of two Wilson lines, we will 
regulate this divergence with a rigid cut-off and performing the derivative with respect this 
rapidity parameter we will obtain the evolution equation. The calculation of diagram in Fig. \ref{BK-type-diagrams} is performed
by first splitting all the fields of the LHS of equ. (\ref{diagram-aBKtype}) in classical and quantum,
then considering only the terms in which operator  $\hat{Q}^{\alpha\beta}_{ij}$ is left classical, and contracting the 
gluon quantum fields using the propagator in the background of a shock-wave given in eq. 

The virtual diagram is given in Fig. (\ref{BK-type-diagrams}b). Following similar procedure as for the real ones 
we obtain
\begin{eqnarray}
&&\hspace{-2cm}
\langle{\rm Tr}\{\tildeQ_{1x} U^\dagger_y\}\rangle_{\small{\rm Fig.}\ref{BK-type-diagrams}b}
= - {\alpha_s\over \pi^2} \int_0^{+\infty}\!{d\alpha\over \alpha}\int d^2z \,
U_z^{ba}\,{\rm Tr}\{\tildeQ_{1x} t^a U^\dagger_y t^b\}\, 
{(y-z,z-x)_\perp\over (x-z)^2_\perp(y-z)^2_\perp}
\end{eqnarray}
Using the symmetry which relates the diagrams in Fig. \ref{BK-type-diagrams} we can also 
obtain the result of the last two \textit{i.e.} diagrams in Fig. \ref{BK-type-diagrams}a and b. Summing all them up 
and working out the color factor we have
\begin{eqnarray}
\hspace{-1cm}&&
\langle{\rm Tr}\{\tildeQ_{1x} U^\dagger_y\}
\rangle_{\small{\rm Fig.}\ref{BK-type-diagrams}}
=
2\langle{\rm Tr}\{\tildeQ_{1x}U^\dagger_y\}
\rangle_{\small{\rm Fig.}\ref{BK-type-diagrams}a+b}
\nonumber\\
\hspace{-1cm}&&=  {\alpha_s\over 2\pi^2}\int_0^{+\infty}\!{d\alpha\over \alpha}\int d^2z \,
\big[2\,U^{ba}_z - 2U_y^{ba}\big]{\rm Tr}\{\tildeQ_{1x} t^a U^\dagger_y t^b\}\, 
{2(y-z,z-x)_\perp\over (y-z)^2_\perp(z-x)^2_\perp}
\nonumber\\
\hspace{-1cm}&&= {\alpha_s\over 2\pi^2}\int_0^{+\infty}\!{d\alpha\over \alpha}\int\!\!d^2z \,
{2(y-z,z-x)_\perp\over (y-z)^2_\perp(z-x)^2_\perp}
\Big[{\rm Tr}\{\tildeQ_{1x} U^\dagger_z \}{\rm Tr}\{U_zU^\dagger_y\} 
- N_c{\rm Tr}\{\tildeQ_{1x}U^\dagger_y\}\Big]
\label{BKtype-qoperator}
\end{eqnarray}

\begin{figure}[htb]
	\begin{center}
		\includegraphics[width=3.0in]{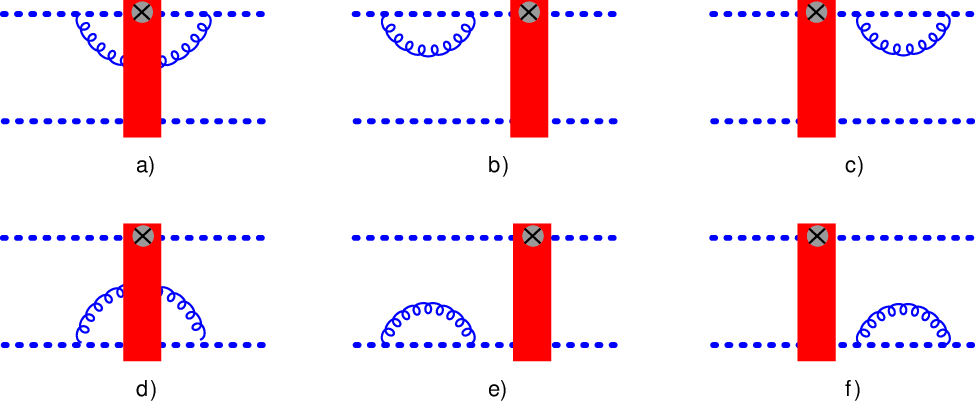}
		\caption{Self-energy diagrams with sub-eikonal correction inside and outside shock-wave.}
		\label{bktype-self}
	\end{center}
\end{figure}

Let us consider self energy diagrams in Fig. \ref{bktype-self}. The result is
\begin{eqnarray}
&&\hspace{-2cm}
\langle{\rm Tr}\{\tildeQ_{1x} \, U^\dagger_y\}\rangle_{\small{\rm Fig.}\ref{bktype-self}a+b+c}
=  {\alpha_s\over 2\pi^2}\int_0^{+\infty}\!\!{d\alpha\over \alpha}\int\!\!{d^2z \over (x-z)^2}
\big[2\,U^{ab}_z - 2U^{ab}_y\big]
{\rm Tr}\{t^a \tildeQ_{1x} t^b U^\dagger_y\}
\end{eqnarray}
and
\begin{eqnarray}
&&\hspace{-2cm}
\langle{\rm Tr}\{\tildeQ_{1x} \, U^\dagger_y\}\rangle_{\small{\rm Fig.}\ref{bktype-self}d+e+f}
=  {\alpha_s\over 2\pi^2}\int_0^{+\infty}\!\!{d\alpha\over \alpha}\int\!\!d^2z
\big[2\,U^{ab}_z -  2U^{ab}_y\big]
{\rm Tr}\{t^a \tildeQ_{1x} t^b U^\dagger_y\}\,{1 \over (y-z)^2}
\end{eqnarray}
Adding up the results for diagrams in Fig. \ref{BK-type-diagrams} and \ref{bktype-self} we have
\begin{eqnarray}
\langle{\rm Tr}\{\tildeQ_{1x}\, U^\dagger_y\}\rangle_{\small{\rm Fig.}\ref{BK-type-diagrams}+\ref{bktype-self}}
=\!\!&& {\alpha_s\over 2\pi^2}\int_0^{+\infty}\!{d\alpha\over \alpha}\int\!\!d^2z\,
{(x-y)^2_\perp \over (x-z)^2_\perp(y-z)^2_\perp}
\big[2\,U^{ab}_z - 2U_y^{ab}\big]
{\rm Tr}\{t^a\tildeQ_{1x}t^b U^\dagger_y \}
\nonumber\\
=\!\!&&{\alpha_s\over 2\pi^2}\int_0^{+\infty}\!{d\alpha\over \alpha}\int\!\!d^2z\,
{(x-y)^2_\perp \over (x-z)^2_\perp(y-z)^2_\perp}
\nonumber\\
&&\times\Big[\Tr\{U^\dagger_z \tildeQ_{1x}\}\Tr\{U^\dagger_y U_z\} - N_c\Tr\{U^\dagger_y \tildeQ_{1x}\}\Big]
\label{BKevolutionQ1}
\end{eqnarray}
Replacing operator $\hat{Q}_{1x}$ with $\calq_{1x}$, $\hat{\tildeQ}_{5x}$, and  $\calf_x$ 
in eq. (\ref{BKevolutionQ1}), we get the respective results for diagrams in Fig. \ref{BK-type-diagrams} and \ref{bktype-self}.

What should be noted in eq. (\ref{BKevolutionQ1}) is that the unitarity property for vanishing 
dipole size is absent because in this limit the term $\Tr\{U^\dagger_z \tildeQ_{1x}\}\Tr\{U^\dagger_y U_z\}$
does not reduce to $N_c\Tr\{U^\dagger_y \tildeQ_{1x}\}$  when $z \to x$.
This will be source of $\alpha_s\ln^2{1\over x_B}$ contributions \cite{Kovchegov:2015pbl}.

\subsection{Evolution equation with $\epsilon^{ij}F_{ij}$ quantum}
\label{sec: calfquantum}

\begin{figure}[htb]
\begin{center}
\includegraphics[width=3.0in]{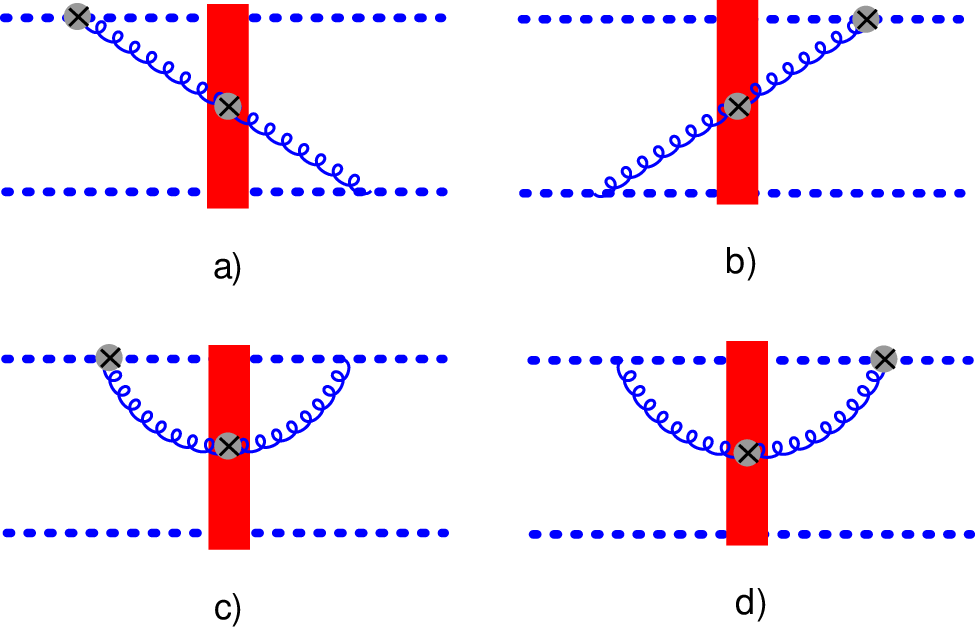}
\caption{Diagrams with $F_{ij}$ outside shock-wave. Here the gluon propagator, which represent the quantum loop,
has gluon sub-eikonal corrections. The gray circle with an x on it symbolizes the $F_{ij}$ operator:
it is quantum when it is on the Wilson-line and it is functionally integrated to create the gluon propagator;
it is classical when it is in the shock-wave red band.}
\label{outshock-2}
\end{center}
\end{figure}

\begin{figure}[htb]
	\begin{center}
		\includegraphics[width=3.0in]{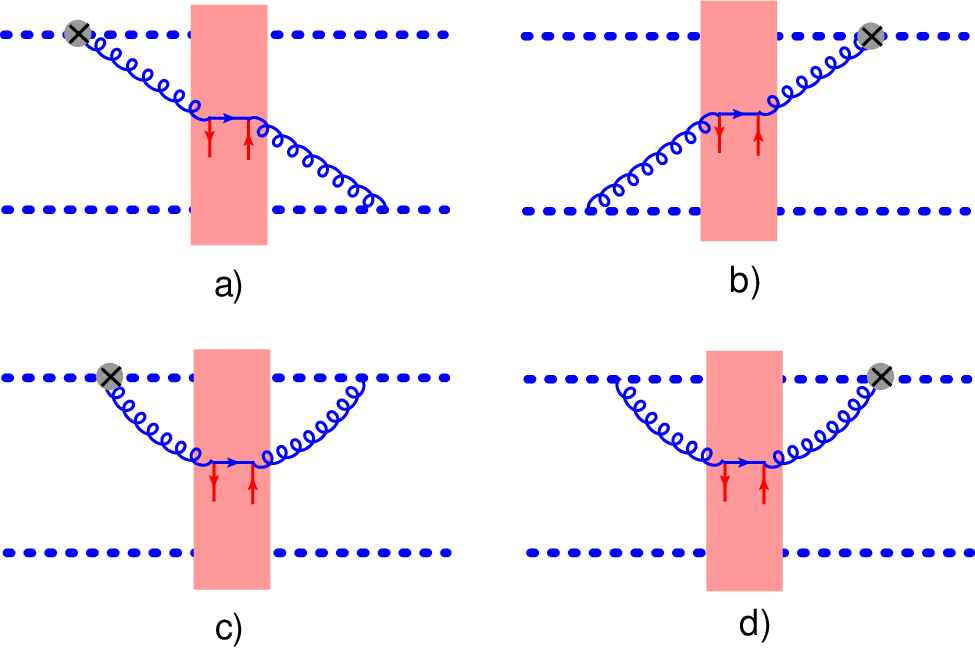}
		\caption{Diagrams with $F_{ij}$ outside shock-wave. Here the gluon propagator, which represent the quantum loop,
			has quark sub-eikonal corrections.}
		\label{outshock-3}
	\end{center}
\end{figure}

In this section we consider the case in which the gluon operator $\calf$ is quantum and, for this reason, it will
be integrated out via functional integration. The diagrams for this case are shown in Fig. \ref{outshock-2} and
in Fig. \ref{outshock-3}. 

In the Diagrams in Fig. \ref{outshock-2} we take the gluon propagator with sub-eikonal correction due to
gluon field. Let us start with diagram in Fig. \ref{outshock-2}a. We have
	
\begin{eqnarray}
&&\langle {\rm Tr}\{g\!\!\int^{+\infty}_{-\infty}\!\!dx_*
\,[\infty p_1,x_*]_x\,\epsilon^{ij}F_{ij}
[x_*,- \infty p_1]_x U^\dagger_y\}\rangle_{\small{\rm Fig.}\ref{outshock-2}\,a}
\nonumber\\
&&~~~~~
= - i g^2\int_{-\infty}^0\!\!dx_*\int_0^{+\infty}\!\!d{2\over s}y_*\,{\rm Tr}\{U_x t^a U^\dagger_y t^b\}
\langle A_\bullet^b(y_*, y_\perp)\,\epsilon^{ij}F^a_{ij}(x_*,x_\perp)\rangle
\label{outshock-2a}
\end{eqnarray}
From (\ref{outshock-2a}), we clearly see that $F_{ij}$ is now quantum and it will be integrated out.
We will perform the calculation in the axial gauge and will use the gluon propagator (\ref{Shwgaxg9}).
For the diagram under consideration the only terms that will survive the
contraction of the Lorentz indexes are the ones given in $\mathfrak{G}_3$, eq.
(\ref{G3}), thus we have
\begin{eqnarray}
&&\langle {\rm Tr}\{g\!\!\int^{+\infty}_{-\infty}\!\!dx_*
\,[\infty p_1,x_*]_x\,\epsilon^{ij}F_{ij}
[x_*,- \infty p_1]_x U^\dagger_y\}\rangle_{\small{\rm Fig.}\ref{outshock-2}\,a}
\nonumber\\
&&=
-ig^2\int_{-\infty}^0\!\!dx_*\int_0^{+\infty}\!\!d{2\over s}y_*\,{\rm Tr}\{U_x t^a U^\dagger_y t^b\}
\langle A_\bullet^b(y_*, y_\perp)\,\epsilon^{ij}F^a_{ij}(x_*,x_\perp)\rangle
\nonumber\\
&&= {ig^2\over s^2}\,{\rm Tr}\{U_x t^a U^\dagger_y t^b\}\int_0^{+\infty}{\dhd\alpha\over \alpha^3}
\int_{-\infty}^0\!\!dx_*\int_0^{+\infty}\!\!dy_*\,
\epsilon^{ij}(\partial_i^x g^l_j - \partial_j^x g^l_i)
\brayp e^{-i{\hatp^2_\perp\over \alpha s}y_*}
\nonumber\\
&&
~~~\times\!g\!\int^{+\infty}_{-\infty}\!\!dz_*\bigg[
\big([\infty p_1, z_*]\,i\,D^k F_{kl}[z_*,-\infty p_1]\big)^{ba}
\nonumber\\
&&
\hspace{3cm} - 2g\!\int_{z_*}^{+\infty}\!\!\!d{2\over s}z'_*
\big([\infty p_1,z'_*]F^i_{~\bullet}[z'_*,z_*]F_{ij}[z_*,-\infty p_1]\big)^{ba}\bigg]
e^{i{\hatp^2_\perp\over \alpha s}x_*}\ketxp
\nonumber\\
&&=  - {ig^2\over 2\pi}{\rm Tr}\{U_x t^a U^\dagger_y t^b\}\int_0^{+\infty}\!{d\alpha\over \alpha}\int d^2z\,\epsilon^{ij}
(\partial_i^x g^l_j - \partial_j^x g^l_i)
\int \!\dhd^2 q_1 {e^{i(q_1,y-z)}\over q^2_{1\perp}}
\int \!\dhd^2 q_2 {e^{i(q_2,z-x)}\over q^2_{2\perp}}
\nonumber\\
&&
~~~\times g\!\int_{-\infty}^{+\infty}\!dz_*\bigg[
\big([\infty p_1, z_*]_{z}\,i\,D^k F_{kl}[z_*,-\infty p_1]_{z}\big)^{ba}
\nonumber\\
&&
\hspace{3cm}- 2g\!\int_{z_*}^{+\infty}\!\!\!d{2\over s}z'_*
\big([\infty p_1,z'_*]_{z}F^i_{~\bullet}[z'_*,z_*]_{z}F_{ij}[z_*,-\infty p_1]_{z}\big)^{ba}\bigg]
\end{eqnarray}
Now we consider diagram \ref{outshock-2}b. 
This time we need the sub-eikonal correction $\mathfrak{G}_4$, eq. 
(\ref{G4}) and obtain
\begin{eqnarray}
&&\langle {\rm Tr}\{g\!\!\int^{+\infty}_{-\infty}\!\!dx_*\,
[\infty p_1,x_*]_x\,\epsilon^{ij}F_{ij}
[x_*,- \infty p_1]_x U^\dagger_y\}\rangle_{\small{\rm Fig.}\ref{outshock-2}\,b}
\nonumber\\
 &&=
-ig^2\int^{+\infty}_0\!\!dx_*\int^0_{-\infty}\!\!d{2\over s}y_*\,{\rm Tr}\{t^aU_x t^b U^\dagger_y \}
\langle F^a_{ij}(x_*,x_\perp)A_\bullet^b(y_*, y_\perp)\,\epsilon^{ij}\rangle
\nonumber\\
&&= - {ig^2\over 2\pi}{\rm Tr}\{t^aU_x t^b U^\dagger_y\}\int_0^{+\infty}\!{d\alpha\over \alpha}\int d^2z\,\epsilon^{ij}
(\partial_i^x g_j^l - \partial_j^x g_i^l)
\int \!\dhd^2 q_1 {e^{i(q_1,x - z)}\over q^2_{1\perp}}
\int \!\dhd^2 q_2 {e^{i(q_2,z-y)}\over q^2_{2\perp}}
\nonumber\\
&&~~~\times g\!\int_{-\infty}^{+\infty}\!dz_*\bigg[
\big([\infty p_1, z_*]_{z}\,i\,D^k F_{kl}[z_*,-\infty p_1]_{z}\big)^{ab}
\nonumber\\
&& \hspace{2.7cm}
-2g\!\int_{z_*}^{+\infty}\!\!\!d{2\over s}z'_*
\big([\infty p_1,z'_*]_{z}F_{ij}[z'_*,z_*]_{z}F^i_{~\bullet}[z_*,-\infty p_1]_{z}\big)^{ab}\bigg]
\end{eqnarray}

In summing diagrams \ref{outshock-2}a and \ref{outshock-2}b, we observe that 
the term with $D^kF_{kl}$ doubles and we may use the identity
\begin{eqnarray}
&&2g\!\int_{-\infty}^{+\infty}\!\!dz_*\,\Big[\hat{P}^k,\big([\infty p_1, z_*]_{z}\, F_{kl}[z_*,-\infty p_1]_{z}\big)^{ba}\Big]
\nonumber\\
&&~~~~~~= 2g\!\int_{-\infty}^{+\infty}\!\!dz_*\,i\,\mathfrak{D}_{z}^k\big([\infty p_1, z_*]_{z}\, F_{kl}[z_*,-\infty p_1]_{z}\big)^{ba}
\nonumber\\
&&~~~~~~= 2g\!\int_{-\infty}^{+\infty}\!dz_*\bigg[
\big([\infty p_1, z_*]_{z}\,i\,D^k F_{kl}[z_*,-\infty p_1]_{z}\big)^{ba}
\nonumber\\
&&\hspace{3.3cm}- g\!\int_{z_*}^{+\infty}\!\!\!d{2\over s}z'_*
\big([\infty p_1,z'_*]_{z}F^i_{~\bullet}[z'_*,z_*]_{z}F_{ij}[z_*,-\infty p_1]_{z}\big)^{ba}
\nonumber\\
&&\hspace{3.3cm}- g\!\int_{z_*}^{+\infty}\!\!\!d{2\over s}z'_*
\big([\infty p_1,z'_*]_{z}F_{ij}[z'_*,z_*]_{z}F^i_{~\bullet}[z_*,-\infty p_1]_{z}\big)^{ba}\bigg]
\label{outshock-2-identity}
\end{eqnarray}
where in the first step we substituted the covariant derivative acting on gauge links and which is defined in
eqs. (\ref{coderiv-glink}) and (\ref{defPi}). Furthermore, we observe that
the operator $P^i = p^i+gA^i$ can be traded with $p^i$ because, as usual in these cases, the transverse gauge field is zero at
the points outside the shock-wave.

So, using (\ref{outshock-2-identity}), the sum of diagrams \ref{outshock-2}a and \ref{outshock-2}b is
\begin{eqnarray}
&&\langle {\rm Tr}\{g\!\!\int^{+\infty}_{-\infty}\!\!dx_*
\,[\infty p_1,x_*]_x\,\epsilon^{ij}F_{ij}[x_*,- \infty p_1]_x U^\dagger_y\}\rangle_{\small{\rm Fig.}\ref{outshock-2}\,a+b}
\nonumber\\
&&= {ig^2\over s^2}\,{\rm Tr}\{U_x t^a U^\dagger_y t^b\}\int_0^{+\infty}{\dhd\alpha\over \alpha^3}
\int_{-\infty}^0\!\!dx_*\int_0^{+\infty}\!\!du_*\,
\epsilon^{ij}(\partial_i^x g^l_j - \partial_j^x g^l_i)
\brazp e^{-i{\hatp^2_\perp\over \alpha s}u_*}
\nonumber\\
&&
~~~\times 2\,g\!\int^{+\infty}_{-\infty}\!\!dz_*
\big[\hat{P}^k, \big([\infty p_1, z_*]\, F_{kl}[z_*,-\infty p_1]\big)^{ba}\big]
e^{i{\hatp^2_\perp\over \alpha s}x_*}\ketop
\nonumber\\
 &&= - {ig^2\over 2\pi}{\rm Tr}\{U_x t^a U^\dagger_y t^b\}\int_0^{+\infty}\!{d\alpha\over \alpha}\int d^2z\,\epsilon^{ij}
(iq_{2\,i} \,g^l_j - iq_{2\,j} \,g^l_i)
\nonumber\\
&&
~~~\times\!\int \!\dhd^2 q_1 {e^{i(q_1,y-z)}\over q^2_{1\perp}}
\int \!\dhd^2 q_2 {e^{i(q_2,z-x)}\over q^2_{2\perp}}(q_1^k - q_2^k)
\nonumber\\
&&
~~~\times 2g
\!\int_{-\infty}^{+\infty}\!\!dz_*\,\big([\infty p_1, z_*]_{z}\, F_{kl}[z_*,-\infty p_1]_{z}\big)^{ba}\,.
\end{eqnarray}
Performing the Fourier transform we arrive at
\begin{eqnarray}
\hspace{-1cm}&&\langle {\rm Tr}\{g\!\!\int^{+\infty}_{-\infty}\!\!dx_*
\,[\infty p_1,x_*]_x\,\epsilon^{ij}F_{ij}(x_*,x_\perp)
[x_*,- \infty p_1]_x U^\dagger_y\}\rangle_{\small{\rm Fig.}\ref{outshock-2}\,a+b}
\nonumber\\
\hspace{-1cm}
&&= - {g^2\over 2\pi}{\rm Tr}\{U_x t^a U^\dagger_y t^b\}\int_0^{+\infty}\!{d\alpha\over \alpha}\int d^2z
\,2g\!\int_{-\infty}^{+\infty}\!\!dz_*\,\big([\infty p_1, z_*]_{z}\,\epsilon^{ij} F_{ij}[z_*,-\infty p_1]_{z}\big)^{ba}
\nonumber\\
\hspace{-1cm}&&
~~~\times \Big[{i(y - z)^i\over 2\pi(y - z)^2}{i(z-x)^i\over 2\pi(z-x)^2} 
- \int \!\dhd^2 q_1 {e^{i(q_1,y-z)}\over q^2_{1\perp}}\,\delta^{(2)}(z-x)\Big]
\end{eqnarray}
where we used $\epsilon^{ij}\big(i\,q_{2i} \, F_{kj} - i\,q_{2j} \, F_{ki}\big)(q^k_1-q^k_2) = 
-i\,\epsilon^{ij}F_{ij}(q_2,q_1 - q_2)_\perp$\,.

At this point it should be clear why we performed all this massaging: the sum of diagrams \ref{outshock-2}a and \ref{outshock-2}b gave back
the same operator we started with, namely $\calf(z_\perp)$ but in the adjoint representation.

Let us consider the self energy diagrams in Fig. \ref{outshock-2}c and \ref{outshock-2}d. 
Integrating over the quantum field we have
\begin{eqnarray}
\hspace{-1cm}&&
\langle{\rm Tr}\{g\!\!\int_{-\infty}^{+\infty}\!\!dx_*\,[\infty p_1, x_*]_x \,\epsilon^{ij}F_{ij}
[x_*,-\infty p_1]_x \, U^\dagger_z\}\rangle_{\small{\rm Fig.}\ref{outshock-2}\,c+d}
\\
\hspace{-1cm}&&~~~~~ = i g^2{2\over s}\!\!\int_{-\infty}^0\!x_*\int^{+\infty}_0\!\!dx'_*\,{\rm Tr}\{t^a U_x t^b U^\dagger_y\}\,\epsilon^{ij}
(\partial_i g^\nu_j - \partial_j g^\nu_i)\langle A_\bullet^a(x'_*, x_\perp)A_\nu^b(x_*, x_\perp)
\rangle
\nonumber
\end{eqnarray}
Repeating similar steps done for the real ones we arrive at
\begin{eqnarray}
&&
\langle{\rm Tr}\{g\!\!\int_{-\infty}^{+\infty}\!\!dx_*\,[\infty p_1, x_*]_x \,\epsilon^{ij}F_{ij}
[x_*,-\infty p_1]_x \, U^\dagger_z\}\rangle_{\small{\rm Fig.}\ref{outshock-2}\,c+d}
\nonumber\\
&&= {ig^2\over 2\pi}\,{\rm Tr}\{t^a U_x t^b U^\dagger_y\}\int_0^{+\infty}\!{d\alpha\over \alpha}\,
\epsilon^{ij}(iq_{2i}\, g^\nu_j - iq_{2j} \,g^\nu_i)\int d^2z\int\!\dhd q_1  {e^{i(q_1,x-z)} \over q^2_{1\perp}}
\int\!\dhd^2 q_2 {e^{ i(q_2,z-x)} \over q^2_{2\perp}}
\nonumber\\
&&
~~~\times 2g\!\int^{+\infty}_{-\infty}\!\!dz_*\bigg[\big([\infty p_1, z_*]_z iD^i F_{ij}[z_*,-\infty p_1]_z\big)^{ab}
\nonumber\\
&&
\hspace{3cm} - g{2\over s}\!\int_{z_*}^{+\infty}\!dz'_*\,\big([\infty p_1, z'_*]_z F^l_{~\bullet}[z'_*,z_*]_z
F_{lm}[z_*,-\infty p_1]_z \big)^{ab}
\nonumber\\
&&
\hspace{3cm} - g{2\over s}\!\int_{z_*}^{+\infty}\!dz'_*\,\big([\infty p_1, z'_*]_z F_{lm}[z'_*,z_*]_z
F^l_{~\bullet}[z_*,-\infty p_1]_z\big)^{ab}
\bigg]
\end{eqnarray}
and using again the identity (\ref{outshock-2-identity}) we obtain
\begin{eqnarray}
\hspace{-1cm}&&
\langle{\rm Tr}\{g\!\!\int_{-\infty}^{+\infty}\!\!dx_*\,[\infty p_1, x_*]_x \,\epsilon^{ij}F_{ij}
[x_*,-\infty p_1]_x \, U^\dagger_y\}\rangle_{\small{\rm Fig.}\ref{outshock-2}\,c+d}
\nonumber\\
\hspace{-1cm}
&&= - {ig^2\over 2\pi}{\rm Tr}\{U_x t^a U^\dagger_y t^b\}\int_0^{+\infty}\!{d\alpha\over \alpha}\int d^2z\,
 2g\!\int_{-\infty}^{+\infty}\!\!dz_*\,\big([\infty p_1, z_*]_{z}\, F_{kl}[z_*,-\infty p_1]_{z}\big)^{ba}
\nonumber\\
\hspace{-1cm}&&~~~\times\!\int \!\dhd^2 q_1 {e^{i(q_1,x-z)}\over q^2_{1\perp}}
\int \!\dhd^2 q_2 {e^{i(q_2,z-x)}\over q^2_{2\perp}}
\,\epsilon^{ij}\,(iq_{2\,i} \,g^l_j - iq_{2\,j} \,g^l_i)(q_1^k - q_2^k)
\,.
\end{eqnarray}
We can now perform the Fourier transform and add to it the result of the real diagrams and arrive at
\begin{eqnarray}
\hspace{-1cm}&&\langle{\rm Tr}\{\calf_x \, U^\dagger_y\}\rangle_{\small{\rm Fig.}\ref{outshock-2}}
\label{calfLOgluonfund}\\
\hspace{-1cm}&&= {\alpha_s\over \pi^2}\int_0^{+\infty}\!{d\alpha\over \alpha}\int d^2z\,
{\rm Tr}\{U_x t^a U^\dagger_y t^b\}\calf^{ba}_z
\nonumber\\
\hspace{-1cm}&&~~~\times\Bigg({(x - z,z - y)\over (y - z)^2_\perp(z-x)^2_\perp}
+ {1\over (x - z)^2_\perp} 
+ 4\pi^2\!\!\int \!\dhd^2 q_1 {e^{i(q_1,y-z)}-e^{i(q_1,x-z)}\over q^2_{1\perp}}\,\delta^{(2)}(z-x)
\Bigg)
\nonumber
\end{eqnarray}
where we defined the adjoint representation operator
\begin{eqnarray}
\calf^{ab}(x_\perp) = 
ig{s\over 2}\!\int_{-\infty}^{+\infty}\!\!dz_*\,\big([\infty p_1, z_*]_x\,\epsilon^{ij} F_{ij}(z_*,x_\perp)[z_*,-\infty p_1]_x\big)^{ab}\,.
\end{eqnarray}

Taking into consideration also quarks in the external field we get diagrams represented in Fig. \ref{outshock-3}. 
The gluon propagator with quarks in the external field given in eq. \ref{gluonprop-inq}
has been calculated in Ref. \cite{Chirilli:2018kkw}.

We start with diagram in Fig. \ref{outshock-3}a and  \ref{outshock-3}b
\begin{eqnarray}
\hspace{-1cm}&&\langle {\rm Tr}\{g\!\!\int^{+\infty}_{-\infty}\!\!dx_*
\,[\infty p_1,x_*]_x\,\epsilon^{ij}F_{ij}
[x_*,- \infty p_1]_x U^\dagger_y\}\rangle_{\small{\rm Fig.}\ref{outshock-3}\,a+b}
\nonumber\\
\hspace{-1cm}&&=
-ig^2\bigg[\int_{-\infty}^0\!\!dx_*\int_0^{+\infty}\!\!d{2\over s}y_*\,{\rm Tr}\{U_x t^a U^\dagger_y t^b\} +
\int^{+\infty}_0\!\!dx_*\int^0_{-\infty}\!\!d{2\over s}y_*\,{\rm Tr}\{t^aU_x t^b U^\dagger_y \}\bigg]
\nonumber\\
\hspace{-1cm}&&
~~\times\!\langle A_\bullet^b(y_*, y_\perp)\,\epsilon^{ij}F^a_{ij}(x_*,x_\perp)\rangle
\label{outshock-3a}
\end{eqnarray}
Using propagator (\ref{gluonprop-inq}) and performing simple spinor algebra we get the result for
diagrams \ref{outshock-3}a and \ref{outshock-3}b
\begin{eqnarray}
&&
\langle{\rm Tr}\{g\!\!\int_{-\infty}^{+\infty}\!\!dx_*\,[\infty p_1, x_*]_x \,\epsilon^{ij}F_{ij}(x_*,x_\perp)
[x_*,-\infty p_1]_x \, U^\dagger_y\}\rangle_{\small{\rm Fig.}\ref{outshock-3}\,a+b}
\nonumber\\
\hspace{-1cm}
&&= 4\,{g^2\over s}{\rm Tr}\{U_x t^a U^\dagger_y t^b\}\int_0^{+\infty}\!{\dhd\alpha\over \alpha}
\int_{-\infty}^{+\infty}\!\!dz_{1*}\int^{z_{1*}}_{-\infty}\!\!\!dz_{2*}\int d^2z
\nonumber\\
\hspace{-1cm}&&
~~~\times \!g^2\Bigg[
\int {\dhd^2 q_1\dhd^2 q_2\over q^2_{1\perp}q^2_{2\perp}}\, e^{i(q_1,y-z)+i(q_2,z-x)}
\bar{\psi}(z_{1*},z_\perp)\big(\vec{q}_2\!\times\!\vec{q}_1 + i (q_1,q_2)\gamma^5\big)\ssp_1
\nonumber\\
\hspace{-1cm}&&~~~\times\!
[z_{1*},\infty p_1]_z\, t^b U_z\, t^a\,[-\infty p_1,z_{2*}]_z\,\psi(z_{2*},z_\perp) + {\rm c.c.}
\Bigg]\,.
\end{eqnarray}
Notice that we used the propagator (\ref{gluonprop-inq}) in the limit of $x^+\to +\infty$ and $y^+\to -\infty$
so that the gauge link gets $[x^+,y^+]_z \to U(z_\perp)$.
After Fourier transform we have

\begin{eqnarray}
&&
\langle\Tr\{\calf_x \, U^\dagger_y\}\rangle_{\small{\rm Fig.}\ref{outshock-3}\,a+b}
\nonumber\\
&&= - {\alpha_s\over \pi^2}\,\Tr\{U_x t^a U^\dagger_y t^b\}\,\int_0^{+\infty}\!{d\alpha\over\alpha}
\int d^2z\int_{-\infty}^{+\infty}\!\!dz_{1*}\!\int_{-\infty}^{z_{1*}}\!\!dz_{2*}
\nonumber\\
&&
\times\! g^2\bigg[
\bar{\psi}(z_{1*},z_\perp)\Big({(\vec{x}-\vec{z})\!\times\!(\vec{z}-\vec{y})\over (x-z)^2_\perp(y-z)^2_\perp}
+ i {(x-z,z-y)\gamma^5\over (x-z)^2_\perp(y-z)^2_\perp}\Big)\,i\ssp_1
\nonumber\\
&&
~~~~~\times\![z_{1*},\infty p_1]_z\, t^b U_z\, t^a\,[-\infty p_1,z_{2*}]_z\,\psi(z_{2*},z_\perp)
\nonumber\\
&&
~~~~~+ \bar{\psi}(z_{2*},z_\perp)\Big({(\vec{x}-\vec{z})\!\times\!(\vec{z}-\vec{y})\over (x-z)^2_\perp(y-z)^2_\perp}
+ i {(x-z,z-y)\gamma^5\over (x-z)^2_\perp(y-z)^2_\perp}\Big)\,i\ssp_1
\nonumber\\
&&
~~~~~\times\![z_{2*}, -\infty p_1]_z\, t^a U^\dagger_z\, t^b\,[\infty p_1,z_{1*}]_z\,\psi(z_{1*},z_\perp)
\bigg]
\end{eqnarray}
where we used again the two dimensional vector product as $\vec{x}\times\vec{y} = \epsilon^{ij}x_iy_j$.

Let us consider self energy diagrams in Fig. \ref{outshock-3}c and \ref{outshock-3}d. Proceeding in a similar way 
we obtain
\begin{eqnarray}
&&
\langle{\rm Tr}\{g\!\!\int_{-\infty}^{+\infty}\!\!dx_*\,[\infty p_1, x_*]_x \,\epsilon^{ij}F_{ij}(x_*,x_\perp)
[x_*,-\infty p_1]_x \, U^\dagger_y\}\rangle_{\small{\rm Fig.}\ref{outshock-3}\,c+d}
\nonumber\\
&&= - {\alpha_s\over \pi^2}{2\over s}\,\Tr\{U_x t^a U^\dagger_y t^b\}\,\int_0^{+\infty}\!{d\alpha\over\alpha}
\int {d^2z\over (x-z)^2_\perp}\int_{-\infty}^{+\infty}\!\!dz_{1*}\!\int_{-\infty}^{z_{1*}}\!\!dz_{2*}
\nonumber\\
&&
~~~\times g^2\bigg[
\bar{\psi}(z_{1*},z_\perp)\,i\,\gamma^5\ssp_1\,
[z_{1*},\infty p_1]_z\, t^b U_z\, t^a\,[-\infty p_1,z_{2*}]_z\,\psi(z_{2*},z_\perp)
\nonumber\\
&&\hspace{1.4cm}
+ \bar{\psi}(z_{2*},z_\perp)\,i\,\gamma^5\ssp_1\,
[z_{2*}, -\infty p_1]_z\, t^a U^\dagger_z\, t^b\,[\infty p_1,z_{1*}]_z\,\psi(z_{1*},z_\perp)
\bigg]
\end{eqnarray}
Let us define the following quark parity odd operators
\begin{eqnarray}
\hspace{-1cm}&&\calq^{ab}_5(z_\perp) 
\label{calq5ab}\\
\hspace{-1cm}&&\equiv g^2\!\int_{-\infty}^{+\infty}\!\!dz_{1*}\!\int_{-\infty}^{z_{1*}}\!\!dz_{2*}
\bar{\psi}(z_{1*},z_\perp)\,\gamma^5\ssp_1\,
[z_{1*},\infty p_1]_z\, t^a U_z\, t^b\,[-\infty p_1,z_{2*}]_z\,\psi(z_{2*},z_\perp)\,,
\nonumber
\\
\hspace{-1cm}&&{\calq^{ab}}^\dagger_5(z_\perp) 
\label{calq5abd}\\
\hspace{-1cm}&&\equiv g^2\!\int_{-\infty}^{+\infty}\!\!dz_{1*}\!\int_{-\infty}^{z_{1*}}\!\!dz_{2*}
\bar{\psi}(z_{2*},z_\perp)\gamma^5\ssp_1\,
[z_{2*}, -\infty p_1]_z\, t^b U^\dagger_z\, t^a\,[\infty p_1,z_{1*}]_z\,\psi(z_{1*},z_\perp)\,,
\nonumber
\end{eqnarray}
and parity even operators
\begin{eqnarray}
\hspace{-1cm}&&\calq^{ab}_1(z_\perp) 
\label{calq1ab}\\
\hspace{-1cm}&&\equiv g^2\!
\int_{-\infty}^{+\infty}\!\!dz_{1*}\!\int_{-\infty}^{z_{1*}}\!\!dz_{2*}
\bar{\psi}(z_{1*},z_\perp)\,i\ssp_1\,
[z_{1*},\infty p_1]_z\, t^a U_z\, t^b\,[-\infty p_1,z_{2*}]_z\,\psi(z_{2*},z_\perp)\,,
\nonumber
\\
\hspace{-1cm}&&{\calq^{ab}}^\dagger_1(z_\perp) 
\label{calq1abd}\\
\hspace{-1cm}&&\equiv - g^2\!\int_{-\infty}^{+\infty}\!\!dz_{1*}\!\int_{-\infty}^{z_{1*}}\!\!dz_{2*}
\bar{\psi}(z_{2*},z_\perp)\,i\,\ssp_1\,
[z_{2*}, -\infty p_1]_z\, t^b U^\dagger_z\, t^a\,[\infty p_1,z_{1*}]_z\,\psi(z_{1*},z_\perp)\,.
\nonumber
\end{eqnarray}
As usual, we will use the short-hand notation $\calq_{1z}=\calq_1(z_\perp)$.
Summing diagrams of Fig. \ref{outshock-3} and using definitions (\ref{calq5ab})-(\ref{calq1abd}) we have

\begin{eqnarray}
&&\langle{\rm Tr}\{\calf_x\, U^\dagger_y\}\rangle_{\small{\rm Fig.}\ref{outshock-3}}
= - {\alpha_s\over \pi^2}\,{\rm Tr}\{U_x t^a U^\dagger_y t^b\}\,\int_0^{+\infty}\!{d\alpha\over\alpha}
\int d^2z\Bigg[
{(\vec{x}-\vec{z})\!\times\!(\vec{z}-\vec{y})\over (x-z)^2_\perp(y-z)^2_\perp}\,
\Big(\calq_{1z}^{ba} - {\calq_{1z}^{ba}}^\dagger\Big)
\nonumber\\
&&\hspace{4.5cm}
 - \Big({(x-z,z-y)\over (x-z)^2_\perp(y-z)^2_\perp}
+ {1\over (x-z)^2_\perp}\Big)
\Big({\calq_{5z}^{ba}} + {\calq_{5z}^{ba}}^\dagger\Big)
\Bigg]
\label{calf-quantumb}
\end{eqnarray}
The interesting thing to notice in result (\ref{calf-quantumb}) is the appearance of operator $\calq_{1x}^{ab}$
which although the operator itself is parity even it is multiplied by the $(\vec{x}-\vec{z})\!\times\!(\vec{z}-\vec{y})$
so parity is preserved.

We can now sum up diagrams in Fig. \ref{outshock-2} and \ref{outshock-3} and obtain
\begin{eqnarray}
&&\langle{\rm Tr}\{\calf_x\, U^\dagger_y\}\rangle_{\small{\rm Figs.}\ref{outshock-2}+\ref{outshock-3}}
\label{calf-quantum}\\
&&=
- {\alpha_s\over \pi^2}\,{\rm Tr}\{U_x t^a U^\dagger_y t^b\}\,\int_0^{+\infty}\!{d\alpha\over\alpha}
\int d^2z\Bigg\{
{(\vec{x}-\vec{z})\!\times\!(\vec{z}-\vec{y})\over (x-z)^2_\perp(y-z)^2_\perp}\,
\Big(\calq_{1z}^{ba} - {\calq_{1z}^{ba}}^\dagger\Big)
\nonumber\\
&&\hspace{5cm} -  \Big({(x-z,z-y)\over (x-z)^2_\perp(y-z)^2_\perp}
+ {1\over (x-z)^2_\perp}\Big)
\Big(\calq_{5z}^{ba} + {\calq_{5z}^{ba}}^\dagger + \calf^{ba}_z\Big)
\nonumber\\
&&\hspace{5cm} - 4\pi^2\!\!\int \!\dhd^2 q_1 {e^{i(q_1,y-z)}-e^{i(q_1,x-z)}\over q^2_{1\perp}}\,\delta^{(2)}(z-x)\calf^{ba}_z
\Bigg\}\,.
\nonumber
\end{eqnarray}

The first thing to notice in eq. (\ref{calf-quantum}) is that 
diagrams \ref{outshock-2} and \ref{outshock-3} have
generated, after one loop evolution, operators $\hat{\calq}_1^{ab}(x_\perp)$, $\hat{\calq}_5 ^{ab}(x_\perp)$, and 
$\hat{\calf}^{ab}(x_\perp)$
which are not present in the OPE (\ref{OPEtot}). This means that, either we have to find the evolution of this new operators or try to
reduce them to the operators we started with \textit{i.e} $\hat{\calf}(x_\perp)$, $\hat{\calq}_1(x_\perp)$ and 
$\hat{\calq}_5(x_\perp)$.
Performing some color algebra we can reduce the RHS of eq.  (\ref{calf-quantum}) in the fundamental representation
\begin{eqnarray}
&&\hspace{-0.6cm}
\langle{\rm Tr}\{\calf_x\, U^\dagger_y\}\rangle_{\small{\rm Figs.}\ref{outshock-2}+\ref{outshock-3}}
\nonumber\\
&&\hspace{-0.6cm}={\alpha_s\over 2\pi^2}\,\int_0^{+\infty}\!{d\alpha\over\alpha}
\int d^2z
\Bigg\{\half{(\vec{x}-\vec{z})\!\times\!(\vec{z}-\vec{y})\over (x-z)^2_\perp(y-z)^2_\perp}\,
\bigg[ {\rm Tr}\{U^\dagger_y \tilde{Q}_{1\,z}\}{\rm Tr}\{U^\dagger_z U_x\}
-  {\rm Tr}\{U_x \tildeQ^\dagger_{1z}\}{\rm Tr}\{U^\dagger_y U_z\}
\nonumber\\
&&+ {1\over N_c} \Big(\Tr\{ U_x U^\dagger_y \tilde{Q}_{1z}U^\dagger_z\}
+ \Tr\{U^\dagger_y U_x U_z^\dagger \tilde{Q}_{1z} \}
-  \Tr\{ U_x U^\dagger_y U_z\tilde{Q}^\dagger_{1z}\}
- \Tr\{U^\dagger_y U_x \tilde{Q}_{1z}^\dagger U_z\}
\Big)
\nonumber\\
&&+ {1\over N^2_c}\Tr\{U^\dagger_y U_x\}\Big(Q^\dagger_{1z} - Q_{1z}\Big)
\bigg]
-  \half\bigg[{(x-z,z-y)\over (x-z)^2_\perp(y-z)^2_\perp}
+ {1\over (x-z)^2_\perp}\bigg]
\nonumber\\
&&
\times\bigg[ {\rm Tr}\{U^\dagger_y \big(\tildeQ_{5z} -2\calf_z\big)\}{\rm Tr}\{U^\dagger_z U_x\}
+  \Tr\{U_x \big(\tilde{Q}^\dagger_{5z}-2\calf^\dagger_z\big)\}{\rm Tr}\{U^\dagger_y U_z\}
\nonumber\\
&&
~~~ - {1\over N_c}\Big( {\rm Tr}\{ U_x U^\dagger_y U_z\tilde{Q}^\dagger_{5z}\}
+ \Tr\{U^\dagger_y U_x \tilde{Q}_{5z}^\dagger U_z\}
+ \Tr\{ U_x U^\dagger_y \tilde{Q}_{5z}U^\dagger_z\}
+ \Tr\{U^\dagger_y U_x U_z^\dagger \tilde{Q}_{5z} \}
\Big)
\nonumber\\
&&
~~~ + {1\over N^2_c}\Tr\{U^\dagger_y U_x\}\Big(Q_{5z}+Q^\dagger_{5z}\Big)
\bigg]
\nonumber\\
&&
~~~+ 4\pi^2\!\!\int \!\dhd^2 q_1 {e^{i(q_1,y-z)}-e^{i(q_1,x-z)}\over q^2_{1\perp}}\,\delta^{(2)}(z-x)
\nonumber\\
&&\times\!\Big[\Tr\{U_x U^\dagger_z\}\Tr\{U^\dagger_y \calf_z\} + \Tr\{U^\dagger_x U_z\}\Tr\{U_x\calf^\dagger_z\}\Big]
\Bigg\}
\label{calf-quantum-fund}
\end{eqnarray}
So, writing the evolution equation for ${\rm Tr}\{\calf_x\, U^\dagger_y\}$ in the fundamental representation,
we eliminated the operators $\hat{\calf}^{ab}_x$, $\hat{\calq}^{ab}_{1x}$, $\hat{\calq}_{5x}^{ab}$ and their adjoint conjugated,
in favor of operators $\hat{\calf}_x$, $\hat{Q}_{1x}$, $\hat{Q}_{5x}$, $\hat{\tildeQ}_{1x}$, $\hat{\tildeQ}_{5x}$ 
and their adjoint conjugated, defined in
eqs. (\ref{Q1})-(\ref{Qtd5}), that are present in the OPE (\ref{OPEtot}) and (\ref{OPEtot-nonsing}).

From eq. (\ref{calf-quantum-fund}) it should be clear why we decided to calculate the diagrams using operators
$\hat{Q}_{1x}$, $\hat{Q}_{5x}$, $\hat{\tildeQ}_{1x}$, $\hat{\tildeQ}_{5x}$, instead of $\hat{\calq}_{1x}$ and $\hat{\calq}_{5x}$.

\subsection{Diagrams with quark-to-gluon propagator}
\label{sec: q2gdiag}

\begin{figure}[htb]
		\begin{center}
		\includegraphics[width=2.9in]{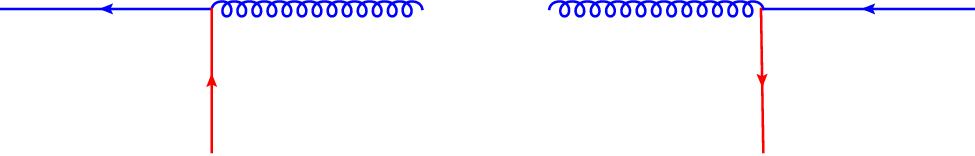}
		\caption{Diagrams for quark-to-gluon propagator.}
		\label{q2g-propagator}
\end{center}
\end{figure}

The diagrams we have calculated until now were obtained using the quark and gluon propagators originally 
calculated in reference \cite{Chirilli:2018kkw} and summarized in Appendix \ref{sec: propagators}.

In this section we are going to calculate the one-loop correction to the quark operators $\hat{Q}_{1x}$, $\hat{Q}_{5x}$,
$\hat{\tildeQ}_{1x}$, and $\hat{\tildeQ}_{5x}$
defined in eqs. (\ref{Q1})-\ref{Qtd5} using the quark-to-gluon propagator shown in Fig. \ref{q2g-propagator}. 

\subsubsection{quark-to-gluon propagator}
First, let us calculate the quark-to-gluon propagator. We have two contributions represented in diagrams 
in Fig. \ref{q2g-propagator}. As usual, the quantum fields are indicated with the superscript (or subscript) $q$ while the classical ones 
are indicated with the superscript (or subscript) $cl$ although the superscript (or subscript) $cl$ will be 
eventually dropped.  

We need the following quark propagators in the eikonal approximation
\begin{eqnarray}
&&\langle\psi(x)\bar{\psi}(y)\rangle 
\nonumber\\
&&=\Big[\int_0^{+\infty}\!\!{\dhd\alpha\over 2\alpha}\theta(x_*-y_*) 
- \int_{-\infty}^0\!\!{\dhd\alpha\over 2\alpha}\theta(y_*-x_*)\Big]e^{-i\alpha(x_\bullet-y_\bullet)}{1\over \alpha s}
\nonumber\\
&&~~~\times\braxp e^{-i{p^2_\perp\over \alpha s}x_*}\Big(
[x_*,y_*]\ssp\ssp_2\ssp - g\!\int_{y_*}^{x_*}\!\!d{2\over s}\omega_*
[x_*,\omega_*]F_{i\bullet}[\omega_*,y_*]\gamma^i\ssp_2\ssp\Big)e^{i{p^2_\perp\over \alpha s}y_*}\ketyp
\label{righteikona}
\end{eqnarray}
and
\begin{eqnarray}
&&
\langle\psi(x)\bar{\psi}(y)\rangle 
\nonumber\\
&&=\Big[\int_0^{+\infty}\!\!{\dhd\alpha\over 2\alpha}\theta(x_*-y_*) 
- \int_{-\infty}^0\!\!{\dhd\alpha\over 2\alpha}\theta(y_*-x_*)\Big]e^{-i\alpha(x_\bullet-y_\bullet)}{1\over \alpha s}
\nonumber\\
&&~~~\times\braxp e^{-i{\hatp^2_\perp\over \alpha s}x_*}\Big(
\hat{\ssp}\ssp_2\hat{\ssp}[x_*,y_*] - g\!\int_{y_*}^{x_*}\!\!d{2\over s}\omega_*
\hat{\ssp}\ssp_2\gamma^i[x_*,\omega_*]F_{i\bullet}[\omega_*,y_*]\Big)e^{i{\hatp^2_\perp\over \alpha s}y_*}\ketyp\,.
\label{lefteikonal}
\end{eqnarray}
Notice that we are not using the usual eikonal quark propagator in the shock-wave which has structure
free propagator-shock-wave-free propagator as given in the eikonal term of quark propagator in 
eq. (\ref{sym-quarksubnoedge2}). The transition from the eikonal propagators
(\ref{righteikona}) and (\ref{lefteikonal}) to the eikonal part of the quark propagator
(\ref{sym-quarksubnoedge2}) is explained ino Ref. \cite{Chirilli:2018kkw}. Here we just mention that
with the help of eqs. (\ref{deriv-glink}), (\ref{coderiv-glink}), and (\ref{defPi}), we can
show that the work propagators (\ref{lefteikonal}), and (\ref{righteikona}) can easily be put in the usual form
\begin{eqnarray}
\brax {i\over \Sp + i\epsilon} \kety
=\!\!\!&&  {1\over s}\left[\int_0^{+\infty}\!{\dhd\alpha\over 2\alpha^2}\,\theta(x_* - y_*) - 
\int^0_{-\infty}\!{\dhd\alpha\over 2\alpha^2}\theta(y_* - x_*)\right]e^{-i\alpha(x_\bullet-y_\bullet)}
\nonumber\\
&&\times\braxp\,e^{-i{\hatp^2_\perp\over \alpha s}x_*} \ssp\,\ketzp
\ssp_2U_z\brazp\ssp\,
e^{i{\hatp^2_\perp\over \alpha s}y_*}\ketyp\,.
\label{qpropa-sw}
\end{eqnarray}
However, as we will soon see, to calculate diagrams in Fig. \ref{q2g-propagator}
we will need the form of the quark propagator given in (\ref{lefteikonal}), and (\ref{righteikona}).

We also need the eikonal gluon propagator in the axial gauge
\begin{eqnarray}
\hspace{-1cm}\langle A^a_\mu(x)A^b_\nu(y)\rangle_A
&&=\left[-\int_0^{+\infty}\!\!{\dhd \alpha\over 2\alpha}\theta(x_*-y_*) 
+ \int_{-\infty}^0\!\!{\dhd\alpha\over 2\alpha}\theta(y_*-x_*) \right]e^{-i\alpha(x_\bullet - y_\bullet)}
\nonumber\\
\hspace{-1cm}&&\times\braxp e^{-i{\hatp^2_\perp\over \alpha s}x_*}\left(\delta_\mu^\xi - {2\over s}{p_{2\mu}\over \alpha}p^\xi\right)\![x_*,y_*]\!
\left(g_{\xi\nu} - 
{2\over s}p_\xi{p_{2\nu}\over \alpha}\right)\!e^{i{\hatp^2_\perp\over \alpha s}y_*} \ketyp^{ab}
\label{gpropaxial}
\end{eqnarray}

Using quark propagator (\ref{lefteikonal}) and gluon propagator (\ref{gpropaxial}), we can calculate the first
diagram for the quark-to-gluon propagator
\begin{eqnarray}
	\hspace{-0.7cm}\langle A_\mu^a(x)\bar{\psi}(y)\rangle =\!\!&& \langle A_\mu^{q,a}(x)\bar{\psi}^q(y)\,ig\int d^4z
	\bar{\psi}^{cl}(z)\Sa^q(z)\psi^q(z)\rangle
	\nonumber\\
	=\!\!&&  ig\!\int d^4z\left[-\int_0^{+\infty}\!\!{\dhd \alpha\over 2\alpha}\theta(x_*-z_*) 
	+ \int_{-\infty}^0\!\!{\dhd\alpha\over 2\alpha}\theta(z_*-x_*) \right]e^{-i\alpha(x_\bullet - z_\bullet)}
	\nonumber\\
	&&\times\braxp e^{-i{\hatp^2_\perp\over \alpha s}x_*}\left(\delta_\mu^\xi - {p_{2\mu}\over p_*}p^\xi\right)\![x_*,z_*]\!
	\left(g_{\xi\nu} - p_\xi{p_{2\nu}\over p_*}\right)\!e^{i{\hatp^2_\perp\over \alpha s}z_*} \ketzp^{ab}\barpsi^{cl}(z)t^b
	\nonumber\\
	&&\times\Big[\int_0^{+\infty}\!\!{\dhd\alpha\over 2\alpha}\theta(z_*-y_*) 
	- \int_{-\infty}^0\!\!{\dhd\alpha\over 2\alpha}\theta(y_*-z_*)\Big]e^{-i\alpha(z_\bullet-y_\bullet)}{1\over \alpha s}
	\nonumber\\
	&&\times\brazp e^{-i{p^2_\perp\over \alpha s}z_*}\Big(
	\ssp\ssp_2\ssp[z_*,y_*] - g\!\int_{y_*}^{z_*}\!\!d{2\over s}\omega_*
	\ssp\ssp_2\gamma^i[z_*,\omega_*]F_{i\bullet}[\omega_*,y_*]\Big)e^{i{p^2_\perp\over \alpha s}y_*}\ketyp
	\nonumber\\
	=\!\!&& ig\!\int_{y_*}^{x_*}d{2\over s}z_*\Big[-\int_0^{+\infty}\!{\dhd\alpha\over 4\alpha^2}\theta(x_*-y_*) 
	+ \int_{-\infty}^0{\dhd\alpha\over 4\alpha^2}\theta(y_*-x_*)\Big]e^{-i\alpha(x_\bullet-y_\bullet)}\int\!d^2z
	\nonumber\\
	&&\times\braxp e^{-i{p^2_\perp\over \alpha s}x_*}\Big(g^\perp_{\mu\nu} - {2\over s}{p^\perp_\nu p_{2\mu}\over \alpha}\Big)\ketzp
	[x_*,z_*]^{ab}_z \bar{\psi}^{cl}(z_*,z_\perp)t^b\gamma^\nu_\perp
	[z_*,y_*]_z\nonumber\\
	&&\times
	\brazp(\alpha \ssp_1 + \ssp_\perp)e^{i{p^2_\perp\over \alpha s}y_*}\ketyp
	+ O(\lambda^{-2})
	\label{q2gpropagator1}
\end{eqnarray}

where we used $\ssp_2\psi = \bar{\psi}\ssp_2 = O(\lambda^{-2})$ and
\begin{eqnarray}
\big(\delta_\mu^\nu - {p_{2\mu}p_\nu\over p_*}\big)[x_*,z_*]^{ab}\bar{\psi}(z_*)\gamma^\nu
= \big(g_{\perp\mu}^\nu - {p_{2\mu}p_{\perp\nu}\over  p_*}\big)[x_*,z_*]^{ab}\bar{\psi}(z_*)\gamma^\nu_\perp
+ O(\lambda^{-2})
\end{eqnarray}
and
\begin{eqnarray}
&&\big(\delta_\mu^\nu - {p_{2\mu}p_\nu\over p_*}\big)[x_*,z_*]^{ab}
\big(g_{\xi\nu} - p_\xi{p_{2\nu}\over p_*}\big)\bar{\psi}(z_*)\gamma^\nu
\nonumber\\
&&= \big(g_{\perp\mu}^\nu - {p_{2\mu}p_{\perp\nu}\over  p_*}\big)[x_*,z_*]^{ab}\bar{\psi}(z_*)\gamma^\nu_\perp
+ O(\lambda^{-2})
\end{eqnarray}
We remind the reader that the parameter $\lambda$ is the large boost-parameter which discriminates between the components of the fields
(see Appendix \ref{sec: notation} for details and Ref. \cite{Chirilli:2018kkw}). 
Similarly, we have
\begin{eqnarray}
\langle \psi(y)A^a_\mu(x)\rangle =\!\!&& \langle \psi^q(y)A_\mu^{q,a}(x)\,
ig\int d^4z\,\bar{\psi}^q(z)\Sa^q(z)\psi^{cl}(z)\rangle
\nonumber\\
=\!\!&& ig\int_{x_*}^{y_*}\!\!d{2\over s}z_*\Big[-\int_0^{+\infty}\!\!{\dhd\alpha\over 4\alpha^2}\theta(y_*-x_*) + 
\int_{-\infty}^0{\dhd\alpha\over 4\alpha^2}\theta(x_*-y_*)\Big]e^{-i\alpha(y_\bullet - x_\bullet)}
\nonumber\\
&&\times \!\int \!d^2z\brayp e^{-i{p^2_\perp\over \alpha s}y_*}(\alpha \ssp_1 + \ssp_\perp)\ketzp[y_*,z_*]_z
\gamma^\nu
t^b\psi^{cl}(z_*,z_\perp)[z_*,x_*]^{ba}_z
\nonumber\\
&&\times
\brazp\Big(g^\perp_{\nu\mu} - {2\over s}{p^\perp_\nu p_{2\mu}\over \alpha}\Big) e^{i{p^2_\perp\over \alpha s}x_*}\ketxp
+ O(\lambda^{-2})
\label{q2gpropagator2}
\end{eqnarray}

\subsubsection{Operators definition}

Before we start the calculation of the diagrams we defined operators that will be useful for the subsequent results. So, we define
\begin{eqnarray}
	&&\hspace{-1cm}\calx_1(x_\perp,y_\perp)  =- g^2\!\!\int_{-\infty}^{+\infty}\!\!dz_*d\omega_*
	\barpsi(z_*,y_\perp)[z_*, -\infty p_1]_y\,i\,\ssp_1\,[\infty p_1,\omega_*]_x\psi(\omega_*,x_\perp)\,,
	\label{calx1}
	\\
	&&\hspace{-1cm}\calx_1^\dagger(x_\perp,y_\perp) =
	g^2\!\!\int_{-\infty}^{+\infty}\!\!dz_*d\omega_*\,
	\barpsi(\omega_*,x_\perp)[\omega_*,\infty p_1]_x\,i\,\ssp_1\,\,[-\infty p_1,z_*]_y\psi(z_*,y_\perp)\,,
	\label{calx1d}
	\\	
	&&\hspace{-1cm}\calx_5(x_\perp,y_\perp) = - g^2\!\!\int_{-\infty}^{+\infty}\!\!dz_*d\omega_*
	\barpsi(z_*,y_\perp)[z_*,-\infty p_1]_y\,\gamma^5\ssp_1\,[\infty p_1,\omega_*]_x\psi(\omega_*,x_\perp)\,,
	\label{calx5}
	\\	
	&&\hspace{-1cm}\calx_5^\dagger(x_\perp,y_\perp) =  - g^2\!\!\int_{-\infty}^{+\infty}\!\!dz_*d\omega_*
	\barpsi(\omega_*,x_\perp)[\omega_*,\infty p_1]_x\,\gamma^5\ssp_1\,[-\infty p_1,z_*]_y\psi(z_*,y_\perp)\,.
	\label{calx5d}
\end{eqnarray}
Operators (\ref{calx1})-(\ref{calx5d}) are not related, to the author's knowledge to any known 
Transverse Momentum Distributions (TMD).

We also define
\begin{eqnarray}
	&&\hspace{-1.3cm}\calh^+_1(x_\perp,y_\perp) = - g^2\!\!\int_{-\infty}^{+\infty}\!\!dz_*d\omega_*\,
	\barpsi(\omega_*,y_\perp)[\omega_*,\infty p_1]_y\,i\ssp_1\,\,[\infty p_1,z_*]_x\psi(z_*,x_\perp)\,,
	\label{calh1plus}
	\\
	&&\hspace{-1.3cm}\calh^+_5(x_\perp,y_\perp) = - g^2\!\!\int_{-\infty}^{+\infty}\!\!dz_*d\omega_*\,
	\barpsi(\omega_*,y_\perp)[\omega_*,\infty p_1]_y\,\gamma^5\ssp_1\,\,[\infty p_1,z_*]_x\psi(z_*,x_\perp)\,,
	\label{calh5plus}
	\\
	&&\hspace{-1.3cm}\calh^-_1(x_\perp,y_\perp) = - g^2\!\!\int_{-\infty}^{+\infty}\!\!dz_*d\omega_*\,
	\barpsi(\omega_*,y_\perp)[\omega_*,-\infty p_1]_y\,i\ssp_1\,\,[-\infty p_1,z_*]_x\psi(z_*,x_\perp)\,,
	\label{calh1minus}
	\\
	&&\hspace{-1.3cm}\calh^-_5(x_\perp,y_\perp) = - g^2\!\!\int_{-\infty}^{+\infty}\!\!dz_*d\omega_*\,
	\barpsi(\omega_*,y_\perp)[\omega_*,-\infty p_1]_y\,\gamma^5\ssp_1\,\,[-\infty p_1,z_*]_x\psi(z_*,x_\perp)\,,
	\label{calh5minus}
\end{eqnarray}
where the superscripts $+$ and $-$ remind that the semi-infinite Wilson lines point to $+\infty p_1$ and $-\infty p_1$ respectively.
We will write subsequent results in terms of the above defined operators.

\subsubsection{quark-to-gluon diagrams for $\tildeQ_{1x}$ and $\tildeQ_{5x}$}

\begin{figure}[thb]
\begin{center}
		\includegraphics[width=4.5in]{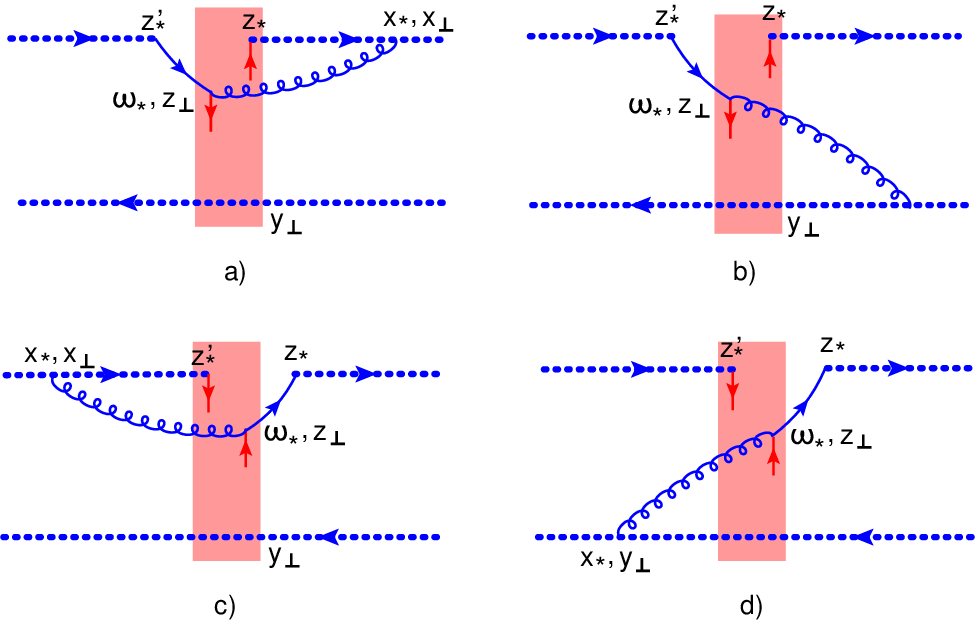}
		\caption{Quark-to-gluon diagrams for $\hat{\tildeQ}_1(x_\perp)$ and $\hat{\tildeQ}_5(x_\perp)$ operators.}
		\label{quark2gluon-diagram1}
\end{center}
\end{figure}

Let us start with the calculation for operators $\tildeQ_{1x}$ and $\tildeQ_{5x}$.
Diagram \ref{quark2gluon-diagram1}a is 
\begin{eqnarray}
\hspace{-1cm}&&\langle{\rm Tr}\{ U^\dagger_y\tilde{Q}_{1x}\}\rangle_{\small{\rm Fig.} \ref{quark2gluon-diagram1}a}
\nonumber\\
&&= g^2\!\!\int_{-\infty}^{+\infty}\!dz_*\!\int_{-\infty}^{z_*}dz'_*\,{\rm Tr}\big\{
U^\dagger_y
\nonumber\\
&&~~~\times ig\!\!\int_{z_*}^{+\infty}\!d{2\over s}x_*\,A^q_\bullet(x_*,x_\perp)[\infty p_1,z_*]_x
{\rm tr}\{\psi^{cl}(z_*,x_\perp)\barpsi^q(z'_*,x_\perp)i\ssp_1\}[z'_*,-\infty p_1]_x\big\}
\nonumber\\
&&= - g^4\!\!\int_{-\infty}^{+\infty}\!dz_*\!\int_{-\infty}^{z_*}\!dz'_*\!
\int_{z_*}^{+\infty}dx_*\!\int_{z'_*}^{x_*}\!d\omega_*
\!\int_0^{+\infty}\!{\dhd\alpha\over s^2\alpha^3}
\nonumber\\
&&~~~\times\Tr\Big\{U^\dagger_y t^a[\infty p_1, z_*]_x \,\tr\{\psi^{cl}(z_*,x_\perp)
\nonumber\\
&&~~~\times
\braxp e^{-i{\hatp^2_\perp\over \alpha s}x_*}p_\nu^\perp[x_*,\omega_*]^{ab}
\barpsi^{cl}(\omega_*)t^b\gamma^\nu_\perp[\omega_*,z'_*]
(\alpha\ssp_1+\ssp_\perp)e^{i{\hatp^2_\perp\over \alpha s}z'_*}\ketxp\,i\ssp_1\}\Big\}
\end{eqnarray}
where we used propagator (\ref{q2gpropagator1}) and used the fact that the classical shock-wave field has
line support only within the interval $[-\epsilon_*,\epsilon_*]$ (with $\epsilon_*>0$), and that, we are working in a gauge 
in which the gauge link made of pure gauge can be set to 1. So, in the limit of $\epsilon_*\to 0$,
the line support of the quantum fields, $x_*$ and $z'_*$ (with the condition that $x_*\ge z'_*$)
extend up to $0$, and, on the other hand, the line support of the classical field can be extend
to infinity, \textit{i.e.} we extend the limit of integration of $\omega^+$
to $]-\infty, \infty[$ and the gauge link becomes semi-infinite Wilson lines. Thus,
\begin{eqnarray}
\langle\Tr\{ U^\dagger_y\tildeQ_{1x}\}\rangle_{\small{\rm Fig.} \ref{quark2gluon-diagram1}a}
=\!\!&& - g^4\!\int_{-\infty}^{+\infty}\!dz_*d\omega_*\!\int_{-\infty}^0\!dz'_*\!
\int_0^{+\infty}dx_*
\!\int_0^{+\infty}\!{\dhd\alpha\over s^2\alpha^3}\int d^2z
\nonumber\\
&&\times
\Tr\big\{U^\dagger_y t^a[\infty p_1, z_*]_x \,\tr\{\psi^{cl}(z_*,x_\perp)
\nonumber\\
&&\times
\braxp e^{-i{\hatp^2_\perp\over \alpha s}x_*}p_\nu^\perp\ketzp[\infty p_1,\omega_*]^{ab}_z
\barpsi^{cl}(\omega_*,z_\perp)t^b\gamma^\nu_\perp
\nonumber\\
&&\times
[\omega_*,-\infty p_1]_z\brazp(\alpha\ssp_1+\ssp_\perp)e^{i{\hatp^2_\perp\over \alpha s}z'_*}\ketxp\,i\ssp_1\}\big\}
\end{eqnarray}
Next, we perform the integration over the light-cone coordinates $z'^+$ and $x^+$, and the Fourier transform
\begin{eqnarray}
&&\hspace{-2cm}\langle{\rm Tr}\{ U^\dagger_y\tilde{Q}_{1x}\}\rangle_{\small{\rm Fig.} \ref{quark2gluon-diagram1}a}
= {g^4\over 4\pi^2}\!\int_{-\infty}^{+\infty}\!dz_*d\omega_*
\!\int_0^{+\infty}\!{\dhd\alpha\over \alpha}\int d^2z\,
\Tr\left\{U^\dagger_y t^a[\infty p_1, z_*]_x \right.
\nonumber\\
&&\hspace{-2cm}
\left.~~~~~\times\tr\left\{\psi^{cl}(z_*,x_\perp)
[\infty p_1,\omega_*]^{ab}_z
\barpsi^{cl}(\omega_*,z_\perp)t^b[\omega_*,-\infty p_1]_z{(\ssx-\ssz)_\perp\over (x-z)^2_\perp}
{(\ssx-\ssz)_\perp\over (x-z)^2_\perp}\,i\ssp_1\right\}\right\}
\label{tQ1q2ga}
\end{eqnarray}
After color and Dirac algebra we arrive at
\begin{eqnarray}
\hspace{-1cm}\langle\Tr\{ U^\dagger_y\tilde{Q}_{1x}\}\rangle_{\small{\rm Fig.} \ref{quark2gluon-diagram1}a}
={\alpha_s\over 4\pi^2}\!\!\int_0^{+\infty}\!{d\alpha\over \alpha}
\!\int{d^2z\over (x-z)^2_\perp}
\Big[{1\over N_c}\Tr\big\{U^\dagger_y\calx_{1xz}\big\}
- {\rm Tr}\big\{U_z U^\dagger_y\}\calh^+_{1xz}\Big]
\label{quark2gluon-diagram1a1}
\end{eqnarray}

From eq. (\ref{tQ1q2ga}), substituting $i\ssp_1$ with $\gamma^5\ssp_1$ we get
\begin{eqnarray}
\hspace{-1cm}\langle{\rm Tr}\{ U^\dagger_y\tilde{Q}_{5x}\}\rangle_{\small{\rm Fig.} \ref{quark2gluon-diagram1}a}
= {\alpha_s\over 4\pi^2}\!\!\int_0^{+\infty}\!{d\alpha\over \alpha}
\!\int{d^2z\over (x-z)^2_\perp}
\Big[{1\over N_c}\Tr\big\{U^\dagger_y\calx_{5xz}\big\} - {\rm Tr}\big\{U_z U^\dagger_y\}\calh^+_{5xz}\Big]
\label{quark2gluon-diagram1a5}
\end{eqnarray}

Let us turn to diagram in Fig. \ref{quark2gluon-diagram1}b and perform similar steps as the ones we performed for
\ref{quark2gluon-diagram1}a
\begin{eqnarray}
&&\langle\Tr\{ U^\dagger_y\tilde{Q}_{1x}\}\rangle_{\small{\rm Fig.} \ref{quark2gluon-diagram1}b}
\\
&&= -ig^3\!\int_{-\infty}^{+\infty}\!dz_*\!\int_{-\infty}^{z_*}\!dz'_*\,
\Tr\big\{U^\dagger_y
\nonumber\\
&&~~~\times{2\over s}\!\int_0^{+\infty}\!dx_* A_\bullet(x_*,y_\perp)[\infty p_1,z_*]_x{\rm tr}\{i\ssp_1
\psi(z_*,x_\perp)\bar{\psi}(z'_*,x_\perp)\}[z'_*,-\infty p_1]_x\big\}
\nonumber\\
\hspace{-1cm}&&= g^4\!\int_{-\infty}^{+\infty}\!dz_*\!\int_{-\infty}^{z_*}\!dz'_*\!\int_0^{+\infty}\!dx_*\!
\int_{z'_*}^{+\infty}\!d\omega_*
\int_0^{+\infty}\!{\dhd\alpha\over s^2\alpha^3}\,{\rm Tr}\big\{U^\dagger_y t^a[\infty p_1,z_*]_x
\nonumber\\
&&
~~~\times{\rm tr}\{i\ssp_1\psi(z_*,x_\perp)
\brayp e^{-i{\hatp^2_\perp\over \alpha s}x_*}p_\nu^\perp[\infty p_1,\omega_*]^{ab}\bar{\psi}(\omega_*)t^b\gamma^\nu_\perp
[\omega_*,-\infty p_1]\ssp_\perp e^{i{\hatp^2_\perp\over \alpha s}z'_*}\ketxp\}\big\}
\nonumber
\end{eqnarray}
Now we perform the integration over the light-cone variable $z'^+$ and $x^+$ 
(taking into consideration the observations done for diagram \ref{quark2gluon-diagram1}a) and get
\begin{eqnarray}
\hspace{-0.8cm}\langle{\rm Tr}\{ U^\dagger_y\tilde{Q}_{1x}\}\rangle_{\small{\rm Fig.} \ref{quark2gluon-diagram1}b}
=\!\!&& - {\alpha_s\over 2\pi^2}\int_0^{+\infty}\!{\dhd\alpha\over \alpha}\!\int_{-\infty}^{+\infty}\!dz_*d\omega_*\!\int d^2z\,
{(y-z)_i(x-z)_j\over (y-z)^2_\perp(z-x)^2_\perp}g^2\Tr\big\{U^\dagger_y t^a
\nonumber\\
&&
\times [\infty p_1,z_*]_x
\tr\{\gamma^i\gamma^j i\,\ssp_1 \psi(z_*,x_\perp)\bar{\psi}(\omega_*,z_\perp)\}[\omega_*,\infty p_1]_z t^aU_z\big\}
\label{tQ1q2gb}
\end{eqnarray}
Performing color and Dirac algebra we obtain
\begin{eqnarray}
\hspace{-1cm}\langle{\rm Tr}\{ U^\dagger_y\tilde{Q}_{1x}\}\rangle_{\small{\rm Fig.} \ref{quark2gluon-diagram1}b}
=&& - {\alpha_s\over 4\pi^2}\int_0^{+\infty}\!{\dhd\alpha\over \alpha}\int d^2z
\label{quark2gluon-diagram1b1}\\
&&
\times\Bigg\{ {(x-z,z-y)\over (y-z)^2_\perp(z-x)^2_\perp} 
\bigg(\Tr\{U_z U^\dagger_y\}\calh^+_{1xz}
- {1\over N_c}\Tr\{U^\dagger_y\calx_{1xz}\}\bigg)
\nonumber\\
&&~~~ - {(\vec{x}-\vec{z})\times(\vec{y}-\vec{z})\over (y-z)^2_\perp(z-x)^2_\perp}
\bigg(\Tr\{U_z U^\dagger_y\}\calh^+_{5xz}
- {1\over N_c}{\rm Tr}\big\{U^\dagger_y \calx_{5xz}\big\}\bigg)	\Bigg\}
\nonumber
\end{eqnarray}
From (\ref{tQ1q2gb}), substituting $i\ssp_1$ with $\gamma^5\ssp_1$ we have
\begin{eqnarray}
\hspace{-1cm}\langle{\rm Tr}\{ U^\dagger_y\tilde{Q}_{5x}\}\rangle_{\small{\rm Fig.} \ref{quark2gluon-diagram1}b}
=\!\!&& - {\alpha_s\over 4\pi^2}\int_0^{+\infty}\!{\dhd\alpha\over \alpha}\!\int d^2z
\label{quark2gluon-diagram1b5}\\
&&
\times\Bigg\{{(x-z,z-y)\over (y-z)^2_\perp(z-x)^2_\perp} 
\bigg(\Tr\{U_z U^\dagger_y\}\calh^+_{5xz}
 - {1\over N_c}{\rm Tr}\big\{U^\dagger_y\calx_{5xz}\big\}\bigg)
\nonumber\\
&&~~~ + {(\vec{x}-\vec{z})\times(\vec{y}-\vec{z})\over (y-z)^2_\perp(z-x)^2_\perp}
\bigg(\Tr\{U_z U^\dagger_y\}\calh^+_{1xz}
- {1\over N_c}{\rm Tr}\big\{U^\dagger_y \calx_{1xz}\big\}\bigg)	\Bigg\}\,.
\nonumber
\end{eqnarray}
The calculation for diagrams \ref{quark2gluon-diagram1}c and \ref{quark2gluon-diagram1}d is the 
same as the previous two ones, so we may write 
\begin{eqnarray}
\hspace{-1cm}\langle{\rm Tr}\{ U^\dagger_y\tilde{Q}_{1x}\}\rangle_{\small{\rm Fig.} \ref{quark2gluon-diagram1}c}
=\!\!&& ig^3\!\int_{-\infty}^{+\infty}\!dz_*\!\int_{-\infty}^{z_*}\!dz'_*\!\int_{-\infty}^{z'_*}\!d{2\over s}x_*
\,\Tr\big\{U^\dagger_y[\infty p_1,z_*]_x
\label{quark2gluon-diagram1c1}\\
&&\times\tr\{\psi(z_*,x_\perp)\bar{\psi}(z'_*,x_\perp)i\ssp_1\}[z'_*,-\infty p_1]_xA_\bullet(x_*,x_\perp)\big\}
\nonumber\\
=\!\!&& - g^4\int_{-\infty}^{+\infty}\!dz'_*\!\int_{z'_*}^{+\infty}\!dz_*\!\int_{-\infty}^{z'_*}\!d{2\over s}x_*\!
\int_{x_*}^{z_*}\!d{2\over s}\omega_*\!\int_0^{+\infty}\!{\dhd\alpha\over 4\alpha^3}\int d^2z
\nonumber\\
&&\times\Tr\big\{U^\dagger_y\tr\{i\ssp_1\braxp e^{-i{\hatp_\perp^2\over \alpha s}z_*}
(\alpha \ssp_1+\ssp_\perp)\ketzp[z_*,\omega_*]\gamma^\nu_\perp t^b
\psi(\omega_*)[\omega_*,x_*]^{ba}
\nonumber\\
&&\times\brazp p_\nu^\perp e^{i{\hatp^2_\perp\over\alpha s}x_*} \ketxp\bar\psi(z'_*,x_\perp)\}
[z'_*,-\infty p_1]_x t^a\big\}
\nonumber\\
=\!\!&& - {\alpha_s\over 4\pi^2}\int_0^{+\infty}\!{d\alpha\over \alpha}
\!\int {d^2z\over (x-z)^2_\perp}
\Big[
 \Tr\{U^\dagger_y U_z\}\calh^-_{1zx}
- {1\over N_c}\Tr\{U^\dagger_y\calx_{1zx}\}\Big]\,.
\nonumber
\end{eqnarray}
For $\tildeQ_{5x}$ we have
\begin{eqnarray}
\hspace{-1cm}\langle{\rm Tr}\{ U^\dagger_y\tilde{Q}_{5x}\}\rangle_{\small{\rm Fig.} \ref{quark2gluon-diagram1}c}
= - {\alpha_s\over 4\pi^2}\int_0^{+\infty}\!{d\alpha\over \alpha}
\!\int {d^2z\over (x-z)^2_\perp}
\Big[
\Tr\{U^\dagger_y U_z\}\calh^-_{5zx} - {1\over N_c}\Tr\{U^\dagger_y\calx_{5zx}\}
\Big]\,.
\label{quark2gluon-diagram1c5}
\end{eqnarray}

Finally, diagram \ref{quark2gluon-diagram1}d is
\begin{eqnarray}
\hspace{-1cm}\langle{\rm Tr}\{ U^\dagger_y\tilde{Q}_{1x}\}\rangle_{\small{\rm Fig.} \ref{quark2gluon-diagram1}d}
=\!\!&& - {\alpha_s\over 4\pi^2}\int_0^{+\infty}\!{\dhd\alpha\over \alpha}\!\int d^2z
\label{quark2gluon-diagram1d1}\\
&&
\times\Bigg\{ {(x-z,z-y)\over (y-z)^2_\perp(z-x)^2_\perp} 
\bigg(\Tr\{U_z U^\dagger_y\}\calh^-_{1zx}
- {1\over N_c}{\rm Tr}\big\{U^\dagger_y \calx_{1zx}\big\}\bigg)
\nonumber\\
&&~~~ + {(\vec{x}-\vec{z})\times(\vec{y}-\vec{z})\over (y-z)^2_\perp(z-x)^2_\perp}
\bigg(\Tr\{U_z U^\dagger_y\}\calh^-_{5zx}
- {1\over N_c}{\rm Tr}\big\{U^\dagger_y \calx_{5zx}\big\}\bigg)	\Bigg\}\,.
\nonumber
\end{eqnarray}
For $\tildeQ_{5x}$ we have
\begin{eqnarray}
\hspace{-1cm}\langle{\rm Tr}\{ U^\dagger_y\tilde{Q}_{5x}\}\rangle_{\small{\rm Fig.} \ref{quark2gluon-diagram1}d}
=\!\!&& - {\alpha_s\over 4\pi^2}\int_0^{+\infty}\!{\dhd\alpha\over \alpha}\!\int d^2z
\label{quark2gluon-diagram1d5}\\
&&\times\Bigg\{ {(x-z,z-y)\over (y-z)^2_\perp(z-x)^2_\perp} 
\bigg(\Tr\{U_z U^\dagger_y\}\calh^-_{5zx}
- {1\over N_c}{\rm Tr}\big\{U^\dagger_y \calx_{5zx}\big\}\bigg)
\nonumber\\
&&~~~ - {(\vec{x}-\vec{z})\times(\vec{y}-\vec{z})\over (y-z)^2_\perp(z-x)^2_\perp}
\bigg(\Tr\{U_z U^\dagger_y\}\calh^-_{1zx}
- {1\over N_c}{\rm Tr}\big\{U^\dagger_y\calx_{1zx}\big\}\bigg)	\Bigg\}\,.
\nonumber
\end{eqnarray}

We now may sum up diagrams in Fig. \ref{quark2gluon-diagram1}.
Starting with $\tildeQ_{1x}$, we sum eqs. (\ref{quark2gluon-diagram1a1}), (\ref{quark2gluon-diagram1b1}),
(\ref{quark2gluon-diagram1c1}), and (\ref{quark2gluon-diagram1d1}) we have
\begin{eqnarray}
&&\langle\Tr\{ U^\dagger_y\tilde{Q}_{1x}\}\rangle_{\small{\rm Fig.} \ref{quark2gluon-diagram1}}
\nonumber\\
&&= - {\alpha_s\over 4\pi^2}\int_0^{+\infty}\!{d\alpha\over \alpha}
\!\int d^2z
\Bigg\{{1\over (x-z)^2_\perp}
\bigg[\Tr\big\{U_z U^\dagger_y\}\Big(\calh^+_{1xz} + \calh^-_{1zx}\Big)
- {1\over N_c}\Tr\big\{U^\dagger_y\big(\calx_{1xz} + \calx_{1zx}\big)\big\} \bigg]
\nonumber\\
&&
~~~ +{(x-z,z-y)\over (y-z)^2_\perp(z-x)^2_\perp} \bigg[
\Tr\{U_z U^\dagger_y\}\Big(\calh^+_{1xz} + \calh^-_{1zx}\Big)
- {1\over N_c}{\rm Tr}\big\{U^\dagger_y \big(\calx_{1xz}+\calx_{1zx}\big)\big\}
\bigg]
\nonumber\\
&& ~~~ + {(\vec{x}-\vec{z})\times(\vec{y}-\vec{z})\over (y-z)^2_\perp(z-x)^2_\perp}
\bigg[
\Tr\{U_z U^\dagger_y\}\Big(\calh^-_{5zx} - \calh^+_{5xz}\Big)
+ {1\over N_c}{\rm Tr}\big\{U^\dagger_y\big(\calx_{5xz}-\calx_{5zx}\big)\big\}\
\bigg]
\Bigg\}\,.
\label{quark2gluon-diagram1asum1}
\end{eqnarray}

The sum of diagrams in Fig. \ref{quark2gluon-diagram1} with $\tildeQ_{5x}$ 
is obtained summing eqs. (\ref{quark2gluon-diagram1a5}), (\ref{quark2gluon-diagram1b5}),
(\ref{quark2gluon-diagram1c5}), and (\ref{quark2gluon-diagram1d5}). So, we have
\begin{eqnarray}
&&\langle\Tr\{ U^\dagger_y\tildeQ_{5x}\}\rangle_{\small{\rm Fig.} \ref{quark2gluon-diagram1}}
\nonumber\\
&&= - {\alpha_s\over 4\pi^2}\int_0^{+\infty}\!{d\alpha\over \alpha}
\!\int d^2z
\Bigg\{{1\over (x-z)^2_\perp}\bigg[\Tr\big\{U_z U^\dagger_y\}\Big(\calh^+_{5xz} + \calh^-_{5zx}\Big)
- {1\over N_c}\Tr\big\{U^\dagger_y\big(\calx_{5xz} + \calx_{5zx}\big)\big\} \bigg]
\nonumber\\
&&
~~~ +{(x-z,z-y)\over (y-z)^2_\perp(z-x)^2_\perp} \bigg[
\Tr\{U_z U^\dagger_y\}\Big(\calh^+_{5xz} + \calh^-_{5zx}\Big)
- {1\over N_c}{\rm Tr}\big\{U^\dagger_y \Big(\calx_{5xz}+\calx_{5zx}\big)\big\}\bigg)
\bigg]
\nonumber\\
&&~~~ + {(\vec{x}-\vec{z})\times(\vec{y}-\vec{z})\over (y-z)^2_\perp(z-x)^2_\perp}
\bigg[
\Tr\{U_z U^\dagger_y\}\Big(\calh^+_{1xz} - \calh^-_{1zx}\Big)
+ {1\over N_c}{\rm Tr}\big\{U^\dagger_y\big(\calx_{1zx} - \calx_{1xz}\big)\big\}\
\bigg]\Bigg\}\,.
\label{quark2gluon-diagram1asum5}
\end{eqnarray}

\subsubsection{quark-to-gluon diagrams for $\hat{Q}_{1x}$ and $\hat{Q}_{5x}$}

\begin{figure}[thb]
		\begin{center}
		\includegraphics[width=4.5in]{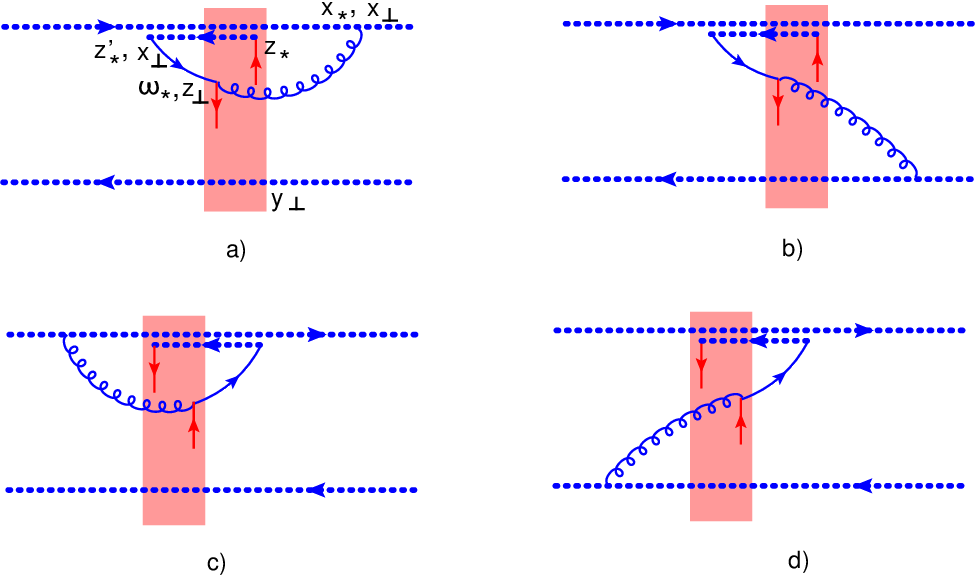}
		\caption{Quark-to-gluon diagrams for $\hat{Q}_1(x_\perp)$ and $\hat{Q}_5(x_\perp)$ operators.}
		\label{quark2gluon-diagram2}
	\end{center}
	\end{figure}

Now we turn our attention to the diagrams with the quark-to-gluon propagator for operators 
$\Tr\{\hat{U}_x\hat{U}^\dagger_y\}\hat{Q}_{1x}$ and $\Tr\{\hat{U}_x\hat{U}^\dagger_y\}\hat{Q}_{5x}$ 
given in Fig. \ref{quark2gluon-diagram2}. 
Note that, these type of diagrams will, after one loop, make the dipole operator $\Tr\{\hat{U}_x\hat{U}^\dagger_y\}$
\textit{talk} with operator $\hat{Q}_{1x}$ (or $\hat{Q}_{5x}$).

Let us start with the calculation of the first diagram, \ref{quark2gluon-diagram2}a. The procedure is similar to the one
adopted in the previous section
\begin{eqnarray}
\hspace{-0.8cm}\langle{\rm Tr}\{U_x U^\dagger_y\}Q_{1x}\rangle_{\small{\rm Fig.} \ref{quark2gluon-diagram2}a}
=\!\!&& ig^3{2\over s}\int_{-\infty}^{+\infty}\!dz_*\int_{-\infty}^{z_*}\!dz'_*\int_0^{+\infty}dx_*
\Tr\{t^aU_x U^\dagger_y\}
\nonumber\\
&&
\times
\Tr\big\{[z'_*,z_*]_x\tr\{i\ssp_1\psi^{cl}(z_*,x_\perp)\big\langle A^q_\bullet(x_*,x_\perp)\bpsi^q(z'_*,x_\perp)\big\rangle\}\big\}\,.
\end{eqnarray}
Using propagator (\ref{q2gpropagator1}) we have
\begin{eqnarray}
\hspace{-1cm}\langle{\rm Tr}\{U_x U^\dagger_y\}Q_{1x}\rangle_{\small{\rm Fig.} \ref{quark2gluon-diagram2}a}
=\!\!&& - g^4\int_{-\infty}^{+\infty}\!dz_*\!\int_{-\infty}^{z_*}\!dz'_*\!\int_0^{+\infty}dx_*\!\int_{z'_*}^{x_*}\!d\omega_*
\int_0^{+\infty}\!{\dhd\alpha\over s^2\alpha^3}\Tr\{t^a U_x U^\dagger_y\}
\nonumber\\
&&
\times\!\int d^2z\Tr\big\{[-\infty p_1,z_*]_x\tr\{i\ssp_1\psi(z_*,x_\perp)
\braxp e^{-i{\hatp^2_\perp\over \alpha s}x_*}p^\perp_\nu\ketzp
\nonumber\\
&&
\times[\infty p_1,\omega_*]^{ab}_z\bpsi(\omega_*,z_\perp)t^b\gamma_\perp^\nu
[\omega_*,-\infty p_1]_z\brazp\ssp_\perp e^{i{\hatp^2_\perp\over \alpha s}z'_*}\ketxp\}\big\}\,.
\end{eqnarray}
Next step is again to recognize the range of the quantum and classical fields as we did for diagram 
in Fig. \ref{quark2gluon-diagram1}. So, we have
\begin{eqnarray}
\hspace{-2cm}
&&\langle{\rm Tr}\{U_x U^\dagger_y\}Q_{1x}\rangle_{\small{\rm Fig.} \ref{quark2gluon-diagram2}a}
\nonumber\\
&&= - {\alpha_s\over 2\pi^2}\!\int_0^{+\infty}\!{d\alpha\over \alpha}\,g^2\!\int_{-\infty}^{+\infty}dz_*d\omega_*\!
\int {d^2z\over (x-z)^2_\perp}
\nonumber\\
&&~~~ \times\Tr\{t^a U_x U^\dagger_y\}
\Tr\big\{[-\infty p_1,z_*]_x {\rm tr}\{i\ssp_1\psi(z_*,x_\perp)
\bpsi(\omega_*,z_\perp)\}[\omega_*,\infty p_1]_z t^a U_z\big\}
\nonumber\\
&&= - {\alpha_s\over 2\pi^2}\!\int_0^{+\infty}\!{d\alpha\over \alpha}g^2\!\int_{-\infty}^{+\infty}dz_*d\omega_*\!
\int {d^2z\over (x-z)^2_\perp}
\nonumber\\
&&
~~~ \times\Big[
\half\Tr\big\{U_x U^\dagger_y U_z [-\infty p_1,z_*]_x 
\tr\{i\ssp_1 \psi(z_*,x_\perp)\bpsi(\omega_*,z_\perp)\}[\omega_*,\infty p_1]_z\big\}
\nonumber\\
&& ~~~ - {1\over 2N_c}\Tr\{U_x U^\dagger_y\}\Tr\big\{[-\infty p_1,z_*]_x
\tr\{i\ssp_1\psi(z_*,x_\perp)\bpsi(\omega_*,z_\perp)\}[\omega_*,-\infty p_1]_z\big\}\Big]\,.
\end{eqnarray}
Using definition of operators (\ref{calx1})-(\ref{calx5d}) and (\ref{calh1plus})-(\ref{calh5minus}) we arrive at
\begin{eqnarray}
\hspace{-1cm}
\langle{\rm Tr}\{U_x U^\dagger_y\}Q_{1x}\rangle_{\small{\rm Fig.} \ref{quark2gluon-diagram2}a}
=\!\! &&  {\alpha_s\over 4\pi^2}\!\int_0^{+\infty}\!{d\alpha\over \alpha}\!
\int {d^2z\over (x-z)^2_\perp}
\nonumber\\
&&\times\Big[\Tr\big\{U_x U^\dagger_y U_z\calx_{1zx}^\dagger\big\}
+ {1\over N_c}\Tr\{U_x U^\dagger_y\}\calh^-_{1xz}\Big]\,.
\label{quark2gluon-diagram2a1}
\end{eqnarray}

Substituting $i\ssp_1$ with $\gamma^5\ssp_1$ we get
\begin{eqnarray}
\hspace{-1cm}
\langle{\rm Tr}\{U_x U^\dagger_y\}Q_{5x}\rangle_{\small{\rm Fig.} \ref{quark2gluon-diagram2}a}
=\!\! && - {\alpha_s\over 4\pi^2}\!\int_0^{+\infty}\!{d\alpha\over \alpha}\!
\int {d^2z\over (x-z)^2_\perp}
\nonumber\\
&&\times\Big[
\Tr\big\{U_x U^\dagger_y U_z \calx_{5zx}^\dagger\big\} - {1\over N_c}\Tr\{U_x U^\dagger_y\}\calh^-_{5xz}\Big]\,.
\label{quark2gluon-diagram2a5}
\end{eqnarray}
Similarly, for diagram \ref{quark2gluon-diagram2}b we have
\begin{eqnarray}
&&\langle{\rm Tr}\{U_x U^\dagger_y\}Q_{1x}\rangle_{\small{\rm Fig.} \ref{quark2gluon-diagram2}b}
\nonumber\\
&&= -ig^3{2\over s}\int_{-\infty}^{+\infty}dz_*\!\int_{-\infty}^{z_*}\!dz'_*\!\int_0^{+\infty}\!dx_*
\nonumber\\
&&~~~\times
\Tr\{U_x U^\dagger_y t^a\}
\Tr\big\{[z'_*,z_*]_x\tr\{i\ssp_1\psi^{cl}(z_*,x_\perp)
\big\langle A^{q,a}_\bullet(x_*,y_\perp)\bpsi^q(z'_*,x_\perp)\big\rangle\}\big\}
\nonumber\\
&&= g^4\! \int_{-\infty}^{+\infty}\!dz_*\int_{-\infty}^{z_*}\!dz'_*\int_0^{+\infty}\!dx_*\int_{z'_*}^{x_*}\!d\omega_*
\int_0^{+\infty}{\dhd\alpha \over s^2\alpha^3}\,\Tr\{U_x U^\dagger_y t^a\}\int\! d^2z
\nonumber\\
&&
~~~\times\Tr\big\{[z'_*,z_*]_x \tr\{i\ssp_1\psi(z_*,x_\perp)
\brayp e^{-i{\hatp^2_\perp\over \alpha s}x_*}p_\perp^\nu\ketzp [x_*,\omega_*]_z^{ab}\bpsi(\omega_*,z_\perp)
\nonumber\\
&&~~~\times t^b\gamma_\nu^\perp
[\omega_*,z'_*]_z\brazp\ssp_\perp e^{i{\hatp^2_\perp\over \alpha s}z'_*}\ketxp\}\big\}\,.
\end{eqnarray}
After Fourier transform and integration over the light-cone coordinates we have
\begin{eqnarray}
\hspace{-0.5cm}\langle{\rm Tr}\{U_x U^\dagger_y\}Q_{1x}\rangle_{\small{\rm Fig.} \ref{quark2gluon-diagram2}b}
=\!\!&& - {\alpha_s\over 2\pi^2} g^2\!\int_{-\infty}^{+\infty}\!dz_*d\omega_*
\int_0^{+\infty}{d\alpha \over \alpha}\,\Tr\{U_x U^\dagger_y t^a\}\int\! d^2z
\nonumber\\
&&\times\Tr\big\{[-\infty p_1,z_*]_x \tr\{i\ssp_1\psi(z_*,x_\perp){(y-z)^\nu_\perp\over (y-z)^2_\perp}
[\infty p_1,\omega_*]_z^{ab}\bpsi(\omega_*,z_\perp)
\nonumber\\
&&\times t^b\gamma_\nu^\perp
[\omega_*,-\infty p_1]_z {(\ssx-\ssz)\over (x-z)^2_\perp}\}\big\}\,.
\end{eqnarray}
Performing color and Dirac algebra we obtain
\begin{eqnarray}
\hspace{-1cm}
\langle{\rm Tr}\{U_x U^\dagger_y\}Q_{1x}\rangle_{\small{\rm Fig.} \ref{quark2gluon-diagram2}b}
=\!\!&&  {\alpha_s\over 4\pi^2}\!\int_0^{+\infty}{d\alpha \over \alpha}\int\! d^2z
\label{quark2gluon-diagram2b1}\\
&&\hspace{0.2cm}\times\Bigg\{
{(x-z,z-y)_\perp\over (y-z)^2_\perp (x-z)^2_\perp} 
\Big(\Tr\{U_xU^\dagger_y U_z\calx_{1zx}^\dagger\}
+{1\over N_c}\Tr\{U_xU^\dagger_y\}\calh^-_{1xz}\Big)
\nonumber\\
&&\hspace{0.2cm} + {(\vec{x}-\vec{z})\times(\vec{y}-\vec{z})\over (y-z)^2_\perp (x-z)^2_\perp}
\Big(\Tr\{U_xU^\dagger_y U_z\calx_{5zx}^\dagger\}
-{1\over N_c}\Tr\{U_xU^\dagger_y\}\calh^-_{5xz}
\Big)\Bigg\}\,.
\nonumber
\end{eqnarray}

For $Q_5(x_\perp)$ we have
\begin{eqnarray}
\langle{\rm Tr}\{U_x U^\dagger_y\}Q_{5x}\rangle_{\small{\rm Fig.} \ref{quark2gluon-diagram2}b} 
=\!\!&& - {\alpha_s\over 4\pi^2}\!
\int_0^{+\infty}{d\alpha \over \alpha}\int\! d^2z
\label{quark2gluon-diagram2b5}\\
&&\hspace{0.2cm}\times\Bigg\{
{(x-z,z-y)_\perp\over (y-z)^2_\perp (x-z)^2_\perp} 
\Big(\Tr\{U_xU^\dagger_y U_z\calx_{5zx}^\dagger\}
-{1\over N_c}\Tr\{U_xU^\dagger_y\}\calh^-_{5xz}
\Big)
\nonumber\\
&&\hspace{0.2cm} - {(\vec{x}-\vec{z})\times(\vec{y}-\vec{z})\over (y-z)^2_\perp (x-z)^2_\perp}
\Big(\Tr\{U_xU^\dagger_y U_z\calx^\dagger_{1zx}\}
+ {1\over N_c}\Tr\{U_xU^\dagger_y\}\calh^-_{1xz}
\Big)\Bigg\}\,.
\nonumber
\end{eqnarray}

Diagram in Fig. \ref{quark2gluon-diagram2}c is
\begin{eqnarray}
\langle{\rm Tr}\{U_x U^\dagger_z\}Q_{1x}\rangle_{\small{\rm Fig.} \ref{quark2gluon-diagram2}c}
=\!\!&& - g^4 \int_{-\infty}^{+\infty}\!\!dz'_*d\omega_*
\!\int_{z'_*}^{+\infty}\!dz_*\!\int_{\omega_*}^0\!dx_*
\!\int_0^{+\infty}\!{\dhd\alpha\over \alpha^3 s^2}
\Tr\{U_x t^aU^\dagger_y\}
\nonumber\\
&&\hspace{-3cm}~~\times\!\int d^2z\,\Tr\big\{{\rm tr}\{i\ssp_1
\braxp e^{-i{\hatp^2_\perp\over \alpha s}z_*}\ssp_\perp\ketzp [\infty p_1,\omega_*]_z\gamma^\nu_\perp t^b
\psi(\omega_*,z_\perp)[\omega_*,x_*]_z^{ba}
\nonumber\\
&&\hspace{-3cm}~~\times\brazp p_\perp^\nu\,e^{i{\hatp^2_\perp\over \alpha s}x_*}\ketxp
\bar{\psi}(z'_*,x_\perp)[z'_*,\infty p_1]_x\}\big\}
\nonumber\\
&&\hspace{-3cm}
= - {\alpha_s\over 2\pi^2}\int_0^{+\infty}\!{d\alpha \over \alpha}\,g^2\!\int_{-\infty}^{+\infty}\!dz'_*d\omega_*
\int {d^2z\over (x-z)^2_\perp}\,\Tr\{U^\dagger_y U_x t^a\}
\nonumber\\
&&\hspace{-3cm}~~\times\Tr\big\{t^a[-\infty p_1,\omega_*]_z
\tr\{i\ssp_1\psi(\omega_*,z_\perp)\bar{\psi}(z'_*,x_\perp)\}[z'_*,\infty p_1]_x U_z\big\}
\nonumber\\
&&\hspace{-3cm}
= {\alpha_s\over 4\pi^2}\int_0^{+\infty}\!{d\alpha \over \alpha}
\int {d^2z\over (x-z)^2_\perp}
\bigg[\Tr\big\{U_zU^\dagger_y U_x \calx^\dagger_{1xz}\big\}
+ {1\over N_c}{\rm Tr}\{U^\dagger_y U_x\}\calh^+_{1zx}
\bigg]\,.
\label{quark2gluon-diagram2c1}
\end{eqnarray}

For operator $Q_{5x}$ we have
\begin{eqnarray}
\langle{\rm Tr}\{U_x U^\dagger_z\}Q_{5x}\rangle_{\small{\rm Fig.} \ref{quark2gluon-diagram2}c} 
=\!\!&&
 - {\alpha_s\over 4\pi^2}\int_0^{+\infty}\!{d\alpha \over \alpha}
\int {d^2z\over (x-z)^2_\perp}
\nonumber\\
&&\times\bigg[\Tr\big\{U_zU^\dagger_y U_x \calx^\dagger_{5xz}\big\}
- {1\over N_c}{\rm Tr}\{U^\dagger_y U_x\}\calh^+_{5zx}\bigg]\,.
\label{quark2gluon-diagram2c5}
\end{eqnarray}

Last diagram of this set of diagrams is \ref{quark2gluon-diagram2}c. The steps we have to perform to calculate
this diagram are again the same as the ones performed above. We have
\begin{eqnarray}
&&\langle\Tr\{U_x U^\dagger_y\}Q_{1x}\rangle_{\small{\rm Fig.} \ref{quark2gluon-diagram2}d}
\nonumber\\
&&= g^4\!\int_{-\infty}^{+\infty}\!dz'_*\int_{-\infty}^{z_*}\!d\omega_*
\!\int_{z'_*}^{+\infty}\!dz_*\int_{-\infty}^0\!dx_*\!\int_0^{+\infty}\!{\dhd\alpha\over \alpha^3 s^2}
\Tr\{U_x t^a U^\dagger_y\}
\nonumber\\
&&
~~~\times\!\int d^2z\,\Tr\big\{[z'_*,z_*]_x \tr\{i\ssp_1
\braxp e^{-i{\hatp^2_\perp\over \alpha s}z_*}\ssp_\perp\ketzp
[z_*,\omega_*]_z\gamma^\nu_\perp t^b\psi(\omega_*,z_\perp)[\omega_*,x_*]^{ba}_z
\nonumber\\
&&
~~~\times\!
\brazp{p^\perp_\nu}e^{i{\hatp^2_\perp\over \alpha s}x_*}\ketyp\bar{\psi}(z'_*,x_\perp)\}\big\}\,.
\end{eqnarray}

Fourier transform and integration over the longitudinal coordinates give
\begin{eqnarray}
&&\langle\Tr\{U_x U^\dagger_y\}Q_{1x}\rangle_{\small{\rm Fig.} \ref{quark2gluon-diagram2}d}
\nonumber\\
&&= {\alpha_s\over 2\pi^2}\int_0^{+\infty}\!{d\alpha\over \alpha}\,g^2\!\int_{-\infty}^{+\infty}\!dz'_*d\omega_*
\!\int d^2z{(x-z)_i(z-y)_j\over (x-z)^2_\perp(z-y)^2_\perp}\Tr\{U^\dagger_yU_x t^a\}
\nonumber\\
&&~~~\times
\Tr\big\{t^a
[-\infty p_1,\omega_*]_z{\rm tr}\{i\ssp_1\gamma^i\gamma^j\psi(\omega_*,z_\perp)\bar{\psi}(z'_*,x_\perp)\}
[z'_*,\infty p_1]_x U_z\big\}\,.
\end{eqnarray}
Color and Dirac algebra give
\begin{eqnarray}
\hspace{-1cm}&&\langle\Tr\{U_x U^\dagger_y\}Q_{1x}\rangle_{\small{\rm Fig.} \ref{quark2gluon-diagram2}d}
\nonumber\\
\hspace{-1cm}&&= {\alpha_s\over 4\pi^2}\int_0^{+\infty}\!{d\alpha\over \alpha}\!\int d^2z
\Bigg\{{(x-z,z-y)_\perp\over (x-z)^2_\perp(z-y)^2_\perp}
\bigg[{\rm Tr}\big\{U_zU^\dagger_y U_x\calx^\dagger_{1xz} \big\}
+ {1\over N_c}\Tr\{U^\dagger_y U_x\}
\calh^+_{1zx}\bigg]
\nonumber\\
\hspace{-1cm}&&
~~~~~~ - {(\vec{x}-\vec{z})\times(\vec{y}-\vec{z})\over (x-z)^2_\perp(z-y)^2_\perp}
\bigg[\Tr\big\{U_zU^\dagger_y U_x\calx_{5xz}^\dagger\big\}
- {1\over N_c}\Tr\{U^\dagger_y U_x\}\calh_{5zx}^+
\bigg]\Bigg\}\,.
\label{quark2gluon-diagram2d1}
\end{eqnarray}
With $Q_{5x}$ we have
\begin{eqnarray}
&&\langle\Tr\{U_x U^\dagger_y\}Q_{5x}\rangle_{\small{\rm Fig.} \ref{quark2gluon-diagram2}d} 
\nonumber\\
&&= - {\alpha_s\over 4\pi^2}\int_0^{+\infty}\!{d\alpha\over \alpha}\!\int d^2z
\Bigg\{{(x-z,z-y)_\perp\over (x-z)^2_\perp(z-y)^2_\perp}
\bigg[\Tr\big\{U_zU^\dagger_y U_x\calx^\dagger_{5xz} \big\}
- {1\over N_c}\Tr\{U^\dagger_y U_x\}
\calh^+_{5zx}\bigg]
\nonumber\\
&&\hspace{1cm}
+ {(\vec{x}-\vec{z})\times(\vec{y}-\vec{z})\over (x-z)^2_\perp(z-y)^2_\perp}
\bigg[\Tr\big\{U_zU^\dagger_y U_x\calx_{1xz}^\dagger\big\}
+ {1\over N_c}\Tr\{U^\dagger_y U_x\}\calh_{1zx}^+
\bigg]\Bigg\}\,.
\label{quark2gluon-diagram2d5}
\end{eqnarray}

Let us sum up diagrams in Fig. \ref{quark2gluon-diagram2}. 
Summing (\ref{quark2gluon-diagram2a1}), (\ref{quark2gluon-diagram2b1}),
(\ref{quark2gluon-diagram2c1}), and (\ref{quark2gluon-diagram2d1}) we have
\begin{eqnarray}
\hspace{-1cm}&&\langle\Tr\{U_x U^\dagger_y\}Q_{1x}\rangle_{\small{\rm Fig.} \ref{quark2gluon-diagram2}}
\nonumber\\
\hspace{-1cm}&&= {\alpha_s\over 4\pi^2}\!\int_0^{+\infty}\!{d\alpha\over \alpha}\!
\int d^2z\,\Bigg\{ {1\over (x-z)^2_\perp}
\bigg[\Tr\big\{U_x U^\dagger_y U_z\calx_{1zx}^\dagger\big\} 
\nonumber\\
\hspace{-1cm}&&
~~~+ \Tr\{U_zU^\dagger_yU_x\calx_{1xz}^\dagger\} + {1\over N_c}\Tr\{U_x U^\dagger_y\}\Big(\calh^-_{1xz} + \calh^+_{1zx}\Big)\bigg]
+ {(x-z,z-y)_\perp\over (y-z)^2_\perp (x-z)^2_\perp} 
\nonumber\\
\hspace{-1cm}&&
~~~\times\bigg[\Tr\{U_xU^\dagger_y U_z\calx_{1zx}^\dagger\} + \Tr\big\{U_zU^\dagger_y U_x\calx^\dagger_{1xz} \big\}
+ {1\over N_c}\Tr\{U_xU^\dagger_y\}\Big(\calh^-_{1xz} + \calh^+_{1zx}\Big)\bigg]
\nonumber\\
\hspace{-1cm}&&
~~~+ {(\vec{x}-\vec{z})\times(\vec{y}-\vec{z})\over (y-z)^2_\perp (x-z)^2_\perp}
\bigg[ \Tr\{U_xU^\dagger_y U_z\calx_{5zx}^\dagger\}
 - \Tr\big\{U_zU^\dagger_y U_x\calx_{5xz}^\dagger\big\}
\nonumber\\
\hspace{-1cm}&&
~~~ - {1\over N_c}\Tr\{U_xU^\dagger_y\}\Big(\calh^-_{5xz} - \calh_{5zx}^+\Big)\bigg]\Bigg\}\,.
\label{quark2gluon-diagram2sum1}
\end{eqnarray}

Summing, instead, (\ref{quark2gluon-diagram2a5}), (\ref{quark2gluon-diagram2b5}), (\ref{quark2gluon-diagram2c5}), and
(\ref{quark2gluon-diagram2d5}) we have
\begin{eqnarray}
\hspace{-1cm}&&\langle\Tr\{U_x U^\dagger_y\}Q_{5x}\rangle_{\small{\rm Fig.} \ref{quark2gluon-diagram2}}
\nonumber\\
\hspace{-1cm}&&= - {\alpha_s\over 4\pi^2}\!\int_0^{+\infty}\!{d\alpha\over \alpha}\!
\int d^2z\,\Bigg\{ {1\over (x-z)^2_\perp}
\bigg[\Tr\big\{U_xU^\dagger_y U_z \calx^\dagger_{5zx}\big\}
\nonumber\\
\hspace{-1cm}&&
+ \Tr\big\{U_zU^\dagger_y U_x \calx^\dagger_{5xz}\big\}  
- {1\over N_c}{\rm Tr}\{U^\dagger_y U_x\}\Big(\calh^-_{5xz} + \calh^+_{5zx}\Big)\bigg]
+ {(x-z,z-y)_\perp\over (y-z)^2_\perp (x-z)^2_\perp} 
\nonumber\\
\hspace{-1cm}&&
\times\bigg[\Tr\{U_xU^\dagger_y U_z\calx_{5zx}^\dagger\} + \Tr\big\{U_zU^\dagger_y U_x\calx^\dagger_{5xz} \big\}
-{1\over N_c}\Tr\{U_xU^\dagger_y\}\Big(\calh^-_{5xz} + \calh^+_{5zx}\Big)\bigg]
\nonumber\\
\hspace{-1cm}&&
+ {(\vec{x}-\vec{z})\times(\vec{y}-\vec{z})\over (y-z)^2_\perp (x-z)^2_\perp}
\bigg[\Tr\{U_zU^\dagger_y U_x\calx_{1xz}^\dagger\big\} - \Tr\{U_xU^\dagger_y U_z\calx_{1zx}^\dagger\}
\nonumber\\
\hspace{-1cm}&&
+ {1\over N_c}\Tr\{U_xU^\dagger_y\}\Big(\calh_{1zx}^+ - \calh^-_{1xz}\Big)\bigg]
\Bigg\}\,.
\label{quark2gluon-diagram2sum5}
\end{eqnarray}

\section{Summary of evolution equations}
\label{sec: summaryevo}

In this section we summarize the evolution equations we have calculated. 
As already discussed in the previous sections, the operators of which we want to find the evolution equations are
the ones proportional to the impact factors. The operator proportional to the impact factor $\cali_1^{\mu\nu}$ is
\begin{eqnarray}
&&\Tr\{\hat{\calq}_{1x}\hat{U}^\dagger_y\}
= \half \Tr\{\hat{U}^\dagger_y \hat{U}_x\}\hat{Q}_{1\,x} - 
{1\over 2N_c} \Tr\{\hat{U}^\dagger_y \hat{\tilde{Q}}_{1x}\}
\end{eqnarray}
The operator proportional to the impact factor $\cali_5^{\mu\nu}$ is, for flavor singlet,
\begin{eqnarray}
&&\Tr\{\big(\hat{\calq}_{5z_2}+\hat{\calf}_{z_2}\big)\hat{U}^\dagger_{z_1}\}
+ \Tr\{\big(\hat{\calq}_{5z_2}+\hat{\calf}^\dagger_{z_2}\big)\hat{U}_{z_1}\}
\nonumber\\
&&~~~~ =
\Tr\{\hat{U}^\dagger_{z_1} \hat{U}_{z_2}\}\hat{Q}_{5z_2} +  \Tr\{\hat{U}_{z_1} \hat{U}_{z_2}\}\hat{Q}_{5z_2}^\dagger
\nonumber\\
&&
~~~~~~~-  {1\over N_c} \Tr\{\hat{U}^\dagger_{z_1} \big(\hat{\tilde{Q}}_{5z_2} - 2N_c\hat{\calf}_{z_2}\big)\}
- {1\over N_c}\Tr\{\hat{U}_{z_1}\big(\hat{\tilde{Q}}^\dagger_{5z_2} - 2N_c\hat{\calf}^\dagger_{z_2}\big)\}
\end{eqnarray}
and for flavor non-singlet case
\begin{eqnarray}
&&\Tr\{\hat{\calq}_{5z_2}\hat{U}^\dagger_{z_1}\} + \Tr\{\hat{\calq}_{5z_2}\hat{U}_{z_1}\}
\nonumber\\
&&=
\Tr\{\hat{U}^\dagger_{z_1} \hat{U}_{z_2}\}\hat{Q}_{5z_2} +  \Tr\{\hat{U}_{z_1} \hat{U}_{z_2}\}\hat{Q}_{5z_2}^\dagger
-  {1\over N_c} \Tr\{\hat{U}^\dagger_{z_1} \hat{\tilde{Q}}_{5z_2}\}
- {1\over N_c}\Tr\{\hat{U}_{z_1}\hat{\tilde{Q}}^\dagger_{5z_2} \}
\end{eqnarray}

The evolution equations that we are going to collect in this sections are not strictly speaking evolution equations
of the type ${d\over d\eta}\calo = K\otimes \calo$ like the BK equation (\ref{balitskyeq}).
The reason lies in the fact that there are divergences of the type $d^2z\over (x-z)^2_\perp$ which
are responsible for the double log contributions peculiar of the polarized structure functions at small-$x_B$
and of the unpolarized quark structure functions as well. In this work we limit ourself to calculate
all the diagrams at one loop level, and properly analyze the double log of energy contributions
in a separate publication. Actually, in Appendix \ref{sec: Diag-calq1claq5} , eq. (\ref{calfpluscalq5}), we obtain the double log of energy
evolution equation which agree with the one calculated in Ref. \cite{Kovchegov:2015pbl, Kovchegov:2016zex, Kovchegov:2018znm}, but
as we will argue in the conclusions, the evolution equations will differ when 
written in terms of the operators $Q_{5x}$ and $\tildeQ_{5x}$.

\subsection{$Q_{1x}$ and $\tildeQ_{1x}$ operators}

Here we collect the evolution equations for $Q_{1x}$ and $\tildeQ_{1x}$.

Summing (\ref{Q1quantum-diagram-aball}), (\ref{Q1bkdiagrams}), and (\ref{quark2gluon-diagram2sum1}) we have
\begin{eqnarray}
\hspace{-0.5cm}&&\langle \Tr\{U^\dagger_y U_x\}Q_{1x}\rangle
\nonumber\\
&&= {\alpha_s\over 4\pi^2}\int_0^{+\infty}\!{d\alpha\over \alpha}\int d^2z\Bigg\{
{\Tr\{U^\dagger_y U_x\}\over (x-z)^2_\perp}\,\bigg[
\Tr\{U^\dagger_x U_z\}\,Q_{1z} - {1\over N_c}{\rm Tr}\{U^\dagger_x \tilde{Q}_{1z}\}\bigg]
\nonumber\\
&&~~~ + {2Q_{1x}(x-y)^2_\perp\over (x-z)^2_\perp(y-z)^2_\perp}
\Big[\Tr\{U_xU^\dagger_z\}\Tr\{U_zU^\dagger_y\} - N_c\Tr\{U_x U^\dagger_y\}\Big]
\nonumber\\
&&~~~+
{1\over (x-z)^2_\perp}
\bigg[\Tr\big\{U_x U^\dagger_y U_z\calx_{1zx}^\dagger\big\} + \Tr\{U_zU^\dagger_yU_x\calx_{1xz}^\dagger\}
+ {1\over N_c}\Tr\{U_x U^\dagger_y\}\Big(\calh^-_{1xz} + \calh^+_{1zx}\Big)\bigg]
\nonumber\\
&&~~~
+ {(x-z,z-y)_\perp\over (y-z)^2_\perp (x-z)^2_\perp} 
\bigg[\Tr\{U_xU^\dagger_y U_z\calx_{1zx}^\dagger\} + \Tr\big\{U_zU^\dagger_y U_x\calx^\dagger_{1xz} \big\}
\nonumber\\
&&~~~
+ {1\over N_c}\Tr\{U_xU^\dagger_y\}\Big(\calh^-_{1xz} + \calh^+_{1zx}\Big)\bigg] 
+ {(\vec{x}-\vec{z})\times(\vec{y}-\vec{z})\over (y-z)^2_\perp (x-z)^2_\perp}
\bigg[ \Tr\{U_xU^\dagger_y U_z\calx_{5zx}^\dagger\}
\nonumber\\
&&~~~ - \Tr\big\{U_zU^\dagger_y U_x\calx_{5xz}^\dagger\big\} 
- {1\over N_c}\Tr\{U_xU^\dagger_y\}\Big(\calh^-_{5xz} - \calh_{5zx}^+\Big)\bigg]\Bigg\}\,.
\label{SumQ1}
\end{eqnarray}
Summing  eqs. (\ref{LOpsibarpsiadj5sum-tildeq1}), (\ref{BKevolutionQ1}),  (\ref{quark2gluon-diagram1asum1})
\begin{eqnarray}
\hspace{-1cm}&&\langle {\rm Tr}\{U^\dagger_y \tildeQ_{1x}\} \rangle
\nonumber\\
&&= {\alpha_s\over 4\pi^2}\!\int^{+\infty}_0\!{d\alpha\over \alpha}\!\int d^2z\,
\Bigg\{{1\over (x-z)^2_\perp}	\bigg[
\Tr\big\{U_z U^\dagger_y\}\Big(Q_{1z} - \calh^+_{1xz} - \calh^-_{1zx}\Big)
\nonumber\\
&&\hspace{5cm} + {1\over N_c}\Tr\big\{U^\dagger_y\big(\calx_{1xz} + \calx_{1zx} - \tildeQ_{1z}\big)\big\}
\bigg]
\nonumber\\
&&
~~~ - {(x-z,z-y)\over (y-z)^2_\perp(z-x)^2_\perp} \bigg[
\Tr\{U_z U^\dagger_y\}\Big(\calh^+_{1xz} + \calh^-_{1zx}\Big)
- {1\over N_c}{\rm Tr}\big\{U^\dagger_y \big(\calx_{1xz}+\calx_{1zx}\big)\big\}\bigg]
\nonumber\\
&&
~~~ - {(\vec{x}-\vec{z})\times(\vec{y}-\vec{z})\over (y-z)^2_\perp(z-x)^2_\perp}
\bigg[
\Tr\{U_z U^\dagger_y\}\Big(\calh^-_{5zx} - \calh^+_{5xz}\Big)
+ {1\over N_c}{\rm Tr}\big\{U^\dagger_y\big(\calx_{5xz}-\calx_{5zx}\big)\big\}
\bigg]
\nonumber\\
&&
~~~ + {2(x-y)^2_\perp\over (x-z)^2_\perp(z-y)^2_\perp}\Big(
{\rm Tr}\{U_z U^\dagger_y\}{\rm Tr}\{U^\dagger_z \tilde{Q}_{1\,x}\} 
- N_c{\rm Tr}\{U^\dagger_y\tilde{Q}_{1\,x}\}\Big)
\Bigg\}\,.
\label{SumtildeQ1}
\end{eqnarray}
For operator $\hat{\calq}_1(x_\perp)$, instead, we have to sum (\ref{BKevolutionQ1}) (with $\tildeQ_1(x_\perp)$ replaced by 
$\calq_1(x_\perp)$),
(\ref{quantum-oneloopcaq1}), and (\ref{calq1-q2g}) and get
\begin{eqnarray}
\hspace{-1cm}&&\big\langle\Tr\{\calq_{1x}\, U^\dagger_y\} 
\big\rangle
\nonumber\\
&&={\alpha_s\over 2\pi^2}
\int_{0}^{+\infty}{d\alpha\over \alpha}\int d^2z\Bigg\{
\half{1\over (x-z)^2_\perp}
\Big(\Tr\{U^\dagger_x \calq_{1x}\}\Tr\{U^\dagger_y U_x\}
-{1\over N_c}\Tr\{\calq_{1z}U^\dagger_y\}\Big)
\nonumber\\
&&~~~ +  {(x-y)^2_\perp \over (x-z)^2_\perp(y-z)^2_\perp}
\Big(\Tr\{U^\dagger_z \calq_{1x}\}\Tr\{U^\dagger_y U_z\} 
- N_c\Tr\{U^\dagger_y \calq_{1x}\} \Big)
\nonumber\\
&&~~~ -  {1\over 4}\bigg[ {(x-z,z-y)\over (x-z)^2_\perp(y-z)^2_\perp} + {1\over (x-z)^2_\perp} 
\bigg]\bigg[
\Tr\{ U_xU^\dagger_y U_z \calx^\dagger_{1zx}\} + \Tr\{U_zU^\dagger_y U_x \calx^\dagger_{1zx} \}
\nonumber\\
&&~~~ + {1\over N_c}\Big(\Tr\{U^\dagger_y U_z\}\big(\calh^+_{1xz} + \calh^-_{1zx}\big)
+ \Tr\{U^\dagger_y U_x\}\big(\calh_{1zx}^+ + \calh^-_{1xz} \big)\Big)
\nonumber\\
&&~~~
- {1\over N^2_c}\Big(\Tr\{U_zU^\dagger_y U_x \calx^\dagger_{1zx}\}  + \Tr\{U_xU^\dagger_y U_z \calx^\dagger_{1xz}\}\Big)
\bigg]
\nonumber\\
&&~~~ + {1\over 4}{(\vec{x}-\vec{z})\times(\vec{y}-\vec{z})\over (x-z)^2_\perp(y-z)^2_\perp}
\bigg[\Tr\{ U_zU^\dagger_y U_x \calx^\dagger_{5xz}\} - \Tr\{ U_xU^\dagger_y U_z \calx^\dagger_{5zx}\}
\nonumber\\
&&~~~
+{1\over N_c}\Big(\Tr\{U^\dagger_y U_z\}\big(\calh^+_{5xz} - \calh^-_{5zx}\big)
+ \Tr\{U^\dagger_y U_x\}\big(\calh^-_{5xz} - \calh^+_{5zx}\big)\Big)
\nonumber\\
&&~~~ 
+ {1\over N^2_c}\Big(\Tr\{U_zU^\dagger_y U_x \calx^\dagger_{5zx}\} - \Tr\{U_xU^\dagger_y U_z \calx^\dagger_{5xz} \}\Big) 
\bigg]
\Bigg\}\,.
\label{SumcalQ1}
\end{eqnarray}

\subsection{$Q_{5x}$, $\tildeQ_{5x}$, and $\calf_x$ operators: flavor singlet}

Here we collect the evolution equations for operators $Q_{5x}$, $\tildeQ_{5x}$, and $\calf_x$ in the flavor singlet case.

Summing (\ref{Q5quantum-diagram-aball}), (\ref{Q1bkdiagrams}) (with $Q_1$ replaced by $Q_5$), and
(\ref{quark2gluon-diagram2sum5}) we obtain
\begin{eqnarray}
&&\langle \Tr\{U^\dagger_y U_x\}Q_{5x}\rangle
\nonumber\\
&&= {\alpha_s\over 4\pi^2}\int_0^{+\infty}\!{d\alpha\over \alpha}\int d^2z
\Bigg\{
{\Tr\{U^\dagger_y U_x\}\over (x-z)^2_\perp}\,\bigg[
\Tr\{U^\dagger_x U_z\}\,Q_{5z} - {1\over N_c}{\rm Tr}\{U^\dagger_x \big(\tilde{Q}_{5z} -2N_c\calf_z\big)\}\bigg]
\nonumber\\
&&~~~ + {2Q_{5x}(x-y)^2_\perp\over (x-z)^2_\perp(y-z)^2_\perp}
\Big[\Tr\{U_xU^\dagger_z\}\Tr\{U_zU^\dagger_y\} - N_c\Tr\{U_x U^\dagger_y\}\Big]
\nonumber\\
&&~~~
+ {1\over (x-z)^2_\perp}
\bigg[\Tr\big\{U_xU^\dagger_y U_z \calx^\dagger_{5zx}\big\} + \Tr\big\{U_zU^\dagger_y U_x \calx^\dagger_{5xz}\big\}  
- {1\over N_c}{\rm Tr}\{U^\dagger_y U_x\}\Big(\calh^-_{5xz} + \calh^+_{5zx}\Big)\bigg]
\nonumber\\
&&~~~
+ {(x-z,z-y)_\perp\over (y-z)^2_\perp (x-z)^2_\perp} 
\bigg[\Tr\{U_xU^\dagger_y U_z\calx_{5zx}^\dagger\} + \Tr\big\{U_zU^\dagger_y U_x\calx^\dagger_{5xz} \big\}
\nonumber\\
&&~~~ - {1\over N_c}\Tr\{U_xU^\dagger_y\}\Big(\calh^-_{5xz} + \calh^+_{5zx}\Big)\bigg]
+ {(\vec{x}-\vec{z})\times(\vec{y}-\vec{z})\over (y-z)^2_\perp (x-z)^2_\perp}
\bigg[ \Tr\{U_zU^\dagger_y U_x\calx_{1xz}^\dagger\big\} 
\nonumber\\
&&~~~ - \Tr\big\{\Tr\{U_xU^\dagger_y U_z\calx_{1zx}^\dagger\}
+ {1\over N_c}\Tr\{U_xU^\dagger_y\}\Big(\calh_{1zx}^+ - \calh^-_{1xz}\Big)\bigg]
\Bigg\}\,.
\label{SumQ5}
\end{eqnarray}

The evolution equation for operator 
$\Tr\{\hat{U}^\dagger_y \hat{\tildeQ}_{5x}\}$ is obtained summing eqs. (\ref{BKevolutionQ1}) (with $\tildeQ_{1x}$ replaced by 
$\tildeQ_{5x}$), (\ref{LOpsibarpsiadj5sum-tildeq5}), and (\ref{quark2gluon-diagram1asum5}). Thus, we have
\begin{eqnarray}
\hspace{-1cm}&&\langle \Tr\{U^\dagger_y \tildeQ_{5x}\} \rangle
\nonumber\\
&&= {\alpha_s\over 4\pi^2}\!\int^{+\infty}_0\!{d\alpha\over \alpha}\!\int d^2z\Bigg\{
{1\over (x-z)^2_\perp}\bigg[
{\rm Tr}\{U^\dagger_y U_z\}\Big( Q_{5z} - \calh^+_{5xz} - \calh^-_{5zx}\Big) 
\nonumber\\
&&
~~~+ {1\over N_c}{\rm Tr}\{U^\dagger_y\big(\calx_{5xz} + \calx_{5zx} -\tilde{Q}_{5z} + N_c\calf_z\big)\}\bigg]
\nonumber\\
&&~~~ - {(x-z,z-y)\over (y-z)^2_\perp(z-x)^2_\perp} \bigg[
\Tr\{U_z U^\dagger_y\}\Big(\calh^+_{5xz} + \calh^-_{5zx}\Big)
- {1\over N_c}{\rm Tr}\big\{U^\dagger_y \Big(\calx_{5xz}+\calx_{5zx}\big)\big\}\bigg)\bigg]
\nonumber\\
&&~~~ - {(\vec{x}-\vec{z})\times(\vec{y}-\vec{z})\over (y-z)^2_\perp(z-x)^2_\perp}
\bigg[
\Tr\{U_z U^\dagger_y\}\Big(\calh^+_{1xz} - \calh^-_{1zx}\Big)
+ {1\over N_c}{\rm Tr}\big\{U^\dagger_y\big(\calx_{1zx} - \calx_{1xz}\big)\big\}\bigg]
\nonumber\\
&&~~~ + {2(x-y)^2_\perp\over (x-z)^2_\perp(z-y)^2_\perp}\Big(
{\rm Tr}\{U_z U^\dagger_y\}{\rm Tr}\{U^\dagger_z \tilde{Q}_{5x}\} 
- N_c{\rm Tr}\{U^\dagger_y\tilde{Q}_{5x}\}\Big)
\Bigg\}
\label{SumtildeQ5}
\end{eqnarray}

Evolution equation for operator $\hat{\calf}(z_\perp)$ is the sum of eqs. (\ref{BKevolutionQ1}) (with $\hat{\tildeQ}_{1x}$ replaced by 
$\hat{\calf}_x$), and (\ref{calf-quantum-fund})
\begin{eqnarray}
\hspace{-0.5cm}&&\langle\Tr\{\calf_x\, U^\dagger_y\}\rangle 
\nonumber\\
&&={\alpha_s\over 4\pi^2}\,\int_0^{+\infty}\!{d\alpha\over\alpha}\int d^2z
\Bigg\{	{(\vec{x}-\vec{z})\!\times\!(\vec{z}-\vec{y})\over (x-z)^2_\perp(y-z)^2_\perp}\,
\bigg[ \Tr\{U^\dagger_y \tilde{Q}_{1\,z}\}\Tr\{U^\dagger_z U_x\}
-  \Tr\{U_x \tilde{Q}^\dagger_{1\,z}\}\Tr\{U^\dagger_y U_z\}
\nonumber\\
&&~~~ + {1\over N_c} \Big(\Tr\{ U_x U^\dagger_y \tilde{Q}_{1z}U^\dagger_z\}
+ \Tr\{U^\dagger_y U_x U_z^\dagger \tilde{Q}_{1z} \}
 -  \Tr\{ U_x U^\dagger_y U_z\tilde{Q}^\dagger_{1z}\}
- \Tr\{U^\dagger_y U_x \tilde{Q}_{1z}^\dagger U_z\}\Big)
\nonumber\\
&&~~~
+ {1\over N^2_c}\Tr\{U^\dagger_y U_x\}\Big(Q^\dagger_{1z}- Q_{1z}\Big)\bigg]
+ \bigg({(x-z,z-y)\over (x-z)^2_\perp(y-z)^2_\perp} + {1\over (x-z)^2_\perp}\bigg)
\nonumber\\	
&&~~~
\times
\bigg[ \Tr\{U^\dagger_y \big(2\calf_z - \tildeQ_{5z}\big)\}\Tr\{U^\dagger_z U_x\}
+  \Tr\{U_x \big(2\calf^\dagger_z - \tildeQ^\dagger_{5z}\big)\}\Tr\{U^\dagger_y U_z\}
\nonumber\\
&&~~~ + {1\over N_c}\Big( \Tr\{ U_x U^\dagger_y U_z\tilde{Q}^\dagger_{5z}\}
+ \Tr\{U^\dagger_y U_x \tilde{Q}_{5z}^\dagger U_z\}
+ \Tr\{ U_x U^\dagger_y \tilde{Q}_{5z}U^\dagger_z\}
+ \Tr\{U^\dagger_y U_x U_z^\dagger \tilde{Q}_{5z} \}\Big)
\nonumber\\
&&~~~ - {1\over N^2_c}\Tr\{U^\dagger_y U_x\}\Big(Q_{5z} + Q^\dagger_{5z} \Big)
\bigg]
+ 8\pi^2\!\!\int \!\dhd^2 q_1 {e^{i(q_1,y-z)}-e^{i(q_1,x-z)}\over q^2_{1\perp}}\,\delta^{(2)}(z-x)
\nonumber\\
&&~~~
\times\!\Big[\Tr\{U_x U^\dagger_z\}\Tr\{U^\dagger_y \calf_z\} + \Tr\{U^\dagger_u U_z\}\Tr\{U_x\calf^\dagger_z\}\Big]
\nonumber\\
&&~~~
+ {2(x-y)^2_\perp\over (x-z)^2_\perp(z-y)^2_\perp}\Big(
\Tr\{U_z U^\dagger_y\}\Tr\{U^\dagger_z \calf_x\} 
- N_c\Tr\{U^\dagger_y\calf_x\}\Big)
\Bigg\}
\label{Sumcalf}
\end{eqnarray}
Evolution equation (\ref{Sumcalf}) should clarify why we needed evolution equations (\ref{SumQ1}), (\ref{SumtildeQ1}),
(\ref{SumQ5}), and (\ref{SumtildeQ5}).

We can also write done the evolution equation for operator $\Tr\{\hat{\calq}_{5x} \hat{U}^\dagger_y\}$ 
summing eqs. (\ref{BKevolutionQ1}) (with $\hat{\tildeQ}_{1x}$ replaced by 
$\hat{\calq}_{5x}$), (\ref{quantum-oneloopcaq5}), and (\ref{calq5-q2g}). We have
\begin{eqnarray}
\hspace{-0.4cm}&&
\big\langle\Tr\{\calq_{5x} U^\dagger_y\}
\big\rangle
\nonumber\\
&&= {\alpha_s\over 2\pi^2}\int_0^{+\infty}\!\!{d\alpha\over \alpha}\int\!\!d^2z\Bigg\{
\half{1\over (x-z)^2_\perp}
\bigg[{\rm Tr}\{U^\dagger_x \big(\calq_{5\,z} + \calf_{z}\big)\}{\rm Tr}\{U^\dagger_y U_x\}
- {1\over N_c}{\rm Tr}\{\big(\calq_{5\,z} + \calf_{z}\big)U^\dagger_y\} 
\bigg]
\nonumber\\
&&~~~ + {(x-y)^2_\perp \over (x-z)^2_\perp(y-z)^2_\perp}
\bigg[\Tr\{\calq_{5x}U^\dagger_z \}\Tr\{ U_zU^\dagger_y\} 
- N_c\Tr\{U^\dagger_y \calq_{5x}\} \bigg]
\nonumber\\
&&~~~
+ {1\over 4}\bigg[ {(x-z,z-y)\over (x-z)^2_\perp(y-z)^2_\perp} + {1\over (x-z)^2_\perp} \bigg]
\bigg[\Tr\{ U_xU^\dagger_y U_z \calx^\dagger_{5zx}\} + \Tr\{ U_zU^\dagger_y U_x \calx^\dagger_{5xz}\}
\nonumber\\
&&~~~  - {1\over N_c}\Big(\Tr\{U^\dagger_y U_z\}\big(\calh^+_{5xz} + \calh_{5zx}^-\big)
 +\Tr\{U^\dagger_y U_x\}\big(\calh_{5xz}^- + \calh_{5zx}^+\big)
  \Big)
\nonumber\\
&&~~~
- {1\over N^2_c}\Big(\Tr\{U_zU^\dagger_y U_x \calx^\dagger_{5zx} \} 
+ \Tr\{U_xU^\dagger_y U_z \calx^\dagger_{5xz}\}\bigg]
\nonumber\\
&&~~~ + {1\over 4}{(\vec{x}-\vec{z})\times(\vec{y}-\vec{z})\over (x-z)^2_\perp(y-z)^2_\perp}
\bigg[
\Tr\{ U_zU^\dagger_y U_x \calx^\dagger_{1xz}\} -\Tr\{ U_xU^\dagger_y U_z \calx^\dagger_{1zx}\}
\nonumber\\
&&~~~
 - {1\over N_c}\Big(\Tr\{U^\dagger_y U_z\}\big(\calh_{1xz}^+  - \calh_{1zx}^- \big)
 + \Tr\{U^\dagger_y U_x\}\big(\calh_{1xz}^-  - \calh_{1zx}^+\big)\Big)
\nonumber\\
&&~~~
+ {1\over N^2_c}\Big(\Tr\{U_zU^\dagger_y U_x \calx^\dagger_{1zx}\} - \Tr\{U_xU^\dagger_y U_z \calx^\dagger_{1xz}\}\Big)
\bigg]
\Bigg\}
\label{Sumcalq5}
\end{eqnarray}

In eqs. (\ref{SumQ5}), (\ref{SumtildeQ5}), (\ref{Sumcalf}), and (\ref{Sumcalq5}) we have again mixing
between odd operators which carry the subscript $5$, and even operators with subscript $1$.

The evolution equation of operators in the adjoint representation $\hat{\calq}_1(x_\perp)^{ab}$,
$\hat{\calq}_5(x_\perp)^{ab}$,  and $\hat{\calf}^{ba}(x_\perp)$ can be found in Appendix \ref{sec: evolutionadjointrep}.

\subsection{$Q_{5x}$, $\tildeQ_{5x}$, and $\calf_x$ operators: flavor non-singlet}

For flavor non-singlet case we do not have the operator $\hat{\calf}(x_\perp)$ because it does not allow flavor exchange.
In this case the evolution equations, using results for the singlet case (neglecting operator $\calf(x_\perp)$), are
\begin{eqnarray}
&&\langle \Tr\{U^\dagger_y U_x\}Q_{5x}\rangle
\nonumber\\
&&= {\alpha_s\over 4\pi^2}\int_0^{+\infty}\!{d\alpha\over \alpha}\int d^2z
\Bigg\{
{\Tr\{U^\dagger_y U_x\}\over (x-z)^2_\perp}\,\bigg[
\Tr\{U^\dagger_x U_z\}\,Q_{5z} - {1\over N_c}{\rm Tr}\{U^\dagger_x \tilde{Q}_{5z}\}\bigg]
\nonumber\\
&&~~~ + {2Q_{5x}(x-y)^2_\perp\over (x-z)^2_\perp(y-z)^2_\perp}
\Big[\Tr\{U_xU^\dagger_z\}\Tr\{U_zU^\dagger_y\} - N_c\Tr\{U_x U^\dagger_y\}\Big]
\nonumber\\
&&~~~
+ {1\over (x-z)^2_\perp}
\bigg[\Tr\big\{U_xU^\dagger_y U_z \calx^\dagger_{5zx}\big\} + \Tr\big\{U_zU^\dagger_y U_x \calx^\dagger_{5xz}\big\}  
- {1\over N_c}{\rm Tr}\{U^\dagger_y U_x\}\Big(\calh^-_{5xz} + \calh^+_{5zx}\Big)\bigg]
\nonumber\\
&&~~~
+ {(x-z,z-y)_\perp\over (y-z)^2_\perp (x-z)^2_\perp} 
\bigg[\Tr\{U_xU^\dagger_y U_z\calx_{5zx}^\dagger\} + \Tr\big\{U_zU^\dagger_y U_x\calx^\dagger_{5xz} \big\}
\nonumber\\
&&~~~ - {1\over N_c}\Tr\{U_xU^\dagger_y\}\Big(\calh^-_{5xz} + \calh^+_{5zx}\Big)\bigg]
+ {(\vec{x}-\vec{z})\times(\vec{y}-\vec{z})\over (y-z)^2_\perp (x-z)^2_\perp}
\bigg[ \Tr\{U_zU^\dagger_y U_x\calx_{1xz}^\dagger\big\}
\nonumber\\
&&~~~  - \Tr\big\{\Tr\{U_xU^\dagger_y U_z\calx_{1zx}^\dagger\}\tilde
+ {1\over N_c}\Tr\{U_xU^\dagger_y\}\Big(\calh_{1zx}^+ - \calh^-_{1xz}\Big)\bigg]
\Bigg\}\,.
\label{SumQ5nonsing}
\end{eqnarray}

\begin{eqnarray}
\hspace{-1cm}&&\langle \Tr\{U^\dagger_y \tildeQ_{5x}\} \rangle
\nonumber\\
&&= {\alpha_s\over 4\pi^2}\!\int^{+\infty}_0\!{d\alpha\over \alpha}\!\int d^2z\Bigg\{
{1\over (x-z)^2_\perp}\bigg[
{\rm Tr}\{U^\dagger_y U_z\}\Big( Q_{5z} - \calh^+_{5xz} - \calh^-_{5zx}\Big) 
\nonumber\\
&&
~~~+ {1\over N_c}{\rm Tr}\{U^\dagger_y\big(\calx_{5xz} + \calx_{5zx} -\tilde{Q}_{5z}\big)\}\bigg]
\nonumber\\
&&~~~ - {(x-z,z-y)\over (y-z)^2_\perp(z-x)^2_\perp} \bigg[
\Tr\{U_z U^\dagger_y\}\Big(\calh^+_{5xz} + \calh^-_{5zx}\Big)
- {1\over N_c}{\rm Tr}\big\{U^\dagger_y \Big(\calx_{5xz}+\calx_{5zx}\big)\big\}\bigg)\bigg]
\nonumber\\
&&~~~ - {(\vec{x}-\vec{z})\times(\vec{y}-\vec{z})\over (y-z)^2_\perp(z-x)^2_\perp}
\bigg[
\Tr\{U_z U^\dagger_y\}\Big(\calh^+_{1xz} - \calh^-_{1zx}\Big)
+ {1\over N_c}{\rm Tr}\big\{U^\dagger_y\big(\calx_{1zx} - \calx_{1xz}\big)\big\}\bigg]
\nonumber\\
&&~~~ + {2(x-y)^2_\perp\over (x-z)^2_\perp(z-y)^2_\perp}\Big(
{\rm Tr}\{U_z U^\dagger_y\}{\rm Tr}\{U^\dagger_z \tilde{Q}_{5x}\} 
- N_c{\rm Tr}\{U^\dagger_y\tilde{Q}_{5x}\}\Big)
\Bigg\}
\label{SumtildeQ5nonsing}
\end{eqnarray}

\begin{eqnarray}
\hspace{-0.7cm}&&
\big\langle\Tr\{\calq_{5x} U^\dagger_y\}
\big\rangle
\nonumber\\
\hspace{-0.7cm}&&= {\alpha_s\over 2\pi^2}\int_0^{+\infty}\!\!{d\alpha\over \alpha}\int\!\!d^2z\Bigg\{
\half{1\over (x-z)^2_\perp}
\bigg[{\rm Tr}\{U^\dagger_x \calq_{5\,z}\}{\rm Tr}\{U^\dagger_y U_x\}
- {1\over N_c}{\rm Tr}\{\calq_{5\,z} U^\dagger_y\} 
\bigg]
\nonumber\\
\hspace{-0.7cm}&&~~~ + {(x-y)^2_\perp \over (x-z)^2_\perp(y-z)^2_\perp}
\bigg[\Tr\{U^\dagger_z \calq_{5z}\}\Tr\{U^\dagger_y U_z\} 
- N_c\Tr\{U^\dagger_y \calq_{5z}\} \bigg]
\nonumber\\
\hspace{-0.7cm}&&~~~
+ {1\over 4}\bigg[ {(x-z,z-y)\over (x-z)^2_\perp(y-z)^2_\perp} + {1\over (x-z)^2_\perp} \bigg]
\bigg[\Tr\{ U_xU^\dagger_y U_z \calx^\dagger_{5zx}\} + \Tr\{ U_zU^\dagger_y U_x \calx^\dagger_{5xz}\}
\nonumber\\
\hspace{-0.7cm}&&~~~  - {1\over N_c}\Big(\Tr\{U^\dagger_y U_z\}\big(\calh^+_{5xz} + \calh_{5zx}^-\big)
+\Tr\{U^\dagger_y U_x\}\big(\calh_{5xz}^- + \calh_{5zx}^+\big)
\Big)
\nonumber\\
\hspace{-0.7cm}&&~~~
- {1\over N^2_c}\Big(\Tr\{U_zU^\dagger_y U_x \calx^\dagger_{5zx} \} 
+ \Tr\{U_xU^\dagger_y U_z \calx^\dagger_{5xz}\}\bigg]
 + {1\over 4}{(\vec{x}-\vec{z})\times(\vec{y}-\vec{z})\over (x-z)^2_\perp(y-z)^2_\perp}
\bigg[
\Tr\{ U_zU^\dagger_y U_x \calx^\dagger_{1xz}\} 
\nonumber\\
\hspace{-0.7cm}&&~~~ -\Tr\{ U_xU^\dagger_y U_z \calx^\dagger_{1zx}\}
- {1\over N_c}\Big(\Tr\{U^\dagger_y U_z\}\big(\calh_{1xz}^+  - \calh_{1zx}^- \big)
+ \Tr\{U^\dagger_y U_x\}\big(\calh_{1xz}^-  - \calh_{1zx}^+\big)\Big)
\nonumber\\
\hspace{-0.7cm}&&~~~ 
+ {1\over N^2_c}\Big(\Tr\{U_zU^\dagger_y U_x \calx^\dagger_{1zx}\} - \Tr\{U_xU^\dagger_y U_z \calx^\dagger_{1xz}\}\Big)
\bigg]
\Bigg\}
\label{Sumcalq5nonsing}
\end{eqnarray}
In the flavor non-singlet case, because of the absence of mixing with operator $\calf(x_\perp)$
the evolution equation for polarized structure functions is eq. (\ref{Sumcalq5nonsing}). This might represent a great simplification
to find its solution.

\section{Conclusions}

We have extended the high-energy OPE of the $T$-product of two
electromagnetic currents at sub-eikonal level. The impact factors $\cali_1^{\mu\nu}$ and $\cali_5^{\mu\nu}$, 
given in eqs. (\ref{I1definition}), and (\ref{I5definition}) respectively, are part of the main results of this paper. 
They satisfy electromagnetic gauge invariance and conformal $SL(2,C)$ invariance.

We have identified several new distribution functions, summarized in section \ref{sec: parametri-matrixele},
which came up from the parametrization of the matrix elements of the operators associated to the impact factors
$I_1^{\mu\nu}$ and $I^{\mu\nu}_5$. In the Appendix \ref{sec: matrielements} we have identified
further distribution functions that, however, will not contribute to $g_1$ structure functions.

We found that the polarized and unpolarized quark distribution functions as well as the polarized
gluon distributions are energy suppressed with respect to the unpolarized gluon ones.
Moreover we have observed that the polarized gluon distributions $G_L$, and $G_T$
enter on the same footing with the polarized distribution functions $q_{5L}$, $q_{5T}$, $\tildeq_{5L}$,
$\tildeq_{5T}$, because they are associated to the same impact factor $I^{\mu\nu}_5$. We also showed that
under one loop evolution we have mixing also with the operators parametrized by the quark distribution functions
$q_1$, $q_{1T}$, $\tildeq_1$, $\tildeq_{1T}$.  

We have also observed the evolution equations are not in a closed form; after one loop we have generated new operators
$\hat{\calx}_{1xy}$, $\hat{\calx}_{5xy}$, $\hat{\calh}^+_{1xy}$, $\hat{\calh}^+_{5xy}$, 
$\hat{\calh}^-_{1xy}$, $\hat{\calh}^-_{5xy}$.

In the evolution equations (see section \ref{sec: summaryevo}), we have found also mixing between operators 
of different parity, \textit{i.e.} operators $\hat{Q}_{5x}$, $\hat{\tildeQ}_{5x}$, $\hat{\calf}_x$,
$\hat{\calx}_{5xy}$, $\hat{\calh}_{5xy}^+$, and $\hat{\calh}_{5xy}^-$
and operators $\hat{Q}_{1x}$, $\hat{\tildeQ}_{1x}$,
$\hat{\calx}_{1xy}$, $\hat{\calh}_{1xy}^+$, and $\hat{\calh}_{1xy}^-$.
These are all new evolution equations that are presented here for the first time.
As already emphasized before, we are aware that the form of the evolution equations summarized in section
\ref{sec: summaryevo} are not written in the usual form ${d\over d\eta}\calo^\eta = K\otimes \calo^\eta$.
The reason is that in all these equations there are contaminations of infrared divergences which remind
us the double log of energy nature of the small-x resummation of quark polarized (and unpolarized) structure functions.
We plan to disentangle the leading log from the double log of energy and to put the evolution equations 
of section \ref{sec: summaryevo} in a conventional form in future work where we will try to reproduce the
result obtained in the double log formalism \cite{Bartels:1995iu, Bartels:1996wc}.

We now compare our results with the ones obtained in recent literature
\cite{Kovchegov:2016zex, Kovchegov:2018znm, Altinoluk:2015gia, Altinoluk:2014oxa, Altinoluk:2020oyd}.

Let us start with the ones obtained in refs. \cite{Kovchegov:2016zex, Kovchegov:2018znm}.
In Appendix \ref{sec: Diag-calq1claq5} we have shown that, in the double logarithm approximation, 
the evolution equation for the operator $\Tr\{\big(\calf_x+\calq_{5x}\big)U^\dagger_y\}$, eq. (\ref{calfpluscalq5})
(see also eq. (\ref{sumcalfcalq-adjrep}) in the adjoint representation), does agree
with the one derived in Refs. \cite{Kovchegov:2016zex, Kovchegov:2018znm} provided that
we neglect the mixing with operator $\hat{\calq}_{1x}$. However, when we consider
all diagrams we find some differences.

In Refs. \cite{Kovchegov:2016zex, Kovchegov:2018znm} the evolution equations
for the operators $\hat{Q}_{1x}$, $\hat{Q}_{5x}$, $\hat{\tildeQ}_{1x}$, $\hat{\tildeQ}_{5x}$ were not calculated. 
Only operator $\hat{\calq}_{5x}$ was considered. With equation (\ref{Sumcalf}), we have shown 
that operator $\hat{\calf}_x$ does mix with operators $\hat{Q}_{1x}$, 
$\hat{Q}_{5x}$, $\hat{\tildeQ}_{1x}$, $\hat{\tildeQ}_{5x}$, thus
justifying the calculation of their evolution equations.

Here, for the first time, we calculated all the diagrams required for the evolution of the polarized (and unpolarized) structure functions 
including the quark-to-gluon propagator in the fundamental 
(see Figs. \ref{quark2gluon-diagram1} and \ref{quark2gluon-diagram2})
and in the adjoint representation (see Fig. \ref{diagrams-q2g-prop}). The quark-to-gluon propagator diagrams have generated new
operators, (\ref{calx1}) - (\ref{calh5minus}), which will affect the spin dynamics at small-x. 

In Refs. \cite{Kovchegov:2016zex, Kovchegov:2018znm} the matrix element 
of operator $\Tr\{\hat{\calf}_x \hat{U}^\dagger_y\}$ was associated only with the longitudinal 
helicity distributions. The same conclusion was reached also in Ref. \cite{Altinoluk:2020oyd}.
In section \ref{sec: parametri-matrixele} (see also Appendix \ref{sec: matrielements}), however,
it was shown that the matrix element of operator
$\Tr\{\hat{\calf}_x \hat{U}^\dagger_y\}$ is, in general, parametrized with longitudinal and transverse distributions.

Sub-eikonal corrections in the framework of Color Glass Condensate (CGC)
have been considered also in Ref. \cite{Altinoluk:2015gia, Altinoluk:2014oxa}. At the moment the result
in Ref. \cite{Altinoluk:2015gia, Altinoluk:2014oxa} is incomplete
because sub-eikonal corrections due to 
pure transverse components of the field strength tensor, like $F_{ij}$, have not been included. As we have seen, such corrections
turned out to be very important and central to the study of spin dynamics.
Recently, the sub-eikonal corrections to the quark propagator 
have been considered in Ref. \cite{Altinoluk:2020oyd}. The authors have reproduced the terms in operator
$\hat{\calo}_1$ of eq. (\ref{O1}), but
their result seem to be missing
some of the terms that are, instead, present in Ref. \cite{Chirilli:2018kkw} and that we
presented here in eq. (\ref{ojobulletstar}).

In Ref. \cite{Boussarie:2019icw} the small-x behavior of the orbital angular momentum distributions
was found as a generalization of the double log of energy for $g_1$ structure function
originally found in Refs. \cite{Bartels:1995iu, Bartels:1996wc}.
On the other hand, in Ref. \cite{Kovchegov:2019rrz} the same calculation was carried on in the frame work
of saturation and Color Glass Condensate formalism. To our knowledge, it is not known whether
the two results are consistent. Our work, among other things, set the foundation of the formalism that
eventually will be able to reconcile the Double Logarithm Asymptotics formalism with the non-linear Wilson-line formalism.

Concluding, we obtained novel evolution equations for polarized structure functions 
(and unpolarized quark structure functions). New operators and new distribution functions
have appeared for the first time in small-x physics. 
First, we have the appearance of the light-ray operators $\hat{Q}_{1x}$ and $\hat{Q}_{5x}$ multiplied
by the usual dipole operator $\Tr\{\hat{U}_x\hat{U}^\dagger_y\}$; we have the operators
$\Tr\{\hat{\tildeQ}_{1x}U^\dagger_y\}$, and  $\Tr\{\hat{\tildeQ}_{5x}U^\dagger_y\}$; the gluon helicity and transverse distribution
is obtained from $\Tr\big\{U^\dagger_{z_1}\calf_{z_2}\big\}
+ {\rm Tr}\big\{U_{z_1}\calf_{z_2}^\dagger\big\}$; and finally we have also
the TMDs operators $\hat{\calx}_{1xy}$, $\hat{\calx}_{5xy}$, 
$\hat{\calh}^+_{1xy}$, $\hat{\calh}^+_{5xy}$, $\hat{\calh}^-_{1xy}$, $\hat{\calh}^-_{5xy}$.

The author is grateful to I. Balitsky for numerous valuable discussions.
He also thanks V.M. Braun for discussions and reading the manuscript, and Y. Kovchegov, and A. Vladimirov  
for discussions. This project has received funding from the European Union's Horizon 2020 research and innovation 
programme under grant No 824093.

\appendix

\section{Notation}
\label{sec: notation}

The notations used throughout this paper are the same as the ones used in Ref.\cite{Chirilli:2018kkw}.

Given two light-cone vectors $p_1^\mu$ and $p_2^\mu$, with $p_1^\mu p_{2\mu} = {s\over 2}$,
we can decompose any coordinate as $x^\mu = {2\over s}x_*p_1^\mu + {2\over s}x_\bullet p_2^\mu + x_\perp^\mu$
with $x_* = x_\mu p_2^\mu= \sqrt{{s\over 2}}x^+$, $x_\bullet = x_\mu p_1^\mu= \sqrt{{s\over 2}}x^-$ and 
$x^\pm = {x^0\pm x^3\over \sqrt{2}}$. Our metric is (1, -1, -1, -1,).
We use the notation $x^\mu_\perp = (0,x^1,x^2,0)$ and $x^i = (x^1,x^2)$ such that $x^ix_i = x^\mu_\perp x^\perp_\mu = - x^2_\perp$. 
So, Latin indexes assume values $1, 2$, while Greek indexes run from $0$ to $3$. We also use notation 
for scalar product $(x,y)_\perp= x^1 y^1 + x^2 y^2$.
For a vector in momentum space we have $k^\mu = \alpha p_1^\mu + \beta p_2^\mu + p_\perp^\mu$
with $\alpha = \sqrt{2\over s}k^+$ and $\beta = \sqrt{2\over s}k^-$.

Under a longitudinal boost, the components of the gauge fields gets rescaled by the large boost parameter $\lambda$ as follows
\begin{eqnarray}
&&A_\bullet(x_\bullet, x_*, x_\perp) \to \lambda\, A_\bullet(\lambda^{-1}x_\bullet, \lambda\, x_*, x_\perp)\,,\nonumber\\
&&A_*(x_\bullet, x_*, x_\perp) \to  \lambda^{-1}A_*(\lambda^{-1}x_\bullet, \lambda\, x_*, x_\perp)\,,
\label{boost}\\
&&A_\perp(x_\bullet, x_*, x_\perp)  \to  A_\perp(\lambda^{-1}x_\bullet, \lambda\, x_*, x_\perp)\,.\nonumber
\end{eqnarray}
while field strength tensor as
\begin{eqnarray}
&&F_{i\bullet}(x_\bullet, x_*, x_\perp) \to  \lambda\, F_{i\bullet}(\lambda^{-1}x_\bullet, \lambda\, x_*, x_\perp)\,,\nonumber\\
&&F_{i *}(x_\bullet, x_*, x_\perp) \to  \lambda^{-1}F_{i *}(\lambda^{-1}x_\bullet, \lambda\, x_*, x_\perp)\,,
\label{Fboost}\nonumber
\\
&&F_{\bullet *}(x_\bullet, x_*, x_\perp)  \to  F_{\bullet *}(\lambda^{-1}x_\bullet, \lambda\, x_*, x_\perp)\,,
\nonumber\\
&&F_{ij}(x_\bullet, x_*, x_\perp)  \to  F_{ij}(\lambda^{-1}x_\bullet, \lambda\, x_*, x_\perp)\,.
\end{eqnarray}

Under the same large longitudinal boost the spinor fields get rescaled as
\begin{eqnarray}
\bar{\psi}t^a\ssp_1\psi \to  \lambda \bar{\psi}t^a\ssp_1\psi\,,~~~~
\bar{\psi}t^a\gamma^\perp_\nu\psi \to  \bar{\psi}t^a\gamma^\perp_\nu\psi\,,~~~~
\bar{\psi}t^a\ssp_2\psi \to  \lambda^{-1} \bar{\psi}t^a\ssp_2\psi\,.
\label{spinorboost}
\end{eqnarray}

In Schwinger representation, which will be frequently used throughout this paper, the free scalar propagator can be written as
\begin{eqnarray}
	\brax {i\over p^2 + i\epsilon}\kety = i\int\!\dhd^4 k \,{e^{-ik\cdot(x-y)}\over k^2 + i\epsilon}\,,
	\label{schwrep}
\end{eqnarray}
with $\langle k\ketx = e^{ix\cdot k}$. In (\ref{schwrep}) we 
used the $\hbar$-inspired notation $\dhd^4 k \equiv {d^4k\over (2\pi)^4}$ and
$\dbar^{(4)}(k) = (2\pi)^4\delta^{(4)}(k)$ so that, $\int\!\dhd^4 k \,\dbar^{(4)}(k) = 1$.

We define the gauge link at fixed transverse position as
\begin{eqnarray}
[up_1,vp_1]_z \equiv [up_1 + z_\perp, vp_1 + z_\perp] 
\equiv {\rm Pexp}\Big\{ig\!\!\int_v^u\!\!d t \,A_\bullet(tp_1+z_\perp)\Big\}\,.
\end{eqnarray}
The derivative of the gauge link with respect to the transverse position is
\begin{eqnarray}
{\partial \over \partial z^i} [up_1, vp_1 ]_z
=&& ig A_i(up_1+z_\perp)[up_1,vp_1]_z - ig [up_1,vp_1]_zA_i(vp_1+z_\perp)
\nonumber\\
&&- ig\!\int^u_v\!\!\! ds\,[up_1,sp_1]_zF_{\bullet i}(p_1s+z_\perp)[p_1s,p_1v]_z\,,
\label{deriv-glink}
\end{eqnarray}
with index $i=1,2$.  
From (\ref{deriv-glink}) we may formally define the transverse covariant derivative 
$\mathfrak{D}_i $ that acts on a non-local operator as
\begin{eqnarray}
\hspace{-1cm}i\mathfrak{D}_i\, [up_1, vp_1]_z &\!\equiv\!& 
i{\partial \over \partial z^i} [up_1, vp_1]_z + g\big[A_i(z_\perp), [up_1,vp_1]_z\big]
\nonumber\\
&\!=\!& g\!\int^u_v\!\!\! ds\,[up_1,sp_1]_zF_{\bullet i}(p_1s+z_\perp)[p_1s,p_1v]_z\,,
\label{coderiv-glink}
\end{eqnarray}
where we have used the implicit notation 
$\big[A_i(z_\perp), [up_1,vp_1]_z\big] = A_i(z_\perp+up_1) [up_1,vp_1]_z - [up_1,vp_1]_z\,A_i(z_\perp + vp_1)$.

Given a gauge link $[x_*,y_*]_z\equiv[{2\over s}x_*p_1 + z_\perp, {2\over s}y_*p_1 + z_\perp]$, in Schwinger notation we have
\begin{eqnarray}
\braxp [x_*,y_*] \ketyp = [x_*,y_*]_x\,\delta^{(2)}(x-y)\,.
\label{schwi-nota}
\end{eqnarray}
The transverse momentum operator $\hat{P}_i = \hat{p}_i + g \hat{A}_i$ acts on the gauge link as
\begin{eqnarray}
\hspace{-1cm}\braxp \hat{P}_i[x_*,y_*]\ketyp \equiv\!\!&& \braxp \big(\hatp_i + \hat{A}_i(x_*)\big)[x_*,y_*]\ketyp 
\label{Pileft}
\nonumber\\=\!\!&& 
\big(i{\partial\over \partial x^i} + A_i(x_*,x_\perp)\big)\braxp [x_*,y_*]\ketyp\\
\hspace{-1cm}\braxp [x_*,y_*]\hat{P}_i\ketyp \equiv\!\!&& \braxp [x_*,y_*]\big(\hatp_i + A_i(y_*)\big)\ketyp 
\nonumber\\=\!\!&& 
\braxp [x_*,y_*]\ketyp\big(- i{\partial\over \partial y^i} + A_i(y_*,y_\perp)\big)\,.
\label{Piright}
\end{eqnarray}
So, from (\ref{Pileft}), and (\ref{Piright}), and using (\ref{coderiv-glink}), and (\ref{schwi-nota}) we finally have
\begin{eqnarray}
&&\hspace{-1.2cm}\braxp \big[\hat{P}_i, [x_*,y_*]\big]\ketyp \!=\! \braxp i\mathfrak{D}_i[x_*,y_*] \ketyp
\!= \!\braxp g{2\over s}\int^{x_*}_{y_*}\!\!\! d\omega_*
\,[x_*,\omega_*]F_{\bullet i}[\omega_*,y_*] \ketyp\,,
\label{defPi}
\end{eqnarray}
where we used again the short-hand notation 
$[x_*,\omega_*]F_{i\bullet}[\omega_*,y_*] = [x_*,\omega_*]F_{i\bullet}(\omega_*)[\omega_*,y_*]$.
From (\ref{defPi}) we observe that the covariant derivative $i\mathfrak{D}_i$ acts on the gauge link even though the 
transverse coordinate has not been specified yet and, as matter of fact, it does not have to
in order to know how it acts on the gauge link. Throughout the paper we will often use
this property of the momentum operator $\hat{P}_i$ and the
covariant derivative $i\mathfrak{D}_i$ without specifying the \textit{bra} $\braxp$ and the \textit{ket} $\ketyp$.

\section{Quark and Gluon Propagators with sub-eikonal corrections}
\label{sec: propagators}

We want to extend the high-energy OPE to include sub-eikonal corrections in order to study polarized structure function at small-x.
The idea is to proceed in a similar way as we did in the unpolarized case with the exception that now the quark propagator (and the gluon one)
has also sub-eikonal corrections. In reference \cite{Chirilli:2018kkw}
a complete analysis of the quark and gluon propagator has been performed. All the 
sub-eikonal corrections (regardless of the twist) that scale as ${1\over \lambda}$ with $\lambda$ a large boost parameter, have been identified.

In what follow we will summarize the all the propagator with sub-eikonal corrections.

\subsection{Quark propagator in the background of gluon fields}

In this sub-section we are going to derive the quark propagator in the gluon background field.
Here, however, we will perform
a different expansion than the one we performed in Ref. \cite{Chirilli:2018kkw}, but
that was only suggested in the Appendix of the same reference. 

If we define $\calb_1 = {4\over s^2}F_{\bullet *}\sigma_{*\bullet} + \half\sigma_{ij}F^{ij}$, 
and $O = \{p^\mu_\perp,A^\perp_\mu\} + \{{2\over s}P_\bullet, A_*\} - gA^2_\perp$, then the quark propagator
up to sub-eikonal corrections is
\begin{eqnarray}
\hspace{-1cm}&&\brax \Sp{i\over P^2 + {g\over 2}F_{\mu\nu}\sigma^{\mu\nu}}\kety
\nonumber\\
\hspace{-1cm}&&= i\Slash{D}\brax\Big[{i\over p^2 + 2g\alpha A_\bullet + gO} - 
{i\over p^2 + 2g\alpha A_\bullet + gO}ig{2\over s}F_{i\bullet}\gamma^i\ssp_2{1\over p^2 + 2g\alpha A_\bullet + gO}
\nonumber\\
\hspace{-1cm}&&
- {i\over p^2 + 2g\alpha A_\bullet }\,g\calb_1\,{1\over p^2 + 2g\alpha A_\bullet }
+ {i\over p^2 + 2g\alpha A_\bullet }ig{2\over s}F_{i\bullet}\gamma^i\ssp_2{1\over p^2 + 2g\alpha A_\bullet }
g\calb_1{1\over p^2 + 2g\alpha A_\bullet }
\nonumber\\
\hspace{-1cm}&&
+ {i\over p^2 + 2g\alpha A_\bullet }\,g\calb_1\,{1\over p^2 + 2g\alpha A_\bullet }
ig{2\over s}F_{i\bullet}\gamma^i\ssp_2{1\over p^2 + 2g\alpha A_\bullet }\Big]\kety\,.
\label{qprop-1st-stepexp}
\end{eqnarray}
Now we need the scalar propagator up to sub-eikonal corrections \cite{Chirilli:2018kkw}
\begin{eqnarray}
&&\brax {i\over P^2+i\epsilon}\kety = \brax {i\over p^2 + 2\alpha g A_\bullet + g O + i\epsilon}\kety
\nonumber\\
&& = \left[\int_0^{+\infty}\!\!{\dhd \alpha\over 2\alpha}\theta(x_*-y_*) - 
\int_{-\infty}^0\!\!{\dhd\alpha\over 2\alpha}\theta(y_*-x_*) \right]e^{-i\alpha(x_\bullet - y_\bullet)}
\nonumber\\
&&~~~\times
\braxp\, e^{-i{\hatp^2_\perp\over \alpha s}x_*}\bigg\{[x_*,y_*] 
+ {ig\over 2\alpha}\Bigg[{2\over s}x_*\Big(\{P_i,A^i(x_*)\} - gA_i(x_*)A^i(x_*)\Big)[x_*,y_*]
\nonumber\\
&&~~~ - [x_*,y_*]{2\over s}y_*\Big(\{P_i,A^i(y_*)\} - gA_i(y_*)A^i(y_*)\Big)
+ \int^{x_*}_{y_*}\!\!d{2\over s}\omega_*\bigg(\big\{P^i,[x_*,\omega_*]\,{2\over s}\,\omega_*\, F_{i\bullet}(\omega_*)\,[\omega_*,y_*]\big\}
\nonumber\\
&&
\hspace{2cm}+ g\!\!\int^{x_*}_{\omega_*}\!\!d{2\over s}\,\omega'_*\,{2\over s}\big(\omega_* - \omega'_*\big)
[x_*,\omega'_*]F^i_{~\bullet}[\omega'_*,\omega_*]\,
\,F_{i\bullet}\,[\omega_*,y_*]\bigg)\Bigg]\bigg\}e^{i{\hatp^2_\perp\over \alpha s}y_*}\ketyp\,.
\label{scalrprop-subeik}
\end{eqnarray}
Now we observe that
\begin{eqnarray}
&&\hspace{-1.2cm}i\Sd\Bigg( e^{-i\alpha(x_\bullet - y_\bullet)}
\braxp\, e^{-i{\hatp^2_\perp\over \alpha s}x_*} [x_*,y_*]\ketzp\Bigg)
\nonumber\\
&&\hspace{-1.2cm}= \big(i\slashd^x + g{2\over s}\ssp_2 A_\bullet(x_*,x_\perp) + 
g\Sa_\perp(x_*,x_\perp) \big)e^{-i\alpha(x_\bullet - y_\bullet)}
\braxp\, e^{-i{\hatp^2_\perp\over \alpha s}x_*} [x_*,y_*]\ketzp
\nonumber\\
&&\hspace{-1.2cm}= 
e^{-i\alpha(x_\bullet - y_\bullet)}\braxp\, e^{-i{\hatp^2_\perp\over \alpha s}x_*}
\nonumber\\
&&\hspace{-1.2cm}~~\times\!
\Big({1\over \alpha s}\hat{\ssp}\ssp_2\hat{\ssp} + i{2\over s}\ssp_2D^x_\bullet
+ {ix_*\over \alpha s}[\hat{p}^2_\perp,g{2\over s}\ssp_2 A_\bullet(x_*)]
+ g\Sa_\perp(x_*) \Big)[x_*,y_*]\ketzp\,.
\label{usefla}
\end{eqnarray}
Note that $i[p^2_\perp, gA_\bullet(x_*)] = g\{p^i, F_{i\bullet}(x_*) + D_\bullet A_i(x_*)\}$. 
The field strength tensor $F_{i\bullet}(x_*) = 0$ since $x_*$ is outside the shock-wave
($x_*$ and $y_*$ here are always point outside the shock-wave which will be sent to $+\infty$
and $-\infty$ respectively). Similarly, we can set all the transverse fields
at the edges of the gauge-link (outside the shock-wave) to zero since they are pure gauge. Moreover,
we make use of $iD_\bullet[x_*,y_*]_x = (i\partial_\bullet^x + g A_\bullet(x_*))[x_*,y_*]_x=0$.
So, we may reduce (\ref{usefla}) to
\begin{eqnarray}
&&\big(i\slashd^x + g{2\over s}\ssp_2 A_\bullet(x_*,x_\perp) + g\Sa_\perp(x_*,x_\perp) \big)e^{-i\alpha(x_\bullet - y_\bullet)}
\braxp\, e^{-i{\hatp^2_\perp\over \alpha s}x_*}\ketzp \nonumber\\
&&~~= 
e^{-i\alpha(x_\bullet - y_\bullet)}\braxp\, e^{-i{\hatp^2_\perp\over \alpha s}x_*}
{1\over \alpha s}\ssp\ssp_2\ssp [x_*,y_*]
\ketzp\,.
\label{usefla2}
\end{eqnarray}
We will also need the following two identities. Given two generic operators $\cala_1$ and $\cala_2$, we have
\begin{eqnarray}
&&\brax {1\over p^2 + 2g\alpha A_\bullet + i\epsilon } A_1 {1\over p^2 + 2g\alpha A_\bullet + i\epsilon}\kety\,
\nonumber\\
&&~~~~~= \left[-\int_0^{+\infty}\!\!{\dhd \alpha\over 4\alpha^2}\theta(x_*-y_*) 
+ \int_{-\infty}^0\!\!{\dhd\alpha\over 4\alpha^2}\theta(y_*-x_*) \right]e^{-i\alpha(x_\bullet - y_\bullet)}
\nonumber\\
&&~~~~~~~~\times
\int_{y_*}^{x_*}\!\!\!d{2\over s}z_{1*}\braxp e^{-i{\hatp^2_\perp\over \alpha s}x_*}
[x_*,z_{1*}]A_1[z_{1*},y_*]e^{i{\hatp^2_\perp\over \alpha s}y_*}\ketyp
\label{A1}
\end{eqnarray}
and
\begin{eqnarray}
&&\brax {1\over p^2 + 2g\alpha A_\bullet + i\epsilon } A_1 {1\over p^2 + 2g\alpha A_\bullet + i\epsilon}
A_2 {1\over p^2 + 2g\alpha A_\bullet + i\epsilon}\kety
\nonumber\\
&&= \left[i\int_0^{+\infty}\!\!{\dhd \alpha\over 8\alpha^3}\theta(x_*-y_*) 
- i\int_{-\infty}^0\!\!{\dhd\alpha\over 8\alpha^3}\theta(y_*-x_*) \right]e^{-i\alpha(x_\bullet - y_\bullet)}
\label{A12}\\
&&~~~\times
\int_{y_*}^{x_*}\!\!\!d{2\over s}z_{2*}\int^{x_*}_{z_{2*}}\!\!\!d{2\over s}z_{1*}\,
\braxp e^{-i{\hatp^2_\perp\over \alpha s}x_*}
[x_*,z_{1*}]A_1[z_{1*},z_{2*}]A_2[z_{2*},y_*]e^{i{\hatp^2_\perp\over \alpha s}y_*}\ketyp
\nonumber
\end{eqnarray}
To obtain (\ref{A1}) and (\ref{A12}), one hae to insert a complite set of states between the operators $A_1$ and $A_2$
and use the eikonal term of the scalar propagator in eq. (\ref{scalrprop-subeik}).
Another identities that we are going to need is
\begin{eqnarray}
\hspace{-1cm}&&{iz_*\over \alpha s}[p^2_\perp,F_{j\bullet}] + {2\over s}z_*{ig\over 2\alpha}
\big(-\{P_i,A^i\}+gA_iA^i\big)F_{j\bullet}
+F_{j\bullet}{2\over s}z_*{ig\over 2\alpha}\big(\{P_i,A^i\}-gA_iA^i\big)
\nonumber\\
\hspace{-1cm}&&~~~~~~= - {iz_*\over \alpha s} \big\{P^i,iD_i F_{j\bullet}\big\}
\label{usefla3}
\end{eqnarray}
So, using eq. (\ref{scalrprop-subeik}), and identities (\ref{usefla}), (\ref{usefla2}), (\ref{A1}), (\ref{A12}), and (\ref{usefla3}),
the first two terms of expansion (\ref{qprop-1st-stepexp}) become
\begin{eqnarray}
&&\hspace{-2cm} i\Slash{D}\brax{i\over p^2 + 2g\alpha A_\bullet + gO} - 
{i\over p^2 + 2g\alpha A_\bullet + gO}ig{2\over s}F_{i\bullet}\gamma^i\ssp_2{1\over p^2 + 2g\alpha A_\bullet + gO}\kety
\nonumber\\
&&\hspace{-2cm}
= \left[\int_0^{+\infty}\!\!{\dhd \alpha\over 2\alpha}\theta(x_*-y_*) - 
\int_{-\infty}^0\!\!{\dhd\alpha\over 2\alpha}\theta(y_*-x_*) \right]e^{-i\alpha(x_\bullet - y_\bullet)}{1\over \alpha s}
\braxp\, e^{-i{\hatp^2_\perp\over \alpha s}x_*}
\ssp\ssp_2[x_*,y_*]\ssp
\nonumber\\
&&\hspace{-2cm}
~~~+ {ig\over 2\alpha}
\int^{x_*}_{y_*}\!\!d{2\over s}\omega_*\,\ssp\ssp_2\bigg(
{2\over s}\omega_*\,\big\{P_i,[x_*,\omega_*]F^i_{~\bullet}[\omega_*,y_*]\big\}
\nonumber\\
&&
+ g\!\!\int^{x_*}_{\omega_*}\!\!d{2\over s}\,\omega'_*\,{2\over s}\big(\omega_* - \omega'_*\big)
[x_*,\omega'_*]F^i_{~\bullet}[\omega'_*,\omega_*]\,
\,F_{i\bullet}\,[\omega_*,y_*]
\bigg)\ssp
\Bigg\}e^{i{\hatp^2_\perp\over \alpha s}y_*}\ketyp ~~~~
\label{qprop-1st-term}
\end{eqnarray}
Note that to get eq. (\ref{qprop-1st-term}) we have \textit{pushed} the operator $\hat{\ssp}$ all the way
to the right so to have a $\hat{\ssp}$ to the left and another one to the right. This is the structure 
of the quark propagator in the background of shock-wave: free propagator
until the interaction with the scock-wave, eikonal plus sub-eikonal interaction with the shock-wave, and again free propagation
after the shock-wave. Note also that, since the operator $\hat{\ssp}$ is outside the shock-wave
\textit{i.e.} is at the point $x_*$, it can be promoted to $\hat{\Sp}$ because the gauge fields are pure gauge outside the shock-wave
(points $x_*$ and $y_*$).
Then, to \textit{push} the operator $\hat{\Sp}$ to the right we repeatedly use
\begin{eqnarray}
\braxp \ssp\ssp_2\Sp[x_*,y_*]\ketyp =\!\!&& \braxp \ssp\ssp_2\big(\alpha\ssp_1 + \Sp_\perp\big)[x_*,y_*]\ketyp
\nonumber\\
=\!\!&& \braxp \ssp\ssp_2\Big(\gamma^i\,i\mathfrak{D}_i[x_*,y_*] + [x_*,y_*](\alpha\ssp_1 + \Sp_\perp\big)\Big)\ketyp
\end{eqnarray}
and the definition of $\mathfrak{D}_i$ eq. (\ref{coderiv-glink}). Recall also that $\alpha = {2\over s}p_* 
= {2\over s}i{\partial\over \partial x_\bullet}$, and it commutes with all the fields because they do not depend on $x_\bullet$.
In a very similar way, we have 
\begin{eqnarray}
&&\hspace{-2cm}
i\Slash{D}\brax\Big[- {i\over p^2 + 2g\alpha A_\bullet }\,g\half\sigma_{ij}F^{ij}\,{1\over p^2 + 2g\alpha A_\bullet }
\nonumber\\
&&\hspace{-2cm} + {i\over p^2 + 2g\alpha A_\bullet }ig{2\over s}F_{i\bullet}\gamma^i\ssp_2{1\over p^2 + 2g\alpha A_\bullet }
\half\sigma_{ij}F^{ij}{1\over p^2 + 2g\alpha A_\bullet}
\nonumber\\
&&\hspace{-2cm}
+ {i\over p^2 + 2g\alpha A_\bullet }\,\half\sigma_{ij}F^{ij}\,{1\over p^2 + 2g\alpha A_\bullet }
ig{2\over s}F_{i\bullet}\gamma^i\ssp_2{1\over p^2 + 2g\alpha A_\bullet }\Big]\kety
\nonumber\\
&&\hspace{-2cm}
= \left[\int_0^{+\infty}\!\!{\dhd \alpha\over 2\alpha}\theta(x_*-y_*) - 
\int_{-\infty}^0\!\!{\dhd\alpha\over 2\alpha}\theta(y_*-x_*) \right]e^{-i\alpha(x_\bullet - y_\bullet)}
{1\over \alpha s}\braxp\, e^{-i{\hatp^2_\perp\over \alpha s}x_*}
\nonumber\\
&&\hspace{-2cm}
~~~\times\!\bigg[{ig\over 2\alpha}\int_{y_*}^{x_*}\!\!\!d{2\over s}z_*\, \ssp\ssp_2\Big(
[x_*,z_*]\half F_{ij}\sigma^{ij}[z_*,y_*]\ssp 
+ \big\{p^k,[x_*,z_*]i F_{kj}\gamma^j[z_*,y_*]\big\}
\nonumber\\
&&\hspace{-2cm}
~~~~~~+ [x_*,z_*]i F_{kj}\gamma^j(iD^k[z_*,y_*]) - (iD^k[x_*,z_*])i F_{kj}\gamma^j[z_*,y_*] 
\Big)\bigg]e^{i{\hatp^2_\perp\over \alpha s}y_*}\ketyp
\label{qprop-2nd-term}
\end{eqnarray}
and
\begin{eqnarray}
&&\hspace{-2cm}
i\Slash{D}\brax\Big[- {i\over p^2 + 2g\alpha A_\bullet }\,g{4\over s^2}F_{\bullet *}\sigma_{*\bullet}\,{1\over p^2 + 2g\alpha A_\bullet }
\nonumber\\
&&\hspace{-2cm} + {i\over p^2 + 2g\alpha A_\bullet }ig{2\over s}F_{i\bullet}\gamma^i\ssp_2{1\over p^2 + 2g\alpha A_\bullet}
{4\over s^2}F_{\bullet *}\sigma_{*\bullet}{1\over p^2 + 2g\alpha A_\bullet}
\nonumber\\
&&\hspace{-2cm}
+ {i\over p^2 + 2g\alpha A_\bullet }\,{4\over s^2}F_{\bullet *}\sigma_{*\bullet}\,{1\over p^2 + 2g\alpha A_\bullet }
ig{2\over s}F_{i\bullet}\gamma^i\ssp_2{1\over p^2 + 2g\alpha A_\bullet }\Big]\kety
\nonumber\\
&&\hspace{-2cm}= 
\left[\int_0^{+\infty}\!\!{\dhd \alpha\over 2\alpha}\theta(x_*-y_*) - 
\int_{-\infty}^0\!\!{\dhd\alpha\over 2\alpha}\theta(y_*-x_*) \right]e^{-i\alpha(x_\bullet - y_\bullet)}
{1\over \alpha s}\braxp\, e^{-i{\hatp^2_\perp\over \alpha s}x_*}
\nonumber\\
&&\hspace{-2cm}
~~~\times\bigg[
{ig\over 2\alpha}\int_{y_*}^{x_*}\!\!\!d{2\over s}z_*
\,\ssp\ssp_2(\alpha\ssp_1-\ssp_\perp)[x_*,z_*]i{2\over s} F_{\bullet *}[z_*,y_*]
+ {ig\over 2\alpha} \int_{y_*}^{x_*}\!\!\!d{2\over s}z_*
\nonumber\\
&&\hspace{-2cm}
~~~~~~\times\ssp\ssp_2\,\Big((i\Slash{D}_\perp [x_*,z_*])i{2\over s} F_{\bullet *}[z_*,y_*] 
- [x_*,z_*]i{2\over s} F_{\bullet *}(i\Slash{D}_\perp[z_*,y_*])\Big)
\bigg]e^{i{\hatp^2_\perp\over \alpha s}y_*}\ketyp~~~~~~~~~
\label{qprop-3rd-term}
\end{eqnarray}

Summing the three terms (\ref{qprop-1st-term}), (\ref{qprop-2nd-term}), and (\ref{qprop-3rd-term})
and symmetrizing the propagator by adding LHS of eq. (\ref{qprop-1st-stepexp}) with $\Sp$ to the right,
we arrive at the quark propagator with sub-eikonal corrections in the background of gluon fields \cite{Chirilli:2018kkw}
\begin{eqnarray}
\hspace{-1cm}\brax{i\over \hat{\Sp} +i\epsilon}\kety
=\!\!&& \left[\int_0^{+\infty}\!\!{\dhd \alpha\over 2\alpha}\theta(x_*-y_*) - 
\int_{-\infty}^0\!\!{\dhd\alpha\over 2\alpha}\theta(y_*-x_*) \right] e^{-i\alpha(x_\bullet - y_\bullet)}
{1\over \alpha s}\,
\nonumber\\
&&\hspace{-0.1cm}
\times\braxp\,e^{-i{\hatp^2_\perp\over \alpha s}x_*}\Bigg\{
\hat{\ssp}\,\ssp_2\,[x_*,y_*]\,\hat{\ssp}
+ \hat{\ssp}\,\ssp_2\,\hat{\mathcal{O}}_1(p_\perp; x_*,y_*)\,\hat{\ssp}
\nonumber\\
&&\hspace{-0.1cm}
+ \hat{\ssp} \,\ssp_2\,\half \hat{\mathcal{O}}_2(p_\perp; x_*,y_*) 
- \half\hat{\mathcal{O}}_2(x_*,y_*;p_\perp)\,\ssp_2\,\hat{\ssp} 
\Bigg\}e^{i{\hatp^2_\perp\over \alpha s}y_*}\ketyp + O(\lambda^{-2})\,.
\label{sym-quarksubnoedge2}
\end{eqnarray}
with
\begin{eqnarray}
\hspace{-0.5cm} &&
\hat{\mathcal{O}}_1(x_*,y_*;p_\perp) 
\hspace{-0.5cm} \nonumber\\
&&=
{ig\over 2\alpha}\int^{x_*}_{y_*}\!\!\!d{2\over s}\omega_*\bigg(
[x_*,\omega_*]\half \sigma^{ij}F_{ij}[\omega_*,y_*]
+ \big\{\hat{p}^i,[x_*,\omega_*]\,{2\over s}\,\omega_*\, F_{i\bullet}(\omega_*)\,[\omega_*,y_*]\big\}
\nonumber\\
&&\hspace{2.5cm} + g\!\!\int^{x_*}_{\omega_*}\!\!\!d{2\over s}\,\omega'_*\,{2\over s}\big(\omega_* - \omega'_*\big)
[x_*,\omega'_*]F^i_{~\bullet}[\omega'_*,\omega_*]\,F_{i\bullet}\,[\omega_*,y_*]\bigg)\,,
\label{O1}
\end{eqnarray}
and
\begin{eqnarray}
\hspace{-0.3cm}&&\hat{\mathcal{O}}_2(p_\perp; x_*,y_*) 
\label{O2}\\
\hspace{-0.3cm}&&=
{ig\over 2\alpha}\int^{x_*}_{y_*}\!\!\!d{2\over s}\omega_*
\Bigg[
\big\{\hat{p}^k,[x_*,\omega_*]i F_{kj}\gamma^j[\omega_*,y_*]\big\}
+ (\alpha\ssp_1-\hat{\ssp}_\perp)[x_*,\omega_*]\,i\,{2\over s} F_{\bullet *}[\omega_*,y_*]
\nonumber\\
\hspace{-0.3cm}&&
~~~+ \int^{x_*}_{\omega_*}\!\!\!d{2\over s}{\omega'_*}
\Bigg([x_*,\omega'_*]gF^k_{~\bullet}[\omega'_*,\omega_*]iF_{kj}\gamma^j[\omega_*,y_*]
- [x_*,\omega'_*]iF_{kj}\gamma^j[\omega'_*,\omega_*]gF^k_{~\bullet}[\omega_*,y_*]
\nonumber\\
\hspace{-0.3cm}&&
~~~~~~+ [x_*,\omega'_*]i{2\over s}F_{\bullet*}[\omega'_*,\omega_*]\gamma^kgF_{k\bullet}[\omega_*,y_*]
- [x_*,\omega'_*]\gamma^kgF_{k\bullet}[\omega'_*,\omega_*]i{2\over s}F_{\bullet*}[\omega_*,y_*]
\Bigg)\Bigg]\,.
\nonumber
\end{eqnarray}
where $\big\{\hat{p}^i,[x_*,\omega_*]\big\} =  \hat{p}^i[x_*,\omega_*] + [x_*,\omega_*]\hat{p}^i$.

The definition of operator $\hat{\mathcal{O}}_2$ given here
differs from the one given in the previous work, Ref. \cite{Chirilli:2018kkw}, because
using the identity $\{\gamma^k,\gamma^i\gamma^j\} = 2\gamma^k g^{ij}$
the term ${i\over 4}\big\{(i$ $\Slash{D}_\perp F_{ij}),\gamma^i\gamma^j\big\}$
is identically zero.

Now let us define the operator $\calo_j$ as follow
\begin{eqnarray}
\hspace{-0.3cm}&&\hat{\mathcal{O}}_j(p_\perp; x_*,y_*) 
\label{Oj}\\
&&\equiv
{ig\over 2\alpha}\int^{x_*}_{y_*}\!\!\!d{2\over s}\omega_*
\Bigg[
\big\{\hat{p}^k,[x_*,\omega_*]i F_{kj}[\omega_*,y_*]\big\}
\nonumber\\
&&
\hspace{1.5cm}+ \int^{x_*}_{\omega_*}\!\!\!d{2\over s}{\omega'_*}
\Bigg([x_*,\omega'_*]gF^k_{~\bullet}[\omega'_*,\omega_*]iF_{kj}[\omega_*,y_*]
- [x_*,\omega'_*]iF_{kj}[\omega'_*,\omega_*]gF^k_{~\bullet}[\omega_*,y_*]
\nonumber\\
&&
\hspace{2.5cm}+ [x_*,\omega'_*]i{2\over s}F_{\bullet*}[\omega'_*,\omega_*]gF_{j\bullet}[\omega_*,y_*]
- [x_*,\omega'_*]gF_{j\bullet}[\omega'_*,\omega_*]i{2\over s}F_{\bullet*}[\omega_*,y_*]
\Bigg)\Bigg]\,.
\nonumber
\end{eqnarray}
and we can write
\begin{eqnarray}
\hat{\ssp}\ssp_2\gamma^j \hat{\mathcal{O}}_j - \hat{\mathcal{O}}_j\gamma^j\ssp_2\hat{\ssp}
= i\,\alpha s\epsilon^{ij}\gamma^5\gamma_i\hat{\mathcal{O}}_j - \ssp_2\big[\hat{p}^j,\hat{\mathcal{O}}_j\big]
- i\epsilon^{ij}\gamma^5\ssp_2\{\hat{p}_i,\hat{\mathcal{O}}_j\}
\end{eqnarray}
(recall $\ssp\ssp_2 = \alpha\ssp_1\ssp_2 + \ssp_\perp \ssp_2$).

Moreover, we may define\footnote{Note that the term of eq. (\ref{Obulletstar}), since fields do not depend on $x_\bullet$,
may be analyzed similarly to the term $\{P_\bullet, A_*\}$ in Appendix C of Ref. \cite{Chirilli:2018kkw}.
The result is that it may be reduced to gauge dependent terms $A_*[x_*,y_*] - [x_*,y_*]A_*$.}
\begin{eqnarray}
\hat{\mathcal{O}}_{\bullet*}(x_*,y_*) \equiv {ig\over 2\alpha}\int^{x_*}_{y_*}\!\!\!d{2\over s}\omega_*[x_*,\omega_*]i{2\over s}F_{\bullet *}[\omega_*,y_*]
\label{Obulletstar}
\end{eqnarray}
so we may write 
\begin{eqnarray}
\hat{\ssp}\ssp_2(\alpha\ssp_1-\hat{\ssp}_\perp)\hat{\mathcal{O}}_{\bullet*} 
- \hat{\mathcal{O}}_{\bullet*}(\alpha\ssp_1 - \hat{\ssp}_\perp)\ssp_2\hat{\ssp}
=  i\,\alpha s\gamma^5\gamma_j\epsilon^{ij}\{p_i,\calo_{\bullet *}\} - \ssp_2[\hat{p}^2_\perp,\hat{\mathcal{O}}_{\bullet*}]
\end{eqnarray}

Using the operators $\calo_j(x_*,y_*)$ and $\calo_{\bullet *}(x_*,y_*)$
the quark propagator (\ref{sym-quarksubnoedge2}) becomes
eq. (\ref{ojobulletstar}).

\subsection{Gluon propagator in the background of gluon field}

The gluon propagator with sub-eikonal corrections in the background of gluon fields is \cite{Chirilli:2018kkw}
\begin{eqnarray}
	&&\langle A^a_\mu(x)A^b_\nu(y)\rangle_A
	\nonumber\\
	&&= \left[-\int_0^{+\infty}\!\!{\dhd \alpha\over 2\alpha}\theta(x_*-y_*) 
	+ \int_{-\infty}^0\!\!{\dhd\alpha\over 2\alpha}\theta(y_*-x_*) \right]e^{-i\alpha(x_\bullet - y_\bullet)}
	\braxp e^{-i{\hatp^2_\perp\over \alpha s}x_*}
	\nonumber\\
	&&~~~\times\left(\delta_\mu^\xi - 
	{2\over s}{p_{2\mu}p^\xi\over \alpha}\right)\!\calo_\alpha(x_*,y_*)\!
	\left(g_{\xi\nu} - {2\over s}{p_\xi\, p_{2\nu}\over \alpha}\right)\!e^{i{\hatp^2_\perp\over \alpha s}y_*} \ketyp^{ab}
	+ i\brax {4\over s^2}{p_{2\mu}p_{2\nu}\over \alpha^2}\kety^{ab}\nonumber\\
	&&~~~ + \left[-\int_0^{+\infty}\!\!{\dhd \alpha\over 2\alpha}\theta(x_*-y_*) 
	+ \int_{-\infty}^0\!\!{\dhd\alpha\over 2\alpha}\theta(y_*-x_*) \right]e^{-i\alpha(x_\bullet - y_\bullet)}
	\braxp e^{-i{\hatp^2_\perp\over \alpha s}x_*}
	\nonumber\\
	&&~~~\times\Big[
	\mathfrak{G}^{ab}_{1\mu\nu}(x_*,y_*; p_\perp) +
	\mathfrak{G}^{ab}_{2\mu\nu}(x_*,y_*; p_\perp) + \mathfrak{G}^{ab}_{3\mu\nu}(x_*,y_*; p_\perp) 
	+ \mathfrak{G}^{ab}_{4\mu\nu}(x_*,y_*; p_\perp)\Big]
	\nonumber\\
	&&~~~\times e^{i{\hatp^2_\perp\over \alpha s}y_*}\ketyp
	+ O(\lambda^{-2})\,,
	\label{Shwgaxg9}
\end{eqnarray}
where we defined 
\begin{eqnarray}
\hspace{-1cm}	{\calo_\alpha(x_*,y_*)} \equiv && [x_*,y_*] 
	+ {ig\over 2\alpha}
	\int^{x_*}_{y_*}\!\!d{2\over s}\omega_*\bigg(
	\big\{p^i,[x_*,\omega_*]\,{2\over s}\,\omega_*\, F_{i\bullet}(\omega_*)\,[\omega_*,y_*]\big\}
	\label{Oalfa}\\
	&&\hspace{2cm} + g\!\!\int^{x_*}_{\omega_*}\!\!d{2\over s}\,\omega'_*\,{2\over s}\big(\omega_* - \omega'_*\big)
	[x_*,\omega'_*]F^i_{~\bullet}[\omega'_*,\omega_*]\,
	\,F_{i\bullet}\,[\omega_*,y_*]\bigg)\,.
	\nonumber
\end{eqnarray}
and
\begin{eqnarray}
	\hspace{-1cm}\mathfrak{G}^{ab}_{1\mu\nu}(x_*,y_*; p_\perp) =\!\!&&
	-{g\,p_{2\mu}p_{2\nu}\over s^2\alpha^3}\int_{y_*}^{x_*}\!\!\!d{2\over s}\omega_*\bigg[
	4p^i[x_*,\omega_*]F_{ij}[\omega_*,y_*]p^j
	\nonumber\\
	&& + ig\int_{\omega'_*}^{x_*}\!\!\!d{2\over s}\omega'_*\,{2\over s}(\omega'_* - \omega_*)
	[x_*,\omega'_*]iD^iF_{i\bullet}[\omega'_*,\omega_*]iD^jF_{j\bullet}[\omega_*,y_*]
	\bigg]^{ab}\,,
	\label{G1}\\
	\hspace{-1cm}\mathfrak{G}^{ab}_{2\mu\nu}(x_*,y_*; p_\perp) =\!\!&&
	- {g\over \alpha}\delta^i_\mu \delta^j_\nu\int_{y_*}^{x_*}\!\!\!d{2\over s}\omega_*
	\big([x_*,\omega_*]F_{ij}[\omega_*,y_*]\big)^{ab}\,,
	\label{G2}\\
	\nonumber\\
	\hspace{-1cm}\mathfrak{G}^{ab}_{3\mu\nu}(x_*,y_*; p_\perp) = \!\!&&
	{g\over  \alpha^2 s}\Big(\delta^j_\mu p_{2\nu} + \delta^j_\nu p_{2\mu}\Big)
	\int_{y_*}^{x_*}\!\!\!d{2\over s}\omega_* \,\big([x_*,\omega_*]iD^i F_{ij}[\omega_*,y_*]\big)^{ab}\,,
	\label{G3}\\
	\nonumber\\
	\hspace{-1cm}\mathfrak{G}^{ab}_{4\mu\nu}(x_*,y_*; p_\perp) = \!\!&& - {2g^2 \over \alpha^2s}
	\int_{y_*}^{x_*}\!\!\!d{2\over s}\omega_*\int_{\omega_*}^{x_*}\!\!\!d{2\over s}\omega'_*
	\Big(\delta^j_\nu p_{2\mu}[x_*,\omega'_*]F^i_{~\bullet}[\omega'_*,\omega_*]
	F_{ij}[\omega_*,y_*]
	\nonumber\\
	\hspace{-1cm}&& +  \delta^j_\mu p_{2\nu}[x_*,\omega'_*]F_{ij}[\omega'_*,\omega_*]
	F^i_{~\bullet}[\omega_*,y_*]\Big)^{ab}\,.
	\label{G4}
\end{eqnarray}

\subsection{Gluon propagator in the background of quark field}

The sub-eikonal corrections to the gluon propagator in the background of quark fields has been calculated in Ref. \cite{Chirilli:2018kkw}
\begin{eqnarray}
	\hspace{-1cm}&&\langle A_\mu^a(x) A_\nu^b(y)\rangle_{\psi,\bar{\psi}}
	\nonumber\\
	\hspace{-1cm}&&=
	\Big[-\int_0^{+\infty}\!\!{\dhd\alpha \over  2\alpha}\theta(x_* - y_*)
	+ \int_{-\infty}^0\!\!{\dhd\alpha\over 2\alpha}\theta(y_* - x_*)\Big]e^{-i\alpha(x_\bullet - y_\bullet)}
	\nonumber\\
	\hspace{-1cm}&&~~~\times g^2\!\int_{y_*}^{x_*}\!\!\! d{2\over s}z_{1*}\int_{y_*}^{z_{1_*}}\!\!\! d{2\over s}z_{2*}
	\,{1\over 4\alpha}\int\!d^2z\Bigg[\braxp
	\,e^{-i{\hatp^2_\perp\over \alpha s}x_*}\Big(g_{\perp\mu}^\xi - {s\over 2}{p_{2\mu}\over \alpha}p^\xi_\perp\Big)\ketzp
	\bar{\psi}(z_{1*},z_\perp)
	\nonumber\\
	\hspace{-1cm}&&~~~
	\times\! \gamma^\perp_\xi\,\ssp_1\,[z_{1*},x_*]_zt^a[x_*,y_*]_zt^b[y_*,z_{2*}]_z
	\gamma^\sigma_\perp\psi(z_{2*},z_\perp)
	\brazp\Big(g^\perp_{\sigma\nu} - p^\perp_\sigma{s\over 2}{p_{2\nu}\over \alpha}\Big)
	e^{i{\hatp^2_\perp\over \alpha s}y_*}\ketyp 
	\nonumber\\
	\hspace{-1cm}&&~~~ +\,\brayp\,e^{-i{\hatp^2_\perp\over \alpha s}y_*}
	\Big(g_{\perp\nu}^\xi - {s\over 2}{p_{2\nu}\over\alpha}p^\xi_\perp\Big)\ketzp
	\bar{\psi}(z_{2*},z_\perp)  \gamma^\perp_\xi\,\ssp_1[z_{2*},y_*]_zt^b[y_*,x_*]_zt^a[x_*,z_{1*}]_z\gamma^\sigma_\perp
	\nonumber\\
	\hspace{-1cm}&&~~~\times\!\psi(z_{1*},z_\perp)\brazp\Big(g^\perp_{\sigma\mu} - p^\perp_\sigma{s\over 2}{p_{2\mu}\over \alpha}\Big)
	e^{i{\hatp^2_\perp\over \alpha s}x_*}\ketxp\Bigg] + O(\lambda^{-2}) ~~~~~~~~~~
	\label{gluonprop-inq}
\end{eqnarray}
Note that the entire sub-eikonal correction is at the transverse position $z_\perp$.
Moreover, in the shock-wave limit we are employing here, we have to send $x_*\to +\infty$ and $y_*\to -\infty$.
In this limit the gauge link $[x_*,y_*]_z$ becomes the usual infinite Wilson line $U(z_\perp)$.

\section{Evaluation of matrix elements}
\label{sec: matrielements}

To understand which of the sub-eikonal term in the quark propagator (\ref{ojobulletstar}) will contribute to the
polarized structure functions we have to consider forward matrix elements with the sub-eikonal operators. 
In other words, we have to analyze the matrix element that will be obtained using the quark propagator (\ref{ojobulletstar})
to calculate diagram in Fig. \ref{subeikonalif} for the impact factor.

The polarization vector is a pseudo-vector which satisfy $S^\mu S_\mu = -1$ and $S\!\cdot\!P = 0$.
Let $S_L^\mu$ be the longitudinal component of the spin vector. In the DIS kinematics we have that
$S^\mu_L \simeq {\lambda\over M} P^\mu$
so, we may write $S^\mu \simeq {\lambda\over M}P^\mu + S^\mu_\perp$.

We chose the proton momentum to be mainly in $p_2^\mu$ direction:  
$P^\mu=p_2^\mu + {M^2\over s}p_1^\mu$, and  the virtual photon has momentum $q^\mu = p_1^\mu - x_Bp^\mu_2$.

Helicity is defined as $h=\lambda{\vec{S}\cdot\vec{P}\over |\vec{P}|} \simeq \lambda g^{3\mu}S_\mu \simeq - \lambda{\sqrt{s}\over 2M}$
with $\lambda = \pm \half$.

The components of the hadronic tensor $W^{\mu\nu}$, eq. (\ref{Htensor}),
which are associated the polarized structure functions $g_1$ and $g_2$, are
\begin{eqnarray}
&&W^\perp_{\mu\nu} = - g_{\mu\nu}^\perp F_1 + i\epsilon^\perp_{\mu\nu} g_1
\\
&&g^i_\mu W_{\bullet i} =  i\, Mx_B\, \epsilon^\perp_{\mu\nu}\, S_\perp^\nu g_T = - g^i_\nu W_{i\bullet}
\\
&&g^i_\mu W_{*i} = i \epsilon_{\mu\nu}^\perp S_\perp^\nu M g_T = - g^i_\nu W_{i*}
\end{eqnarray}
where $g_T= g_1+g_2$ is the transverse polarized structure function.
We see that the transverse hadronic tensor is associated to the $g_1$ structure function with longitudinal polarization,
while transverse polarized structure function $g_T$ is obviously associated to the transverse spin $S_\perp^\mu$.

To evaluate the matrix elements we will use spin vector $S^\mu$, the target momentum $P^\mu$,
the direction of the Wilson line ${2\over s}p_1^\mu$ and the transverse momentum $k^\mu_\perp$ 
conjugated to the size of the dipole $\Delta^\mu_\perp = (x-y)^\mu_\perp$.
Note also that all distribution functions that we are going to introduce
have dimensions $[M^{-2}]$.

\subsection{Matrix element with $\hat{\calo}_1$ operator}

Here we are going to evaluate the matrix element that would be generated using the operator $\hat{\calo}_1$
defined in eq. (\ref{O1}).
All matrix elements will be of dipole type with the insertion of the sub-eikonal correction (see Fig. \ref{subeikonalif}).

Let us consider the following matrix elements 
\begin{eqnarray}
&&\int \!d^2\Delta e^{i(\Delta,k)_\perp}\int_{-\infty}^{+\infty}\!\! d z_*
\langle\langle P,S|\Big[\Tr\{[\infty p_1, z_*]_x \,igF^{ij}(z_*,x_\perp)[z_*,-\infty p_1]_x U^\dagger_y\}+{\rm a.c}\Big]|P,S\rangle\rangle
\nonumber\\
&&=  a{4\over s^2}\epsilon^{\bullet * \alpha i}(S_* P_\bullet - S_\bullet P_*) {k^\perp_\alpha k^j\over M ^5}
+ b {4\over s^2}\epsilon^{\bullet * \alpha i}S_\bullet {k^\perp_\alpha k^j\over M^3}
+ c {4\over s^2}\epsilon^{\bullet * \alpha i}S_\alpha^\perp P_\bullet {k^j\over M^3}
+ d {k^i k^j\over M^4}
\nonumber\\
&&=  {2\over s M^2}\epsilon^{k i} k^j\Big({a\over M^3}(S_* P_\bullet - S_\bullet P_*) k_k
+ {b\over M} \,S_\bullet k_k
+ {c\over M} \,S_k P_\bullet \Big) + d \, {k^i k^j\over M^4}\,.
\label{matrixelementFij-1}
\end{eqnarray}
with $a, b, c, d$ dimensionless coefficients.
We now use $P^\mu \simeq (P^0, 0, 0, P^3) = {\sqrt{s}\over 2}(1,0,0,-1)$, and $S\!\cdot\!P =0$, and 
\begin{eqnarray}
S_* P_\bullet - S_\bullet P_* &=& {s\over 2}\big(S^3 P^0 - S^0 P^3\big) 
= {s\over 2}\,{M^2 S^3\over P^0}\simeq {s\over 2} {\lambda MP^3\over P^0}
= - {s\over 2}\lambda M\,,\\
S_\bullet &\simeq& {1\over M}\lambda P_\bullet = \lambda{s\over 2M}\,,
\end{eqnarray}
where we also used $MS^3\simeq \lambda P^3$. 
The matrix element (\ref{matrixelementFij-1}) becomes
\begin{eqnarray}
&&\int \!d^2\Delta e^{i(\Delta,k)_\perp}\!\int_{-\infty}^{+\infty}\!\! d z_*
\langle\langle P,S|\Big[\Tr\{[\infty p_1, z_*]_x \,igF^{ij}(z_*,x_\perp)[z_*,-\infty p_1]_x U^\dagger_y\}+{\rm a.c.}\Big]|P,S\rangle\rangle
\nonumber\\
\nonumber\\
&&=  {\epsilon^{k i} k^j\over  M^2}\Big(\lambda G_L(k^2_\perp, x) k_k+ G_T(k^2_\perp,x) MS_k\Big)\,.
\end{eqnarray}
where we have introduced the polarized longitudinal (helicity) distribution function
$G_L(k^2_\perp, x)$, the transverse gluon distribution function $G_T(k^2_\perp, x)$.
As usual, Latin indexes assume values $1, 2$, while Greek ones run from $0$ to $3$, $(x,y)_\perp = x^1y^1+x^2y^2$.

Let us consider the matrix element generated by the term $\epsilon^{ij}F_{ij}$ in operator (\ref{O1})
\begin{eqnarray}
&&\int \!d^2\Delta e^{i(\Delta,k)_\perp}\!\!\int_{-\infty}^{+\infty}\!\! d z_*
\langle\langle P,S|\Big[\Tr\{[\infty p_1, z_*]_x \,ig{s\over 2}\epsilon^{ij}F_{ij}(z_*,x_\perp)[z_*,-\infty p_1]_x U^\dagger_y\}
+ {\rm a.c}\Big]|P,S\rangle\rangle
\nonumber\\
&& = {s\over 2}\Big[\lambda {k^2_\perp\over M^2}
G_L(k^2_\perp,x)  + {(S,k)_\perp\over M} G_T(k^2_\perp,x) \Big]
\label{examparity}
\end{eqnarray}
So, the matrix element (\ref{examparity}) will contribute to longitudinal polarized structure function, $g_1$, and to
the transverse polarized structure function $g_T$.

Consider
\begin{eqnarray}
&&\int \!d^2\Delta e^{i(\Delta,k)_\perp}\!
\int_{-\infty}^{+\infty}\!\!\!d\omega_*\int^{+\infty}_{\omega_*}\!\!\! d\omega'_*\,\big(\omega_* - \omega'_*\big)
\nonumber\\
&&\times \langle\langle P,S|\Big[\Tr\{[\infty p_1,\omega'_*]_xF^i_{~\bullet}[\omega'_*,\omega_*]_x
\,F_{i\bullet}\,[\omega_*,-\infty p_1]_xU^\dagger_y\}+ {\rm a.c}\Big]|P,S\rangle\rangle
\nonumber\\
&& ~~~~~~=  {s\over 2}\hat{G}(k^2_\perp,x)
\end{eqnarray}
This matrix element will contribute to unpolarized structure function and $\hat{G}(k^2_\perp,x)$ is an
unpolarized gluon distribution function.

Now we consider 
\begin{eqnarray}
&&\int \!d^2\Delta e^{i(\Delta,k)_\perp}\!
\int_{-\infty}^{+\infty}\!\!\!d\omega_*\,\omega_*\,
k_i\langle\langle P,S|\Big[\Tr\{[\infty p_1,\omega'_*]_xF^i_{~\bullet}(\omega_*,x_\perp)[\omega_*,-\infty p_1]_xU^\dagger_y\}
+ {\rm a.c.}\Big]|P,S\rangle\rangle
\nonumber\\
&& =  {s\over 2}G^-(k^2_\perp,x) + {s\over 2}{\hat{S}\times \vec{k}\over M}G^-_T(k^2_\perp,x)
\end{eqnarray}
Also this matrix element will contribute to unpolarized structure function with 
and $G^-(k^2_\perp,x)$ is an unpolarized gluon distribution function (recall that ${F^i}_\bullet = \sqrt{s\over 2}F^{i-}$). 

\subsection{Matrix element with $\hat{\calo}_j$ operator}

Here we are going to evaluate the matrix element that would be generated using the operator $\hat{\calo}_j$
defined in eq. (\ref{Oj}).

Let us consider the term $\half\ssp_2[\hat{p}^j,\hat{\calo}_j]$. One of the matrix element that it will generate is
\begin{eqnarray}
&&\int \!d^2\Delta e^{i(\Delta,k)_\perp}\!\int_{-\infty}^{+\infty}\!\! d \omega_*\!\int_{\omega_*}^{+\infty}\!\!d\omega'_*
\nonumber\\
&&\times
k^j\langle\langle P,S| \Big[\Tr\{[\infty p_1,\omega'_*]_x{F^k}_\bullet [\omega'_*,\omega_*]_x
F_{kj}[\omega_*,-\infty p_1]_x U^\dagger_y\}+ {\rm a.c.}\Big]|P,S\rangle\rangle
\nonumber\\
&&
~~~~~~
= {s\over 2}k^2_\perp \barG(k^2_\perp, x) + {s\over 2}M\big(\vec{S}\times \vec{k}\big)\barG_T(k^2_\perp,x)\,.
\end{eqnarray}
This matrix element will contribute to the unpolarized structure function with gluon distribution
$\barG(k^2_\perp, x)$ and to the transverse polarized structure function with transversely polarized gluon distribution $\barG_T(k^2_\perp,x)$.

We also have matrix element
\begin{eqnarray}
&&\int \!d^2\Delta e^{i(\Delta,k)_\perp}\!\int_{-\infty}^{+\infty}\!\! d \omega_*\!\int_{\omega_*}^{+\infty}\!\!d\omega'_*
\nonumber\\
&&\times\langle\langle P,S| \Big[\Tr\{[\infty p_1,\omega'_*]_xF_{\bullet *} [\omega'_*,\omega_*]_x
F_{j\bullet}[\omega_*,-\infty p_1]_x U^\dagger_y\}+ {\rm a.c.}\Big]|P,S\rangle\rangle
\nonumber\\
&&~~~~~~ = {s\over 2}k_j\breve{G}(k^2_\perp, x) + \lambda {s\over 2}\epsilon^{ij}MS_i\breve{G}_L(k^2_\perp,x)\,.
\end{eqnarray}
This matrix element will contribute to unpolarized structure function with 
gluon distribution $\breve{G}(k^2_\perp,x)$ and to the transverse polarized structure function with transversely polarized 
gluon distribution $\breve{G}_T(k^2_\perp,x)$.

\subsection{Matrix element with $\hat{\calo}_{\bullet *}$ operator}

Let us now consider the forward matrix element obtained with operator  $\hat{\calo}_{\bullet *}$
\begin{eqnarray}
&&\int \!d^2\Delta \, e^{i(\Delta, k)} \langle\langle P,S| \Big[\Tr\{[\infty p_1,\omega_*]_xF_{\bullet *}(\omega_*,x_\perp)[\omega_*,-\infty p_1]_x U^\dagger_y\} + {\rm a.c.}\Big]|P,S\rangle\rangle
\nonumber\\
&&~~~~~~ = {s\over 2}\tildeG(k^2_\perp,x) + {s\over 2}{\vec{S}\times \vec{k}\over M}\, \tildeG_T(k^2_\perp,x)\,,
\end{eqnarray}
where we recall that $F_{\bullet *} = {s\over 2}F^{-+}= {s\over 2}F_{+-}$.
From this matrix element we have extracted an unpolarized gluon distribution $\tildeG(k^2_\perp, x)$,
and a transversally polarized gluon distribution function $\tildeG_T(k^2_\perp,x)$.
This matrix element will not contribute to $g_1$, but to $g_T$.

\subsection{Quark propagator for $g_1$ structure function}

From the analysis just performed we may conclude that
the gluon field sub-eikonal contribution to the quark propagator that we have to use to calculate 
the impact factor for polarized DIS and relevant for $g_1$ structure function is
\begin{eqnarray}
\hspace{-1cm}\brax{i\over \hat{\Sp} +i\epsilon}\kety
\ni && \left[\int_0^{+\infty}\!\!{\dhd \alpha\over 2\alpha}\theta(x_*-y_*) - 
\int_{-\infty}^0\!\!{\dhd\alpha\over 2\alpha}\theta(y_*-x_*) \right] e^{-i\alpha(x_\bullet - y_\bullet)}
{1\over \alpha s}\,
\nonumber\\
&&\times\!\int d^2z\braxp\,e^{-i{\hatp^2_\perp\over \alpha s}x_*}\ssp\ketzp
\nonumber\\
&&\hspace{-0.1cm}
\times\Bigg\{
\ssp_2\,[x_*,y_*]_z
+ {ig\over 2\alpha}\int^{x_*}_{y_*}\!\!\!d{2\over s}\omega_*[x_*,\omega_*]_z\half \ssp_2\gamma^5\epsilon^{ij}F_{ij}[\omega_*,y_*]_z
\Bigg\}
\nonumber\\
&&\times\!\brazp\ssp\, e^{i{\hatp^2_\perp\over \alpha s}y_*}\ketyp
\label{quark-subFij}
\end{eqnarray}
where the symbol $\ni$ in eq. (\ref{quark-subFij}) means that the terms in the right-hand-side (RHS) are 
only part of all the terms of the quark propagators; the terms we have left out will not contribute
to the calculation of the impact factor for polarized $g_1$ structure function. 
It should be stressed that here we are only concerned with the sub-eikonal correction with only gluon  background field.
Propagator (\ref{quark-subFij}) will be used to calculate the impact factor diagram in Fig. \ref{loif} b. 
To calculate the impact factor diagram in Fig. \ref{loif} b we need the quark propagator
with quarks in the background given in eq. (\ref{qprpinquark}).

\section{Derivation of the OPE with quark-sub-eikonal corrections}
\label{sec: qsub-correctionIF}

Here we provide some calculation details of the derivation of the OPE with quark-sub-eikonal corrections eq. (\ref{OPEquark}).
Let us consider the $T$-product of two electromagnetic currents am perform the functional integration over the spinor fields.
As usual, we start with the case in which $x^+>0>y^+$. Since the sub-eikonal correction can be included either in the 
quark fermion line or in the anti-quark fermion line, at sub-eikonal level we have, without exceeding our precision, two terms 
\begin{eqnarray}
&&\hspace{-2cm}\langle {\rm T}\{\hat{j}^{\mu}(x) \hat{j}^{\nu}(y)\}\rangle_{A,\psi,\bar{\psi}}
\stackrel{x^+>0>y^+}{=} - {\rm tr}\{\gamma^\mu \langle \psi(x)\bar{\psi}(y)\,\rangle_{A,\psi,\bar{\psi}}
\gamma^\nu \langle \psi(y)\bar{\psi}(x)\rangle_{A,\psi,\bar{\psi}}\}
\end{eqnarray}
Now we are concerned only with the contribution coming from quarks in the external field, so using the quark propagator
(\ref{qprop-backqua-coord}) we have
\begin{eqnarray}
&&\langle {\rm T}\{\hat{j}^{\mu}(x) \hat{j}^{\nu}(y)\}\rangle_{A,\psi,\bar{\psi}}
\nonumber\\
&&\stackrel{x^+>0>y^+}{\ni}
- \Tr\,\tr\Big\{\gamma^\mu {-g^2\over 16\pi^3x_*^2y_*^2}\int^{x_*}_{y_*}\!\!\!dz_*\!\!\int_{y_*}^{z_*}\!\!\!dz'_* 
\!\!\int d^2z_2
\nonumber\\
&&~~~~~~\times{\Sx_2[x_*,z_*]_{z_2} t^a\gamma_\perp^\alpha
\psi(z_*,z_{2\perp})[z_*,z'_*]^{ab}_{z_2}
\bar{\psi}(z'_*,z_{2\perp})\gamma^\perp_\alpha t^b[z'_*,y_*]_{z_2} \,\Sy_2\over 
\left(\calz_2 + i\epsilon \right)^2}
\nonumber\\
&&~~~~~~\times\gamma^\nu { i \over 2\pi^3x_*^2y_*^2 }\int\!\!d^2z_1{\Sy_1\,\ssp_2[y_*,x_*]_{z_1} \,\Sx_1\over 
\left(\calz_1 + i\epsilon\right)^3}\Big\}
- \Tr\,\tr\Big\{\gamma^\mu 
{- i\over 2\pi^3x_*^2y_*^2 }
\nonumber\\
&&~~~~~~\times\!
\int\!\!d^2z_1{\Sx_1\,\ssp_2[x_*,y_*]_{z_1} \,\Sy_1\over \left(\calz_1 + i\epsilon\right)^3}
\gamma^\nu{-g^2\over 16\pi^3x_*^2y_*^2}\int^{y_*}_{x_*}\!\!\!dz'_*\!\!\int_{x_*}^{z'_*}\!\!\!dz_* 
\!\!\int d^2z_2\,{1\over \left(\calz_2 + i\epsilon\right)^2}
\nonumber\\
&&~~~~~~\times\Sy_2[y_*,z'_*]_{z_2} t^a\gamma_\perp^\alpha
\psi(z'_*,z_{2\perp})[z'_*,z_*]^{ab}_{z_2}
\barpsi(z_*,z_{2\perp})\gamma^\perp_\alpha t^b[z_*,x_*]_{z_2} \,\Sx_2\Big\}
\label{ifdetails1}
\end{eqnarray}
The two terms in (\ref{ifdetails1}) represent the sub-eikonal correction for quark and anti-quark propagator.
We can further simplify (\ref{ifdetails1}) by renaming dummy variables
\begin{eqnarray}
&&	\langle {\rm T}\{\hat{j}^{\mu}(x) \hat{j}^{\nu}(y)\}\rangle_{A,\psi,\bar{\psi}}
\nonumber\\
&&\stackrel{x^+>0>y^+}{\ni}
{ig^2\over 32\pi^6 x_*^4y_*^4} \!\!\int d^2z_2 d^2z_1
\left(\calz_2 + i\epsilon\right)^{-2}\left(\calz_1 + i\epsilon\right)^{-3}
\nonumber\\
&&~~~~~~\times\!
\Bigg[\int^{x_*}_{y_*}\!\!\!dz_*\!\!\int_{y_*}^{z_*}\!\!\!dz'_*{\rm Tr\,tr}\{
\gamma^\mu\Sx_2[x_*,z_*]_{z_2} t^a\gamma_\perp^\alpha\psi(z_*,z_{2\perp})[z_*,z'_*]^{ab}_{z_2}
\barpsi(z'_*,z_{2\perp})\nonumber\\
&&~~~~~~\times\!\gamma^\perp_\alpha t^b[z'_*,y_*]_{z_2} \,\Sy_2
\gamma^\nu\,\Sy_1\,\ssp_2[y_*,x_*]_{z_1} \,\Sx_1\}
\nonumber\\
&&~~~~~~ - \int^{y_*}_{x_*}\!\!\!dz_*\!\!\int_{x_*}^{z_*}\!\!\!dz'_*
{\rm Tr\,tr}\{
\gamma^\mu\Sx_1\ssp_2[x_*,y_*]_{z_1}\,\Sy_1\gamma^\nu\,
\Sy_2[y_*,z_*]_{z_2} t^a\gamma_\perp^\alpha\psi(z_*,z_{2\perp})
\bar{\psi}(z'_*,z_{2\perp})\nonumber\\
&&~~~~~~\times\![z_*,z'_*]^{ab}_{z_2}\gamma^\perp_\alpha t^b[z'_*,x_*]_{z_2} \,\Sx_2\}\Bigg]
\label{ifdetails2}
\end{eqnarray}
Observing that last line in (\ref{ifdetails2}) can be written as the adjoint conjugation of the second, we have
\begin{eqnarray}
&&\langle {\rm T}\{\hat{j}^{\mu}(x) \hat{j}^{\nu}(y)\}\rangle_{A,\psi,\bar{\psi}}
\nonumber\\
&&\stackrel{x^+>0>y^+}{\ni}
{ig^2\over 32\pi^6 x_*^4y_*^4} \!\!\int d^2\omega d^2z
\left(\calz_2 + i\epsilon\right)^{-2}\left(\calz_1 + i\epsilon\right)^{-3}
\nonumber\\
&&\hspace{1cm}\times\!
\Bigg[\int^{x_*}_{y_*}\!\!\!dz_*\!\!\int_{y_*}^{z_*}\!\!\!dz'_*{\rm Tr\,tr}\{
\gamma^\mu\Sx_2[x_*,z_*]_{z_2} t^a\gamma_\perp^\alpha\psi(z_*,z_{2\perp})[z_*,z'_*]^{ab}_{z_2}
\bar{\psi}(z'_*,z_{2\perp})\nonumber\\
&&\hspace{1.5cm}\times\!\gamma^\perp_\alpha t^b[z'_*,y_*]_{z_2} \,\Sy_2
\gamma^\nu\,\Sy_1\,\ssp_2[y_*,x_*]_{z_1} \,\Sx_1\}
\nonumber\\
&&\hspace{1.5cm}
- \Big(\int^{x_*}_{y_*}\!\!\!dz_*\!\!\int_{y_*}^{z_*}\!\!\!dz'_*{\rm Tr\,tr}\{
\gamma^\mu\Sx_2[x_*,z_*]_{z_2} t^a\gamma_\perp^\alpha\psi(z_*,z_{2\perp})[z_*,z'_*]^{ab}_{z_2}
\bar{\psi}(z'_*,z_{2\perp})\nonumber\\
&&\hspace{5.5cm}\times\!\gamma^\perp_\alpha t^b[z'_*,y_*]_{z_2} \,\Sy_2
\gamma^\nu\,\Sy_1\,\ssp_2[y_*,x_*]_{z_1} \,\Sx_1\}\Big)^\dagger\Bigg]\,.
\label{IFquarkintermidiate}
\end{eqnarray}
Now, using definition of operator $Q^{\alpha\beta}_{ij}$ in eq. (\ref{Qoper}) we finally get eq. (\ref{OPEquark})
\begin{eqnarray}
\langle {\rm T}\{j^{\mu}(x) j^{\nu}(y)\}\rangle_{A,\psi,\bar{\psi}}
\stackrel{x_*>0>y_*}{\ni}\!\!&&
{i\over 32\pi^6 x_*^4y_*^4} \!\!\int {d^2z_1 d^2z_2\over [\calz_1+i\epsilon]^3[\calz_2+i\epsilon]^2}
\\
&&\times\!
\Bigg[{\rm Tr\,tr}\{\gamma^\mu \Sx_2 
\gamma^\rho_\perp Q(z_{2\perp})\gamma_\rho^\perp\,\Sy_2\gamma^\nu\,\Sy_1\,\ssp_2U^\dagger_{z_1} \,\Sx_1\}
\nonumber\\
&&~~ - \Big({\rm Tr\,tr}\{\gamma^\mu \Sx_2 \gamma^\rho_\perp Q(z_{2\perp})\gamma_\rho^\perp
\,\Sy_2\gamma^\nu\,\Sy_1\,\ssp_2U^\dagger_{z_1} \,\Sx_1\}\Big)^\dagger
\Bigg]
\nonumber
\end{eqnarray}

\section{Coefficients $I_1^{\mu\nu}$ and $I_5^{\mu\nu}$}
\label{sec: explicitIFs}

Here we provide explicit expressions for the coefficients $I_1^{\mu\nu}$ and $I_5^{\mu\nu}$.
We remind that $X^\mu_i = X^\mu_{i\perp} + {2\over s}x_*p_1^\mu$ and the same for $y$ with $i=1,2$,
$X^\mu_{1\perp} = (x-z_1)^\mu_\perp$ and $X^\mu_{2\perp} = (x-z_2)^\mu_\perp$
and similar expression with $y$.

Coefficient $I_1^{\mu\nu}$ is
\begin{eqnarray}
	I_1^{\mu\nu} =\!\!&&
	\half\Bigg[2x_*y_* \Big(z^2_{12\perp}g^{\mu\nu} + z_{12}^\mu z_{12}^\nu - (X_1+X_2)^\mu(Y_1+Y_2)^\nu\Big)
	\nonumber\\
	&&
	- 2p_2^\mu p_2^\nu\Big((X_1\!\cdot \!Y_1)( X_2\!\cdot\! Y_2)
	- (X_1\!\cdot\! Y_2)( X_2\!\cdot \!Y_1) + (Y_1\!\cdot \!Y_2)( X_1\!\cdot \!X_2)\Big)
	\nonumber\\
	&&
	- x_*p_2^\nu\Big(- (Y^2_1 - Y^2_2)z_{12\perp}^\mu + z_{12\perp}^2(Y_1^\mu + Y_2^\mu)
	- 2Y_1\cdot Y_2 (X^{\mu}_1 + X^{\mu}_2) \Big) 
	\nonumber\\
	&&
	- y_*p_2^\mu\Big(- (X^2_1 - X^2_2)z_{12\perp}^\nu + z_{12\perp}^2(X_1^\nu + X_2^\nu)
	- 2X_1\cdot X_2(Y^{\nu}_1 + Y^{\nu}_2) \Big) \Bigg]
	\nonumber\\
	=\!\!&& \half x_*^2y_*^2{\partial^2\over \partial x_\mu\partial y_\nu}\left(\calz_1\calz_2 - z_{12\perp}^2 {(x-y)^2\over x_*y_*}\right)
\end{eqnarray}

Coefficient $I_5^{\mu\nu}$ is
\begin{eqnarray}
	\hspace{-2cm}
	I_5^{\mu\nu} =\!\!&& \Bigg[x_*y_*\Big(Y_{1j}X_{1k}
	-  Y_{2j}X_{2k}\Big) (g^{\mu\nu}_\perp\epsilon^{jk} + \epsilon^{\mu\nu}g^{jk})
	\nonumber\\
	&&+ p_2^\mu p_2^\nu \Big((Y_1,Y_2)\vec{X}_1\times \vec{X}_2 - (X_1,X_2) \vec{Y}_1\times \vec{Y}_2\Big)
	\nonumber\\
	&&
	+ x_*y_*\Big((Y_{1\perp}^\nu X_{2j} - Y_{2\perp}^\nu X_{1j})\epsilon^{j\mu} 
	- (X_{1\perp}^\mu Y_{2j} - X_{2\perp}^\mu Y_{1j})\epsilon^{j\nu}\Big)
	\nonumber\\
	&&
	- x_*\,p_2^\nu\Big((Y_1,Y_2)(X_{1k} - X_{2k})\epsilon^{k\mu} 
	+ (\vec{Y}_1\times \vec{Y}_2)(X_{1\perp}^\mu + X_{2\perp}^\mu) \Big)
	\nonumber\\
	&&+ y_*\,p_2^\mu\Big( (X_1,X_2)\big(Y_{1k}- Y_{2k}\big)\epsilon^{k\nu}
	+ (\vec{X}_1\times \vec{X}_2)(Y_{1\perp}^\nu + Y_{2\perp}^\nu)\Big)
	\nonumber\\
	&&+ {4\over s}\Big[x_*y_*^2\,p_1^\nu \Big(X_{2j} - X_{1j}\Big)\epsilon^{j\mu}
	- x_*^2y_*\,p_1^\mu\Big( Y_{2j} - Y_{1j}\Big)\epsilon^{j\nu}
	\nonumber\\
	&&~~~~~~~
	+ y_*^2\,p_1^\nu p_2^\mu\,  \vec{X}_1\times \vec{X}_2 - x_*^2\,p_1^\mu p_2^\nu\,  \vec{Y}_1\times \vec{Y}_2\Big] \Bigg]
	\nonumber\\
	=\!\!&& \big(x_*\partial^\mu_x - p_2^\mu\big)\big(y_*\partial^\nu_y - p_2^\nu\big)
	\big[(\vec{Y}_1\times\vec{Y}_2) X_1\cdot X_2 - (\vec{X}_1\times\vec{X}_2)Y_1\cdot Y_2\big]
\end{eqnarray}

\section{Evolution equation for operators $\calq_{1x}$ and $\calq_{5x}$}
\label{sec: Diag-calq1claq5}

\subsection{Diagrams with $\calq_{1x}$ and $\calq_{5x}$ quantum}

	\begin{figure}[thb]
		\begin{center}
		\includegraphics[width=4.5in]{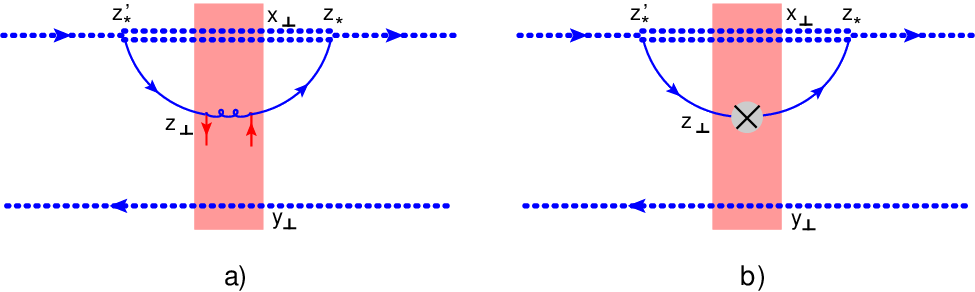}
		\caption{Diagrams with quantum quarks. Blue fermion lines are quantum fields, while red fermion lines are 
			classical fields. The gray circle with a cross on it represents here the $\epsilon^{ij}F_{ij}$ classical field.}
		\label{LO-psibarpsiadj}
	\end{center}
	\end{figure}

It is convenient to calculate the evolution of the operator $Q^{\alpha\beta}_{ij}$ given in eq. (\ref{Qoper}).
Using the following relations
\begin{eqnarray}
	&&\calq_5(x_\perp) = {\rm tr}\{\gamma^5\ssp_1Q_x\}
	\label{trgamma5}\\
	&&\calq_1(x_\perp) = {\rm tr}\{i\,\ssp_1Q_x\}
	\label{trp1}
\end{eqnarray}
we will obtain the evolution equations for $\Tr\{\calq_{1x}U^\dagger_y\}$ and $\Tr\{\calq_{5x}U^\dagger_y\}$.
We will use $\tr$ for traces over Dirac matrices, $\Tr$ for traces over color matrices in the fundamental representation, and
$\Tra$ for traces in the adjoint representation.

In this section, we will calculate the evolution of the operator $Q^{\alpha\beta}_{ij}$ as defined in eq. (\ref{Qoper}).
Let us start calculating diagram in Fig. \ref{LO-psibarpsiadj}a. Performing the contraction over the quark field we have
\begin{eqnarray}
	&&\hspace{-2cm}
	\langle{\rm Tr}\{Q^{\beta\alpha}_x U^\dagger_y\}\rangle_{\small{\rm Fig.} \ref{LO-psibarpsiadj}a}
	= g^2\int_0^{+\infty}\!\!d z_*\int^0_{-\infty}\!\!dz'_*\,t^a_{ij}U^{ab}_\omega\,(t^bU^\dagger_z)_{ki}
	\langle\,\psi_j^{\beta}(z_*,\omega_\perp)\bar{\psi}_k^{\alpha}(z'_*,\omega_\perp)\rangle\,.
\end{eqnarray}
To proceed with the calculation for diagram in Fig. \ref{LO-psibarpsiadj}a we need the quark propagator with quark sub-eikonal correction
(\ref{qprop-backqua-coord}), and obtain
\begin{eqnarray}
&&\langle{\rm Tr}\{Q^{\beta\alpha}_x U^\dagger_y\}\rangle_{\small{\rm Fig.} \ref{LO-psibarpsiadj}a}
\nonumber\\
&&= - \alpha_s\int_0^{+\infty}\!\!{d\alpha\over \alpha}\int d^2z\dhd^2q_1\dhd^2q_2 \, 
e^{i(q_1,x-z)+i(q_2,z-x)}\,
U_x^{ab}\,{(\ssq_{1\perp}\gamma_\rho^\perp)_{\beta\beta'}\over 
q^2_{1\perp}}\Tr\{t^a Q_{z}^{\beta'\alpha'}t^bU^\dagger_y\}
{(\gamma^\rho_\perp\ssq_{2\perp})_{\alpha'\alpha}\over q^2_{2\perp}}
\nonumber\\
&&= {\alpha_s\over 4\pi^2}\int_0^{+\infty}\!\!{d\alpha\over \alpha}\int d^2z \,
U_x^{ab}{[(\ssx - \ssz)^\perp\gamma_\rho^\perp]_{\beta\beta'}\over (x-z)^2_\perp}\,
{\rm Tr}\{t^a Q_{z}^{\beta'\alpha'}t^bU^\dagger_y\}
{[\gamma^\rho_\perp(\ssz - \ssx)_\perp]_{\alpha'\alpha}\over (x-z)^2_\perp}
\nonumber\\
&&= {\alpha_s\over 4\pi^2}\int_0^{+\infty}\!\!{d\alpha\over \alpha}\int d^2z \,
{[(\ssx - \ssz)_\perp\gamma_\rho^\perp]_{\beta\beta'}\over (x-z)^2_\perp}\,
\bigg[\half{\rm Tr}\{U^\dagger_y U_x\}{\rm Tr}\{U^\dagger_x Q_{z}^{\beta'\alpha'}\}
\nonumber\\
&&\hspace{6.5cm} - {1\over 2N_c}{\rm Tr}\{U^\dagger_y Q_{z}^{\beta'\alpha'}\}\bigg]
{[\gamma^\rho_\perp(\ssz - \ssx)_\perp]_{\alpha'\alpha}\over (x-z)^2_\perp}
\label{quantum-q1}
\end{eqnarray} 
We perform same steps for diagram in Fig. \ref{LO-psibarpsiadj}b 
but this time we need the gluon sub-eikonal contribution to the quark propagator. Thus, we have
\begin{eqnarray}
\langle{\rm Tr}\{Q^{\beta\alpha}_x U^\dagger_y\}\rangle_{\small{\rm Fig.} \ref{LO-psibarpsiadj}b}
=\!\!&& {\alpha_s \over 4\pi^2}\int_0^{+\infty}\!\!{d\alpha\over \alpha}\!\int\!d^2z\,
{[(\ssx-\ssz)_\perp\gamma^\rho_\perp]_{\beta\beta'}\over(z-x)^2_\perp}
\,{1\over 2s}[\ssp_2\gamma^5]^{\beta'\alpha'}
\nonumber\\
&&\times\Tr\{t^a \,\calf_{z}t^bU^\dagger_y\}U^{ab}_x\,
{[\gamma_\rho^\perp(\ssz-\ssx)_\perp]_{\alpha'\alpha}\over(z-x)^2_\perp}
\label{quantum-q2}
\end{eqnarray}

Contracting eqs. (\ref{quantum-q1}) and (\ref{quantum-q2}) with $\ssp_1\gamma^5$ and summing them up we have
\begin{eqnarray}
&&
\langle{\rm Tr}\{ \calq_{5\,x}\,U^\dagger_y\}\rangle_{\small{\rm Fig.} \ref{LO-psibarpsiadj}}
\label{quantum-oneloopcaq5}\\
&&= \langle\Tr\{\tr\{\gamma^5\ssp_1Q_x\} \, U^\dagger_y\}\rangle_{\small{\rm Fig.} \ref{LO-psibarpsiadj}}
\nonumber\\
&&= {\alpha_s\over 4\pi^2}\int_0^{+\infty}\!\!{d\alpha\over \alpha}\int\!\!d^2z\,
\tr\left\{\gamma^5\ssp_1{[(\ssx - \ssz)_\perp\gamma_\rho^\perp]\over (x-z)^2_\perp}\,
U_x^{ab}\right.
\nonumber\\
&&\hspace{6cm}\left.\times\Tr\{t^a \Big(Q_{z} + 
{1\over 2s}[\ssp_2\gamma^5]\calf_{z}\Big)t^bU^\dagger_y\}
{[\gamma^\rho_\perp(\ssz - \ssx)_\perp]\over (x-z)^2_\perp}\right\}
\nonumber\\
&&= {\alpha_s\over 2\pi^2}\!\!\int_0^{+\infty}\!\!{d\alpha\over \alpha}\int\!\!d^2z
{U_x^{ab}\over (x-z)^2_\perp}
\Tr\{t^a \big( \calq_{5z} +\calf_z\big)t^bU^\dagger_y\}
\nonumber\\
\hspace{-1.5cm}&&= {\alpha_s\over 4\pi^2}\!\!\int_0^{+\infty}\!\!{d\alpha\over \alpha}\!\!\int\!\!
{d^2z\over (x-z)^2_\perp}
\Big[\Tr\{\big(\calq_{5z} + \calf_z\big)U^\dagger_x\}\Tr\{U^\dagger_y U_x\} 
- {1\over N_c}\Tr\{\big(\calq_{5z} + \calf_z\big)U^\dagger_y\}\Big]
\nonumber
\end{eqnarray}
While contracting eqs. (\ref{quantum-q1}) and (\ref{quantum-q2}) with $i\ssp_1$ and again summing them up we obtain
\begin{eqnarray}
&&
\langle{\rm Tr}\{ \calq_{1x}  \, U^\dagger_y\}\rangle_{\small{\rm Fig.} \ref{LO-psibarpsiadj}}
\\
&&\hspace{0.7cm}= \langle\Tr\{\tr \{i\ssp_1Q_x\} \, U^\dagger_y\}\rangle_{\small{\rm Fig.} \ref{LO-psibarpsiadj}}
\nonumber\\
&&\hspace{0.7cm}
= {\alpha_s\over 4\pi^2}\int_0^{+\infty}\!\!{d\alpha\over \alpha}\int\!\!d^2z\,
\tr\left\{i\,\ssp_1\,{[(\ssx - \ssz)_\perp\gamma_\rho^\perp]\over (x-z)^2_\perp}\right.
\nonumber\\
&&\hspace{6cm}\left.\times
U_x^{ab}\,{\rm Tr}\{t^a \Big(Q_z + {1\over 2s}[\ssp_2\gamma^5]\calf_z\Big)t^bU^\dagger_y\}
{[\gamma^\rho_\perp(\ssz - \ssx)_\perp]\over (x-z)^2_\perp}\,
\right\}
\nonumber
\end{eqnarray}
Using ${\rm tr}\{(\ssx-\ssz)_\perp\gamma^\perp_\rho\ssp_2\sigma^{ij}\gamma^\rho_\perp(\ssz -\ssx)_\perp\ssp_1\}= 0$,
we see that the term proportional to $\calf_z$ cancels, so we do not have mixing of operator of different parity.
Thus, we arrive at

\begin{eqnarray}	
&&\langle{\rm Tr}\{ \calq_{1x}  \, U^\dagger_y\}\rangle_{\small{\rm Fig.} \ref{LO-psibarpsiadj}}
\nonumber\\
&&\hspace{0.7cm}= {\alpha_s\over 2\pi^2}\int_0^{+\infty}\!\!{d\alpha\over \alpha}\int\!\!d^2z
{U_x^{ab}\over (x-z)^2_\perp}\,
{\rm Tr}\{t^a  \calq_{1z}t^bU^\dagger_y\}
\nonumber\\
&&\hspace{0.7cm}= {\alpha_s\over 4\pi^2}\int_0^{+\infty}\!\!{d\alpha\over \alpha}\int\!\!d^2z
{1\over (x-z)^2_\perp}
\Big[\Tr\{U^\dagger_x \calq_{1z}\}\Tr\{U^\dagger_y U_x\} - {1\over N_c}\Tr\{\calq_{1z}U^\dagger_y\}\Big]
\label{quantum-oneloopcaq1}
\end{eqnarray}

To arrive at results (\ref{quantum-oneloopcaq5}) and (\ref{quantum-oneloopcaq1}) we used identities
\begin{eqnarray}
\ssp_2\sigma^{ij} &=& \epsilon^{ij}\ssp_2\gamma^5\,,
\label{idgamDirac1}
\\
{\rm tr}\{(\ssx-\ssz)_\perp\gamma^\perp_\rho\ssp_2\sigma^{ij}\gamma^\rho_\perp(\ssz -\ssx)_\perp\gamma^5\ssp_1\}
&=& 4\,s\,\epsilon^{ij}\,(x-z)^2_\perp\,,
\label{idgamDirac2}
\\
\tr\{\gamma^5\ssp_1(\ssx-\ssz)_\perp\gamma^\perp_\rho\,\psi\,\bar{\psi}\,\gamma^\rho_\perp(\ssz-\ssx)_\perp\}
&=&  2(x-z)^2_\perp\,\tr\{\gamma^5\ssp_1\psi \bar{\psi}\}\,,
\label{idgamDirac3}
\\
\tr\{\ssp_1(\ssx-\ssz)_\perp\gamma^\perp_\rho\,\psi\,\bar{\psi}\,\gamma^\rho_\perp(\ssz-\ssx)_\perp\}
&=& 2(x-z)^2_\perp\,\tr\{\ssp_1\psi\bar{\psi}\}\,,
\label{idgamDirac4}
\end{eqnarray}

The evolution equations (\ref{quantum-oneloopcaq5}) and (\ref{quantum-oneloopcaq1}) are not closed evolution equations because
after one loop evolution we have generated new operators. Consequently,
to solve them we should find the evolution equations of the operators generated after one loop,
thus generating a hierarchy of evolution equations similar to the Balitsky-hierarchy equations for dipoles.
Alternatively, one can try to truncate the hierarchy of evolution equations employing some approximation.
For example, it is known that in the large $N_c$ limit the Balitsky-hierarchy of evolution equations is truncated to the
BK equation. In order to perform a similar truncation in eqs. (\ref{quantum-oneloopcaq5}) and (\ref{quantum-oneloopcaq1})
it is probably convenient to work out the color algebra and reduce all the operators in the fundamental representation,
as it is done in eq. (\ref{calf-quantum-fund}), and only then one can try to find a way to
linearize and solve the evolution equations. 

It is interesting to notice that if we consider the evolution equation of the sum 
of operator $\hat{\calq}_{5x}$, eq. (\ref{Sumcalq5}), of operator $\hat{\calf}_x$, eq. (\ref{calf-quantum})
and their adjoint conjugated we obtain and neglecting the quark-to-gluon diagrams and consider only the
terms contributing to the  double logarithm, we have
\begin{eqnarray}
	&&\Big\langle\Tr\{\big(\calf_x+\calq_{5x}\big) U^\dagger_y\} + \Tr\{\big(\calf^\dagger_x+\calq^\dagger_{5x}\big) U_y\}
	\Big\rangle_{\small{\rm Figs.}\ref{outshock-2}+\ref{outshock-3}+\ref{LO-psibarpsiadj}}
	\nonumber\\
	&&=
	{\alpha_s\over \pi^2}\,\int_0^{+\infty}\!{d\alpha\over\alpha}
	\int {d^2z\over (x-z)^2_\perp}\Bigg\{\Tr\{U_x t^a U^\dagger_y t^b\}
	\Big(\calq_{5z}^{ba} + {\calq_{5z}^{ba}}^\dagger + \calf^{ba}_z\Big)
	\nonumber\\
	&&~~~+ \Tr\{t^bU_yt^aU^\dagger_x\}\Big({\calq_{5z}^{ba}}^\dagger + \calq_{5z}^{ba} + {\calf^{ba}}^\dagger_z\Big)
	\nonumber\\
	&&~~~	+ {1\over 4}\Big[\Tr\{\big(\calq_{5z} + \calf_z\big)U^\dagger_x\}\Tr\{U^\dagger_y U_x\} 
	- {1\over N_c}\Tr\{\big(\calq_{5z} + \calf_z\big)U^\dagger_y\}
	\nonumber\\
	&&~~~ + \Tr\{\big(\calq^\dagger_{5z} + \calf^\dagger_z\big)U_x\}\Tr\{U_y U^\dagger_x\} 
- {1\over N_c}\Tr\{\big(\calq^\dagger_{5z} + \calf^\dagger_z\big)U_y\}\Big]
	\Bigg\}\,.
	\label{calfpluscalq5}
\end{eqnarray}
So, if we neglect the quark-to-gluon diagrams and the 
mixing with the operator $\calq_1$ as is shown in eq. (\ref{calf-quantum}), 
then (\ref{calfpluscalq5}) does agree with the evolution equation calculated in Refs.
\cite{Kovchegov:2016zex, Kovchegov:2018znm}.

\subsection{Diagrams quark-to-gluon for $\calq_{1x}$ and $\calq_{5x}$}

\begin{figure}[htb]
		\begin{center}
		\includegraphics[width=3.7in]{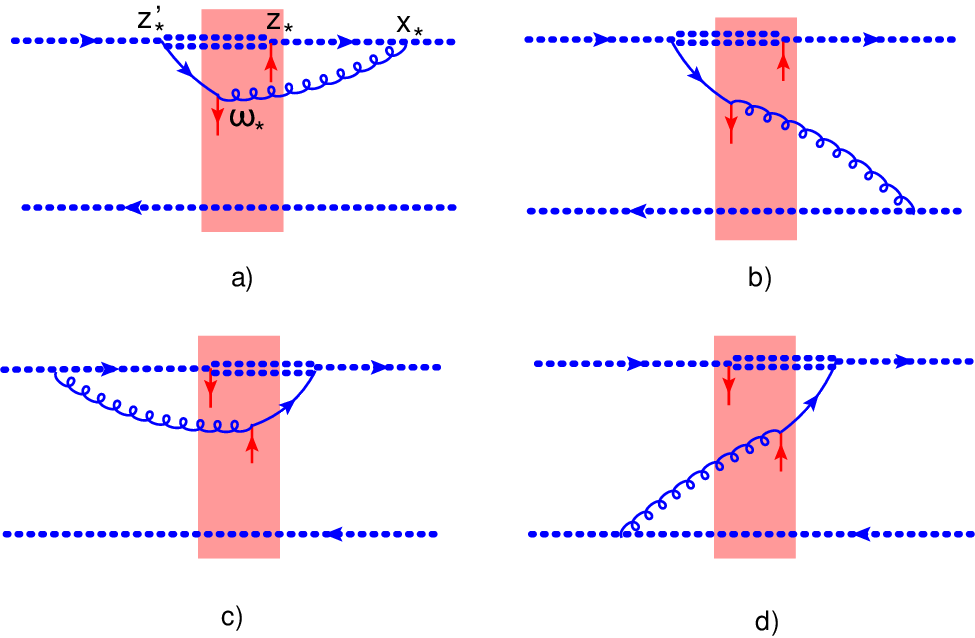}
		\caption{One loop correction diagrams to the operator $Q^{\alpha\beta}_{ij}$ defined in eq. 
			(\ref{Qoper}) using the quark-to-gluon propagator. Single dotted lines are Wilson-line in the fundamental representation, while 
			double dotted lines are Wilson-line in the adjoint representation.}
		\label{diagrams-q2g-prop}
	\end{center}
	\end{figure}

We calculate diagrams in Fig. \ref{diagrams-q2g-prop} using again operator $Q^{\alpha\beta}_{ij}$ and then 
with the help of relations (\ref{trgamma5}) and (\ref{trp1}), we will obtain the results for operators $\calq_{5x}$ and $\calq_{1x}$
respectively.

The first diagram we will calculate is in Fig. \ref{diagrams-q2g-prop}a
\begin{eqnarray}
&&\langle \Tr\{Q^{\alpha\beta}_x U^\dagger_y\}\rangle_{\small{\rm Fig.} \ref{diagrams-q2g-prop}a}
\nonumber\\
	&&=\Big\langle\Tr\Big\{
	g^2\int_{-\infty}^{+\infty}\!dz_*\int_{-\infty}^{z_*}\!dz'_*[\infty p_1,z_*]_x t^a\psi^\alpha(z_*,x_\perp)
	[z_*,z'_*]_x^{ab}\bar{\psi}^\beta(z'_*,x_\perp)t^b[z'_*,-\infty p_1]_x U^\dagger_y\Big\}\Big\rangle
	\nonumber\\
	&&= 
	g^2\!\int_{-\infty}^{+\infty}\!dz_*\int_{-\infty}^{z_*}\!dz'_*ig{2\over s}\int_{z_*}^{+\infty}\!d\xi_*
	\Big\langle\Tr\Big\{[\infty p_1,\xi_*]_x A^q_\bullet(\xi_*,x_\perp)[\xi_*,z_*]_xt^a
	\psi^\alpha_{cl}(z_*,x_\perp)
	\nonumber\\
	&&\hspace{5.6cm}\times[z_*,z'_*]^{ab}_x\bar{\psi}^\beta_q(z'_*,x_\perp)t^b[z'_*,-\infty p_1]_x U^\dagger_y
	\Big\}\Big\rangle\,.
\end{eqnarray}
To proceed, we should remember that the shock-wave has support only within the infinitesimal interval $[-\epsilon_*,\epsilon_*]$
and that, in a particular gauge, we can set to $1$ the gauge links 
made of classical field with support outside the interval $[-\epsilon_*,\epsilon_*]$.
Moreover, we can change
the extremes of integration of the longitudinal variable $\xi^+$ from $[z^+,\infty]$ to $[0,\infty]$
and those of $z'^+$ from $[-\infty,z^+]$ to $[-\infty,0]$
because the support of the quantum field gets up to zero.
Thus, using propagator in eq. (\ref{q2gpropagator1}) we have
\begin{eqnarray}
	&&\langle \Tr\{Q^{\alpha\beta}_x U^\dagger_y\}\rangle_{\small{\rm Fig.} \ref{diagrams-q2g-prop}a}
	\nonumber\\
	&&= g^2\!\int_{-\infty}^{+\infty}\!dz_*\int_{-\infty}^{z_*}\!dz'_*ig{2\over s}\int_{z_*}^{+\infty}\!d\xi_*
	\nonumber\\
	&&~~~\times \Big\langle\Tr\Big\{t^c[\infty p_1,\xi_*]_xt^a
	\psi^\alpha_{cl}(z_*,x_\perp)[z_*,z'_*]^{ab}_x
	\langle  A^{q,c}_\bullet(\xi_*,x_\perp)\bar{\psi}^{q,\beta}(z'_*,x_\perp)\rangle t^b[z'_*,-\infty p_1]_x U^\dagger_y
	\Big\}\Big\rangle
	\nonumber\\
    &&= - {g^4\over s^2}\int_{-\infty}^{+\infty}dz_*d\omega_*\int_{-\infty}^0 dz'_*\int_0^{+\infty}d\xi_*
	\int_{0}^{+\infty}{\dhd\alpha\over \alpha^3}\int d^2z\,\braxp e^{-i{\hatp^2_\perp\over \alpha s}\xi_*}\hat{p}_\nu^\perp\ketzp
	\nonumber\\
	&&~~~\times\Tr\Big\{
	t^c[\infty p_1,z_*]_x t^a\psi^\alpha_{cl}(z_*,x_\perp)[z_*,-\infty p_1 ]_x^{ab}
	[\infty p_1,\omega_*]^{cd}_z
	\bar{\psi}^{\beta'}_{cl}(\omega_*,z_\perp)t^d
	[\omega_*,-\infty p_1]_zt^bU^\dagger_y
	\nonumber\\
	&&\hspace{1.3cm}\times\brazp \big[\gamma_\perp^\nu(\alpha \ssp_1+\ssp_\perp)\big]^{\beta'\beta}
	e^{i{\hatp^2_\perp\over \alpha s}z'_*}\ketxp 
	\Big\}
	\nonumber\\
	&&= - {g^4\over s^2}\int_{-\infty}^{+\infty}dz_*d\omega_*\int_{-\infty}^0 dz'_*\int_0^{+\infty}d\xi_*
	\int_{0}^{+\infty}{\dhd\alpha\over \alpha^3}\int d^2z\,\braxp e^{-i{\hatp^2_\perp\over \alpha s}\xi_*}\hat{p}_\nu^\perp\ketzp
	\nonumber\\
	&&~~~\times\Tr\Big\{
	t^cU_zt^bU^\dagger_yt^cU_x t^b[-\infty p_1,z_*]_x\psi^\alpha_{cl}(z_*,x_\perp)
	\bar{\psi}^{\beta'}_{cl}(\omega_*,z_\perp)[\omega_*,\infty p_1]_z
	\nonumber\\
	&&\hspace{1.3cm}\times\brazp\big[\gamma_\perp^\nu(\alpha \ssp_1+\ssp_\perp)]^{\beta'\beta}
	e^{i{\hatp^2_\perp\over \alpha s}z'_*}\ketxp \Big\}
	\label{Qq2gdiagrama}
\end{eqnarray}

Proceeding in the same way, with the exception that $d\xi_*$ starts from $0$ and not from $z_*$, 
for diagram \ref{diagrams-q2g-prop}b  we have
\begin{eqnarray}
&&\langle \Tr\{Q^{\alpha\beta}_x U^\dagger_y\}\rangle_{\small{\rm Fig.} \ref{diagrams-q2g-prop}b}
\nonumber\\
	&&=\Big\langle\Tr\Big\{
	g^2\int_{-\infty}^{+\infty}\!dz_*\int_{-\infty}^{z_*}\!dz'_*[\infty p_1,z_*]_x t^a\psi^\alpha(z_*,x_\perp)
	[z_*,z'_*]_x^{ab}\bar{\psi}^\beta(z'_*,x_\perp)t^b[z'_*,-\infty p_1]_x U^\dagger_y
	\Big\}\Big\rangle
	\nonumber\\
	&&= {g^4\over s^2}\int_{-\infty}^{+\infty}dz_*d\omega_*\int_{-\infty}^{z_*} dz'_*\int_{z_*}^{+\infty}d\xi_*
	\int_{0}^{+\infty}{\dhd\alpha\over \alpha^3}\int d^2z\,\brayp e^{-i{\hatp^2_\perp\over \alpha s}\xi_*}\hat{p}_\nu^\perp\ketzp
	\nonumber\\
	&&~~~\times\Tr\Big\{
	t^c[\infty p_1,z_*]_x t^a\psi^\alpha_{cl}(z_*,x_\perp)[z_*,-\infty p_1 ]_x^{ab}
	[\infty p_1,\omega_*]^{cd}_z
	\bar{\psi}^{\beta'}_{cl}(\omega_*,z_\perp)t^d
	[\omega_*,-\infty p_1]_zt^bU^\dagger_y
	\nonumber\\
	&&\hspace{1cm}
	\times\brazp\big[\gamma_\perp^\nu(\alpha \ssp_1+\ssp_\perp)\big]^{\beta'\beta}e^{i{\hatp^2_\perp\over \alpha s}z'_*}\ketxp 
	\Big\}
	\nonumber\\
	&&={g^4\over s^2}\int_{-\infty}^{+\infty}dz_*d\omega_*\int_{-\infty}^{z_*} dz'_*\int_{z_*}^{+\infty}d\xi_*
	\int_{0}^{+\infty}{\dhd\alpha\over \alpha^3}\int d^2z\,\brayp e^{-i{\hatp^2_\perp\over \alpha s}\xi_*}\hat{p}_\nu^\perp\ketzp
	\nonumber\\
	&&~~~\times\Tr\Big\{
	t^cU_zt^bU^\dagger_yt^cU_x t^b[-\infty p_1,z_*]_x\psi^\alpha_{cl}(z_*,x_\perp)
	\bar{\psi}^{\beta'}_{cl}(\omega_*,z_\perp)[\omega_*,\infty p_1]_z
	\nonumber\\
	&&\hspace{1cm}\times\brazp\big[\gamma_\perp^\nu(\alpha \ssp_1+\ssp_\perp)\big]^{\beta'\beta}
	e^{i{\hatp^2_\perp\over \alpha s}z'_*}\ketxp 
	\Big\}\,.
	\label{Qq2gdiagramb}
\end{eqnarray}

The second self diagram, Fig. \ref{diagrams-q2g-prop}c is also calculated in a similar way. We have
\begin{eqnarray}
&&\langle \Tr\{Q^{\alpha\beta}_x U^\dagger_y\} \rangle_{\small{\rm Fig.} \ref{diagrams-q2g-prop}c}
\nonumber\\
	&&=\Big\langle\Tr\left\{
	g^2\int_{-\infty}^{+\infty}\!dz_*\int_{-\infty}^{z_*}\!dz'_*[\infty p_1,z_*]_x t^a\psi^\alpha(z_*,x_\perp)
	[z_*,z'_*]_x^{ab}\bar{\psi}^\beta(z'_*,x_\perp)t^b[z'_*,-\infty p_1]_x U^\dagger_y\right\}\Big\rangle
	\nonumber\\
	&&=\Tr\left\{g^2\int_{-\infty}^{+\infty}dz_*\int_{-\infty}^{z_*}dz'_* \,t^a\psi^\alpha_q(z_*,x_\perp)[z_*,z'_*]^{ab}\bar{\psi}^\beta_{cl}(z'_*,x_\perp)t^b\right.
	\nonumber\\
	&&\hspace{1.3cm}\left.\times\!\left(ig{2\over s}\right)\int_{-\infty}^{z'_*}d\xi_*[z'_*,\xi_*]_x A_\bullet^q(\xi_*,x_\perp)U^\dagger_y\right\}
\end{eqnarray}
Now we use propagator (\ref{q2gpropagator2}) and obtain
\begin{eqnarray}
&&\langle \Tr\{Q^{\alpha\beta}_x U^\dagger_y \}\rangle_{\small{\rm Fig.} \ref{diagrams-q2g-prop}c}
\nonumber\\
&&= ig^3{2\over s}\int_{-\infty}^{+\infty}\!\!dz'_*\int_{z'_*}^{+\infty}\!\!dz_*\int_{-\infty}^{z'_*}\!\!d\xi_*
\nonumber\\
&&~~~\times\Tr\{t^a\langle \psi^\alpha_q(z_*,x_\perp)A_\bullet^{q\,,c}(\xi_*,x_\perp)\rangle[\infty p_1,z'_*]^{ab}_x
\bar{\psi}_{cl}^\beta(z'_*,x_\perp)t^b[z'_*,-\infty p_1]_x t^c U^\dagger_y\}
\nonumber\\
&&= - {g^4\over s^2}\int_{-\infty}^{+\infty}\!\!dz'_*d\omega_*\int_0^{+\infty}{\dhd\alpha\over \alpha^3}
\int_0^{+\infty}dz_*\int_{-\infty}^0d\xi_*\int d^2z
\nonumber\\
&&~~~\times\braxp e^{-i{\hatp^2_\perp\over \alpha s}z_*}
\big[(\alpha\ssp_1+\ssp_\perp)\gamma_\perp^\nu\big]^{\alpha\alpha'}\ketzp\brazp p_\nu^\perp 
e^{i{\hatp^2_\perp\over \alpha s}\xi_*}\ketxp
\nonumber\\
&&~~~\times\Tr\{t^a[\infty p_1,\omega_*]_z t^d\psi^{\alpha'}(\omega_*,z_\perp)[\omega_*,-\infty p_1]^{dc}_z
[\infty p_1, z'_*]^{ab}_x\bar{\psi}^\beta(z'_*,x_\perp)t^b[z'_*,-\infty p_1]_x t^c U^\dagger_y\}
\nonumber\\
&&= - {g^4\over s^2}\int_{-\infty}^{+\infty}\!\!dz'_*d\omega_*\int_0^{+\infty}{\dhd\alpha\over \alpha^3}
\int_0^{+\infty}dz_*\int_{-\infty}^0d\xi_*\int d^2z
\nonumber\\
&&~~~\times\braxp e^{-i{\hatp^2_\perp\over \alpha s}z_*}
\big[(\alpha\ssp_1+\ssp_\perp)\gamma_\perp^\nu\big]^{\alpha\alpha'}\ketzp\brazp p_\nu^\perp 
e^{i{\hatp^2_\perp\over \alpha s}\xi_*}\ketxp
\nonumber\\
&&~~~\times\Tr\{t^aU_xt^cU^\dagger_yt^aU_zt^c[-\infty p_1,\omega_*]_z\psi^{\alpha'}(\omega_*,z_\perp)
\bar{\psi}^\beta(z'_*,x_\perp)[z'_*,\infty p_1]_x \}\,.
\label{Qq2gdiagramc}
\end{eqnarray}

Diagram in {\rm Fig.} \ref{diagrams-q2g-prop}d is similar to \ref{diagrams-q2g-prop}c, so we have
\begin{eqnarray}
&&\langle \Tr\{Q^{\alpha\beta}_x U^\dagger_y\} \rangle_{\small{\rm Fig.} \ref{diagrams-q2g-prop}d}
\nonumber\\
	&&=\Big\langle\Tr\Big\{
	g^2\int_{-\infty}^{+\infty}\!dz_*\int_{-\infty}^{z_*}\!dz'_*[\infty p_1,z_*]_x t^a\psi^\alpha(z_*,x_\perp)
	[z_*,z'_*]_x^{ab}\bar{\psi}^\beta(z'_*,x_\perp)t^b[z'_*,-\infty p_1]_x U^\dagger_y\Big\}\Big\rangle
	\nonumber\\
	&&= {g^4\over s^2}\int_{-\infty}^{+\infty}\!\!dz'_*d\omega_*\int_0^{+\infty}{\dhd\alpha\over \alpha^3}
	\int_0^{+\infty}\!\!dz_*\int_{-\infty}^0d\xi_*\int d^2z
	\nonumber\\
	&&~~~\times\braxp e^{-i{\hatp^2_\perp\over \alpha s}z_*}
	\big[(\alpha\ssp_1+\ssp_\perp)\gamma_\perp^\nu\big]^{\alpha\alpha'}\ketzp\brazp p_\nu^\perp e^{i{\hatp^2_\perp\over \alpha s}\xi_*}\ketyp
	\nonumber\\
	&&~~~\times\Tr\{t^aU_xt^cU^\dagger_yt^aU_zt^c[-\infty p_1,\omega_*]_z\psi^{\alpha'}(\omega_*,z_\perp)
	\bar{\psi}^\beta(z'_*,x_\perp)[z'_*,\infty p_1]_x \}\,.
	\label{Qq2gdiagramd}
\end{eqnarray}

The contribution of diagrams in Fig. \ref{diagrams-q2g-prop} to the evolution of operators $\calq_1$ and $\calq_5$ defined
in eqs. (\ref{calq1}) and (\ref{calq5}) respectively, can be obtained by taking the Dirac trace of the operator $Q^{\alpha\beta}_{ij}$
with $i\ssp_1$ and $\gamma^5\ssp_1$. 

\subsubsection{quark-to-gluon diagrams for $\calq_{1x}$}

Summing eqs. (\ref{Qq2gdiagrama}) and (\ref{Qq2gdiagramb}) and taking trace with $i\ssp_1$
we have
\begin{eqnarray}
	&&\langle \Tr\{\calq_{1x}U^\dagger_y\}\rangle_{\small{\rm Fig.} \ref{diagrams-q2g-prop}a+b} 
	\nonumber\\
	&&= \langle \Tr\,\tr\{i\ssp_1Q_xU^\dagger_y\}\rangle_{\small{\rm Fig.} \ref{diagrams-q2g-prop}a+b}
	\nonumber\\
	&&=
	{g^4\over s^2}\int_{-\infty}^{+\infty}\!\!dz_*d\omega_*\int_{-\infty}^0 dz'_*\int_0^{+\infty}d\xi_*
	\int_{0}^{+\infty}{\dhd\alpha\over \alpha^3}\int d^2z
		\nonumber\\
	&&~~~\times
	\Big(\brayp e^{-i{\hatp^2_\perp\over \alpha s}\xi_*}\hat{p}_\nu^\perp\ketzp
	- \braxp e^{-i{\hatp^2_\perp\over \alpha s}\xi_*}\hat{p}_\nu^\perp\ketzp
	\Big)
	\nonumber\\
	&&~~~\times\Tr\,\tr\Big\{
	t^cU_zt^bU^\dagger_yt^cU_x t^b[-\infty p_1,z_*]_x\,i\ssp_1\,\psi^\alpha(z_*,x_\perp)
	\bar{\psi}^{\beta'}(\omega_*,z_\perp)[\omega_*,\infty p_1]_z
		\nonumber\\
	&&~~~~~~~~~\times\!\brazp\gamma_\perp^\nu(\alpha \ssp_1+\ssp_\perp)
	e^{i{\hatp^2_\perp\over \alpha s}z'_*}\ketxp\Big\}
	\nonumber\\
	&&={\alpha_s\over 2\pi^2}
	\int_{0}^{+\infty}{d\alpha\over \alpha}\,g^2\!\!\int_{-\infty}^{+\infty}\!\!dz_*d\omega_*\int d^2z
	\nonumber\\
	&&~~~\times\!\Tr\,\tr\Big\{t^cU_zt^bU^\dagger_yt^cU_x t^b[-\infty p_1,z_*]_x\,i\ssp_1\,\psi^\alpha(z_*,x_\perp)
	\barpsi^{\beta'}(\omega_*,z_\perp)[\omega_*,\infty p_1]_z
	\nonumber\\
	&&~~~~~~~~~\times\bigg[ {(x-z,z-y)\over (x-z)^2_\perp(y-z)^2_\perp} + {1\over (x-z)^2_\perp} 
	+ i\gamma^5{(\vec{x}-\vec{z})\times(\vec{y}-\vec{z})\over (x-z)^2_\perp(y-z)^2_\perp}\bigg] \Big\}
	\label{calq1q2gdiagramab}
\end{eqnarray}
where in the last step we have integrated over the 	longitudinal variables $\xi^+$and $z'^+$ and performed the 
Fourier transform. 

Summing up the next two diagrams, eqs (\ref{Qq2gdiagramc}) and (\ref{Qq2gdiagramd}) we have
\begin{eqnarray}
	&&\langle \Tr\{\calq_{1x}U^\dagger_y\}\rangle_{\small{\rm Fig.} \ref{diagrams-q2g-prop}c+d} 
	\nonumber\\
	&&= \langle \Tr\,\tr\{i\ssp_1Q_xU^\dagger_y\}\rangle_{\small{\rm Fig.} \ref{diagrams-q2g-prop}c+d}
	\nonumber\\
	&&= {g^4\over s^2}\int_{-\infty}^{+\infty}\!\!dz'_*d\omega_*\int_0^{+\infty}{\dhd\alpha\over \alpha^3}
	\int_0^{+\infty}\!\!dz_*\int_{-\infty}^0d\xi_*\int d^2z
	\nonumber\\
	&&~~~\times\tr\Big\{\braxp e^{-i{\hatp^2_\perp\over \alpha s}z_*}
	i\ssp_1\big[(\alpha\ssp_1+\ssp_\perp)\gamma_\perp^\nu\big]\ketzp
	\Big(\brazp p_\nu^\perp e^{i{\hatp^2_\perp\over \alpha s}\xi_*}\ketyp
	- \brazp p_\nu^\perp e^{i{\hatp^2_\perp\over \alpha s}\xi_*}\ketxp\Big)
	\nonumber\\
	&&~~~~~~~~\times\Tr\{t^aU_xt^cU^\dagger_yt^aU_zt^c[-\infty p_1,\omega_*]\psi(\omega_*,z_\perp)
	\bar{\psi}(z'_*,x_\perp)[z'_*,\infty p_1]_x \}\Big\}
	\nonumber\\
	&&= {\alpha_s\over 2\pi^2}\int_0^{+\infty}{d\alpha\over \alpha}\,g^2\!\int_{-\infty}^{+\infty}\!\!dz'_*d\omega_*
	\int d^2z\,
	\nonumber\\
	&&~~~\times\tr\Big\{
	\bigg[{(x-z,z-y)\over (x-z)^2_\perp(y-z)^2_\perp} + {1\over (x-z)^2_\perp} 
	- i\gamma^5{(\vec{x}-\vec{z})\times(\vec{y}-\vec{z})\over (x-z)^2_\perp(y-z)^2_\perp}\bigg]
	\nonumber\\
	&&~~~~~~~~\times\Tr\{t^aU_xt^cU^\dagger_yt^aU_zt^c[-\infty p_1,\omega_*]_z
	\,i\ssp_1\,\psi(\omega_*,z_\perp)
	\bar{\psi}(z'_*,x_\perp)[z'_*,\infty p_1]_x \}\Big\}\,.
	\label{calq1q2gdiagramcd}
\end{eqnarray}

We use definition of the operators (\ref{calx1}) and (\ref{calx5}),
and the sum of eq. (\ref{calq1q2gdiagramab}) and (\ref{calq1q2gdiagramcd}) traced with $i\ssp_1$ is
\begin{eqnarray}
&&\langle \Tr\{\calq_{1x}U^\dagger_y\}\rangle_{\small{\rm Fig.} \ref{diagrams-q2g-prop}} 
\label{calq1-q2g}\\
 &&= - {\alpha_s\over 2\pi^2}
\int_{0}^{+\infty}{d\alpha\over \alpha}\int d^2z
\Bigg\{\bigg[ {(x-z,z-y)\over (x-z)^2_\perp(y-z)^2_\perp} + {1\over (x-z)^2_\perp} \bigg]
\nonumber\\
&&~~~ \times
\Big(\Tr\{
t^aU_zt^bU^\dagger_yt^aU_x t^b\calx^\dagger_{1zx}\} + \Tr\{t^aU_xt^bU^\dagger_yt^aU_zt^b\calx^\dagger_{1xz}\}\Big)
\nonumber\\
&& ~~~ - {(\vec{x}-\vec{z})\times(\vec{y}-\vec{z})\over (x-z)^2_\perp(y-z)^2_\perp}
\Big(\Tr\{t^aU_xt^bU^\dagger_yt^aU_zt^b\calx^\dagger_{5xz}\} - \Tr\{t^aU_zt^bU^\dagger_yt^aU_x t^b\calx^\dagger_{5zx}\}\Big)
\Bigg\}
\nonumber
\end{eqnarray}

The color trace in eq. (\ref{calq1q2gdiagramab}) can be simplified. For example, we have
\begin{eqnarray}
&&\Tr\{t^aU_zt^bU^\dagger_y t^a U_x t^b \calx^\dagger_{1zx}\} 
\nonumber\\
&&\hspace{1cm}= {1\over 4}\Tr\{ U_xU^\dagger_y U_z \calx^\dagger_{1zx}\} - {1\over 4N_c}\Tr\{U^\dagger_y U_z\}\Tr\{U_x\calx^\dagger_{1zx}\}
\nonumber\\
&&\hspace{1.3cm} - {1\over 4N_c}\Tr\{U^\dagger_y U_x\}\Tr\{\calx^\dagger_{1zx} U_z\} 
- {1\over 4N^2_c}\Tr\{U_zU^\dagger_y U_x \calx^\dagger_{1zx} \}\,.
\end{eqnarray}
Note also that the products like $\calx^\dagger_{1zx} U_z$ or $U_x\calx^\dagger_{1zx}$ 
may modify the operator $\calx^\dagger_{1zx}$ to
\begin{eqnarray}
\hspace{-1cm}\Tr\{\calx^\dagger_{1zx}U_z\}  
=\!\!&& - g^2\!\!\int_{-\infty}^{+\infty}\!\!dz_*d\omega_*
\Tr\{[-\infty p_1,z_*]_x\tr\{i\ssp_1\,\psi(z_*,x_\perp)\barpsi(\omega_*,z_\perp)\}[\omega_*, -\infty p_1]_z\}
\nonumber\\
=\!\!&& - \calh^-_1(x_\perp,z_\perp) 
\end{eqnarray}
and
\begin{eqnarray}
\hspace{-1cm}\Tr\{U_x\calx^\dagger_{1zx}\}  
=\!\!&& - g^2\!\!\int_{-\infty}^{+\infty}\!\!dz_*d\omega_*
	\Tr\{[\infty p_1,z_*]_x\tr\{i\ssp_1\,\psi(z_*,x_\perp)\barpsi(\omega_*,z_\perp)\}[\omega_*, \infty p_1]_z\}
	\nonumber\\
=\!\!&& - \calh^+_1(x_\perp,z_\perp)
\end{eqnarray}
where we used definition of operators (\ref{calh1minus}) and (\ref{calh1plus}) respectively.
It is also easy to find the operators $\calh^{-\dagger}_{1xz} =  - \Tr\{U^\dagger_z\calx_{1zx}\}  
= - \calh^-_{1zx}$, and similarly, $\calh^{+\dagger}_{1xz} = - \calh_{1zx}^+$. With $\calx_{5xz}$, instead, we have
\begin{eqnarray}
\hspace{-1cm}\Tr\{\calx^\dagger_{5zx}U_z\}  
=\!\!&& g^2\!\!\int_{-\infty}^{+\infty}\!\!dz_*d\omega_*
\Tr\{[-\infty p_1,z_*]_x\tr\{i\ssp_1\,\psi(z_*,x_\perp)\barpsi(\omega_*,z_\perp)\}[\omega_*, -\infty p_1]_z\}
\nonumber\\
=\!\!&& \calh^-_5(x_\perp,z_\perp) 
\end{eqnarray}
and
\begin{eqnarray}
\hspace{-1cm}\Tr\{U_x\calx^\dagger_{5zx}\}  
=\!\!&& g^2\!\!\int_{-\infty}^{+\infty}\!\!dz_*d\omega_*
\Tr\{[\infty p_1,z_*]_x\tr\{i\ssp_1\,\psi(z_*,x_\perp)\barpsi(\omega_*,z_\perp)\}[\omega_*, \infty p_1]_z\}
\nonumber\\
=\!\!&& \calh^+_5(x_\perp,z_\perp)
\end{eqnarray}
where we used definition of operators (\ref{calh5minus}) and (\ref{calh5plus}) respectively.
We also have $\calh^{-\dagger}_{5xz} = \Tr\{\calx_{5zx}U^\dagger_z\} = \calh^{-}_{5zx}$ and similarly
we get $\calh^{+\dagger}_{5xz} = \calh^+_{5zx}$

Evolution of operator $\Tr\{\hat{\calq}_{1x}\hat{U}^\dagger_y\}$, when diagrams in Fig. \ref{diagrams-q2g-prop} are taken into account,
introduces new operators, $\calx_{1xy}$, $\calx_{5xy}$, $\calh^-_{1xy}$, $\calh^-_{5xy}$,
$\calh^+_{1xy}$, $\calh^+_{5xy}$, which have never been considered before
in the study of spin dynamics at small-$x$.

\subsubsection{quark-to-gluon diagrams for $\calq_{5x}$}

Summing eqs. (\ref{Qq2gdiagrama}) and (\ref{Qq2gdiagramb}) and taking trace with $\gamma^5\ssp_1$, we have
\begin{eqnarray}
	&&\langle \Tr\{\calq_{5x}U^\dagger_y\}\rangle_{\small{\rm Fig.} \ref{diagrams-q2g-prop}a+b}
	\nonumber\\ 
	&&= \langle \Tr\tr\{\gamma^5\ssp_1Q_xU^\dagger_y\}\rangle_{\small{\rm Fig.} \ref{diagrams-q2g-prop}a+b}
	\nonumber\\
	&&= {g^4\over s^2}\int_{-\infty}^{+\infty}dz_*d\omega_*\int_{-\infty}^0 dz'_*\int_0^{+\infty}d\xi_*
	\int_{0}^{+\infty}{\dhd\alpha\over \alpha^3}\int d^2z\,
	\nonumber\\
	&&~~~\times\!\Big(\brayp e^{-i{\hatp^2_\perp\over \alpha s}\xi_*}\hat{p}_\nu^\perp\ketzp
	- \braxp e^{-i{\hatp^2_\perp\over \alpha s}\xi_*}\hat{p}_\nu^\perp\ketzp\Big)
	\Tr\,\tr\Big\{
	t^aU_zt^bU^\dagger_yt^aU_x t^b
	\nonumber\\
	&&~~~\times[-\infty p_1,z_*]_x\,\gamma^5\ssp_1\,\psi_{cl}(z_*,x_\perp)
	\barpsi_{cl}(\omega_*,z_\perp)[\omega_*,\infty p_1]_z
	\brazp\gamma_\perp^\nu(\alpha\ssp_1+\ssp_\perp) e^{i{\hatp^2_\perp\over \alpha s}z'_*}\ketxp \Big\}
	\nonumber\\
	&&={\alpha_s\over 2\pi^2}g^2\int_{-\infty}^{+\infty}dz_*d\omega_*
	\int_{0}^{+\infty}{\dhd\alpha\over \alpha}\int d^2z
	\nonumber\\
	&&~~~\times\Tr\,\tr\bigg\{
	t^aU_zt^bU^\dagger_yt^aU_x t^b[-\infty p_1,z_*]_x\psi_{cl}(z_*,x_\perp)
	\barpsi_{cl}(\omega_*,z_\perp)[\omega_*,\infty p_1]_z
	\nonumber\\
	&&\hspace{2cm}\times\bigg[\gamma^5\ssp_1\bigg( {(x-z,z-y)\over (x-z)^2_\perp(y-z)^2_\perp} + {1\over (x-z)^2_\perp} \bigg)
	+ i\ssp_1{(\vec{x}-\vec{z})\times(\vec{y}-\vec{z})\over (x-z)^2_\perp(y-z)^2_\perp}\Bigg] \bigg\}
	\label{calq5q2gdiagramab}
\end{eqnarray}
Now we sum eqs. (\ref{Qq2gdiagramc}) and (\ref{Qq2gdiagramd}) and trace them with $\gamma^5\ssp_1$ we have
\begin{eqnarray}
	&&\langle \Tr\{\calq_{5x}U^\dagger_y\}\rangle_{\small{\rm Fig.} \ref{diagrams-q2g-prop}c+d} 
	\nonumber\\
	&&= \langle \Tr\,\tr\{\gamma^5\ssp_1Q_xU^\dagger_y\}\rangle_{\small{\rm Fig.} \ref{diagrams-q2g-prop}c+d}
	\nonumber\\
	&&= {g^4\over s^2}\int_{-\infty}^{+\infty}\!\!dz'_*d\omega_*\int_0^{+\infty}{\dhd\alpha\over \alpha^3}
	\int_0^{+\infty}\!\!dz_*\int_{-\infty}^0d\xi_*\int d^2z
	\nonumber\\
	&&~~~\times\tr\bigg\{\braxp e^{-i{\hatp^2_\perp\over \alpha s}z_*}
	\gamma^5\ssp_1\big[(\alpha\ssp_1+\ssp_\perp)\gamma_\perp^\nu\big]\ketzp
	\Big(\brazp p_\nu^\perp e^{i{\hatp^2_\perp\over \alpha s}\xi_*}\ketyp
	- \brazp p_\nu^\perp e^{i{\hatp^2_\perp\over \alpha s}\xi_*}\ketxp\Big)
	\nonumber\\
	&&~~~~~~~~~~\times\Tr\{t^aU_xt^bU^\dagger_yt^aU_zt^b[-\infty p_1,\omega_*]\psi(\omega_*,z_\perp)
	\bar{\psi}(z'_*,x_\perp)[z'_*,\infty p_1]_x \}\bigg\}
	\nonumber\\
    &&= {\alpha_s\over 2\pi^2}\int_0^{+\infty}{d\alpha\over \alpha}\,g^2\!\int_{-\infty}^{+\infty}\!\!dz'_*d\omega_*\int d^2z
	\nonumber\\
	&&~~~\times\!\tr\bigg\{
	\bigg[\gamma^5\ssp_1\bigg({(x-z,z-y)\over (x-z)^2_\perp(y-z)^2_\perp} + {1\over (x-z)^2_\perp} \bigg)
	- i\ssp_1{(\vec{x}-\vec{z})\times(\vec{y}-\vec{z})\over (x-z)^2_\perp(y-z)^2_\perp}\bigg]
	\nonumber\\
	&&~~~~~~~~~~\times\Tr\{t^aU_xt^bU^\dagger_yt^aU_zt^b
	[-\infty p_1,\omega_*]\psi(\omega_*,z_\perp)\bar{\psi}(z'_*,x_\perp)[z'_*,\infty p_1]_x \}\bigg\}\,.
	\label{calq5q2gdiagramcd}
\end{eqnarray}
We can now sum eqs. (\ref{calq5q2gdiagramab}) and (\ref{calq5q2gdiagramcd}) and obtain
\begin{eqnarray}
&&\langle \Tr\{\calq_{5x}U^\dagger_y\}\rangle_{\small{\rm Fig.} \ref{diagrams-q2g-prop}} 
\label{calq5-q2g}\\
&&={\alpha_s\over 2\pi^2}\int_{0}^{+\infty}{\dhd\alpha\over \alpha}\int d^2z
\Bigg\{\bigg[ {(x-z,z-y)\over (x-z)^2_\perp(y-z)^2_\perp} + {1\over (x-z)^2_\perp} \bigg]
\nonumber\\
&&~~~~\times\!
\Big(\Tr\{t^aU_zt^bU^\dagger_yt^aU_x t^b\calx^\dagger_{5zx}\} + \Tr\{t^aU_xt^bU^\dagger_yt^aU_z t^b\calx^\dagger_{5xz}\}\Big)
\nonumber\\
&&~~~~ +{(\vec{x}-\vec{z})\times(\vec{y}-\vec{z})\over (x-z)^2_\perp(y-z)^2_\perp}
\Big(\Tr\{t^aU_xt^bU^\dagger_yt^aU_z t^b\calx^\dagger_{1xz}\} - \Tr\{t^aU_zt^bU^\dagger_yt^aU_x t^b\calx^\dagger_{1zx}\}\Big)
\Bigg\}
\nonumber
\end{eqnarray}
where we used the operators $\calx_{1xy}$ and $\calx_{5xy}$ defined in eqs. (\ref{calx1}) and (\ref{calx5}) respectively.

\subsection{Evolution equation with operators in the adjoint representation}
\label{sec: evolutionadjointrep}

\begin{figure}[thb]
\begin{center}
\includegraphics[width=4.5in]{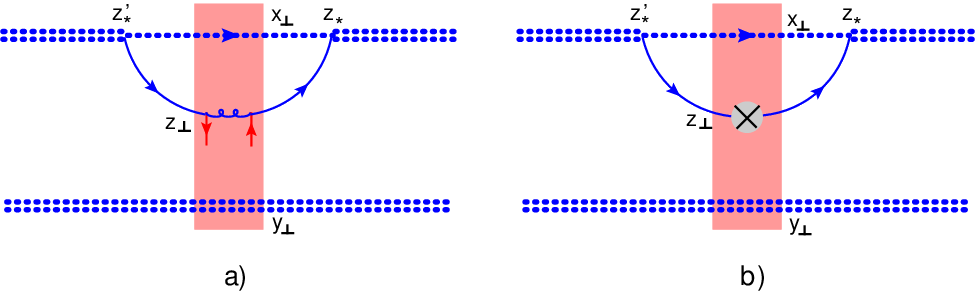}
\caption{Diagrams for the quark operators in the adjoint representation eqs. (\ref{calq5ab})-(\ref{calq1abd}). 
Double dotted lines represent Wilson lines in the adjoint representation.}
\label{LO-psibarpsiadjadj}
\end{center}
\end{figure}

Let $T^a$ be the matrix in the adjoint representation, and $\Tra$ the trace in the adjoint representation.
From result (\ref{calf-quantum}) we can easily deduce
\begin{eqnarray}
&&\langle\Tra\{\calf_x\, U^\dagger_y\}\rangle_{\small{\rm Figs.}\ref{outshock-2}+\ref{outshock-3}}
\nonumber\\
&&= {\alpha_s\over \pi^2}\,\Tra\{U_x T^a U^\dagger_y T^b\}\,\int_0^{+\infty}\!{d\alpha\over\alpha}
\int d^2z
\nonumber\\
&&~~~\times\!\Bigg\{\Bigg[
- {(\vec{x}-\vec{z})\!\times\!(\vec{z}-\vec{y})\over (x-z)^2_\perp(y-z)^2_\perp}\,
\Big(\calq_{1z}^{ba} - {\calq_{1z}^{ba}}^\dagger\Big)
\nonumber\\
&&~~~~~~~ +  \Big({(x-z,z-y)\over (x-z)^2_\perp(y-z)^2_\perp}
+ {1\over (x-z)^2_\perp}\Big)
\Big(\calq_{5z}^{ba} + {\calq_{5z}^{ba}}^\dagger + \calf^{ba}_z\Big)
\Bigg]
\nonumber\\
&&~~~~~~~ + 4\pi^2\!\!\int \!\dhd^2 q_1 {e^{i(q_1,y-z)}-e^{i(q_1,x-z)}\over q^2_{1\perp}}\,\delta^{(2)}(z-x)\calf^{ba}_z
\Bigg\}
\label{calf-quantum-adj}
\end{eqnarray}
Let us consider the evolution of $\calq_{1x}^{ab}U^{ba}_y$ and $\calq_{5x}^{ab}U^{ba}_y$
starting with the diagrams with quark in the background shock-wave given in Fig. \ref{LO-psibarpsiadjadj}a
\begin{eqnarray}
	&&\langle{\calq^{ab}_5}^\dagger(x_\perp)\rangle_{\small{\rm Fig.} \ref{LO-psibarpsiadjadj}a}
	\nonumber\\
	&&=\langle g^2\!\int_{-\infty}^{+\infty}\!\!dz_{1*}\!\int_{-\infty}^{z_{1*}}\!\!dz_{2*}\,
	[-\infty p_1, z_{2*}]^{bc}_x\bar{\psi}(z_{2*})\gamma^5\ssp_1t^c[z_{2*},z_{1*}]_xt^d\psi(z_{1*})[z_{1*},\infty p_1]^{da}_x
	\rangle 
	\nonumber\\
	&&= - g^2\!\int^{+\infty}_0\!dz_{1*}\int^0_{-\infty}\!dz_{2*}(t^bU^\dagger_xt^a)_{ij}\tr\{\gamma^5\ssp_1
	\Big\langle \psi_j(z_{1*},x_\perp)\bar{\psi}_i(z_{2*},x_\perp)\Big\rangle\}
\end{eqnarray}
Using the quark propagator in the background of quark filed given in eq. (\ref{qprpinquark}) we obtain
\begin{eqnarray}
	&&\langle{\calq^{ab}_5}^\dagger(x_\perp)\rangle_{\small{\rm Fig.} \ref{LO-psibarpsiadjadj}a}
	\nonumber\\
	&&=- g^4\! \int^{+\infty}_0\!dz_{1*}\int^0_{-\infty}\!dz_{2*}
	\int_0^{+\infty}\!{\dhd\alpha\over 2\alpha^3 s^2}
	\!\int_{z_{2*}}^{z_{1*}}\!dz_*\!\int_{z_{2*}}^{z_{1*}}\!dz'_*\int d^2z\,(t^bU^\dagger_xt^a)_{ij}
	\nonumber\\
	&&~~~\times 
	\tr\{\gamma^5\ssp_1
	\langle x_\perp|e^{-i{p^2_\perp\over \alpha s}z_{1*}}(\alpha\ssp_1+\ssp_\perp)\ketzp
	[\infty p_1,z_*]_z\gamma^\mu_\perp t^c\psi(z_*,z_\perp)[z_*,z'_*]_z^{cd}
	\nonumber\\
	&&~~~~~~~~\times \bar{\psi}(z'_*,z_\perp)t^d\gamma^\perp_\mu[z'_*,-\infty p_1]_z
	\brazp(\alpha\ssp_1+\ssp_\perp)e^{i{p^2_\perp\over \alpha s}z_{2*}}|x_\perp\rangle\}_{ji}
\end{eqnarray}
The integration over the longitudinal directions $z_1^+$ and $z_2^+$ yields
\begin{eqnarray}
&&\langle{\calq^{ab}_5}^\dagger(x_\perp)\rangle_{\small{\rm Fig.} \ref{LO-psibarpsiadjadj}a}
\\
&&~~~~= - g^4\int_0^{+\infty}\!{\dhd\alpha\over 2\alpha^3s^2}
\int d^2z 
\!\int_{z_{2*}}^{z_{1*}}\!dz_*\!\int_{z_{2*}}^{z_{1*}}\!dz'_*(t^bU^\dagger_xt^a)_{ij}
\tr\{\gamma^5\ssp_1\langle x_\perp | {-i\alpha s \,\ssp_\perp\over p^2_\perp}|z_\perp\rangle
\nonumber\\
&&~~~~~~~\times
[\infty p_1,z_*]_z\gamma^\mu_\perp t^c\psi(z_*)[z_*,z'_*]^{cd}_z
\bar{\psi}(z'_*)t^d\gamma^\perp_\mu[z'_* ,-\infty p_1]_z
\langle z_\perp | {-i\alpha s \,\ssp_\perp\over p^2_\perp}| x_\perp\rangle\}_{ji}
\nonumber
\end{eqnarray}
We can now perform the Fourier transform in coordinate space and arrive at
\begin{eqnarray}
&&\langle{\calq^{ab}_5}^\dagger(x_\perp)\rangle_{\small{\rm Fig.} \ref{LO-psibarpsiadjadj}a}
\nonumber\\
&&~~~= g^2 \int_0^{+\infty}\!{\dhd\alpha\over 2\alpha}
\int d^2z\,(t^bU^\dagger_xt^a)_{ij}\tr\{\gamma^5\ssp_1
{i(\ssx-\ssz)_\perp\over 2\pi(x-z)^2_\perp}
g^2\!\int_{-\infty}^{+\infty}\!dz_*\!\int_{-\infty}^{z_*}\!dz'_*
\nonumber\\
&&~~~~~~\!\times
[\infty p_1,z_*]_z\gamma^\mu_\perp t^c\psi(z_*)[z_*,z'_*]^{cd}_z
\bar{\psi}(z'_*)t^d\gamma^\perp_\mu[z'_* ,z_*]_z
{i(\ssz-\ssx)_\perp\over 2\pi(x-z)^2_\perp}\}_{ji}
\end{eqnarray}
performing the trace over the Dirac gamma-matrices and using $(\ssx-\ssz)_\perp (\ssx-\ssz)_\perp=-(x-z)^2_\perp$ we obtain
\begin{eqnarray}
&&\langle{\calq^{ab}_5}^\dagger(x_\perp)\rangle_{\small{\rm Fig.} \ref{LO-psibarpsiadjadj}a}
\nonumber\\
&&~~~= - {g^2\over 8\pi^3}\int_0^{+\infty}\!{d\alpha \over \alpha}\int d^2 z\,{1\over (x-z)^2_\perp}
(t^bU^\dagger_xt^a)_{ij}\,g^2\!\int_{-\infty}^{+\infty}\!dz_*\int_{-\infty}^{z_*}\!dz'_*
\nonumber\\
&&~~~~~~\times\!
([\infty p_1,z_*]_z t^c\tr\{\gamma^5\ssp_1\psi(z_*)[z_*,z'_*]^{cd}\bar{\psi}(z'_*)t^d[z'_*,-\infty p_1]_z\})_{ji}
\nonumber\\
&&~~~= - {\alpha_s\over 2\pi^2}\int_0^{+\infty}\!{d\alpha \over \alpha}\int d^2 z\,{1\over (x-z)^2_\perp}\,
(t^bU^\dagger_xt^a)_{ij}\,\calq_{5\,ji}(z_\perp)
\end{eqnarray}
where in the last step we inserted the $\calq_{5x}$ definition eq. (\ref{calq5}).
Until now the Wilson line operator $U^{ab}_y$ was only a spectator, inserting it back we finally obtain
\begin{eqnarray}
&&\langle U_y^{ab}{\calq^{ab}_{5x}}^\dagger\rangle_{\small{\rm Fig.} \ref{LO-psibarpsiadjadj}a}
\label{calq5adjevolution}\\
&&~~~~~= - {\alpha_s\over 2\pi^2}\int_0^{+\infty}\!{d\alpha \over \alpha}\int d^2 z\,{1\over (x-z)^2_\perp}\,
U_y^{ab}\Tr\{t^bU^\dagger_xt^a\calq_{5z}\}
\nonumber\\
&&~~~~~= - {\alpha_s\over 2\pi^2}\int_0^{+\infty}\!{d\alpha \over \alpha}\int d^2 z\,{1\over (x-z)^2_\perp}\,
\half\Big[\Tr\{U_yU^\dagger_x\}\Tr\{\calq_{5\,z}U^\dagger_y\} - {1\over N_c}\Tr\{U^\dagger_x \calq_{5\, z}\}\Big]
\nonumber
\end{eqnarray}
Notice that on the LHS of eq. (\ref{calq5adjevolution}) we have trace in the adjoint representation
and operator $\calq^{ab}$ defined in (\ref{calq5abd}), while on the RHS we have
trace over the fundamental representation with operator $\calq_{5x}$ defined in eq. (\ref{calq5}).
Following similar steps, the result of diagram in Fig. \ref{LO-psibarpsiadjadj}a for the ${\calq_{1x}^{ab}}^\dagger$ operator is
\begin{eqnarray}
	&&\langle{\calq^{ab}_{1x}}^\dagger\rangle_{\small{\rm Fig.} \ref{LO-psibarpsiadjadj}a}
	\nonumber\\
	&&=\langle - g^2\!\int_{-\infty}^{+\infty}\!\!dz_{1*}\!\int_{-\infty}^{z_{1*}}\!\!dz_{2*}\,
	[-\infty p_1, z_{2*}]^{bc}_x\bar{\psi}(z_{2*})i\ssp_1t^c[z_{2*},z_{1*}]_xt^d\psi(z_{1*})[z_{1*},\infty p_1]^{da}_x
	\rangle 
	\nonumber\\
	&&=  g^2\!\int^{+\infty}_0\!dz_{1*}\int^0_{-\infty}\!dz_{2*}(t^bU^\dagger_xt^a)_{ij}\tr\{i\ssp_1
	\langle \psi_j(z_{1*},x_\perp)\bar{\psi}_i(z_{2*},x_\perp)\}\rangle
	\nonumber\\
	&&= - {\alpha_s\over 2\pi^2}\int_0^{+\infty}\!{d\alpha \over \alpha}\int d^2 z\,{1\over (x-z)^2_\perp}\,
	\Tr\{t^bU^\dagger_xt^a\,\calq_{1z}\}\,.
\end{eqnarray}
Thus we have
\begin{eqnarray}
&&\langle U_y^{ab}{\calq^{ab}_{1x}}^\dagger\rangle_{\small{\rm Fig.} \ref{LO-psibarpsiadjadj}a}
\\
&&~~~~~~= - {\alpha_s\over 4\pi^2}\int_0^{+\infty}\!{d\alpha \over \alpha}\int d^2 z\,{1\over (x-z)^2_\perp}\,
\Big[\Tr\{U_yU^\dagger_x\}\Tr\{\calq_{1\,z}U^\dagger_y\} - {1\over N_c}\Tr\{U^\dagger_x \calq_{1\, z}\}\Big]
\nonumber
\end{eqnarray}
We now consider diagram in Fig. \ref{LO-psibarpsiadjadj}b which is the one
with the gluon sub-eikonal correction $\epsilon^{ij}F_{ij}$ in the quark propagator.
Let us start again with operator ${\calq_{5x}^{ab}}^\dagger$
\begin{eqnarray}
&&\langle{\calq^{ab}_{5x}}^\dagger\rangle_{\small{\rm Fig.} \ref{LO-psibarpsiadjadj}b}
\\
&&~~~~~~=\langle g^2\!\int_{-\infty}^{+\infty}\!\!dz_{1*}\!\int_{-\infty}^{z_{1*}}\!\!dz_{2*}\,
[-\infty p_1, z_{2*}]^{bc}_x\bar{\psi}(z_{2*})\gamma^5\ssp_1t^c[z_{2*},z_{1*}]_xt^d\psi(z_{1*})[z_{1*},\infty p_1]^{da}_x
\rangle 
\nonumber\\
&&~~~~~~= - g^2\!\int^{+\infty}_0\!dz_{1*}\int^0_{-\infty}\!dz_{2*}(t^bU^\dagger_xt^a)_{ij}\tr\{\gamma^5\ssp_1
\big\langle \psi_j(z_{1*},x_\perp)\bar{\psi}_i(z_{2*},x_\perp)\big\rangle\}
\nonumber
\end{eqnarray}
we now need quark propagator (\ref{qpropFijterm}) and proceeding as before we integrate over the longitudinal coordinates $z_{1*}$
and $z_{2*}$ and  perform the Fourier transform we arrive at
\begin{eqnarray}
&&\langle{\calq^{ab}_{5x}}^\dagger\rangle_{\small{\rm Fig.} \ref{LO-psibarpsiadjadj}b} 
\nonumber\\
&&= - g^2(t^bU^\dagger_xt^a)_{ij}\!\int_0^{+\infty}\!dz_{1*}\!\int^0_{-\infty}\!dz_{2*}\,
\tr\{\gamma^5\ssp_1\!\!\int_0^{+\infty}\!{\dhd\alpha\over 2\alpha}{1\over \alpha s}
\langle x_\perp| e^{-i{p^2_\perp\over \alpha s}z_{1*}}(\alpha \ssp_1 + \ssp_\perp)|z_\perp\rangle
\ssp_2\gamma^5
\nonumber\\
&&~~~\times{ig\over 4\alpha}\int_{-\infty}^{+\infty}\!d{2\over s}z_*[\infty p_1,z_*]_z\epsilon^{ij}F_{ij}[z_*,-\infty p_1]_z
\langle z_\perp |(\alpha p_1+\ssp_\perp)e^{i{p^2_\perp\over \alpha s}z_{2*}}|x_\perp \rangle
\nonumber\\
&&= {g^2\over 8\pi}\int_0^{+\infty}\!{d\alpha\over \alpha}\,\tr\{\gamma^5\ssp_1{i(\ssx-\ssz)_\perp\over 2\pi(x-z)^2_\perp}
\ssp_2\gamma^5 {i(\ssz-\ssx)_\perp\over 2\pi(x-z)^2_\perp}\}
\nonumber\\
&&~~~\times ig\!\int_{-\infty}^{+\infty}\!dz_*([\infty p_1, z_*]_z\epsilon^{ij}F_{ij}[z_*,-\infty p_1]_z)_{ij}(t^bU^\dagger_xt^a)_{ij}
\nonumber\\
&&= - {\alpha_s\over 2\pi^2}\int_0^{+\infty}\!{d\alpha\over \alpha}\int {d^2z\over (x-z)^2_\perp}\,
g\!\int_{-\infty}^{+\infty}\!dz_*([\infty p_1, z_*]_z\epsilon^{ij}i{s\over 2}F_{ij}[z_*,-\infty p_1]_z)_{ij}(t^bU^\dagger_xt^a)_{ij}
\nonumber\\
&&= - {\alpha_s\over 2\pi^2}\int_0^{+\infty}\!{d\alpha\over \alpha}\int {d^2 z\over (x-z)^2_\perp}
\Tr\{\calf_z t^bU^\dagger_xt^a\}
\end{eqnarray}
where we used $\tr\{\gamma^5\ssp_2(\ssx-\ssz)_\perp\ssp_2\gamma^5(\ssz-\ssx)_\perp\} = 2 s (x-z)^2_\perp$
and definition of operator $\hat{\calf_x}$, eq. (\ref{calfoperator}).

Next, we consider ${\calq^{ab}_{1x}}^\dagger$ and proceed in the same way
\begin{eqnarray}
\langle{\calq^{ab}_{1x}}^\dagger\rangle_{\small{\rm Fig.} \ref{LO-psibarpsiadjadj}b} 
=\!\!&& {g^2\over 8\pi}\int_0^{+\infty}\!{d\alpha\over \alpha}\,\tr\{i\ssp_1{i(\ssx-\ssz)_\perp\over 2\pi(x-z)^2_\perp}
\ssp_2\gamma^5 {i(\ssz-\ssx)_\perp\over 2\pi(x-z)^2_\perp}\}
\nonumber\\
&&\times ig\!\int_{-\infty}^{+\infty}\!dz_*([\infty p_1, z_*]_z\epsilon^{ij}F_{ij}[z_*,-\infty p_1]_z)_{ij}(t^bU^\dagger_xt^a)_{ij} =0
\end{eqnarray}
where we used $\tr\{i\ssp_1(\ssx-\ssz)_\perp\ssp_2\gamma^5 (\ssz-\ssx)_\perp\} =0$, so
we find that the quark operator $\calq_1^{ab}$ does not mix with the gluon operator $\calf$ under one loop evolution.
We have obtained the same result for the quark operator $\calq_1$ defined in (\ref{calq1}).

Summing up diagrams inf Fig. \ref{LO-psibarpsiadjadj}a and b for operator ${\calq^{ab}_{5x}}^\dagger$, we have
\begin{eqnarray}
&&\hspace{-2cm}\langle{\calq^{ab}_{5x}}^\dagger\rangle_{\small{\rm Fig.} \ref{LO-psibarpsiadjadj}}  =
- {\alpha_s\over 2\pi^2}\int_0^{+\infty}\!{d\alpha\over \alpha}\int {d^2z\over (x-z)^2_\perp}\,
\Tr\{(\calf_z + \calq_{5z}) t^bU^\dagger_xt^a\}
\end{eqnarray}
from which we can easily deduce
\begin{eqnarray}
&&\hspace{-2cm}\langle\calq^{ab}_{5x}\rangle_{\small{\rm Fig.} \ref{LO-psibarpsiadjadj}}  =
- {\alpha_s\over 2\pi^2}\int_0^{+\infty}\!{d\alpha\over \alpha}\int {d^2z\over (x-z)^2_\perp}\,
\Tr\{(\calf_z^\dagger + \calq^\dagger_{5z}) t^aU_xt^b\}\,.
\end{eqnarray}

The evolution of operator $\Tra\{U^\dagger_y\calf_x\}$, diagrams in Fig. \ref{outshock-2} and {\rm Fig.} \ref{outshock-3}
can be deduced from $\Tr\{U^\dagger_y\calf_x\}$, by changing $t^a\to T^a$ and $\Tr\to \Tra$
in eq. (\ref{calf-quantum}). 

We can conclude that the evolution of $\Tra\{U^\dagger_y(\calf_x + \calq_{5x})\}$, excluding the BK-type diagrams, is
\begin{eqnarray}
	&&\langle \Tra\{U^\dagger_y(\calf_x + \calq_{5x})\}
	\rangle_{\small{\rm Figs.}\ref{outshock-2}+\ref{outshock-3}+\ref{LO-psibarpsiadjadj}}
	\nonumber\\
	&&\hspace{1cm}= - {\alpha_s\over 2\pi^2}\int_0^{+\infty}\!{d\alpha\over \alpha}\int d^2 z\Bigg\{ {1\over (x-z)^2_\perp}
	\Tr\{(\calf_z^\dagger + \calq^\dagger_{5z}) t^aU_xt^b\}U^{ab}_y
	\nonumber\\
	&&\hspace{1.5cm} + 2\,\Tra\{U_x T^a U^\dagger_y T^b\}
	\Bigg[
	 {(\vec{x}-\vec{z})\!\times\!(\vec{z}-\vec{y})\over (x-z)^2_\perp(y-z)^2_\perp}\,
	\Big(\calq_{1z}^{ba} - {\calq_{1z}^{ba}}^\dagger\Big)
	\nonumber\\
	&&\hspace{1.5cm}	-  \Big({(x-z,z-y)\over (x-z)^2_\perp(y-z)^2_\perp} + {1\over (x-z)^2_\perp}\Big)
	\Big(\calq_{5z}^{ba} + {\calq_{5z}^{ba}}^\dagger + \calf^{ba}_z\Big)
	\nonumber\\
	&&\hspace{1.5cm} - 4\pi^2\!\!\int \!\dhd^2 q_1 {e^{i(q_1,y-z)}-e^{i(q_1,x-z)}\over q^2_{1\perp}}\,\delta^{(2)}(z-x)\calf^{ba}_z
	\Bigg]
	\Bigg\}
	\label{sumcalfcalq-adjrep}
\end{eqnarray}
Note, that in the RHS we have trace in the adjoint representation, while in the LHS we have 
trace in the fundamental representation in the first term, and trace in the adjoint representation in the second term.
In eq. (\ref{sumcalfcalq-adjrep}) we have mixing with operator $\hat{\calq}_1(x_\perp)$
which may contribute to spin in the non forward case. 
Evolution equation (\ref{sumcalfcalq-adjrep}) agrees with the one calculated in Refs. \cite{Kovchegov:2016zex, Kovchegov:2018znm}
except for the mixing term proportional to $\hat{\calq}_1(x_\perp)$.
However, when we perform the color algebra, the evolution equation (\ref{sumcalfcalq-adjrep})
is entirely written in terms of operators $\hat{Q}_1(x_\perp)$ $\hat{Q}_5(x_\perp)$, $\hat{\tildeQ}_1(x_\perp)$,
and $\hat{\tildeQ}_5(x_\perp)$.


\begin{thebibliography}{99}

\bibitem{Boer:2011fh}
D.~Boer, M.~Diehl, R.~Milner, R.~Venugopalan, W.~Vogelsang, D.~Kaplan, H.~Montgomery, S.~Vigdor, A.~Accardi 
and E.~C.~Aschenauer, \textit{et al.}
[arXiv:1108.1713 [nucl-th]].

\bibitem{Nocera:2014gqa}
E.~R.~Nocera \textit{et al.} [NNPDF],
Nucl. Phys. B \textbf{887}, 276-308 (2014)
doi:10.1016/j.nuclphysb.2014.08.008
[arXiv:1406.5539 [hep-ph]].

\bibitem{deFlorian:2008mr}
D.~de Florian, R.~Sassot, M.~Stratmann and W.~Vogelsang,
Phys. Rev. Lett. \textbf{101}, 072001 (2008)
doi:10.1103/PhysRevLett.101.072001
[arXiv:0804.0422 [hep-ph]].

\bibitem{deFlorian:2009vb}
D.~de Florian, R.~Sassot, M.~Stratmann and W.~Vogelsang,
Phys. Rev. D \textbf{80}, 034030 (2009)
doi:10.1103/PhysRevD.80.034030
[arXiv:0904.3821 [hep-ph]].
LaTeX (US)

\bibitem{Kovchegov:2015pbl}
Y.~V.~Kovchegov, D.~Pitonyak and M.~D.~Sievert,
JHEP \textbf{01}, 072 (2016)
[erratum: JHEP \textbf{10}, 148 (2016)]
doi:10.1007/JHEP01(2016)072
[arXiv:1511.06737 [hep-ph]].

\bibitem{Kovchegov:2017lsr}
Y.~V.~Kovchegov, D.~Pitonyak and M.~D.~Sievert,
JHEP \textbf{10}, 198 (2017)
doi:10.1007/JHEP10(2017)198
[arXiv:1706.04236 [nucl-th]].

\bibitem{Kovchegov:2016zex}
Y.~V.~Kovchegov, D.~Pitonyak and M.~D.~Sievert,
Phys. Rev. D \textbf{95}, no.1, 014033 (2017)
doi:10.1103/PhysRevD.95.014033
[arXiv:1610.06197 [hep-ph]].


\bibitem{Kovchegov:2018znm}
Y.~V.~Kovchegov and M.~D.~Sievert,
Phys. Rev. D \textbf{99}, no.5, 054032 (2019)
doi:10.1103/PhysRevD.99.054032
[arXiv:1808.09010 [hep-ph]].

\bibitem{Boussarie:2019icw}
R.~Boussarie, Y.~Hatta and F.~Yuan,
Phys. Lett. B \textbf{797}, 134817 (2019)
doi:10.1016/j.physletb.2019.134817
[arXiv:1904.02693 [hep-ph]].

\bibitem{Tarasov:2020cwl}
A.~Tarasov and R.~Venugopalan,
Phys. Rev. D \textbf{102}, 114022 (2020)
doi:10.1103/PhysRevD.102.114022
[arXiv:2008.08104 [hep-ph]].


\bibitem{Hatta:2016aoc}
Y.~Hatta, Y.~Nakagawa, F.~Yuan, Y.~Zhao and B.~Xiao,
Phys. Rev. D \textbf{95}, no.11, 114032 (2017)
doi:10.1103/PhysRevD.95.114032
[arXiv:1612.02445 [hep-ph]].


\bibitem{Hatta:2020riw}
Y.~Hatta,
[arXiv:2012.01865 [hep-ph]].

\bibitem{Hatta:2020ltd}
Y.~Hatta,
Phys. Rev. D \textbf{102}, no.9, 094004 (2020)
doi:10.1103/PhysRevD.102.094004
[arXiv:2009.03657 [hep-ph]].
	
	
\bibitem{Kuraev:1977fs}
E.~A. Kuraev, L.~N. Lipatov and V.~S. Fadin, \emph{{The Pomeranchuk
singlularity in non-Abelian gauge theories}}, {\emph{Sov. Phys. JETP}
{\bfseries 45} (1977) 199--204}.

\bibitem{Balitsky:1978ic}
I.~Balitsky and L.~Lipatov, \emph{{The Pomeranchuk Singularity in Quantum
Chromodynamics}}, {\emph{Sov.J.Nucl.Phys.} {\bfseries 28} (1978) 822--829}.
	
		
\bibitem{Aaron:2009aa}
F.~D.~Aaron \textit{et al.} [H1 and ZEUS],
JHEP \textbf{01}, 109 (2010)
doi:10.1007/JHEP01(2010)109
[arXiv:0911.0884 [hep-ex]].

\bibitem{Abramowicz:2015mha}
H.~Abramowicz \textit{et al.} [H1 and ZEUS],
Eur. Phys. J. C \textbf{75}, no.12, 580 (2015)
doi:10.1140/epjc/s10052-015-3710-4
[arXiv:1506.06042 [hep-ex]].
	
\bibitem{Balitsky:1987bk}
I.~I.~Balitsky and V.~M.~Braun,
Nucl. Phys. B \textbf{311}, 541-584 (1989)
doi:10.1016/0550-3213(89)90168-5

\bibitem{Balitsky:1995ub}
I.~Balitsky,
Nucl. Phys. B \textbf{463}, 99-160 (1996)
doi:10.1016/0550-3213(95)00638-9
[arXiv:hep-ph/9509348 [hep-ph]].

\bibitem{Balitsky:2008zza}
I.~Balitsky and G.~A.~Chirilli,
Phys. Rev. D \textbf{77}, 014019 (2008)
doi:10.1103/PhysRevD.77.014019
[arXiv:0710.4330 [hep-ph]].

\bibitem{Balitsky:2009xg}
I.~Balitsky and G.~A.~Chirilli,
Nucl. Phys. B \textbf{822}, 45-87 (2009)
doi:10.1016/j.nuclphysb.2009.07.003
[arXiv:0903.5326 [hep-ph]].

\bibitem{Balitsky:2010ze}
I.~Balitsky and G.~A.~Chirilli,
Phys. Rev. D \textbf{83}, 031502 (2011)
doi:10.1103/PhysRevD.83.031502
[arXiv:1009.4729 [hep-ph]].

\bibitem{Balitsky:2012bs}
I.~Balitsky and G.~A.~Chirilli,
Phys. Rev. D \textbf{87}, no.1, 014013 (2013)
doi:10.1103/PhysRevD.87.014013
[arXiv:1207.3844 [hep-ph]].


\bibitem{JalilianMarian:1997gr}
J.~Jalilian-Marian, A.~Kovner, A.~Leonidov and H.~Weigert,
Phys. Rev. D \textbf{59}, 014014 (1998)
doi:10.1103/PhysRevD.59.014014
[arXiv:hep-ph/9706377 [hep-ph]].

\bibitem{Ferreiro:2001qy}
E.~Ferreiro, E.~Iancu, A.~Leonidov and L.~McLerran,
Nucl. Phys. A \textbf{703}, 489-538 (2002)
doi:10.1016/S0375-9474(01)01329-X
[arXiv:hep-ph/0109115 [hep-ph]].

\bibitem{Iancu:2000hn}
E.~Iancu, A.~Leonidov and L.~D.~McLerran,
Nucl. Phys. A \textbf{692}, 583-645 (2001)
doi:10.1016/S0375-9474(01)00642-X
[arXiv:hep-ph/0011241 [hep-ph]].


\bibitem{Kovchegov:1999yj}
Y.~V.~Kovchegov,
Phys. Rev. D \textbf{60}, 034008 (1999)
doi:10.1103/PhysRevD.60.034008
[arXiv:hep-ph/9901281 [hep-ph]].

\bibitem{Kovchegov:1999ua}
Y.~V.~Kovchegov,
Phys. Rev. D \textbf{61}, 074018 (2000)
doi:10.1103/PhysRevD.61.074018
[arXiv:hep-ph/9905214 [hep-ph]].

\bibitem{Balitsky:2001gj}
I.~Balitsky,
doi:10.1142/9789812810458\_0030
[arXiv:hep-ph/0101042 [hep-ph]].

\bibitem{Kovchegov:2012mbw}
Y.~V.~Kovchegov and E.~Levin,
Camb. Monogr. Part. Phys. Nucl. Phys. Cosmol. \textbf{33}, 1-350 (2012)
doi:10.1017/CBO9781139022187

\bibitem{Fadin:1998py}
V.~S.~Fadin and L.~N.~Lipatov,
Phys. Lett. B \textbf{429}, 127-134 (1998)
doi:10.1016/S0370-2693(98)00473-0
[arXiv:hep-ph/9802290 [hep-ph]].


\bibitem{Beuf:2017bpd}
G.~Beuf,
Phys. Rev. D \textbf{96}, no.7, 074033 (2017)
doi:10.1103/PhysRevD.96.074033
[arXiv:1708.06557 [hep-ph]].

\bibitem{Beuf:2020dxl}
G.~Beuf, H.~H\"anninen, T.~Lappi and H.~M\"antysaari,
Phys. Rev. D \textbf{102}, 074028 (2020)
doi:10.1103/PhysRevD.102.074028
[arXiv:2007.01645 [hep-ph]].



\bibitem{Kirschner:1983di}
R.~Kirschner and L.~n.~Lipatov,
Nucl. Phys. B \textbf{213}, 122-148 (1983)
doi:10.1016/0550-3213(83)90178-5

\bibitem{Ermolaev:1995fx}
B.~I.~Ermolaev, S.~I.~Manaenkov and M.~G.~Ryskin,
Z. Phys. C \textbf{69}, 259-267 (1996)
doi:10.1007/s002880050026
[arXiv:hep-ph/9502262 [hep-ph]].

\bibitem{Bartels:1995iu}
J.~Bartels, B.~I.~Ermolaev and M.~G.~Ryskin,
Z. Phys. C \textbf{70}, 273-280 (1996)
[arXiv:hep-ph/9507271 [hep-ph]].

\bibitem{Bartels:1996wc}
J.~Bartels, B.~I.~Ermolaev and M.~G.~Ryskin,
Z. Phys. C \textbf{72}, 627-635 (1996)
doi:10.1007/BF02909194
[arXiv:hep-ph/9603204 [hep-ph]].

\bibitem{Jaroszewicz:1982gr}
T.~Jaroszewicz,
Phys. Lett. B \textbf{116}, 291-294 (1982)
doi:10.1016/0370-2693(82)90345-8

\bibitem{BALITSKY:2014zza}
I.~Balitsky,
Int. J. Mod. Phys. Conf. Ser. \textbf{25}, 1460024 (2014)
doi:10.1142/S2010194514600246

\bibitem{Altarelli:1977zs}
G.~Altarelli and G.~Parisi,
Nucl. Phys. B \textbf{126}, 298-318 (1977)
doi:10.1016/0550-3213(77)90384-4

\bibitem{Mertig:1995ny}
R.~Mertig and W.~L.~van Neerven,
Z. Phys. C \textbf{70}, 637-654 (1996)
doi:10.1007/s002880050138
[arXiv:hep-ph/9506451 [hep-ph]].

\bibitem{Moch:2014sna}
S.~Moch, J.~A.~M.~Vermaseren and A.~Vogt,
Case,''
Nucl. Phys. B \textbf{889} (2014), 351-400
doi:10.1016/j.nuclphysb.2014.10.016
[arXiv:1409.5131 [hep-ph]].

\bibitem{Behring:2019tus}
A.~Behring, J.~Bl\"umlein, A.~De Freitas, A.~Goedicke, S.~Klein, A.~von
Manteuffel, C.~Schneider and K.~Sch\"onwald,
Operator Matrix Elements,''
Nucl. Phys. B \textbf{948} (2019), 114753
doi:10.1016/j.nuclphysb.2019.114753
[arXiv:1908.03779 [hep-ph]].

\bibitem{Blumlein:1995jp}
J.~Blumlein and A.~Vogt,
Phys. Lett. B \textbf{370} (1996), 149-155
doi:10.1016/0370-2693(95)01568-X
[arXiv:hep-ph/9510410 [hep-ph]].

\bibitem{Blumlein:1996hb}
J.~Blumlein and A.~Vogt,
small x,''
Phys. Lett. B \textbf{386} (1996), 350-358
doi:10.1016/0370-2693(96)00958-6
[arXiv:hep-ph/9606254 [hep-ph]].


\bibitem{Chirilli:2018kkw} 
G.~A.~Chirilli,
JHEP {\bf 1901}, 118 (2019)
doi:10.1007/JHEP01(2019)118
[arXiv:1807.11435 [hep-ph]].


\bibitem{Balitsky:2015qba}
I.~Balitsky and A.~Tarasov,
JHEP \textbf{10}, 017 (2015)
doi:10.1007/JHEP10(2015)017
[arXiv:1505.02151 [hep-ph]].

\bibitem{Balitsky:2016dgz}
I.~Balitsky and A.~Tarasov,
JHEP \textbf{06}, 164 (2016)
doi:10.1007/JHEP06(2016)164
[arXiv:1603.06548 [hep-ph]].


\bibitem{Sivers:1989cc}
D.~W.~Sivers,
Phys. Rev. D \textbf{41}, 83 (1990)
doi:10.1103/PhysRevD.41.83

\bibitem{Sivers:1990fh}
D.~W.~Sivers,
Phys. Rev. D \textbf{43}, 261-263 (1991)
doi:10.1103/PhysRevD.43.261


\bibitem{Mulders:1995dh}
P.~J.~Mulders and R.~D.~Tangerman,
Nucl. Phys. B \textbf{461}, 197-237 (1996)
[erratum: Nucl. Phys. B \textbf{484}, 538-540 (1997)]
doi:10.1016/0550-3213(95)00632-X
[arXiv:hep-ph/9510301 [hep-ph]].

\bibitem{Goeke:2005hb}
K.~Goeke, A.~Metz and M.~Schlegel,
Phys. Lett. B \textbf{618}, 90-96 (2005)
doi:10.1016/j.physletb.2005.05.037
[arXiv:hep-ph/0504130 [hep-ph]].

\bibitem{Bacchetta:2006tn}
A.~Bacchetta, M.~Diehl, K.~Goeke, A.~Metz, P.~J.~Mulders and M.~Schlegel,
JHEP \textbf{02}, 093 (2007)
doi:10.1088/1126-6708/2007/02/093
[arXiv:hep-ph/0611265 [hep-ph]].

\bibitem{Kovchegov:2019rrz}
Y.~V.~Kovchegov,
JHEP \textbf{03}, 174 (2019)
doi:10.1007/JHEP03(2019)174
[arXiv:1901.07453 [hep-ph]].

\bibitem{fcollins}
J.C. Collins, Foundations of Perturbative QCD, Cambridge University Press, Cambridge
U.K. (2011).

\bibitem{Altinoluk:2014oxa}
T.~Altinoluk, N.~Armesto, G.~Beuf, M.~Mart\'\i{}nez and C.~A.~Salgado,
JHEP \textbf{07}, 068 (2014)
doi:10.1007/JHEP07(2014)068
[arXiv:1404.2219 [hep-ph]].

\bibitem{Altinoluk:2015gia}
T.~Altinoluk, N.~Armesto, G.~Beuf and A.~Moscoso,
JHEP \textbf{01}, 114 (2016)
doi:10.1007/JHEP01(2016)114
[arXiv:1505.01400 [hep-ph]].

\bibitem{Altinoluk:2020oyd}
T.~Altinoluk, G.~Beuf, A.~Czajka and A.~Tymowska,
[arXiv:2012.03886 [hep-ph]].




\end{thebibliography}
\end{document}